\documentclass[fleqn,usenatbib]{mnras}

\usepackage{float}
\usepackage[caption = false]{subfig}

\usepackage{longtable,rotating}
\usepackage{booktabs}
\usepackage{newtxtext,newtxmath}

\usepackage[T1]{fontenc}

\DeclareRobustCommand{\VAN}[3]{#2}
\let\VANthebibliography\thebibliography
\def\thebibliography{\DeclareRobustCommand{\VAN}[3]{##3}\VANthebibliography}

\usepackage{graphicx}	
\usepackage{amsmath}	
\usepackage{tablefootnote}

\def\msun{\mbox{$M_\odot$}}

\def\degr{$^{\circ}$}

\title[Low-mass hub-filament in NGC2071-N]{A low-mass hub-filament with double centre revealed in NGC2071-North}

\author[V. K\"onyves et al.]{
Vera K\"onyves,$^{1}$\thanks{E-mail: vera.konyves@gmail.com}
D. Ward-Thompson,$^{1}$
Y. Shimajiri,$^{2, 3}$
P. Palmeirim,$^{4}$
Ph. Andr\'e$^{5}$
\\
$^{1}$Jeremiah Horrocks Institute, University of Central Lancashire, Preston PR1 2HE, UK\\
$^{2}$National Astronomical Observatory of Japan, National Institutes of Natural Sciences, 2-21-1 Osawa, Mitaka, Tokyo 181-8588, Japan\\
$^{3}$Kyushu Kyoritsu University, Jiyugaoka 1-8, Yahatanishi-ku, Kitakyushu, Fukuoka, 807-8585, Japan\\
$^{4}$Instituto de Astrof\'isica e Ci{\^e}ncias do Espa\c{c}o, Universidade do Porto, CAUP, Rua das Estrelas, PT4150-762 Porto, Portugal\\
$^{5}$Laboratoire d'Astrophysique (AIM), CEA, CNRS, Universit\'e Paris-Saclay, Universit\'e Paris Diderot, Sorbonne Paris Cit\'e, 91191 Gif-sur-Yvette, France
}

\date{Accepted XXX. Received YYY; in original form ZZZ}

\pubyear{2021}

\begin{document}
\label{firstpage}
\pagerange{\pageref{firstpage}--\pageref{lastpage}}
\maketitle

\begin{abstract}
We present the first analysis in NGC2071-North as a resolved hub-filament featuring a double centre. 
This $\sim 1.5 \times 1.5$ parsec-scale filament hub contains $\sim$500\,\msun. Seen from {\it Planck}, 
magnetic field lines may have facilitated the gathering of material at this isolated location. 
The energy balance analysis, supported by infalling gas signatures, reveal that these filaments are 
currently forming stars.
{\it Herschel} 100\,$\mu$m emission concentrates in the hub, at 
IRAS~05451+0037 and LkH$\alpha$~316, and presents diffuse lobes and loops around them. We suggest that 
such a double centre could be formed, because the converging locations of filament pairs are offset, 
by 2.3\arcmin~($0.27$\,pc). This distance also matches the diameter of a hub-ring, seen 
in column density and molecular tracers, such as HCO$^+$(1--0) and HCN(1--0),  
that may indicate a transition and the connection between the hub and the radiating filaments. 
We argue that all of the three components of the emission star LkH$\alpha$~316 are in 
physical association. We find that a $\sim$0.06\,pc-sized gas loop, attached to IRAS~05451+0037, 
can be seen at wavelengths all the way from Pan-STARRS-i to {\it Herschel}-100\,$\mu$m. These 
observations suggest that both protostars at the double hub centre are interacting with the cloud material. 
In our $^{13}$CO data, we do not seem to find the outflow of this region that was identified in the 
80s with much lower resolution.
\end{abstract}
\begin{keywords}
ISM: clouds -- ISM: structure -- ISM: individual objects (NGC2071-North) -- Stars: formation
\end{keywords}



\section{Introduction}\label{sec:intro}

Thanks to recent space-based (e.g., {\it Herschel}) and sensitive ground-based 
(e.g., ALMA, EVLA) observations, much progress is being made on the early stages of 
star formation. This is being achieved by not only looking at the very peak of such sources, 
but also considering their near and far surroundings in the molecular clouds that are forming stars.
In particular, many studies have shown that the cold material of molecular clouds is often 
organized in networks of filaments, whether these clouds are currently forming stars or not 
\citep[e.g.,][]{Arzoumanian+2011, Arzoumanian+2019}. Most of the clumps and 
cores are seen forming in these filaments \citep[e.g.,][]{Andre+2014, Konyves+2015, Konyves+2020},
the formation of which may be due to various mechanisms, invoking one or more of the turbulent-, 
gravitational-, and magnetic forces. Summaries on the origin of interstellar filaments can be found in 
\citet{Andre+2014}, \citet{Hacar+2022}, and \citet{Pineda+2022}.

Nearby {\it Herschel} filaments -- up to $\sim$0.5\,kpc distance -- appear to be characterized
by a narrow distribution of transverse half-power widths with a typical FWHM value of 
0.1\,pc \citep[e.g.,][]{Arzoumanian+2011, Arzoumanian+2019, KochRosolowsky2015}.
While there has been some debate about the reliability of this finding \citep[cf.,][]{Panopoulou+2022}, 
tests performed on synthetic data suggest that published {\it Herschel} width measurements 
are not affected by significant biases, at least in the case of nearby, high-contrast filamentary 
structures \citep{Arzoumanian+2019, Andre+2022}.

\begin{figure*}
 \begin{center}
  \begin{minipage}{1.\linewidth}
   \resizebox{0.448\hsize}{!}{\includegraphics[width=\columnwidth]{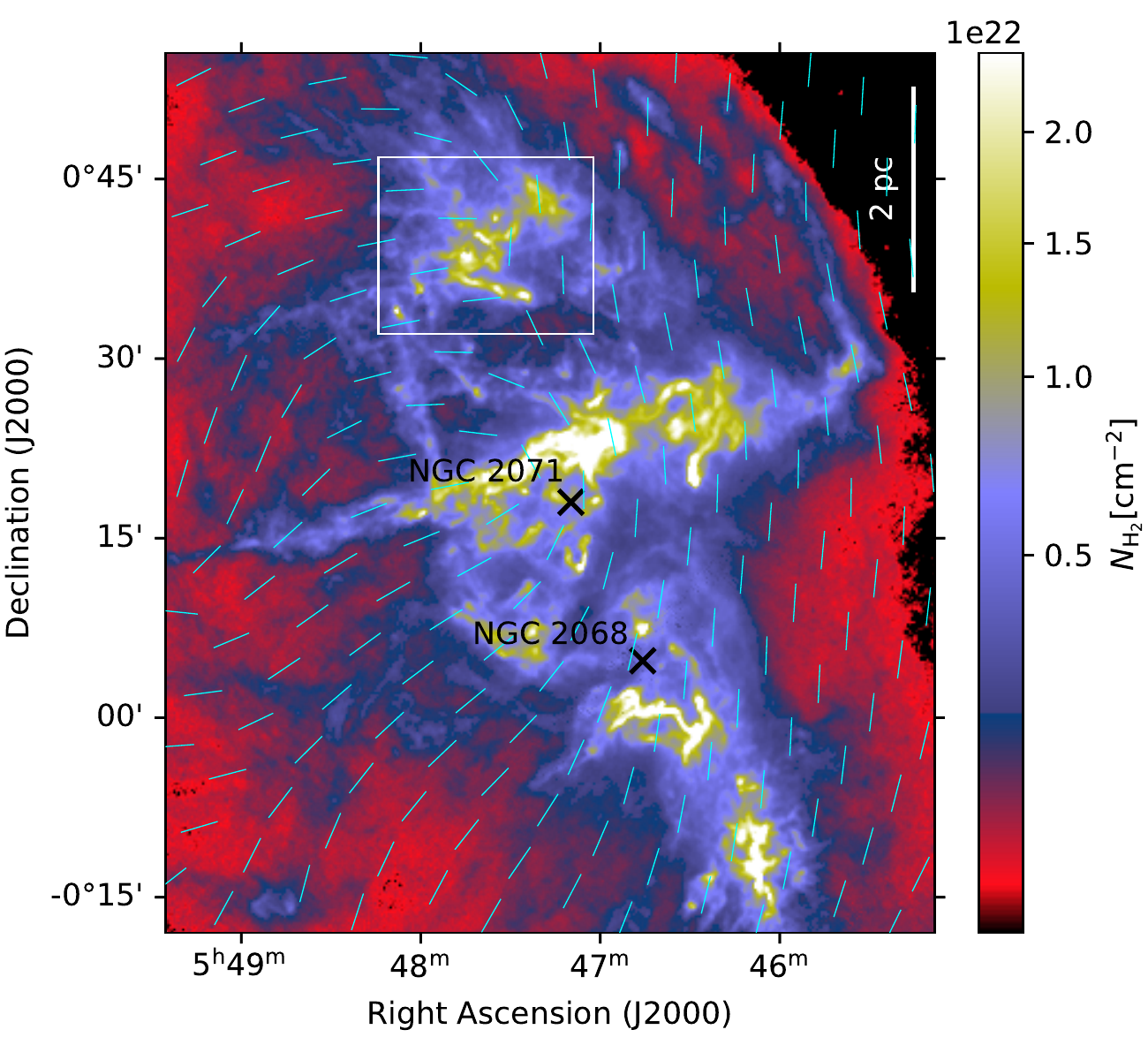}}
   \resizebox{0.552\hsize}{!}{\includegraphics[width=\columnwidth]{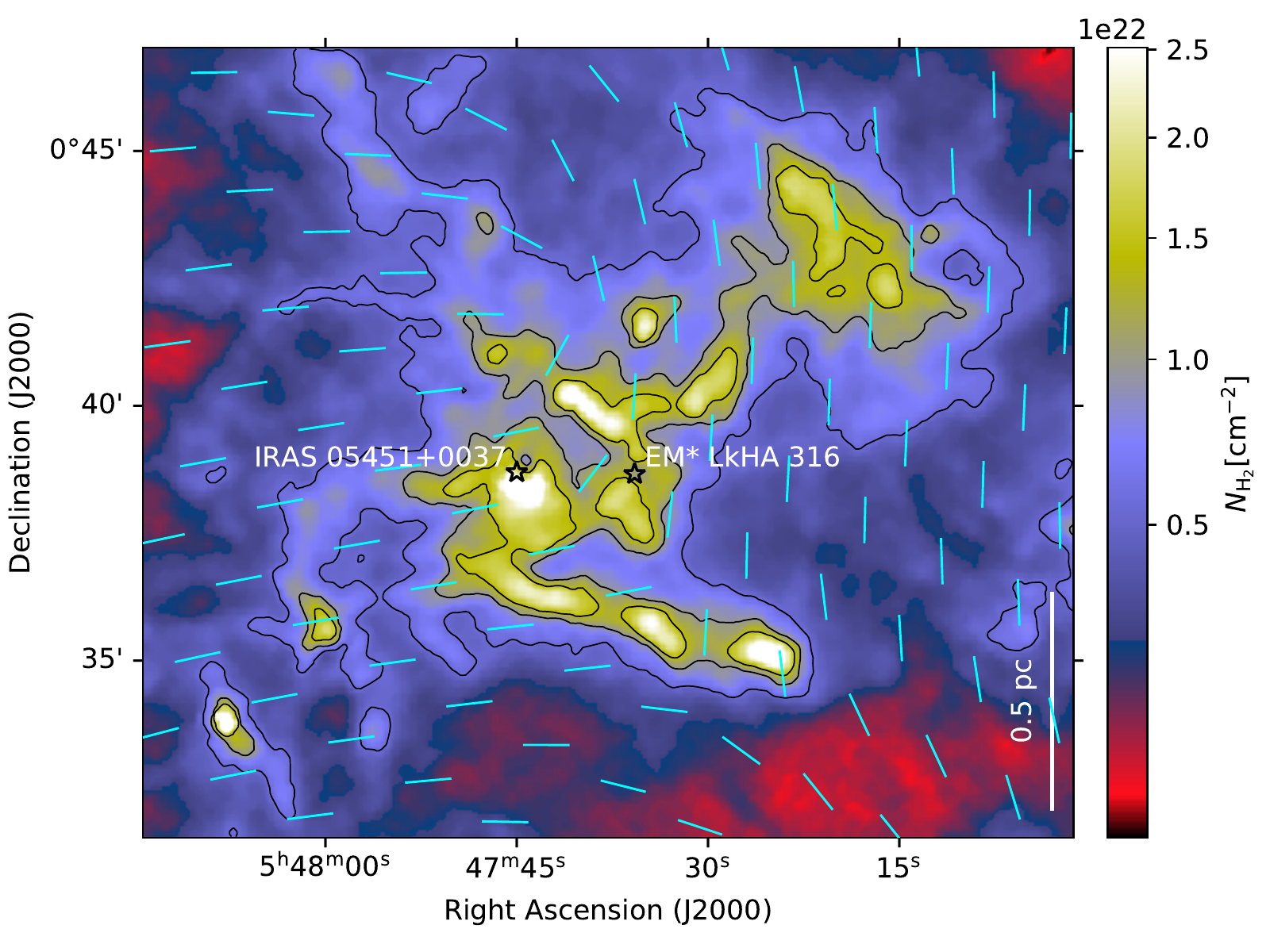}}
  \end{minipage}
 \end{center}
  \caption{{\bf Left:} Column density map of Orion~B showing the region around NGC2071 \citep[see][]{Konyves+2020}.
          The white square outlines a $\sim 18\arcmin \times 15\arcmin$ field centred on RA=05:47:37.8, 
          Dec=+00:39:16. The column density peak within this box, somewhat lower left from the centre, corresponds 
          to $\sim 4 \times 10^{22}$ cm$^{\rm -2}$, where the dust filaments exhibit a multi-arm hub morhology.
	  {\bf Right:} Zoomed column density map on the filament hub, marking the locations of two embedded
	  protostars.
	  Black contours correspond to $N_{\rm H_{2}}$ = [5.5 $\times 10^{21}$, 9.7 $\times 10^{21}$, 
	  1.4 $\times 10^{22}$] cm$^{-2}$, which are equivalent to $\sim$6, 10, and 15 magnitudes
	  of visual extinction. 
	  The same contours are used in subsequent figures.
	  In both panels Planck B-field-oriented vectors are overplotted in cyan on a 5\arcmin-scale (left),
	  and 1.71\arcmin-scale (right).
	  }
  \label{fig1}
\end{figure*}

The Orion~B cloud complex at $d\sim 400$\,pc \citep[][]{Menten+2007, Lallement+2014, Zucker+2019} 
was studied by the {\it Herschel} Gould Belt survey \citep{Andre+2010}.
Here, \citet{Konyves+2020} confirmed the physical existence of a transition in prestellar core 
formation efficiency (CFE) around a fiducial threshold of $A_V^{\rm bg} \sim 7$\,mag 
in background visual extinction. This is similar to the trend observed with {\it Herschel} in other 
regions, such as the Aquila cloud \citep{Konyves+2015}.
Between $A_V^{\rm bg} \sim 5-10$\,mag the CFE goes steeply from low to high, and the bulk of core and 
star formation is occurring above this threshold that had already been suspected earlier 
\citep[e.g.,][]{Onishi+1998, Johnstone+2004, KirkH+2006}.
In the filamentary regions of NGC2023/24, NGC2068/71 (see Fig.~\ref{fig1} left) and 
the L1622 cometary cloud \citet{Konyves+2020} found a total of 1768 starless cores ($\sim 28-45$\% 
of which are gravitationally bound prestellar cores) and an additional 76 protostellar (Class 0-I) 
cores. In Orion~B, the mass in prestellar cores above the mentioned threshold represents only a 
moderate fraction ($\sim 20$\%); and $\sim$60--80\% of the gravitationally bound cores are 
associated with filaments. 

Interstellar filaments also play an important role in the formation of massive stars, where these 
dense elongated features are often organized in a ``hub-filament'' structure with converging arms 
\citep[e.g.,][]{Myers2009, Peretto+2013}. Similar structures were called ``junctions of filaments'' by 
\citet{Schneider+2012}, who found that massive clusters more likely lie in the proximity of junctions 
of filaments in high column density regions, as \citet{DaleBonnell2011} proposed from simulations. 

Clumps with radiating multiple filaments can be seen in low-mass star-forming fields as well 
\citep[e.g., in Pipe Nebula:][]{Peretto+2012}. However, single filaments that do not cross each other 
\citep[e.g., in Taurus:][]{Palmeirim+2013} appear to provide enough material to form solar-type stars.
The hub-filament mode may be more typical, and its role may be more important, in regions of massive 
star formation, where the hub centre represents a deep potential well, able 
to accrete much more material from the surrounding filaments. 

As for the role of the magnetic field ($B$-field) in the formation and evolution 
of such structures, a bimodal distribution of the relative orientations between the 
filaments and the mean magnetic field directions was found observationally, that is
these relative orientations change from parallel to perpendicular with 
increasing (column) density. In other words, the relative orientations between the 
$B$-fields and the elongated structures were found to be mostly parallel in low-density 
filaments, and mostly perpendicular in dense filaments. These structures can be matched
to the lower density ``parallel'' filaments and the high-density elongated hubs
they are connected to in the hub-filament model by \citet{Myers2009}.
\citet{Li+2013} showed from observations that uniform (i.e., dynamically important) 
$B$-fields on larger scales can give rise to typical hub-filament cloud morphology.
A strong interplay and the bimodality between interstellar $B$-fields and filaments 
were also shown by \citet{PlanckXXXV2016}, and \citet{Alina+2019}, after 
these have been predicted by MHD simulations 
\citep{NakamuraLi2008, Soler+2013, Chen+2016, SolerHennebelle2017}.  

The region of interest of this work is a so far poorly studied sub-region north of NGC2071 that 
was named ``NGC2071-North'' (see Fig.~\ref{fig1} right) by \citet{Iwata+1988}. 
They made molecular observations in $^{12}$CO, $^{13}$CO, C$^{18}$0(1--0), and NH$_3$ (1, 1) 
and (2, 2) lines, and followed up a CO outflow, discovered by \citet{Fukui+1986}.
This relatively old ($t \sim 1.7 \times 10^{5}$\,yr) outflow shows a U-shape, and is apparently 
driven by IRAS~05451+0037. 
This was confirmed by \citet{Goldsmith+1992} with 3\,mm wavelength molecular observations. 
They argued that the molecular abundances in NGC2071-North (or NGC2071-N) have been 
significantly affected by this outflow and the presence of YSOs (young stellar objects).
With this study we follow on from \citet{Gibb2008}, who also summarized the findings 
in NGC2071-N up to 2008. 

As we mentioned above, cloud material may be channeled more efficiently through filaments into the seeds 
of star formation. However, newly-formed luminous sources can also significantly alter the 
geometry and composition of the surrounding material. While massive young stars can have dramatic
effects on their neighbourhood by ionisation and major dynamical impact 
\citep[HII regions, e.g.,][]{Tenorio-Tagle1982}, winds and outflows of lower mass young stars can also 
sweep up gas and dust by injecting considerable mechanical energy into the ISM \citep[cf.,][]{Snell1989}.  
All of these feedback effects, which are an integral part of the star formation process, can trigger 
temperature and density changes in the surrounding matter.

\citet{AspinReipurth2000} performed optical CCD imaging around compact reflection nebulae and 
embedded IRAS sources in order to search for new Herbig-Haro (HH) jets and flows. They also 
identified a cluster of new HH objects associated with IRAS 05451+0037 and the nearby young 
star LkH$\alpha$ 316 in the centre of NGC2071-N. 

\citet{Hillenbrand+2012} detected strong emission-line features in the TiO and VO bands at the 
position of the
optically-faint, flat-spectrum protostar IRAS 05451+0037 too, which suggests that it may be
surrounded by dense and warm circumstellar gas.

Due to its relative isolation north of NGC2071 (see Fig.~\ref{fig1} left), this sub-region 
is still not well-studied. We present here the first analysis of NGC2071-N 
with high-resolution data sets of both the extended molecular material and the compact 
star-forming cores. 
It was hypothesized that it was once an elongated molecular clump along the E-W, SE-NW 
directions \citep{Iwata+1988, Goldsmith+1992}, and that NGC2071-North has
now been refined into a `filamentary hub' structure.  

This paper is organized as follows. Section 2 provides details about the used 
data sets. In Section 3 we show distance results from Gaia EDR3 measurements,
$Herschel$-derived properties, molecular line-derived properties, and we also estimate
the energy balance based on energy densities and pressures. 
In Section 4 we discuss the large-scale magnetic field over NGC2071-N, we reason that
star formation is on-going here, we discuss the double center, and revisit the existence
of a CO outflow in the region. Finally, Section 5 presents our conclusions.

\section{Observations and Data}\label{sec:obs}

\subsection{Herschel data}\label{sec:obsH} 

In this study we used part of
the SPIRE and PACS `parallel-mode' {\it Herschel} Gould Belt survey (HGBS) 
observations of the Orion~B complex region that includes NGC2071-N \citep[see][]{Andre+2010}.
The details of the data reduction process are discussed by \citet{Konyves+2020}.
In this work we used the SPIRE 250-$\mu$m data, calibrated in MJy sr$^{-1}$, on 
3\arcsec-pixel scales. We also used the H$_2$ column density map that was created from 
the HGBS observations\footnote{http://gouldbelt-herschel.cea.fr/archives} 
\citep[see][]{Konyves+2020}.

In addition, we used 
PACS-only mode 100-$\mu$m, and 160-$\mu$m data, also from the HGBS project 
(OBSIDs: 1342206054, 1342206055), observed on 08 October 2010 with a scanning speed 
of 20$\arcsec$s$^{-1}$ (as opposed to the parallel-mode's 60$\arcsec$s$^{-1}$). 
They were reduced in the same way as the parallel-mode PACS observations
\citep[see][]{Konyves+2020}.

\subsection{IRAM 30\,m observations}\label{sec:obsIRAM} 

In the 2013 summer semester we carried out IRAM 30\,m observations under project number 030--13. 
We used the EMIR receiver
\citep{Carter+2012} at 3\,mm to take fully-sampled C$^{18}$O(1--0) and $^{13}$CO(1--0) maps 
simultaneously in a $\sim 108$ arcmin$^2$ region toward NGC2071-N. The on-the-fly mapping mode 
was used with position-switching.
At 109.782 GHz, the 30\,m telescope has a beam size of 23.6$\arcsec$~and the forward and 
main beam (MB) efficiencies ($F_{\rm eff}$ and $B_{\rm eff}$) are 95\% and 79\%, respectively. 
The backend used was the VESPA auto-correlator providing 
a frequency resolution of 20\,kHz that corresponds to $\sim$ 0.055\,km s$^{-1}$) in velocity 
resolution.

For the position-switching mode, the reference position was offsetted by $\sim$20\arcmin~from the map centre.
The off position was selected from {\it Herschel} column density images, 
and we made sure that no significant CO emission is appearing there by observations in 
frequency-switching mode.

During the observations, calibration was performed every $\sim$30 minutes, and the
telescope pointing was checked and adjusted every $\sim$2~hours. The pointing accuracy was 
found to be better than 3\arcsec. 
All of the data were reduced with the GILDAS/CLASS software package\footnote{http://www.iram.fr/IRAMFR/GILDAS}.
 
We smoothed the data spatially with a Gaussian function resulting in an effective beam size 
of 28$\arcsec$ ($\sim$0.05\,pc at $\sim$400\,pc). 
The 1$\sigma$ noise level of the final mosaicked data cube is $\sim$0.11 K in T$_{\rm MB}$,
at an effective angular resolution of 28$\arcsec$ and a velocity resolution of $\sim$0.1~km~s$^{-1}$.

\subsection{NRO 45\,m observations}\label{sec:obsNRO}

In 2015 we carried out observations on the 45-m telescope of the Nobeyama Radio Observatory
toward a 0.14 square degree region in the Orion B cloud, including the densest portions of NGC2071-North,
with the TZ receiver \citep{Shimajiri+2017}. All molecular line data (HCN(1--0), H13CN(1--0), HCO+(1--0), 
and H13CO+(1--0)) were obtained simultaneously. At 86 GHz, the telescope has a beam size of 19.1\arcsec 
(HPBW) and a main beam efficiency of $\sim$50\%. As a backend, we used the SAM45 spectrometer, which 
provides a bandwidth of 31~MHz and a frequency resolution of 7.63~kHz. The latter corresponds to a 
velocity resolution of $\sim$0.02~km~s$^{-1}$ at 86~GHz. We applied spatial smoothing to the data 
with a Gaussian function resulting in an effective beam size of 30\arcsec. The 1-sigma noise level 
of the final data is $\sim$0.35 K in T$_{\rm MB}$ at an effective resolution of 30\arcsec and a 
velocity resolution of 0.1~km~s$^{-1}$.
More details of these observations and data reduction are described by \citet{Shimajiri+2017}.

Some of the image operations, such as $moment$ and $smooth$, were performed using the MIRIAD 
software package \citep{Sault+1995}, for both the NRO and the IRAM observations.

\subsection{Archival data}\label{sec:obsArchive} 

In order to trace and visualize NGC2071-North, in particular the central part of the hub,
we have displayed and superimposed various data sets and catalogues within the interactive 
software Aladin sky atlas\footnote{http://aladin.u-strasbg.fr}.

In addition to the above data sets, then we downloaded high-resolution short-wavelength 
SDSS, Pan-STARRS, and 2MASS images to further trace the interesting structures that we
first caught in the 100\,$\mu$m map (see Fig.~\ref{fig_100mu}). 

SDSS images were retrieved from the SkyView Query Form\footnote{https://skyview.gsfc.nasa.gov} 
which service resampled the data from the Sloan Digital Sky Survey\footnote{www.sdss3.org}.
In this work we visualize SDSS data observed with the g, r, i colour filters.  

2MASS\footnote{https://irsa.ipac.caltech.edu/Missions/2mass.html} J, H, K$_{\rm s}$ infrared 
images were also queried from NASA's SkyView service.

Pan-STARRS is a system for wide-field astronomical imaging developed and operated by the 
Institute for Astronomy at the University of Hawaii. We downloaded DR2 data of the first part
of the project to be completed through their Image Cutout Server\footnote{https://ps1images.stsci.edu/cgi-bin/ps1cutouts}
Out of the five broadband filters (g, r, i, z, y), we used i, z, y.

To double-check the distance measurements in this region, we used Gaia Early Data 
Release 3 (Gaia EDR3) sources downloaded from the Gaia Archive\footnote{http://gea.esac.esa.int/archive}.

\begin{figure*}
 \begin{center}
  \begin{minipage}{1.\linewidth}
  \hspace{-1mm}
   \resizebox{0.49\hsize}{!}{\includegraphics[width=\columnwidth]{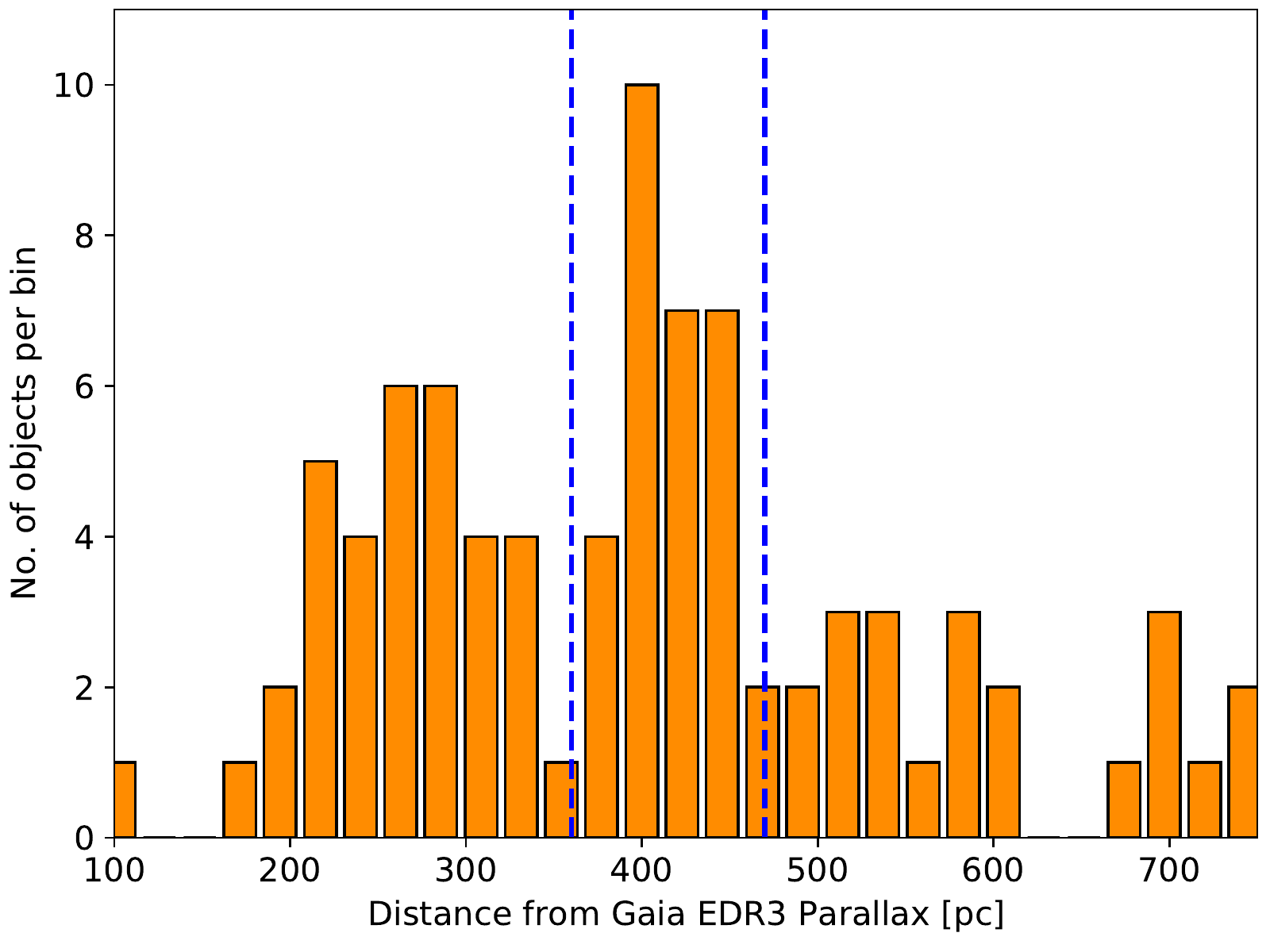}}
  \hspace{0.5mm}
   \resizebox{0.51\hsize}{!}{\includegraphics[width=\columnwidth]{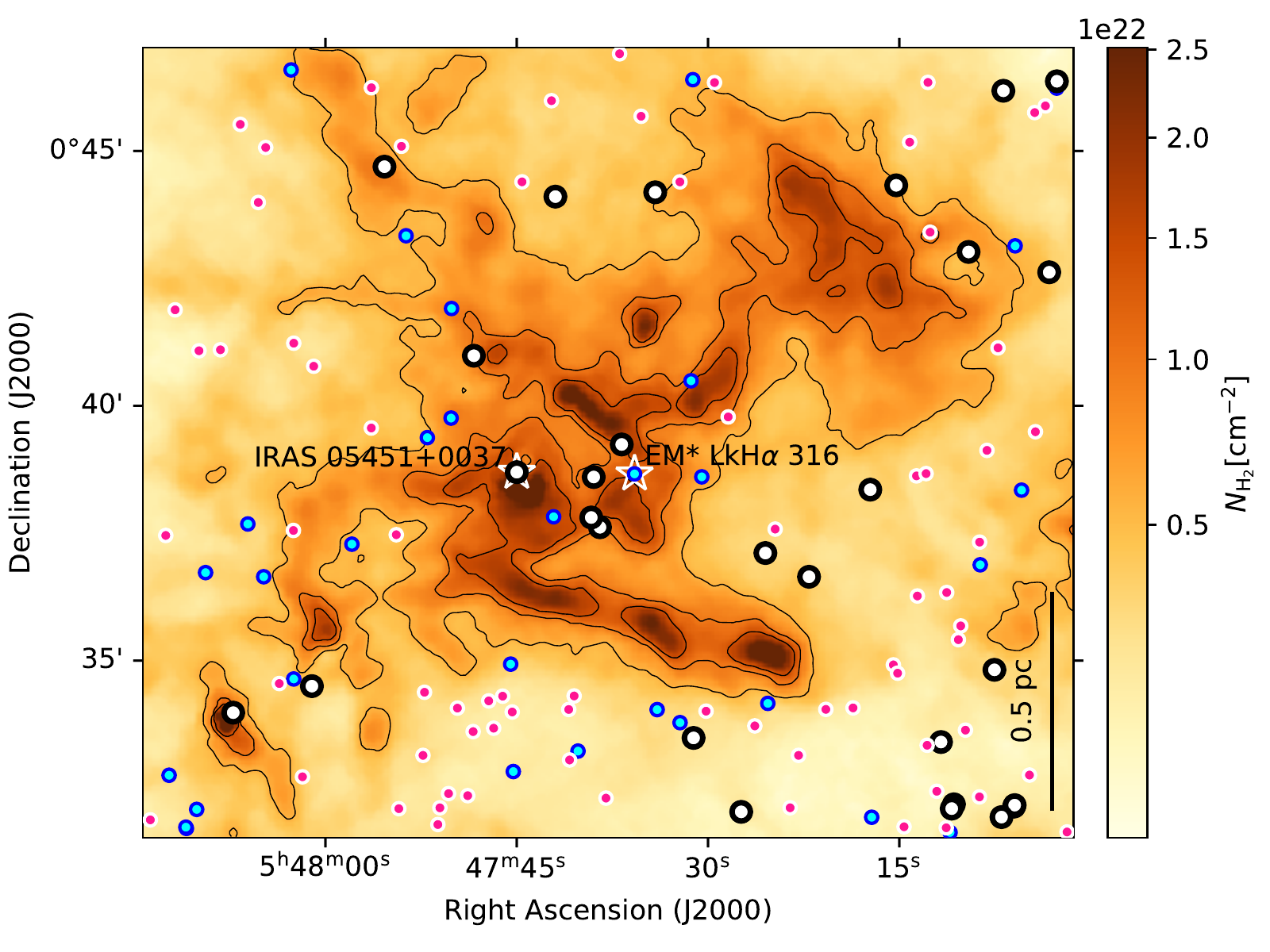}}
  \end{minipage}
 \end{center}
  \caption{{\bf Left:} Histogram of distances converted from Gaia EDR3 parallax measurements over the region shown in the right panel. 
          Dashed blue lines mark the distances at 360\,pc and 470\,pc (see text for details). 
          {\bf Right:} Overlaid on column density, cyan/blue dots mark the Gaia measurement points at $d < 360$\,pc, white/black 
          dots show the positions of $d = 360-470$\,pc measurements, which are a priori considered Orion~B sources, and magenta/white
          dots mark the locations of Gaia EDR3 sources beyond 470\,pc. Again, white stars mark the two named protostars. 
          }
  \label{fig_gaia}
\end{figure*}

\section{Results and Analysis}\label{sec:res}

\subsection{Distance from Gaia EDR3 data}\label{sec:dist}

The most prominent features within Orion~B are the Horsehead Nebula, the NGC~2023/24, NGC~2068/71 
nebulae, as well as the Lynds~1622 (L1622) cometary cloud north-east of NGC2071-North 
\citep[see Fig.~2 of][]{Konyves+2020}. In the Lynds catalogue L1630 covers Orion~B 
without L1622.
As it was mentioned in Sect.~\ref{sec:intro}, we consider $d\sim 400$\,pc for the distance to 
most of the Orion~B clouds \citep[][]{Anthony-Twarog1982, Menten+2007, Gibb2008, Lallement+2014, 
Schlafly+2014, Zucker+2019}, while there is indication that the cometary, trunk-like features
at and around L1622 may not be at this same distance. The alignment of the trunks and comets
suggests that this northern part of Orion~B interacts with the Barnard's Loop
\citep[see Fig.~1 of][]{Konyves+2020}. Thus, the L1622 region may also be at a closer distance 
($\sim$170--180\,pc), which is tentatively found for Barnard's Loop by \citet{Bally2008} and \citet{Lallement+2014}.
However, its distance is still uncertain \citep[see e.g.,][and references therein]{Ochsendorf+2015}.

Revisiting Fig.~1 of \citet{Konyves+2020} gave the idea to check the distances at NGC2071-N with 
available Gaia EDR3 data, as the H$\alpha$ shell of Barnard's Loop seems to cut through Orion~B 
between NGC2071 and L1622. 

Gaia EDR3 measurements have been downloaded for the coverage shown in, for example, Fig.~\ref{fig1}
right, 
from which we only exploited RA(J2000) and Dec(J2000) coordinates, parallax, and parallax errors. 
Distances in parsec from the parallax data (in arcseconds) have been converted with 
Astropy's $to()$ unit conversion method \citep{Greenfield+2013}. For the analysis, 
we ignored data points with negative parallax, as well as data where the parallax error was 
larger than half of the parallax value.

The histogram in the left-hand panel of Fig.~\ref{fig_gaia} shows the distances for the region 
in the right-hand panel.
A significant group of distances around 400\,pc is apparent. However a very wide range of distances 
is found in this $\sim$18$\times$15 arcmin region. 
From the histogram, we set lower and upper distance limits 
around Orion~B, at 360\,pc and 470\,pc, 
which seem to be reasonable choices, and they may also indicate a cloud depth, at NGC2071-N, of about 100\,pc.
The Gaia data points in these three distance ranges are plotted in Fig.~\ref{fig_gaia} right.
The cyan/blue dots mark the positions of Gaia measurements up to 360\,pc, the white/black dots
show the a priori Orion~B sources between 360\,pc and 470\,pc, and the sources beyond 470\,pc
are marked with magenta/white dots. These latter ones most probably belong to the background, 
as they almost only appear where the cloud is more transparent (i.e., less column density). 
A group of white/black dots at around 400\,pc is concentrated in the central portion of 
the hub and/or at higher column densities, which is reassuring. More also show up
in the lower right corner that is in the direction of NGC2071.      
However the cyan/blue dots appear everywhere in the line of sight, and apparently the 
nebulous star LkH$\alpha$ 316, around the centre of 
NGC2071-N, also seems to lie closer to us
than 360\,pc. 

Distances from EDR3 parallaxes of the two central objects, IRAS 05451+0037 and LkH$\alpha$ 316,
gave us $d_{\rm IRAS} = 383.4_{-14.2}^{+15.3}\,$pc, and $d_{\rm LkH\alpha} = 300.6_{-36.5}^{+48.2}\,$pc,
respectively, at which distances they may still belong to the same cloud.  

After the above filtering of Gaia EDR3 measurement points, the distances range from $\sim$100\,pc, 
to 6300\,pc. They all are plotted in Fig.~\ref{fig_gaia} right, while only
those up to $\sim$800\,pc are shown in the left-hand side histogram.

\begin{figure*}
 \begin{center}
  \begin{minipage}{1.\linewidth}
   \resizebox{0.5\hsize}{!}{\includegraphics[width=\columnwidth]{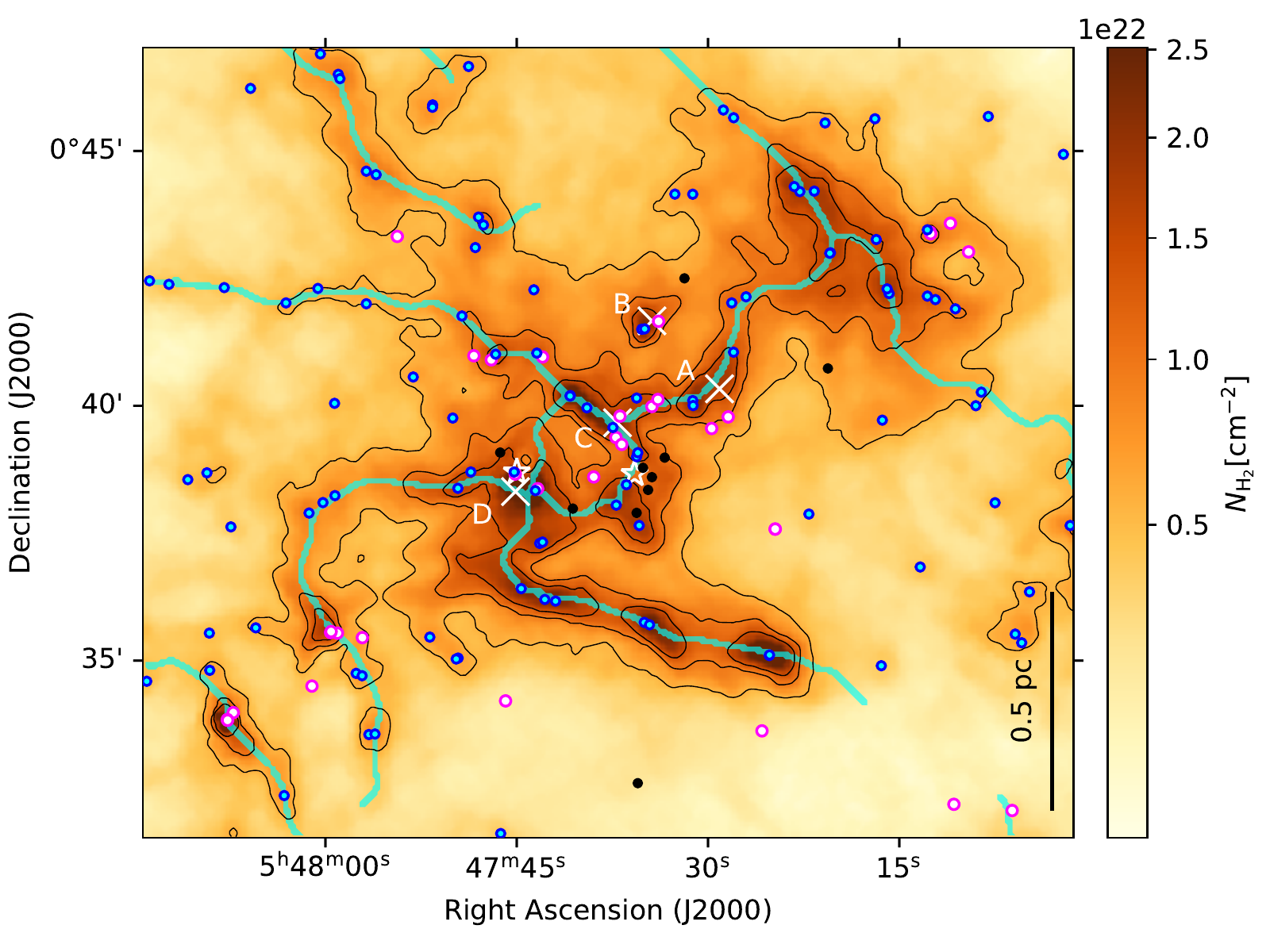}}   
   \resizebox{0.5\hsize}{!}{\includegraphics[width=\columnwidth]{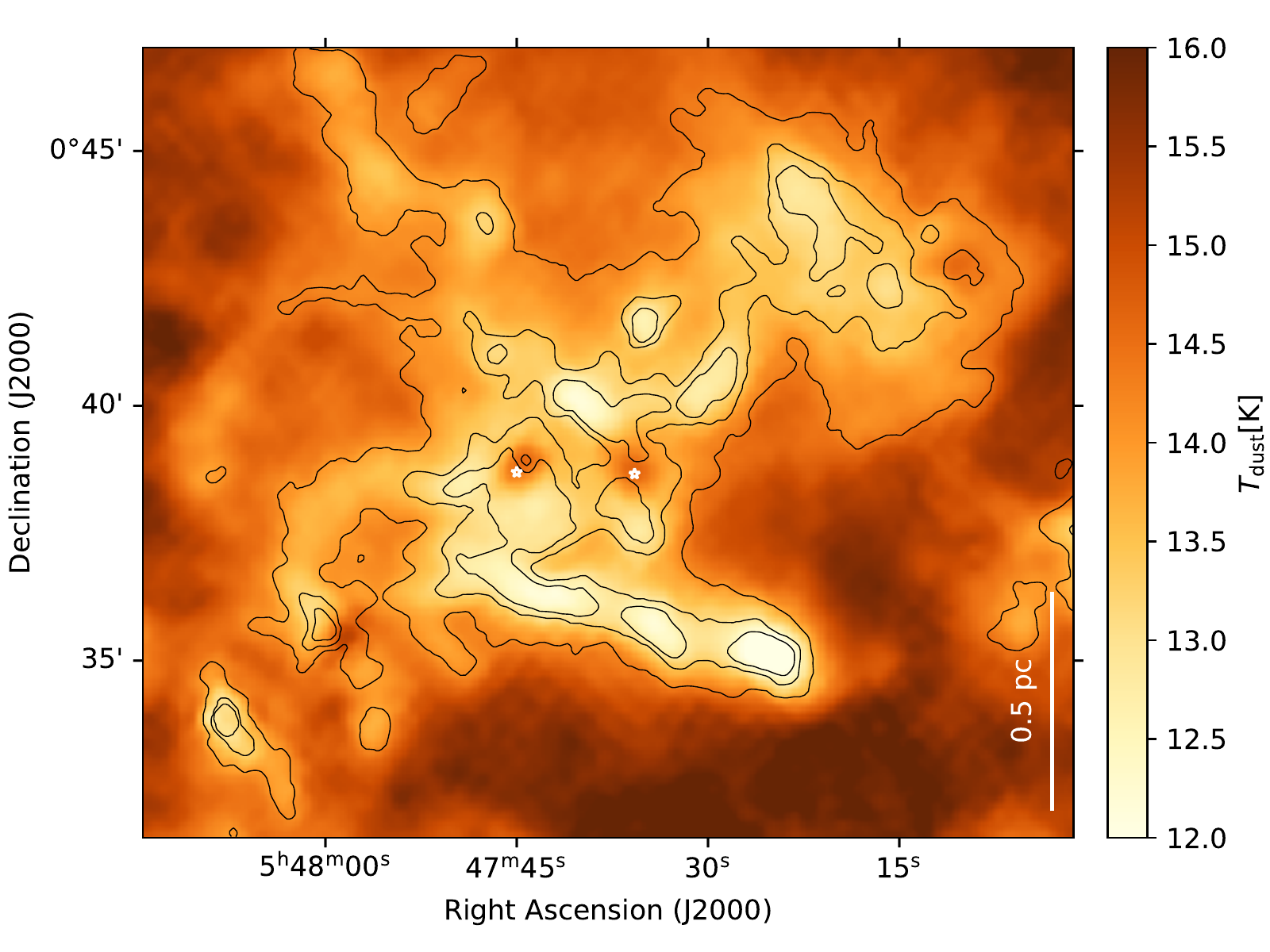}}
  \end{minipage}
 \end{center}
  \caption{{\bf Left:} Column density map of the NGC2071-N filament hub, with DisPerSE \citep{Sousbie2011} filaments.            
          Dense cores (bound prestellar and unbound starless) from \citet{HKirk+2016a} and \citet{Konyves+2020} are overplotted
          with cyan/blue dots. White/magenta dots mark YSOs/protostars from SIMBAD together with a few embedded protostars from 
          \citet{Konyves+2020}. Black dots mark Herbig-Haro objects from SIMBAD. Four white crosses show the positions of
          NH$_3$(1, 1) cores defined by \citet{Iwata+1988}. {\bf Right:} Dust temperature map of the same region.
          Maps and filaments are from \citet{Konyves+2020}. In both panels the left and right white stars (also protostars) 
          mark the locations of IRAS 05451+0037 and LkH$\alpha$ 316, resp.
          }
  \label{fig_fil_obj}
\end{figure*}

\begin{figure}
 \begin{center}
  \begin{minipage}{1.\linewidth}
   \resizebox{1.0\hsize}{!}{\includegraphics[width=\columnwidth]{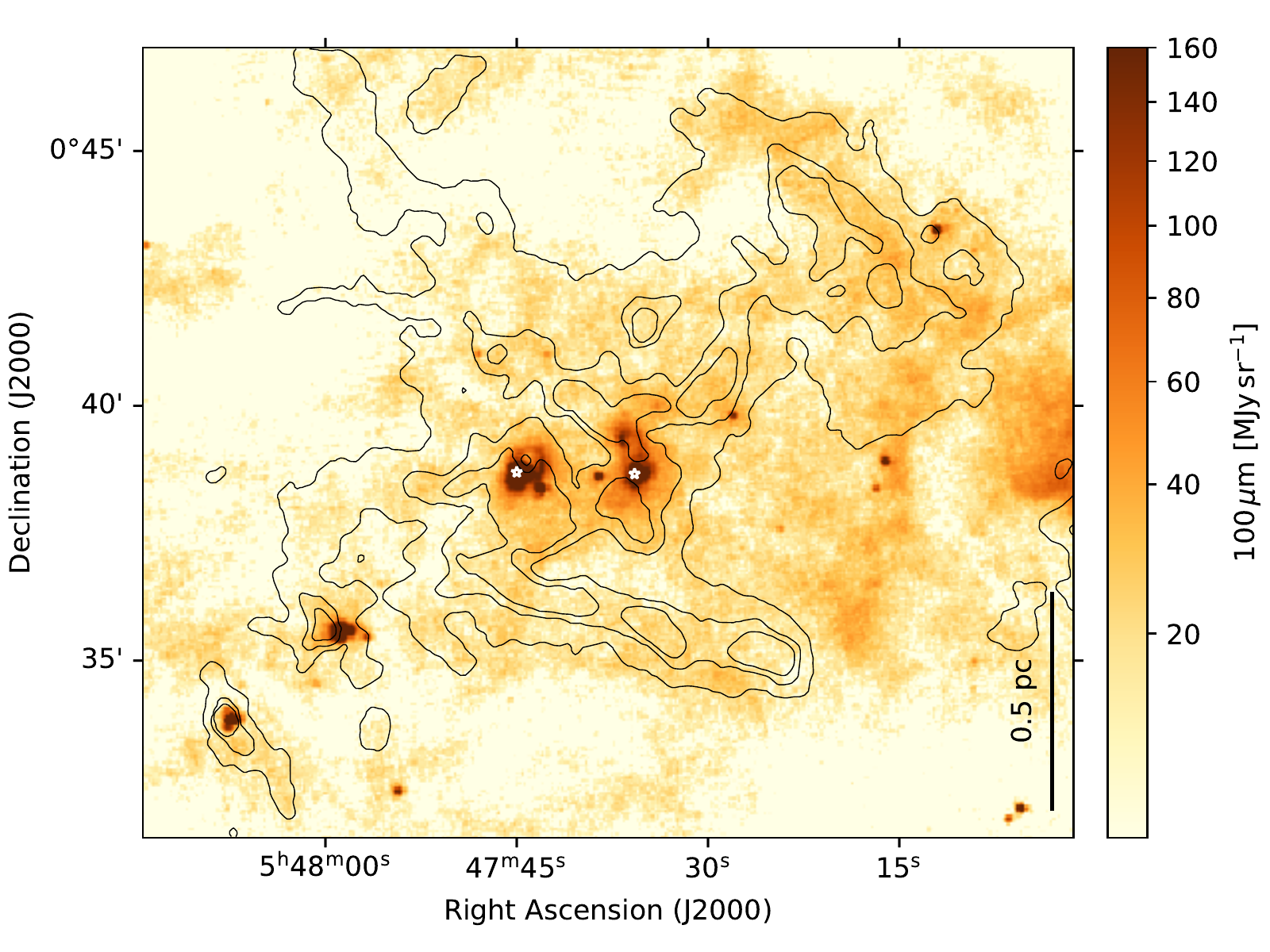}}
  \end{minipage}
 \end{center}
  \caption{The same region as above in 100\,$\mu$m emission. 
          The left and right white stars mark the locations of IRAS 05451+0037 and LkH$\alpha$ 316, resp.
          }
  \label{fig_100mu}
\end{figure}

\begin{table*}\small\setlength{\tabcolsep}{3.8pt}
 \caption{Physical properties of 42 selected locations within 25\arcsec radius in NGC2071-N, shown in Fig.~\ref{fig_circ}. 
          See Sects.~\ref{sec:props} and \ref{sec:energy} for details.
          {\bf (1)}: Spot numbers, as in Fig.~\ref{fig_circ}; {\bf (2)} and {\bf (3)}: Right ascension and declination of spot centres; 
	  {\bf (4)}: Visual extinction calculated from median column density assuming 
	  $N_{\rm H_2}\, ({\rm cm}^{-2}) = 0.94 \times 10^{21}\, A_{\rm V} \,({\rm mag }) $ \citep{Bohlin+1978}; 
	  {\bf (5)}: Dust mass within these circles, no 20\% error is included; {\bf (6)} and {\bf (7)}: Dust temperature with stdev uncertainty; 
	  {\bf (8)}: average volume density; {\bf (9)}: gravitational energy density; 
	  {\bf (10)}: volume-averaged gravitational pressure;
	  {\bf (11)}: Type of central object taken from the following studies: 
	  $^1$ \citet{Konyves+2020}; $^2$ \citet{Megeath+2012}; $^3$ \citet{KirkH+2016a}; 
	  $^4$ \citet{Gaia2018}; $^5$ \citet{Cutri+2003}. 
         }
 \label{tab_uCalcG}      
 {\renewcommand{\arraystretch}{1.2}        
  \begin{tabular}{l c c c c c c c c r}  
  \hline
  \hline
  Spot no.	& RA$_{\rm 2000}$	& Dec$_{\rm 2000}$		& $A_{\rm V}$	& Mass		& $T_{\rm dust}$	& $n^{\rm ave}_{\rm H_2}$	& $u_{\rm grav}$		& $\overline{P}_{\rm G} / k_{\rm B}$	& obj$_{\rm type}$	\\
		& (h m s)		& (\degr~\arcmin~\arcsec)	& (mag)		& ($M_\odot$)	& (K)			& (10$^{4}$ cm$^{-3}$)		& (10$^{-10}$ erg cm$^{-3}$)	& (10$^{5}$ K cm$^{-3}$)		&			\\   
  (1)		& (2)			& (3)				& (4)		& (5)		& (6) $\pm$ (7)		& (8)				& (9)				& (10)				& (11)			\\   
  \hline
   1 		& 05:47:37.00		& +00:38:10.3			& 17.7 	& 2.75		& 13.4 ~ 0.3		& 8.33				& 5.70				& 18.71			    & --		    \\
   2 		& 05:47:40.72		& +00:38:16.4		 	& 12.7		& 2.06		& 13.3 ~ 0.1		& 6.23				& 3.19				& 10.50			    & --		    \\
   3 		& 05:47:44.50		& +00:38:22.2		 	& 28.7		& 4.34		& 13.4 ~ 0.5		& 13.16			& 14.21			& 46.59			    & --		    \\
   4 		& 05:47:49.62		& +00:38:22.9		 	& 14.2		& 2.18		& 12.8 ~ 0.1		& 6.61				& 3.58				& 11.75			    & prestellar core $^1$  \\  	     
   5 		& 05:47:55.55		& +00:38:22.1		 	&  8.6		& 1.28		& 13.6 ~ 0.1		& 3.88				& 1.23				& 4.05			& --		    \\
   6 		& 05:48:01.29		& +00:37:53.8		 	&  8.2		& 1.29		& 13.7 ~ 0.1		& 3.93				& 1.26				& 4.12			    & prestellar core $^1$  \\
   7 		& 05:48:01.75		& +00:36:38.9		 	&  8.2		& 1.28		& 13.5 ~ 0.2		& 3.87				& 1.23				& 4.05			    & --		    \\
   8 		& 05:47:59.56		& +00:35:34.3		 	& 12.9		& 1.91  	& 14.1 ~ 0.6		& 5.80				& 2.77				& 9.02			    & protostellar core $^1$\\
   9 		& 05:47:56.12		& +00:33:33.7		 	&  6.5		& 1.02  	& 14.0 ~ 0.2		& 3.10				& 0.79				& 2.57			    & prestellar core $^1$  \\
  10 		& 05:48:07.68		& +00:33:49.9		 	& 13.6		& 2.46  	& 13.1 ~ 0.2		& 7.46				& 4.56				& 14.97			    & protostellar core $^1$\\
  11 		& 05:47:42.98		& +00:37:19.6		 	& 16.7		& 2.45  	& 13.0 ~ 0.2		& 7.44				& 4.55				& 14.85			    & prestellar core $^1$  \\
  12 		& 05:47:46.43		& +00:36:39.6		 	& 17.3		& 2.68  	& 12.6 ~ 0.2		& 8.12				& 5.41				& 17.77			    & --		    \\
  13 		& 05:47:41.96		& +00:36:10.2		 	& 18.5		& 2.81  	& 12.4 ~ 0.3		& 8.51				& 5.94				& 19.53			    & prestellar core $^1$  \\
  14 		& 05:47:38.18		& +00:35:57.0		 	& 13.2		& 2.05  	& 12.9 ~ 0.2		& 6.22				& 3.18				& 10.39			    & --		    \\
  15 		& 05:47:34.59		& +00:35:42.3		 	& 19.0		& 2.98  	& 12.4 ~ 0.2		& 9.04				& 6.71				& 21.97			    & prestellar core $^1$  \\
  16 		& 05:47:30.05		& +00:35:21.4		 	& 12.1		& 1.89  	& 13.0 ~ 0.1		& 5.72				& 2.69				& 8.84			    & --		    \\  
  17 		& 05:47:25.18		& +00:35:06.8		 	& 24.8		& 3.97  	& 11.7 ~ 0.3		& 12.04			& 11.90				& 38.98			    & prestellar core $^1$  \\  
  18 		& 05:47:05.91		& +00:35:31.3		 	&  6.6		& 1.01  	& 14.1 ~ 0.2		& 3.07				& 0.77				& 2.52			    & starless core $^1$    \\  
  19 		& 05:47:40.36		& +00:39:07.4		 	&  9.5		& 1.50  	& 13.5 ~ 0.1		& 4.54				& 1.69				& 5.57			    & --		    \\  
  20 		& 05:47:43.73		& +00:39:26.2		 	& 12.1		& 1.82  	& 13.5 ~ 0.3		& 5.53				& 2.51				& 8.19			    & --		    \\
  21 		& 05:47:35.51		& +00:39:04.8		 	& 13.6		& 2.11  	& 14.1 ~ 0.3		& 6.40				& 3.36				& 11.01			    & prestellar core $^1$  \\  
  22 		& 05:47:37.46		& +00:39:34.5		 	& 18.8		& 2.98  	& 13.2 ~ 0.4		& 9.03				& 6.69				& 21.97			    & prestellar core $^1$  \\
  23		& 05:47:40.82		& +00:40:11.5		 	& 18.6		& 3.11  	& 12.5 ~ 0.3		& 9.43				& 7.30				& 23.92			    & prestellar core $^1$  \\  						  
  24		& 05:47:42.97		& +00:40:57.4		 	& 12.0		& 1.86  	& 13.3 ~ 0.1		& 5.65				& 2.62				& 8.56			    & YSO $^{2, 4}$	    \\
  25		& 05:47:46.66		& +00:41:00.7		 	& 12.9		& 2.04  	& 13.3 ~ 0.1		& 6.18				& 3.14				& 10.29			    & prestellar core $^1$  \\
  26		& 05:47:49.30		& +00:41:45.7		 	&  9.4		& 1.44  	& 13.6 ~ 0.1		& 4.38				& 1.57				& 5.13			    & prestellar core $^1$  \\
  27		& 05:47:53.00		& +00:41:55.3		 	&  6.3		& 0.98  	& 14.1 ~ 0.1		& 2.98				& 0.73				& 2.38			    & --		    \\
  28		& 05:47:56.80		& +00:42:00.1		 	&  5.9		& 0.90  	& 14.2 ~ 0.1		& 2.73				& 0.61				& 2.00			    & dense core $^3$	    \\  	    
  29		& 05:48:03.10		& +00:42:01.1		 	&  5.1		& 0.79  	& 14.4 ~ 0.1		& 2.39				& 0.47				& 1.54			    & prestellar core $^1$  \\
  30		& 05:47:47.61		& +00:43:32.8		 	&  9.6		& 1.48  	& 13.4 ~ 0.1		& 4.49				& 1.66				& 5.42			    & prestellar core $^1$  \\
  31		& 05:47:56.03		& +00:44:32.0		 	&  9.2		& 1.38  	& 13.6 ~ 0.1		& 4.19				& 1.44				& 4.71			    & prestellar core $^1$  \\
  32		& 05:47:51.61		& +00:45:51.3		 	&  6.8		& 1.07  	& 14.2 ~ 0.1		& 3.24				& 0.86				& 2.83			    & starless core $^1$    \\
  33		& 05:47:58.87		& +00:46:25.0		 	&  8.5		& 1.31  	& 13.9 ~ 0.1		& 3.97				& 1.29				& 4.24			    & prestellar core $^1$  \\
  34		& 05:47:34.40		& +00:39:59.1		 	& 15.0		& 2.31  	& 13.2 ~ 0.1		& 7.02				& 4.04				& 13.2			    & YSO $^{2, 5}$	    \\
  35		& 05:47:31.17		& +00:40:00.3		 	& 16.0		& 2.58  	& 13.0 ~ 0.2		& 7.81				& 5.01				& 16.46			    & prestellar core $^1$  \\
  36		& 05:47:34.97		& +00:41:30.5		 	& 14.7		& 2.39  	& 12.9 ~ 0.2		& 7.23				& 4.30				& 14.13			    & prestellar core $^1$  \\
  37		& 05:47:28.00		& +00:41:03.1		 	& 13.0		& 2.12  	& 13.2 ~ 0.2		& 6.42				& 3.38				& 11.12			    & dense core $^3$	    \\
  38		& 05:47:28.14		& +00:42:01.1		 	& 10.9		& 1.66  	& 13.5 ~ 0.1		& 5.03				& 2.07				& 6.82			    & prestellar core $^1$  \\
  39		& 05:47:24.34		& +00:42:31.9		 	& 11.3		& 1.73  	& 13.5 ~ 0.1		& 5.26				& 2.27				& 7.40			    & --		    \\
  40		& 05:47:20.45		& +00:42:59.5		 	& 16.0		& 2.50  	& 13.1 ~ 0.1		& 7.58				& 4.72				& 15.46			    & prestellar core $^1$  \\
  41		& 05:47:15.96		& +00:42:17.6		 	& 16.0		& 2.55  	& 13.2 ~ 0.1		& 7.74				& 4.91				& 16.08			    & prestellar core $^1$  \\
  42		& 05:47:23.24		& +00:44:18.0		 	& 17.5		& 2.69  	& 12.9 ~ 0.1		& 8.16				& 5.46				& 17.90			    & prestellar core $^1$  \\  										   
  \hline
  \hline  
  \end{tabular}
 }   
\end{table*}

\subsection{{\it Herschel} properties of the hub-filament}\label{sec:props}

The spectacular hub-filament structure of NGC2071-North, seen in Figures 1 \& 2, 
is made up of high column density multi-arms, 
corresponding to equivalent visual extinctions of $A_{\rm V} \ga 5$.
It is relatively isolated from NGC2071, thus this sub-region is still not much studied.
It occupies a $\sim 1.5 \times 1.5$\,pc projected area at a distance of
$\sim$ 400\,pc, and it was identified as such a structure in HGBS images \citep{Konyves+2020}.  

We have studied the star-forming properties of these filaments in \citet{Konyves+2020}, where 
the dense core population of the whole Orion~B cloud was discussed, also with respect
of filaments (see Fig.~\ref{fig_fil_obj} left).
The mass of the NGC2071-North structure is found to be about 500\,$\msun$~above $A_V \sim 5$\,mag 
which is less than that of the Serpens-South filament hub; $\sim 750\,\msun$~in a
$\sim 1 \times 2$\,pc region, most of which is even above $A_V \sim 10$\,mag, assuming a 
distance of 260\,pc \citep{Konyves+2015}. On the other hand, the B59 filament hub in Pipe 
\citep{Peretto+2012} is at lower extinction, its mass above $A_V \sim 5$ is only $\sim 30\,\msun$
(covering the hub centre and one filament arm).

Dust temperatures, also derived from HGBS data, within the lowest contours in the column density
map at 5.5 $\times 10^{21}$\,cm$^{-2}$ are found to be $\sim$14\,K or less 
(see Fig.~\ref{fig_fil_obj} right).
Within the densest portions (southern filament, clumps A, B, C, NW corner) the
temperature drops below 13\,K, while the direct surroundings (within $\sim$20\arcsec) of the 
two central protostars exhibit temperatures between 14 and 15\,K.

The bound prestellar core masses are in the range of $\sim$0.2--10\,\msun~with a median mass of 
$\sim$1\,\msun, which would eventually collapse to low-mass stars. These sources are also among 
the overplotted ones in Fig.~\ref{fig_fil_obj} left. The properties around them, 
together with further derived properties from Sect.~\ref{sec:energy}, are listed in 
Table~\ref{tab_uCalcG}. 
In Fig.~\ref{fig_fil_obj} left it is clear that most of the dense cores (cyan/blue dots) 
are located along filaments and elongated features. This tendency has been
noted and discussed in several HGBS papers \citep[e.g.,][]{Andre+2014, Marsh+2016, 
Benedettini+2018, Ladjelate+2020, Fiorellino+2021}, and this result does not depend significantly 
on the method of filament extraction \citep{Konyves+2020}. 
This same figure panel also shows that YSOs and protostars (white/magenta dots) tend to appear 
at locations which are at the crossing points of filaments. For instance, at core C and D of 
\citet{Iwata+1988} (identified from NH$_3$(1, 1) observations), which are at the junction of 
extracted filaments (in cyan). 
A similar effect on larger scales, that infrared clusters are found at the junction 
of filaments, has been found by several studies 
\citep[][]{Myers2009, Schneider+2012, Peretto+2013, Dewangan+2015}.

We also see diffuse and nebulous features at the apparent double-centre in the HGBS 
100\,$\mu$m map (see Fig.~\ref{fig_100mu}).  
At 100\,$\mu$m the emission is concentrated in the central part of the filament hub, at 
IRAS 05451+0037 and EM* LkH$\alpha$ 316, and features diffuse lobes and loops. 
This kind of activity at 100\,$\mu$m (and also at 70\,$\mu$m) cannot be found in the
neighbourhood; at least within $\sim$17\arcmin~to the south, and up to $\sim$1.8\degr~to 
the north, north-east (as far as L1622).
For a better visualization of the central part of NGC2071-N, at various
wavelengths, see Sect.~\ref{sec:double}.

\begin{figure}
 \begin{center}
  \begin{minipage}{1.0\linewidth}
    \hspace{-1mm}
   \resizebox{1.018\hsize}{!}{\includegraphics[width=\columnwidth]{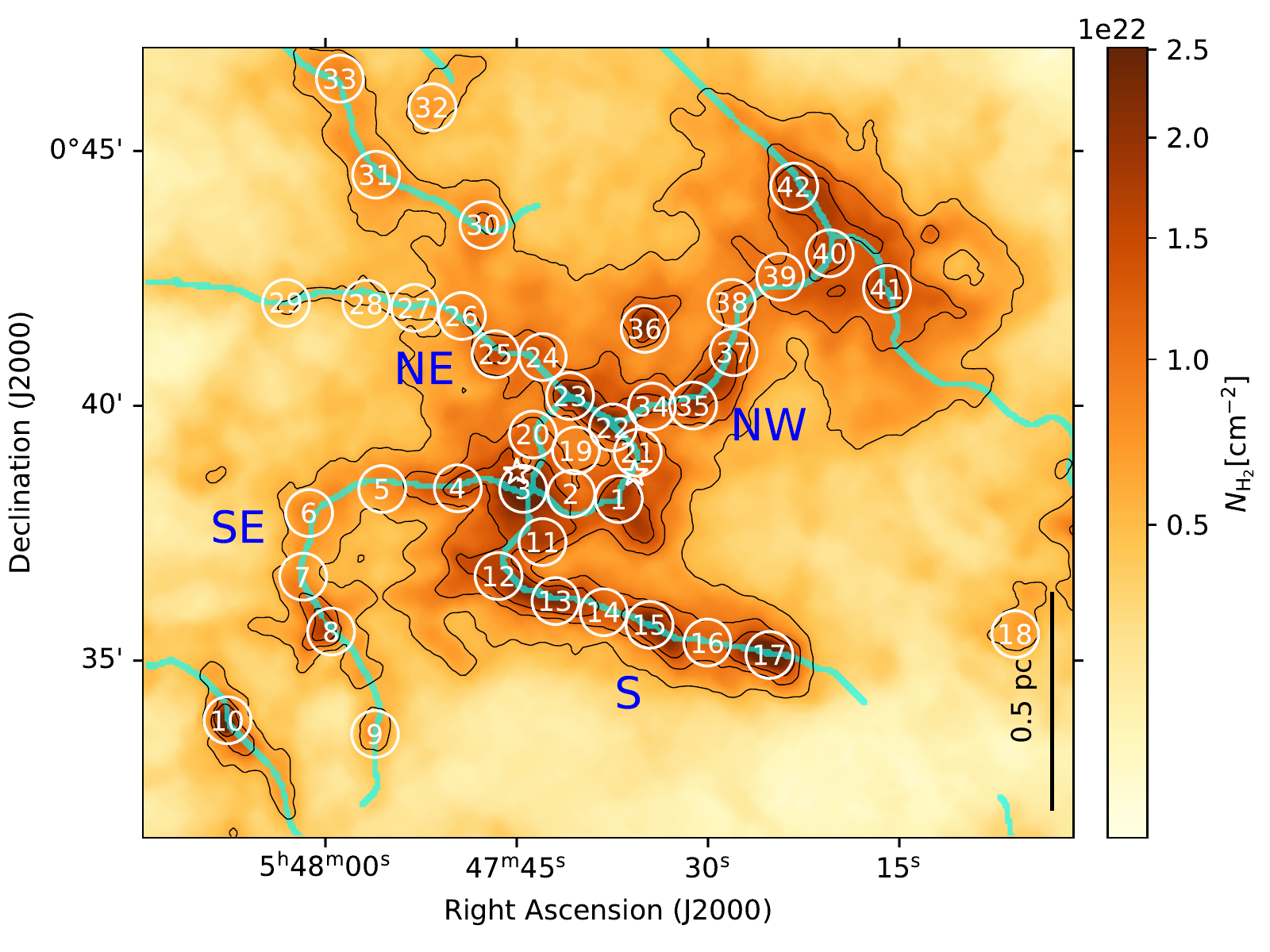}}   
  \end{minipage}
 \end{center}
  \caption{Column density map and filaments in NGC2071-N, as in Fig.~\ref{fig_fil_obj} left. 
          White numbered circles with 25\arcsec radius indicate the spots within which we estimated the 
          energy balance (see Sects.~\ref{sec:energy} and \ref{sec:coherent}, and 
          Tables~\ref{tab_uCalcG} and \ref{tab_uCalcT}). Filament designations by their locations are also shown:
          southern (S); south-east (SE); north-east (NE); north-west (NW).
          }
  \label{fig_circ}
\end{figure}

For the following calculations we mainly choose core positions, displayed in 
Fig.~\ref{fig_fil_obj} left, that are sampling the filaments and the hub centre. 
Around them, we defined spots/circles (see Fig.~\ref{fig_circ}) within which we then 
performed the measurements and calculations.
For their sizes we defined a uniform 25\arcsec~(i.e, $\sim$0.05\,pc) radius, knowing that the 
median FWHM size of the HGBS cores in this subregion is 0.05\,pc. We may take the FHWM size 
as a radius, because 1) the column density profile of a critical Bonnor-Ebert sphere \citep[e.g.,][]{Bonnor1956} 
with outer radius $R_{BE}$ is very well approximated by a Gaussian profile of FWHM $\sim R_{BE}$
\citep[see in][]{Konyves+2015}, and 2) for a Gaussian 2D circular distribution, $\sim$94\% of the 
emission is contained within a circular aperture of radius FWHM \citep[see also in][]{Peretto+2006}.
Thus, closely 100\% of the flux and mass of the cores is contained in the --preferably non-overlapping-- 
drawn circles in Fig.~\ref{fig_circ}, within which we can also sample the molecular emission with 
independent beams. 
The two circles \#1 and \#3 were placed over column density peaks, 
and most of the rest of the positions correspond to starless, prestellar, protostellar dense cores, 
or YSOs. In these spheres we assume uniform densities. 

\begin{figure} 
 \begin{center}
  \subfloat{\includegraphics[width=0.24\textwidth]{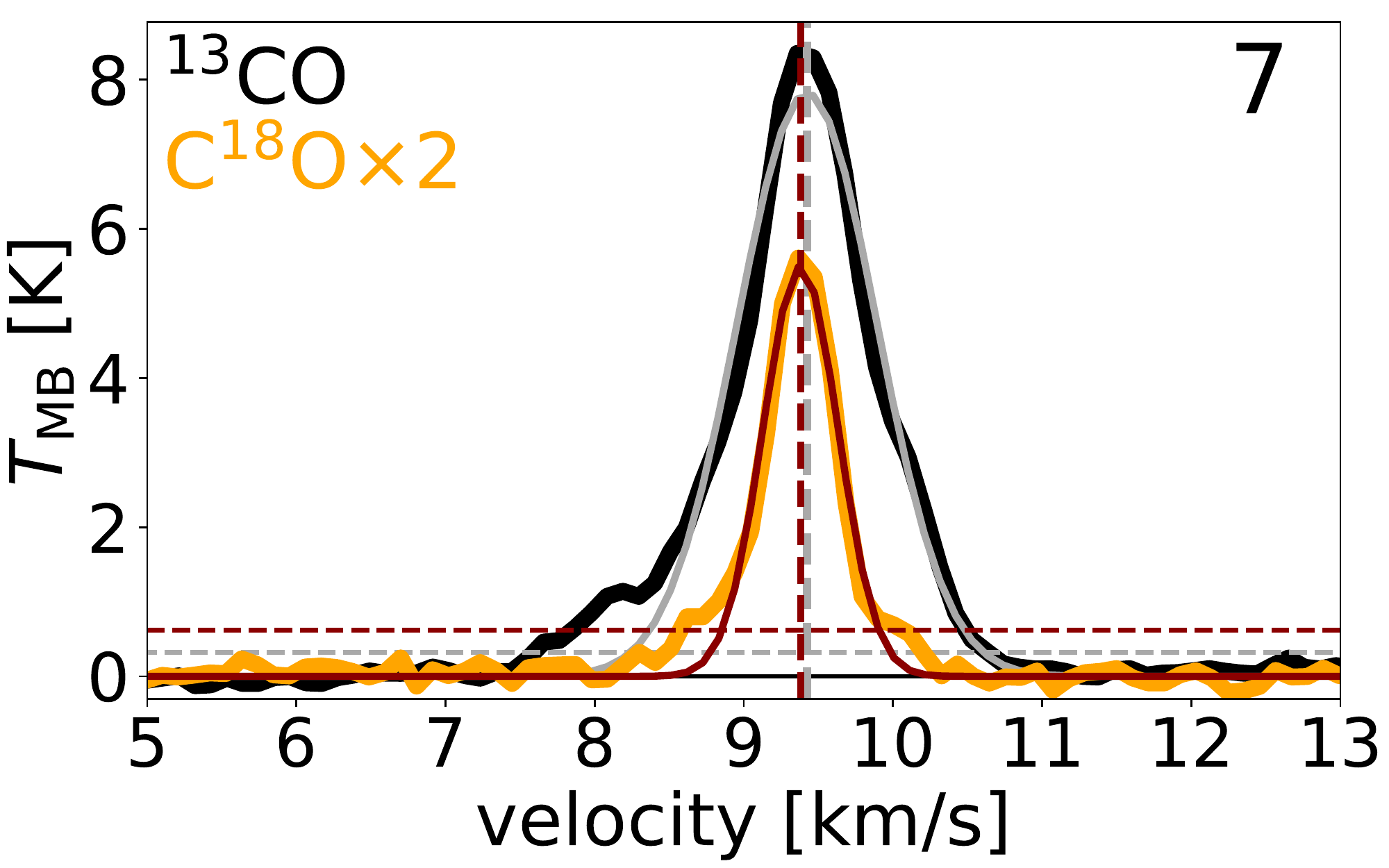}} 
  \subfloat{\includegraphics[width=0.24\textwidth]{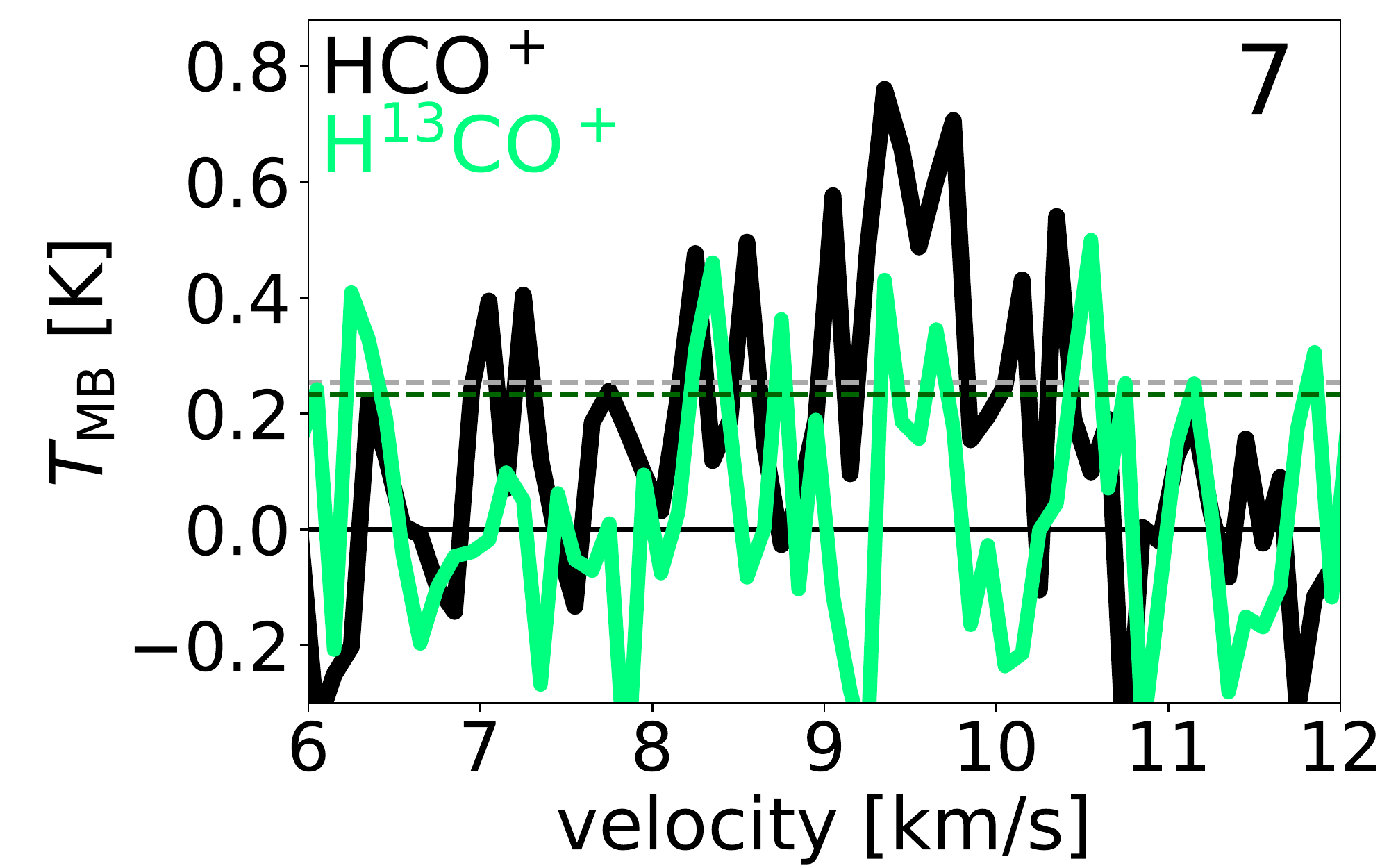}} 
  \begin{center}
   \subfloat{\includegraphics[width=0.365\textwidth]{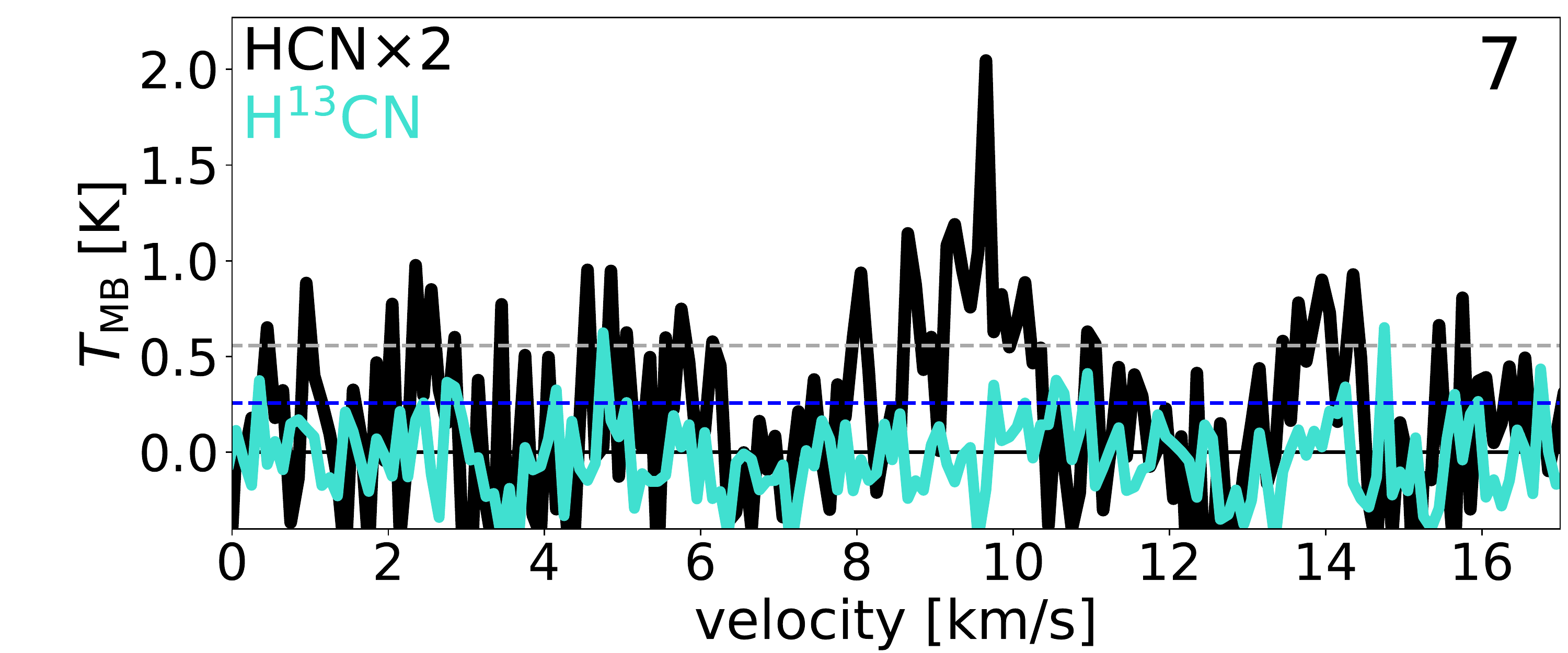}}  
  \end{center}
  \subfloat{\includegraphics[width=0.24\textwidth]{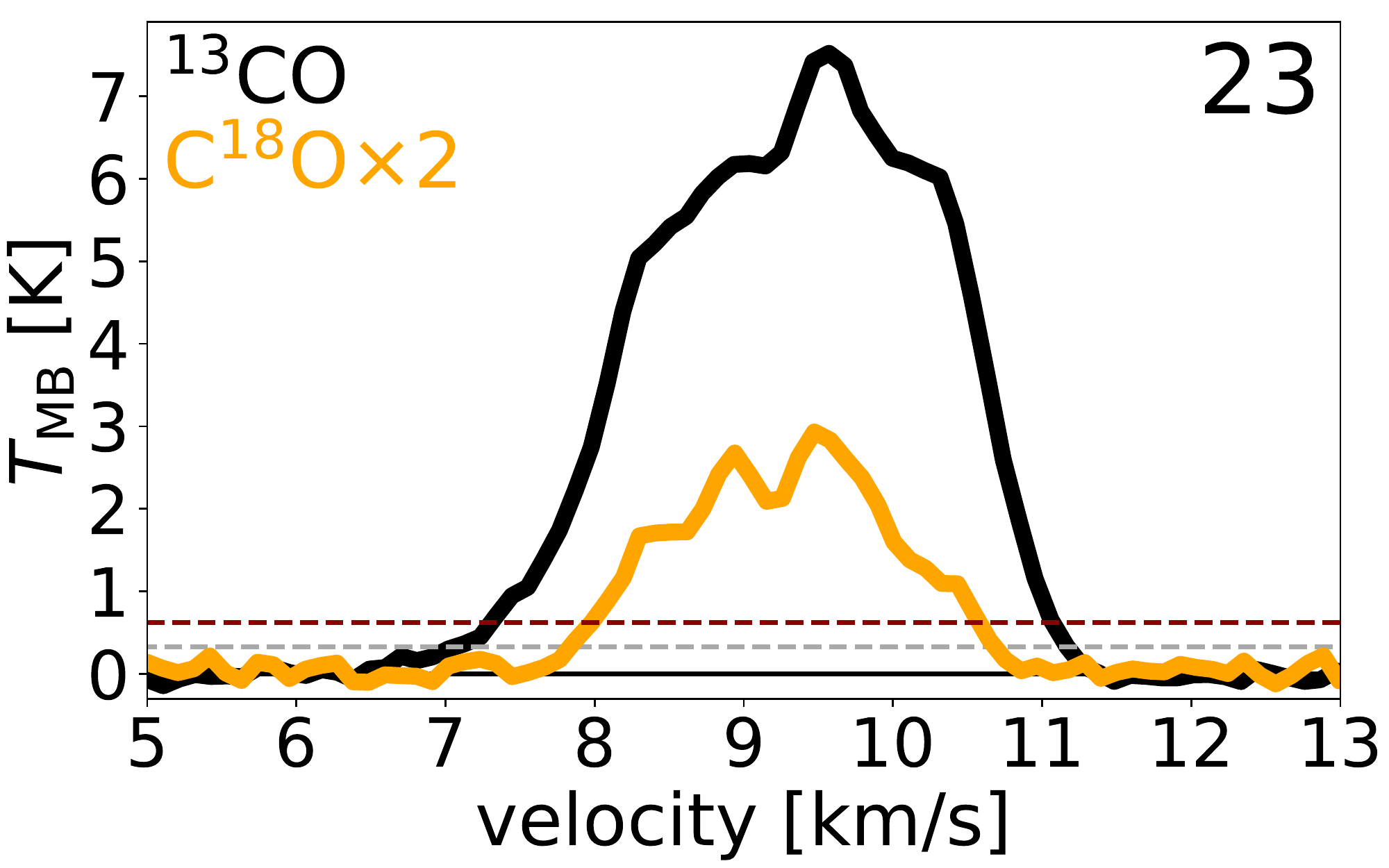}} 
  \subfloat{\includegraphics[width=0.24\textwidth]{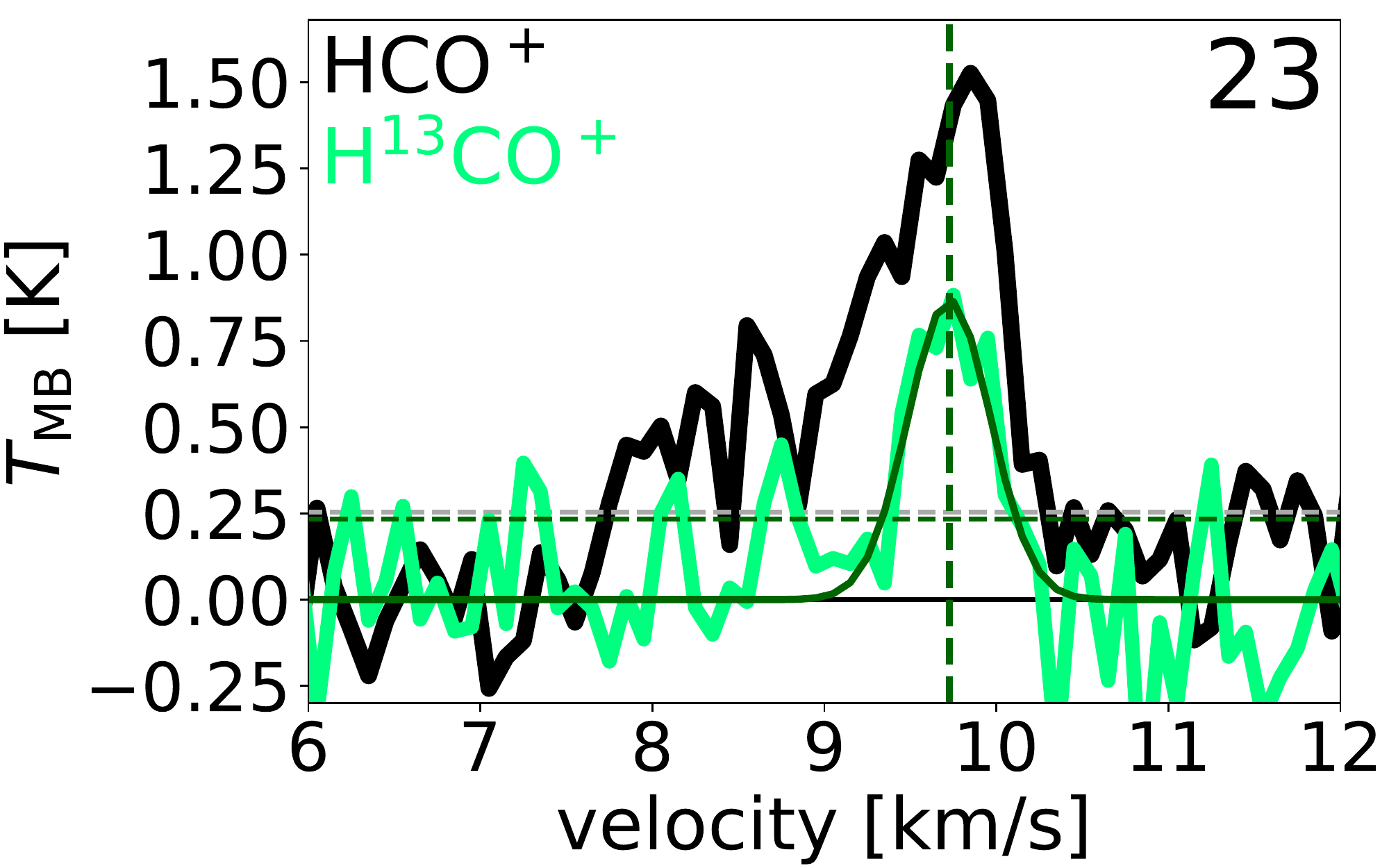}}\\ 
  \begin{center}
   \subfloat{\includegraphics[width=0.365\textwidth]{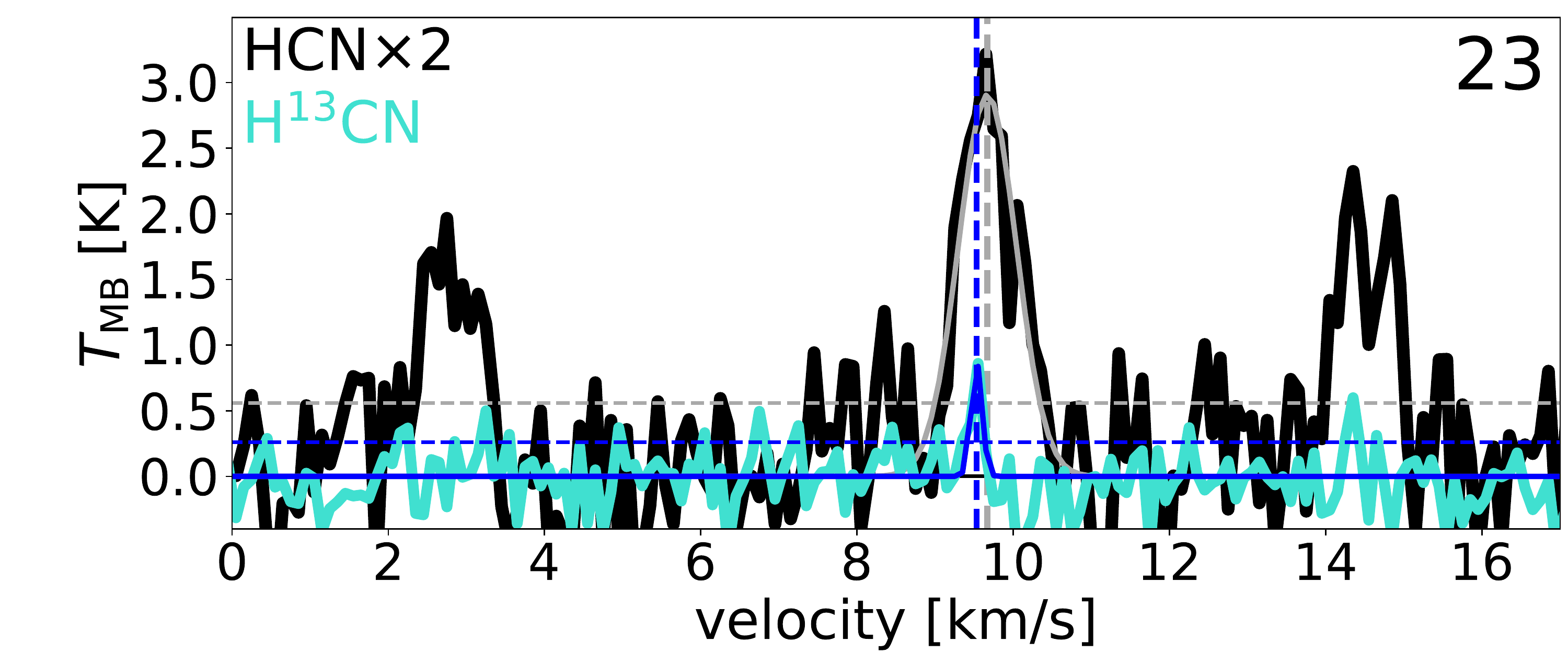}} 
  \end{center}
 \end{center}
\caption{Spectra averaged over circles \#7 \& \#23. All of them are $J = 1\rightarrow 0$ transitions. When it is reasonable, a Gaussian fit 
         and its peak position are also included. Dashed horizontal lines mark rms thresholds; grey for the black spectra, in colour for the
         overplotted spectra (6$\sigma$ rms level is indicated for the CO species, 3$\sigma$ for the HCX species).   
         All of the avaraged spectra over the analysis circles (see Fig.~\ref{fig_circ}) are included in Appendix~\ref{app:spectra}, when 
         available. Derived properties from these molecular lines are listed in Table~\ref{tab_uCalcT}. }  
\label{fig_sampleSpec}
\end{figure}
\subsection{Molecular line data}\label{sec:molec}

Low- and high-density molecular line tracers of gas kinematics were also analysed in our 
NGC2071-North hub.
$^{13}$CO(1--0) and C$^{18}$O(1--0) observations were done with the IRAM~30\,m telescope, 
HCN(1--0), H$^{13}$CN(1--0) and HCO$^+$(1--0), H$^{13}$CO$^+$(1--0) mapping were performed with 
the Nobeyama 45\,m antenna. 

Sample spectra of all these observed lines are displayed in Fig.~\ref{fig_sampleSpec}. As example 
spot \#7 has $A_{\rm V} \sim$8\,mag, while the visual extinction is higher ($\sim$19\,mag) in spot \#23, 
which is also located at the junction of filaments in the central part of the hub.
Among other properties, $A_{\rm V}$ values are listed for each analysis spot in Table~\ref{tab_uCalcG}.   
All of the avaraged spectra over these spots, where available, are included in Appendix~\ref{app:spectra}.
When we find a reasonable Gaussian fit of the main component of the line, it is also overplotted on 
the spectra. The goodness of the fits has been evaluated with residual sum of squares, then we also
eye-inspected the results.

Given the high column density values across NGC2071-N, we first evaluated the optical depth of the 
HCO$^+$ and HCN lines with the help of their $^{13}$C-isotopes. We considered the same assumptions
as \citet{Shimajiri+2017} and used their Eq.~8 for deriving $\tau^{\rm HCO+}$ and $\tau^{\rm HCN}$. 
The peak intensity of the rare isotopic species we could fit only in limited cases (see 
Fig.~\ref{fig_HCO+_H13CO+_spectra} for H$^{13}$CO$^+$(1--0), and Fig.~\ref{fig_HCN_H13CN_spectra} for
H$^{13}$CN(1--0)). At the velocity position of the peak of the fit, we recorded the observed intensity 
(in $T_{\rm MB}$) of the averaged H$^{13}$CO$^+$ and H$^{13}$CN lines, then that of the main species, 
HCO$^+$ and HCN, at the same position. The resulting optical depths are listed in 
Table~\ref{tab_uCalcT}. The actual number results suggest that the HCO$^+$ and HCN lines are optically
thick, however $\tau^{\rm HCO+}$ at circle \#8, and $\tau^{\rm HCN}$ at spot \#24 may be upper limits. 
At the locations where the intensity of the main species is weaker than that of the rare species,
most probably due to self-absorption, we indicate $\gg$1 as optical depth. In the rest of the cases,
the emission of the rare species was not detected, or not well characterized by the fit.
At the same time, based on the spectra, however noisy, we assume that the rare species, H$^{13}$CO$^+$ 
and H$^{13}$CN, remain optically thin, or nearly so. 
Considering the critical densities of these rare $J = 1\rightarrow 0$ lines at 10\,K for local 
thermodynamic equilibrium (LTE), that is $n_{\rm crit}^{\rm H13CO+} = 1.5 \times 10^5$\,cm$^{-3}$ and 
$n_{\rm crit}^{\rm H13CN} = 2.0 \times 10^6$\,cm$^{-3}$ \citep{Dhabal+2018}, our estimated volume densities
are everywhere lower (see Table~\ref{tab_uCalcG}), with a median value of 6.23 $\times$ 10$^{4}$ cm$^{-3}$.

High-density molecular tracers, especially when combining an optically thick and a thin line, can also 
indicate infalling gas, and thus global collapse of parts of the cloud \citep[e.g.,][]{Myers+1996, Evans1999, 
Schneider+2010, Rygl+2013, He+2015, Traficante+2017}.
This shows up in the thick tracer (i.e., HCO$^+$) as a blue-red asymmetry, with brighter blue-shifted peak,
whereas the thin line (H$^{13}$CO$^+$) peaks at the velocity of the self-absorption dip. 
Such configuration of the thin and thick lines we find in a couple of positions along the filaments (see
Fig.~\ref{fig_HCO+_H13CO+_spectra}) that we note with an upper index ``c'' in column 8 of 
Table~\ref{tab_uCalcT}. 
At these spots we have a strong hint that self-gravity plays a significant role that further
analysis may confirm (see Sects.~\ref{sec:energy} and \ref{sec:coherent}). 
These locations, except one, lie along the southern filament.
Profiles of this shape also provide confirmation that the thick double peaks are not coming from two 
velocity components along the line of sight. Among these, the peaks of the thin lines (and the dips of the 
thick ones) are found at $V_{\rm LSR} \sim$9\,km~s$^{-1}$.

In addition to the presented line profiles and calculations, we are also providing the integrated
intensity (moment 0) maps of the observed molecules in Appendix~\ref{app:mom0}, however
they may be somewhat influenced by the optically thick conditions. In the case of the thin lines
(H$^{13}$CO$^+$, H$^{13}$CN), the integrated intensity of this region is much weaker and rather sparse.

\subsection{Energy balance in NGC2071-North}\label{sec:energy}

From column density and optically thin molecular line data we have estimated gravitational 
and turbulent kinetic energy densities, as well as related gravitational and internal pressures  
in several spots within the region.
These pressure terms contribute to the energy densities, nevertheless they are often equivalently 
used to evaluate the stability of clumps and clouds. We will separately probe the interplay between 
gravity and turbulence with the energy density and pressure properties, estimated with standard 
formulae and assumpsions.

Throughout the calculations we assume that the cores have spherical shape, uniform density, 
no contribution of external pressure or support of magnetic field, and no rotation either.
In NGC2071-N there are no available polarisation observations (with comparable resolution 
to the above data) that we could derive magnetic energy densities, therefore we did not attempt 
to derive it from {\it Planck} measurements on 5\arcmin-scale, which would most probably 
characterize different phenomena.

The gravitational energy density was estimated from $u_{\rm grav} = 4/5 \pi G \rho^{2} R^{2}$ 
\citep[see, e.g.,][]{Lyo+2021}, where $G$ is the gravitational constant, and 
$\rho = \mu m_{\rm H} n_{\rm H_2} $ is the uniform density for which the volume density we 
calculated from the $Herschel$ column density map of \citet{Konyves+2020} (column
(8) of Table~\ref{tab_uCalcG}). $R$ is the sphere 
radius (25\arcsec), $\mu = 2.8$ is the mean molecular weight per H$_2$ molecule, and $m_{\rm H}$ is 
the hydrogen atom mass. 

These calculations gave a mean value of 
$u_{\rm grav} = 3.6 \times 10^{-10}$\,erg cm$^{-3}$ with a standard deviation 
of $2.8 \times 10^{-10}$\,erg cm$^{-3}$,  
for the whole range of results at the 42 locations in Fig.~\ref{fig_circ}.

For a later comparison, the volume-averaged gravitational pressure can be estimated as 
$\overline{P}_{\rm G} / k_{\rm B} \approx 1.01 a_1 \langle \phi_{\rm G} \rangle M^2 R^{-4} $, 
expressed by \citet{BertoldiMcKee1992}, and used, for example, by \citet{Sadavoy+2015}.
$k_{\rm B}$ is the Boltzmann constant, $a_1 = 1.3$ is a scaling factor that measures the 
effects of a nonuniform density distribution; our value is appropriate for a self-gravitating
cloud. $\phi_{\rm G}$ is a scaling factor describing the cloud geometry; we consider $\phi_{\rm G} = 1$
for perfect spheres. $M$ (in $M_\odot$) is the mass of the clumps (column (5) of Table~\ref{tab_uCalcG}),
$R=0.048$\,pc (used in pc) is the radius of our spots/spheres. 
We thus calculated the volume-averaged gravitational pressures that give 
a mean value of 
$\overline{P}_{\rm G} / k_{\rm B} = 11.8 \times 10^{5}$\,K\,cm$^{-3}$ and a standard deviation of
$9.3 \times 10^{5}$\,K\,cm$^{-3}$ for the whole range of results.

The individual values are listed in Table~\ref{tab_uCalcG}, along with additional physical properties.
We assumed 20\% error on the column densities, mass, then on the volume densities, energy densities, 
and pressures in order to avoid the propagation of the typical factor of about 2 systematic errors mainly 
due to the uncertainties in the dust opacity law. 

The turbulent kinetic energy densities were calculated from 
$u_{\rm turb} = 3/2 \rho \sigma^2_{\rm NT}$, where $\sigma_{\rm NT}$ is the non-thermal 
component of the velocity dispersion. For this, first we converted the linewidths, as 
$\sigma = \Delta v / \sqrt{8~{\rm ln}2}$, that we 
estimated by Gaussian fitting to the observed profiles of H$^{13}$CO$^+$(1--0) and H$^{13}$CN(1--0), 
where it was possible (see Sect.~\ref{sec:molec}). Then, we separated the non-thermal 
component $\sigma_{\rm NT}$ using a similar relation to eqn.~6 of \citet{Dunham+2011}, 
and found that the observed velocity dispersions (or linewidths) are almost entirely due 
to non-thermal motions.
These, we assume, may represent random turbulent motions 
(and infall motions), which are independent of the gas temperature.
The resulting turbulent kinetic energy densities (for H$^{13}$CO$^+$, where it was
possible to derive) 
yield a mean of $u_{\rm turb}^{\rm H13CO+} = 3.23 \times 10^{-10}$\,erg cm$^{-3}$ with a high standard 
deviation of $3.20 \times 10^{-10}$\,erg cm$^{-3}$.
From H$^{13}$CN data, we could calculate $u_{\rm turb}^{\rm H13CN}$ only in two cases. 
See Figs.~\ref{fig_HCO+_H13CO+_spectra} and \ref{fig_HCN_H13CN_spectra} for the fitted line profiles, 
and Table~\ref{tab_uCalcT} for the derived line properties.

In comparison to the gravitational pressure, we estimate the internal pressure as well
in the spots, where the quality of the H$^{13}$CO$^+$(1--0) and H$^{13}$CN(1--0) spectra
allowed us to derive turbulent kinetic energy densities too.
The internal pressure can be approximated from the ideal gas law, as
$P_{\rm int} / k_{\rm B} \approx n^{\rm ave}_{\rm H_2} \sigma^2_{\rm NT} $, where
$n^{\rm ave}_{\rm H_2}$ is the average volume density (column (8) of Table~\ref{tab_uCalcG}),
and $\sigma_{\rm NT}$ is the non-thermal part of the velocity dispersion, as above.
In this or similar form the internal pressure has been calculated, for example, by 
\citet{Hatchell+2005, Sadavoy+2015, Pattle+2015, MAMD+2017}.
The gravitational pressure estimates from H$^{13}$CO$^+$ can be summarized with a 
mean value of 
$P_{\rm int} / k_{\rm B} = 5.4 \times 10^5$\,K\,cm$^{-3}$ with an as large standard deviation 
of $5.4 \times 10^5$\,K\,cm$^{-3}$.

Apart from the 20\% uncertainties inherited from the column density measurements, 
we consider 1 standard deviation uncertainties on the measured median dust temperatures 
within the sampling spots. Along with the calculations of $\sigma_{\rm NT}$, $u_{\rm turb}$,
and $P_{\rm int}$ we propagated the corresponding individual errors using the maximum 
error formula (see Tables~\ref{tab_uCalcG} and \ref{tab_uCalcT}, and Fig.~\ref{fig_u} for the 
visualization of the uncertainties).

The left panel of Fig.~\ref{fig_u} shows the estimated energy densities, 
where both $u_{\rm grav}$ and $u_{\rm turb}$ (latter from H$^{13}$CO$^+$ or H$^{13}$CN) 
could be calculated. Such spots are denoted on the horizontal axis.
In the right panel of Fig.~\ref{fig_u}, black and red data points display the volume-averaged 
gravitational pressure and internal pressure of these same locations, respectively, using the 
same molecular lines as in the left-hand panel.
In both panels yellow stars indicate the positions where we observed possible infall signatures 
based on the blue-red asymmetry of the averaged HCO$^+$ profiles 
(see Fig.~\ref{fig_HCO+_H13CO+_spectra}). For a discussion involving Fig.~\ref{fig_u}, see
Sect.~\ref{sec:coherent}.

\begin{table*}\small\setlength{\tabcolsep}{1.5pt}
 \caption{Physical properties of 42 selected locations within 25\arcsec radius in NGC2071-N, shown in Fig.~\ref{fig_circ}. 
          See Sects.~\ref{sec:molec} and \ref{sec:energy} for details.
          {\bf (1)}: Spot numbers, as in Fig.~\ref{fig_circ}; {\bf (2) $\pm$ (3)}, {\bf (4) $\pm$ (5)}, {\bf (6) $\pm$ (7)}, {\bf (9) $\pm$ (10)}, 
          {\bf (15) $\pm$ (16)}, {\bf (18) $\pm$ (19)}: non-thermal velocity dispersion and uncertainties of the given (1--0) lines, $^{13}$CO, C$^{18}$O,
          HCO$^+$, H$^{13}$CO$^+$, HCN, H$^{13}$CN, resp.; {\bf (8)} and {\bf (17)}: optical depth of HCO$^+$ and HCN, resp.;   
          {\bf (11) $\pm$ (12)}, and {\bf (20) $\pm$ (21)}: turbulent kinetic energy densities and errors from H$^{13}$CO$^+$ and H$^{13}$CN
          linewidths, resp.; {\bf (13) $\pm$ (14)}, and {\bf (22) $\pm$ (23)}: internal pressure and uncertainties at the latter locations.          
          An upper index ``c'' in column {\bf (8)} marks the spots where HCO$^+$ speactra show infall signatures 
          (see Fig.~\ref{fig_HCO+_H13CO+_spectra}).
          }
 \label{tab_uCalcT}      
 {\renewcommand{\arraystretch}{1.2}        
  \begin{tabular}{l c c c c c c c c c c c c c c c c c c c}  
  \hline
  \hline
		& $^{13}$CO				& C$^{18}$O				& \multicolumn{2}{c}{HCO$^+$}					& \multicolumn{3}{c}{H$^{13}$CO$^+$}									& \multicolumn{2}{c}{HCN}				& \multicolumn{3}{c}{H$^{13}$CN}									\\
  \hline
  Spot \#	& $\sigma_{\rm NT}^{\rm 13CO}$	& $\sigma_{\rm NT}^{\rm C18O}$		& $\sigma_{\rm NT}^{\rm HCO+}$		& $\tau^{\rm HCO+}$	& $\sigma_{\rm NT}^{\rm H13CO+}$	& $u_{\rm turb}^{\rm H13CO+}$	& $P^{\rm H13CO+}_{\rm int}/k_{\rm B}$ & $\sigma_{\rm NT}^{\rm HCN}$	& $\tau^{\rm HCN}$	& $\sigma_{\rm NT}^{\rm H13CN}$	& $u_{\rm turb}^{\rm H13CN}$	& $P^{\rm H13CN}_{\rm int}/k_{\rm B}$	\\			    
  \hline
		& (km\,s$^{-1}$)			& (km\,s$^{-1}$)			& (km\,s$^{-1}$)			& --			& (km\,s$^{-1}$)			& (10$^{-10}$ erg cm$^{-3}$)	& (10$^{5}$ K cm$^{-3}$)	& (km\,s$^{-1}$)		& --			& (km\,s$^{-1}$)			& (10$^{-10}$ erg cm$^{-3}$)	& (10$^{5}$ K cm$^{-3}$)	\\
  (1)		& (2) $\pm$ (3)				& (4) $\pm$ (5)				& (6) $\pm$ (7)				& (8)			& (9) $\pm$ (10)			& (11) $\pm$ (12)		& (13) $\pm$ (14)		& (15) $\pm$ (16)		& (17)			& (18) $\pm$ (19)			& (20) $\pm$ (21)		& (22) $\pm$ (23)		\\	
  \hline
   1 		& --					& --					& 0.48 ~ 0.03				& 11			& 0.19 ~ 0.07				&  2.09 $\pm$ 1.91		& 3.38 $\pm$ 3.10		& --				& --			& --					& --				& --			       \\		       
   2 		& --					& --					& --					& --			& --					& --				& --				& --				& --			& --					& --				& --			       \\		       
   3 		& --					& --					& --					& $\gg$1		& 0.39 ~ 0.05				& 14.32 $\pm$ 6.47		& 24.38 $\pm$ 11.02		& --				& --			& --					& --				& --			       \\		       
   4 		& --					& --					& --					& --			& --					& --				& --				& --				& --			& --					& --				& --			       \\		       
   5 		& --					& --					& --					& --			& --					& --				& --				& --				& --			& --					& --				& --			       \\		       
   6 		& 0.60 ~ 0.01				& 0.39 ~ 0.01 				& --					& --			& --					& --				& --				& --				& --			& --					& --				& --			       \\	       
   7 		& 0.47 ~ 0.01				& 0.24 ~ 0.01 				& --					& --			& -- 					& --				& --				& --				& --			& --					& --				& --			       \\				       
   8 		& 0.43 ~ 0.02				& 0.27 ~ 0.01 				& --					& 90			& 0.33 ~ 0.07 				&  4.52 $\pm$ 2.92 		& 7.64 $\pm$ 4.93 		& 0.54 ~ 0.08			& --			& --					& --				& --			       \\	       
   9 		& 0.31 ~ 0.01				& 0.18 ~ 0.01 				& --					& --			& --					& --				& --				& --				& --			& --					& --				& --			       \\	       
  10 		& --					& --					& --					& --			& --					& --				& --				& --				& --			& --					& --				& --			       \\		       
  11 		& --					& --					& --					& --			& --					& --				& --				& --				& --			& --					& --				& --			       \\		       
  12 		& --					& --					& --					& $\gg$1$\rm ^c$	& 0.24 ~ 0.05				&  3.21 $\pm$ 2.01		& 5.33 $\pm$ 3.34		& --				& --			& --					& --				& -- 			  \\			  
  13 		& --					& --					& --					& $\gg$1$\rm ^c$	& 0.23 ~ 0.04				&  3.09 $\pm$ 1.60		& 5.12 $\pm$ 2.66		& --				& --			& --					& --				& --			     \\ 		     
  14 		& --					& --					& --					& $\gg$1$\rm ^c$	& 0.21 ~ 0.05				&  1.88 $\pm$ 1.28		& 3.08 $\pm$ 2.10		& --				& --			& --					& --				& --			     \\ 		     
  15 		& --					& --					& --					& $\gg$1$\rm ^c$	& 0.22 ~ 0.06				&  3.09 $\pm$ 2.21		& 5.11 $\pm$ 3.66		& --				& --			& --					& --				& --			     \\ 		     
  16 		& --					& --					& --					& $\gg$1$\rm ^c$	& 0.21 ~ 0.05				&  1.77 $\pm$ 1.20		& 2.91 $\pm$ 1.97		& --				& --			& --					& --				& --			     \\ 		     
  17 		& --					& --					& --					& $\gg$1$\rm ^c$	& 0.24 ~ 0.04				&  5.05 $\pm$ 2.72		& 8.45 $\pm$ 4.55		& 0.23 ~ 0.03			& --			& --					& --				& --			     \\ 		     
  18 		& --					& --					& --					& --			& --					& --				& --				& --				& --			& --					& --				& --			       \\		       
  19 		& --					& --					& --					& --			& --					& --				& --				& --				& --			& --					& --				& --			       \\		       
  20 		& --					& --					& --					& --			& --					& --				& --				& --				& --			& --					& --				& --			       \\		       
  21 		& --					& --					& 0.43 ~ 0.02				& --			& --					& --				& --				& --				& --			& --					& --				& --			       \\		       
  22 		& --					& --					& --					& 14			& 0.26 ~ 0.06				&  4.23 $\pm$ 2.97		& 7.07 $\pm$ 4.97		& --				& --			& --					& --				& --			     \\ 		     
  23		& --					& --					& --					& 59			& 0.24 ~ 0.04				&  3.76 $\pm$ 2.02		& 6.27 $\pm$ 3.36		& 0.37 ~ 0.04			& 61			& 0.07 ~ 0.02				& 0.29 $\pm$ 0.23		& 0.25 $\pm$ 0.20 	     \\ 		     
  24		& --					& --					& --					& --			& --					& --				& --				& --				& 89			& 0.11 ~ 0.06				& 0.46 $\pm$ 0.45		& 0.64 $\pm$ 0.60	       \\		       
  25		& 0.77 ~ 0.02 				& 0.39 ~ 0.01				& --					& --			& --					& --				& --				& --				& --			& --					& --				& --			       \\		       
  26		& 0.66 ~ 0.01 				& 0.31 ~ 0.01 				& --					& --			& --					& --				& --				& --				& --			& --					& --				& --			       \\		       
  27		& 0.65 ~ 0.02 				& 0.27 ~ 0.01				& --					& --			& --					& --				& --				& --				& --			& --					& --				& --			       \\		       
  28		& --					& --					& --					& --			& --					& --				& --				& --				& --			& --					& --				& --			       \\		       
  29		& --					& --					& --					& --			& --					& --				& --				& --				& --			& --					& --				& --			       \\		       
  30		& --					& --					& --					& --			& --					& --				& --				& --				& --			& --					& --				& --			       \\		       
  31		& --					& --					& --					& --			& --					& --				& --				& --				& --			& --					& --				& --			       \\		       
  32		& --					& --					& --					& --			& --					& --				& --				& --				& --			& --					& --				& --			       \\		       
  33		& --					& --					& --					& --			& --					& --				& --				& --				& --			& --					& --				& --			       \\		       
  34		& --					& --					& --					& 30			& 0.19 ~ 0.06				&  1.73 $\pm$ 1.48		& 2.81 $\pm$ 2.40		& --				& --			& --					& --				& --			      \\		      
  35		& --					& --					& --					& 34			& 0.19 ~ 0.07				&  2.03 $\pm$ 1.86		& 3.30 $\pm$ 3.03		& --				& --			& --					& --				& --			      \\		      
  36		& --					& --					& --					& $\gg$1		& 0.11 ~ 0.03				&  0.62 $\pm$ 0.42		& 0.88 $\pm$ 0.60		& --				& --			& --					& --				& --			      \\		      
  37		& --					& --					& --					& --			& --					& --				& --				& --				& --			& --					& --				& --			       \\		       
  38		& --					& --					& --					& $\gg$1		& 0.08 ~ 0.03				&  0.23 $\pm$ 0.22		& 0.26 $\pm$ 0.24		& --				& --			& --					& --				& --			       \\		       
  39		& --					& --					& --					& $\gg$1$\rm ^c$	& 0.05 ~ 0.03				&  0.11 $\pm$ 0.10		& 0.05 $\pm$ 0.04		& --				& --			& --					& --				& --			       \\		       
  40		& --					& --					& --					& --			& --					& --				& --				& --				& --			& --					& --				& --			       \\		       
  41		& --					& --					& --					& --			& --					& --				& --				& --				& --			& --					& --				& --			       \\		       
  42		& --					& --					& --					& --			& --					& --				& --				& --				& --			& --					& --				& --			       \\		       
  \hline
  \hline  
  \end{tabular}
 }   
\end{table*}

\section{Discussion}\label{sec:discuss}

\subsection{Large-scale magnetic field pattern}\label{sec:magn}

Revisiting the panels of Fig.~\ref{fig1}, we can see the large-scale structure of the 
POS magnetic field that was discussed in \citet{Soler2019} for the 
whole Orion A and B clouds too. 
What shows up in Fig.~\ref{fig1} right is that the POS magnetic field (on the scale
of 1/3 of the 5\arcmin~{\it Planck} beam) is largely horizontal in the eastern part,
and vertical in the western part, as it would turn and change orientation 
across the hub. 

Comparing Fig.~\ref{fig1} left with Fig.~4 of \citet{Soler2019} and Fig.~8 of 
\citet{Tahani+2018}, we can see that there is a magnetic loop structure apparently 
starting or ending at NGC2071, extending up to NGC2071-N, while the B-field lines 
run mostly along the right ascension lines in the western side of all the clouds 
shown in Fig.~\ref{fig1} left.

Providing that the magnetic field is carrying material, a significant
change in the B-field orientation (i.e., from horizontal to perpendicular)
may be an efficient configuration for deposing and gathering gas and dust,
which could also explain the location of this relatively small and isolated cloud. 
Not discussing here other physical forces, indeed it can be seen in the ISM that 
the B-field shows a bend and turns at high-density (elongated) molecular clouds, 
such as Orion~A, Perseus, unlike in Musca and the Chamaeleon, for example. See these
maps of {\it Planck} POS magnetic field and column density in \citet{Soler2019}.  

We note that in the case of Orion~A, such a simple visualization helped 
\citet{Tahani+2018} to conclude that a bow-shaped magnetic field is surrounding 
that filament.

However, how the large-scale magnetic field is cascading down, and most likely
affecting the morphology of the centre of the hub, is not yet known.

\begin{figure*}
 \begin{center}
  \begin{minipage}{1.0\linewidth}
   \resizebox{0.5\hsize}{!}{\includegraphics[width=\columnwidth]{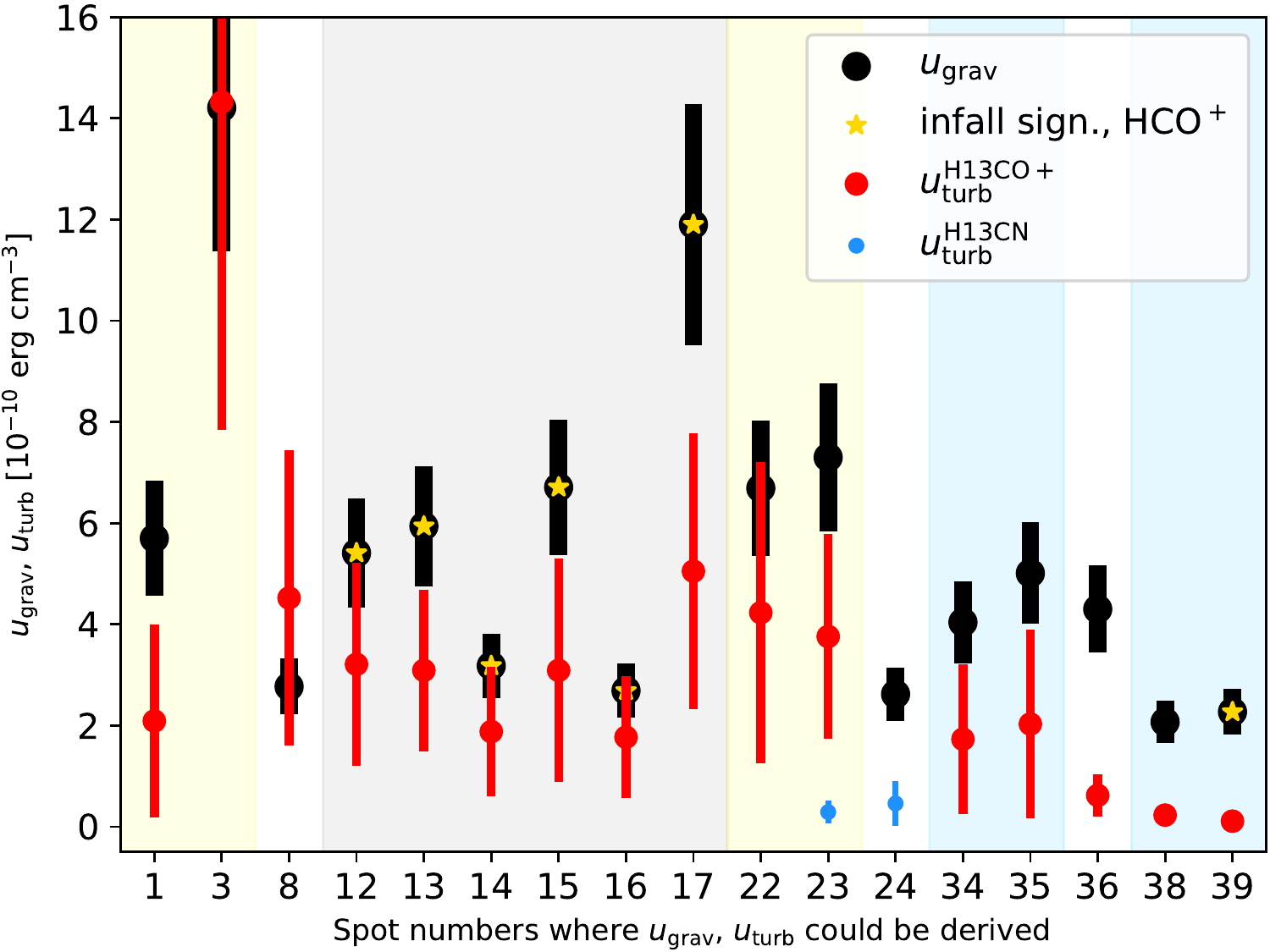}}   
   \resizebox{0.5\hsize}{!}{\includegraphics[width=\columnwidth]{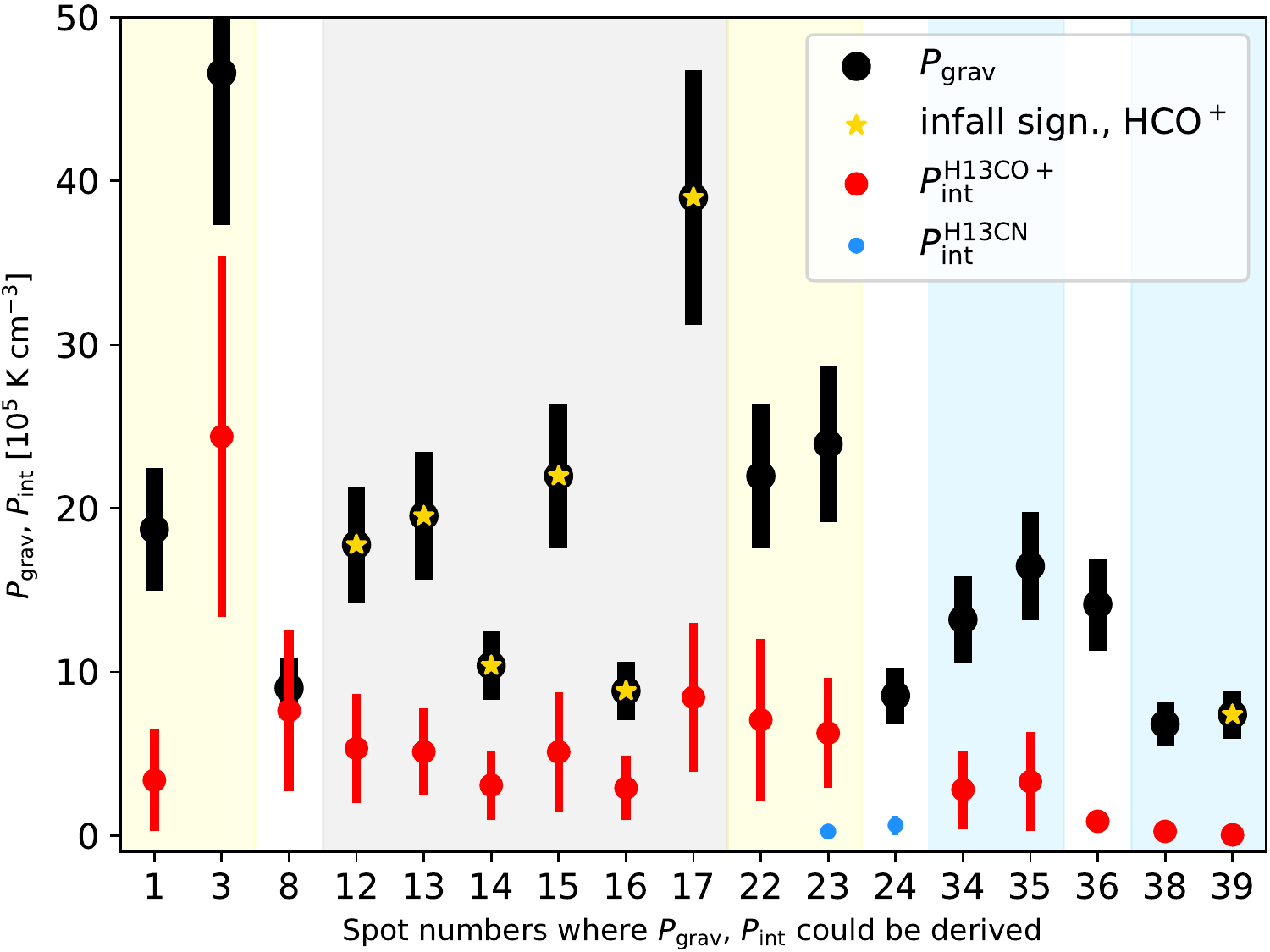}}   
  \end{minipage}
 \end{center}
  \caption{{\bf Left:} Gravitational and turbulent energy densities (in 10$^{-10}$ erg cm$^{-3}$) with uncertainties.
           {\bf Right:} Volume-averaged gravitational pressure and internal pressure (in 10$^{5}$ K cm$^{-3}$), more precisely
           in the form of $P / k_{\rm B}$.
           They are indicated per spot number, where both pairs could be calculated; see the legends. 
           The yellow stars mark the positions where the blue-red asymmetry of HCO$^+$ lines revealed possible 
           infall signatures.          
           See Tables~\ref{tab_uCalcG} and \ref{tab_uCalcT} for details, and Fig.~\ref{fig_circ} for the numbered 
           analysis spots. Among the latters, light yellow background marks the spots which are along the central
           filament loop. Light grey background shows the spots in the southern filament, and light blue background
           indicates the ones along the north-western filament. See Sects.~\ref{sec:energy} and \ref{sec:coherent} 
           for details.
          }
  \label{fig_u}
\end{figure*}
\subsection{Coherent structure with contracting cores}\label{sec:coherent}

\citet{Gibb2008} has noticed that NGC2071-N attracted very little study since the 
1980s \citep{Fukui+1986, Iwata+1988}, which continues to hold ever since 2008.
It is located at $\sim$20\arcmin~north of the NGC2071 reflection nebula, and its 
structures turned out to be more extended than the 0.7\,pc-diameter clump first 
seen in C$^{18}$O \citep{Iwata+1988}.
While NGC2071-N seems isolated from the rest of the L1630-North complex at the
$\rm H_2$ column density level of $\sim 1 \times 10^{22}$ cm$^{\rm -2}$, it is 
well part of it within the contours at $\sim 2 \times 10^{21}$ cm$^{\rm -2}$.
At this higher value, NGC2071-N looks more fragmented than, for example, the 
region of HH24--26, south of the NGC2068 reflection nebula (see Fig.~\ref{fig1} 
left at RA$\sim$05h46m, Dec$\sim$--00d15m). 
As we go down from this higher to the lower value, the column density contours
become more extended around NGC2071-N, than around HH24--26. 
In other words, the same mass lies in a somewhat larger area in NGC2071-N.
The high-column density fragments in our sub-region are the two centres, portions 
of surrounding filaments, and the extension at the north-west. (see e.g., 
Fig.~\ref{fig1} right).
The basis of this simple comparison is the similar projected location of 
both sub-regions, i.e., north/south of a reflection nebula, respectively, 
on either side of the L1630-North complex (see Fig.~\ref{fig1} left).
The above may mean that NGC2071-N still has a great potential to form more 
solar-type stars over a longer period of time.

At the time of \citet{Iwata+1988}, on the other hand, they inferred that this 
cloud is nearly in dynamical equilibrium, having calculated the same figures both
for the gravitational energy and for total kinetic energy of internal turbulence.
Then, \citet{Goldsmith+1992} found their C$^{18}$O clumps gravitationally unbound and 
deemed as probably transient structures.

We infer the current activity of this cloud from Fig.~\ref{fig_u}, the 
left panel of which shows that the gravitational energy densities dominate over 
the turbulent kinetic energy densities in most of the displayed positions. 
This same trend is also supported by the right panel of Fig.~\ref{fig_u},
that is that the volume-averaged gravitational pressures are higher than 
the internal turbulent pressures.

In spot \#8, at the dense tip of the SE filament, the energy balance is less 
conclusive. Here, we measure one of the largest velocity dispersions based
on the H$^{13}$CO$^+$ and HCN lines; $\sigma_{\rm NT}^{\rm H13CO+}$ is larger 
than that of the large-scale C$^{18}$O, and $\sigma_{\rm NT}^{\rm HCN}$ is 
even larger than $\sigma_{\rm NT}^{\rm 13CO}$.
However, the velocity dispersions of H$^{13}$CO$^+$ and HCN at \#8 are fairly 
noisy that probably led to their overestimations, and therefore to higher
turbulent energy densities and internal pressures.

Among the clumps along the central hub-ring, in particular \#3 
(around IRAS 05451+0037), \#22, \#23 
that are located at junctions of filaments, we also observe higher estimated 
$u_{\rm turb}^{\rm H13CO+}$ and $P^{\rm H13CO+}_{\rm int}$ (Fig.~\ref{fig_u}). 
Besides the noisy lines, this might also arise from infalling material along 
the filamets, however more investigations on this are out of the scope of 
the present paper. 

In addition to the marked HCO$^+$ infall candidates in Fig.~\ref{fig_u}, we 
can also identify blue-shifted asymmetric line profiles in the optically 
thick C$^{18}$O (Fig.~\ref{fig_13CO_C18O_spectra}), in more than the 
already marked positions. 
Together with the indications that inside-out collapse has been detected 
in these clumps \citep{Evans1999}, the overall higher gravitational 
energy densities and pressures support the picture that gravity is acting 
stronger on spatial scales of $\sim$0.05\,pc and is currently forming 
new stars in several spots of this region.

\subsection{Hub with a double centre}\label{sec:double}

The HGBS column density map shows a detailed filamentary structure,
both at $A_{\rm V} \sim$ 5 and 10\,mag contours 
(see Fig.~\ref{fig_fil_obj} left), which was not known before.
We suggest that a double centre may be formed in this hub, because the converging 
locations of the filaments S and SE, {\it and} NE and NW are somewhat offset
along the hub-ring that is traced by a filament loop.
In other words, the converging locations of filament pairs are offset. 
This offset is about 2.3\arcmin~($0.27$\,pc) measured from spot \#3 to \#22, 
(see Fig.~\ref{fig_circ}), while there is almost the same projected 
distance between the IRAS and LkH$\alpha$ sources.
Circle positions \#3 and \#22 correspond to the ammonia cores D and C, resp., 
defined by \citet{Iwata+1988}, while \#3 contains IRAS 05451+0037 too.
This span also matches the diameter of a ring-like structure that 
can be approximated from the combined emission at HCO$^+$ and HCN 
(Fig.~\ref{fig_all_mom0}), given that the maxima of their integrated intensities 
are scattered along the hub-ring. This ring may indicate a transition and 
the connection between the hub and the filaments. 
We note that the junction of filament NE and that of filament NW
with the hub-ring are not appearing as one projected point (as apparently the 
junction of filaments S and SE on the central ring), however the western section
of the hub-ring, $\sim$0.1\,pc around the nominal position of core C along the 
filaments, seems to be an active spot of star formation with dense cores and protostars
(see Fig.~\ref{fig_fil_obj} left).

Hub-filament systems are expected to feature one single centre on a larger scale, 
while they may break up after a closer look. 
Based on molecular line data, \citet{Trevino-Morales+2019} extracted a complex 
filamentary network in Monoceros R2 ($d \sim$830\,pc) also with a ring at the hub 
centre, where radially oriented filaments are joining from larger scales 
(see their Fig.~4.). Their hub-ring has a radius of approximately 0.30\,pc,
that we read from their figures, 
while that of ours is $\sim$0.13\,pc that was estimated from a fitted
circle to the closed loop of filament skeletons.
There is a massive cluster forming in the Mon R2 filament hub \citep{Rayner+2017}, 
unlike in NGC2071-N. Its most massive star IRS 1 ($\sim$12\,$M_\odot$) appears 
close in projection to the location where multiple filaments arrive onto the 
Mon R2 hub-ring. IRS 1 is also associated with an ultra-compact 
HII region that may have helped clean the innermost ring area of the $^{13}$CO and 
C$^{18}$O emission, while our hub centre is not devoid of the large-scale emission
(see Fig.~\ref{fig_all_mom0}). 

Their central hub (without the extending filament arms) has a radius of $\sim$1\,pc 
within which the mass is about 1000\,$M_\odot$, and the visual extinction is much higher 
than in NGC2071-N, above $\sim$20\,mag, also derived from {\it Herschel} observations 
\citep{Rayner+2017}. The latter authors suggest that star-formation is an on-going 
process in the Mon R2 hub that has already been regulated by feedback, which is again a 
substantial difference between our subregions. However, probably more such central hub-ring 
structures may exist on a larger scale of cloud properties.

\subsection{At the centre of the hub}\label{sec:centre}

The sources in the central part of the hub, while it was not known that it is a hub, 
have been discussed in more detail by \citet{AspinReipurth2000} and \citet{Hillenbrand+2012}.  

As it was mentioned before that this region is somewhat isolated, there are no such 
compact sources with nebulous emission in the vicinity either, which strengthens the 
confidence that IRAS 05451+0037 and LkH$\alpha$ 316 belong to the same neighbourhood 
(see also Sect.~\ref{sec:dist}).

\begin{figure*}
 \begin{center}
  \begin{minipage}{1.\linewidth}
   \resizebox{0.5\hsize}{!}{\includegraphics[width=\columnwidth]{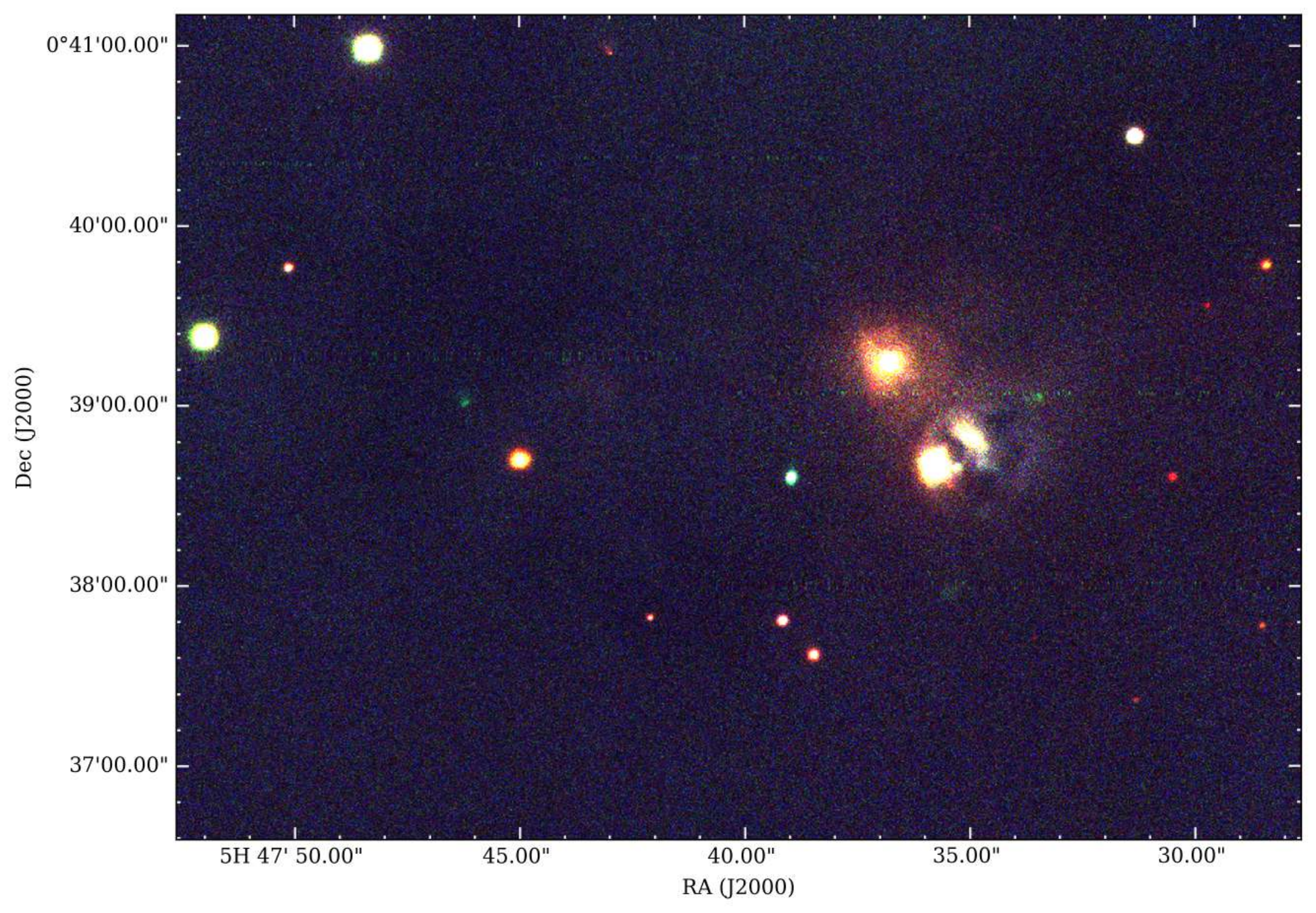}}
   \hfill
   \resizebox{0.5\hsize}{!}{\includegraphics[width=\columnwidth]{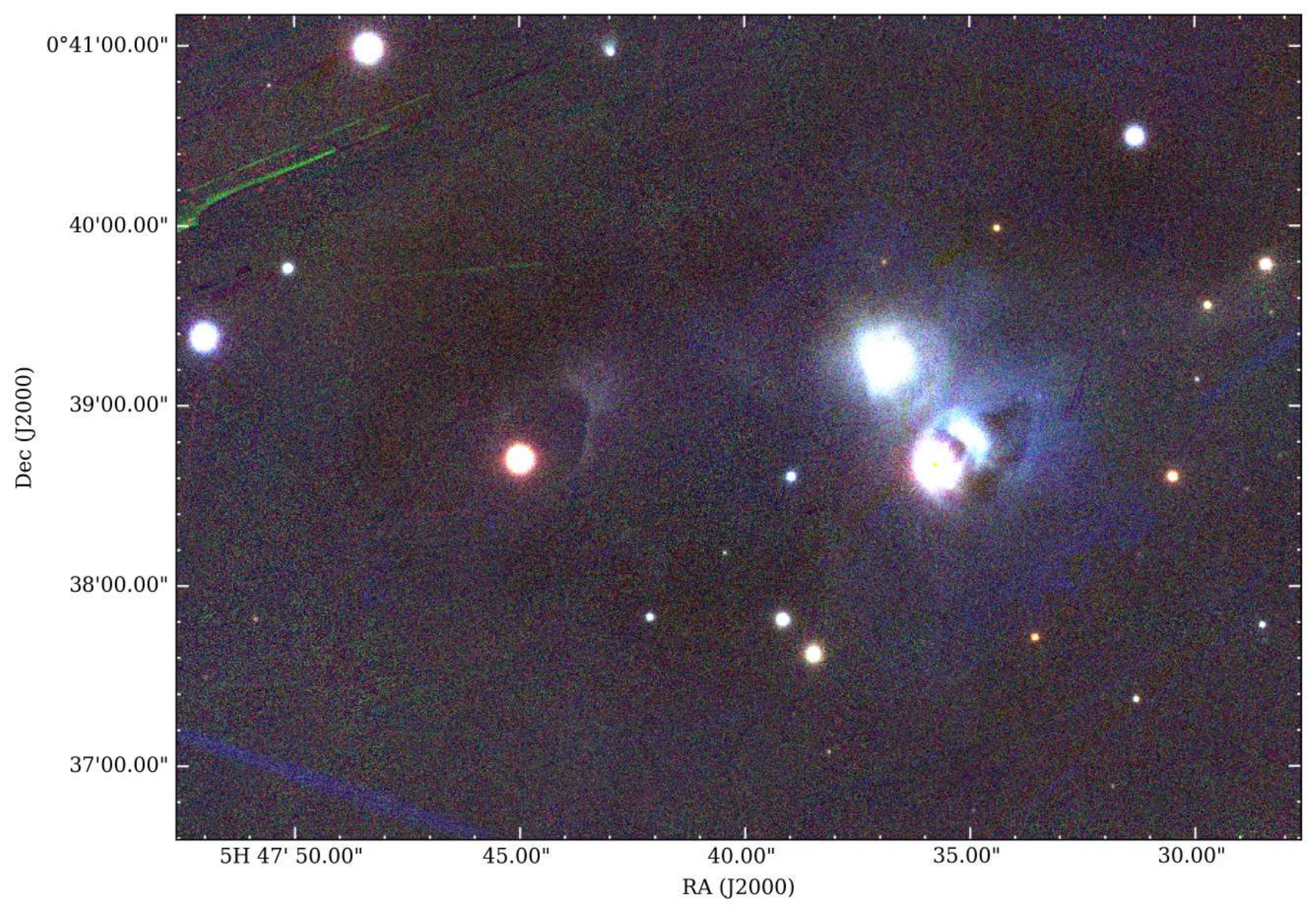}}
   \vspace{3mm}
   \resizebox{0.5\hsize}{!}{\includegraphics[width=\columnwidth]{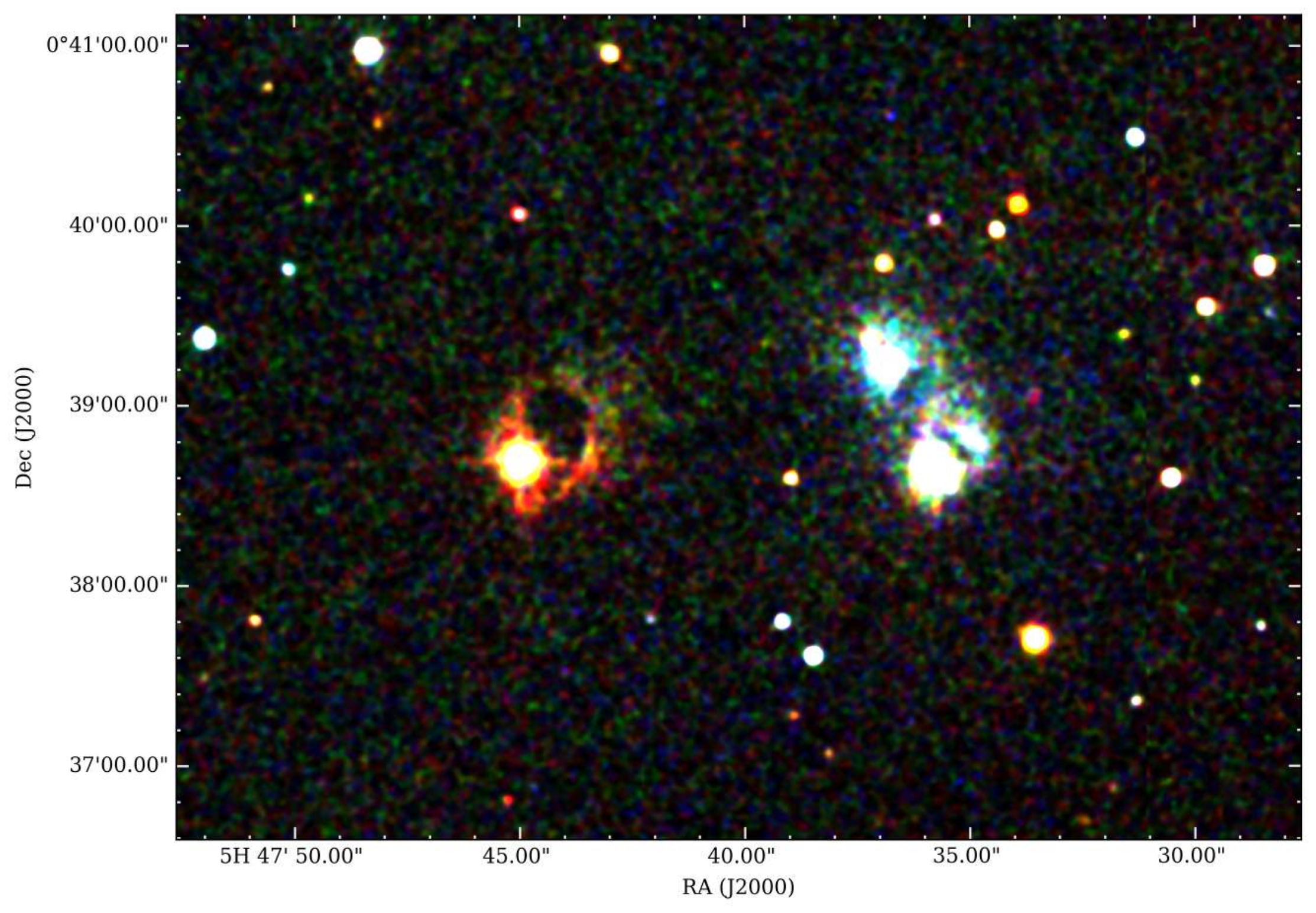}}   
   \hfill
   \resizebox{0.5\hsize}{!}{\includegraphics[width=\columnwidth]{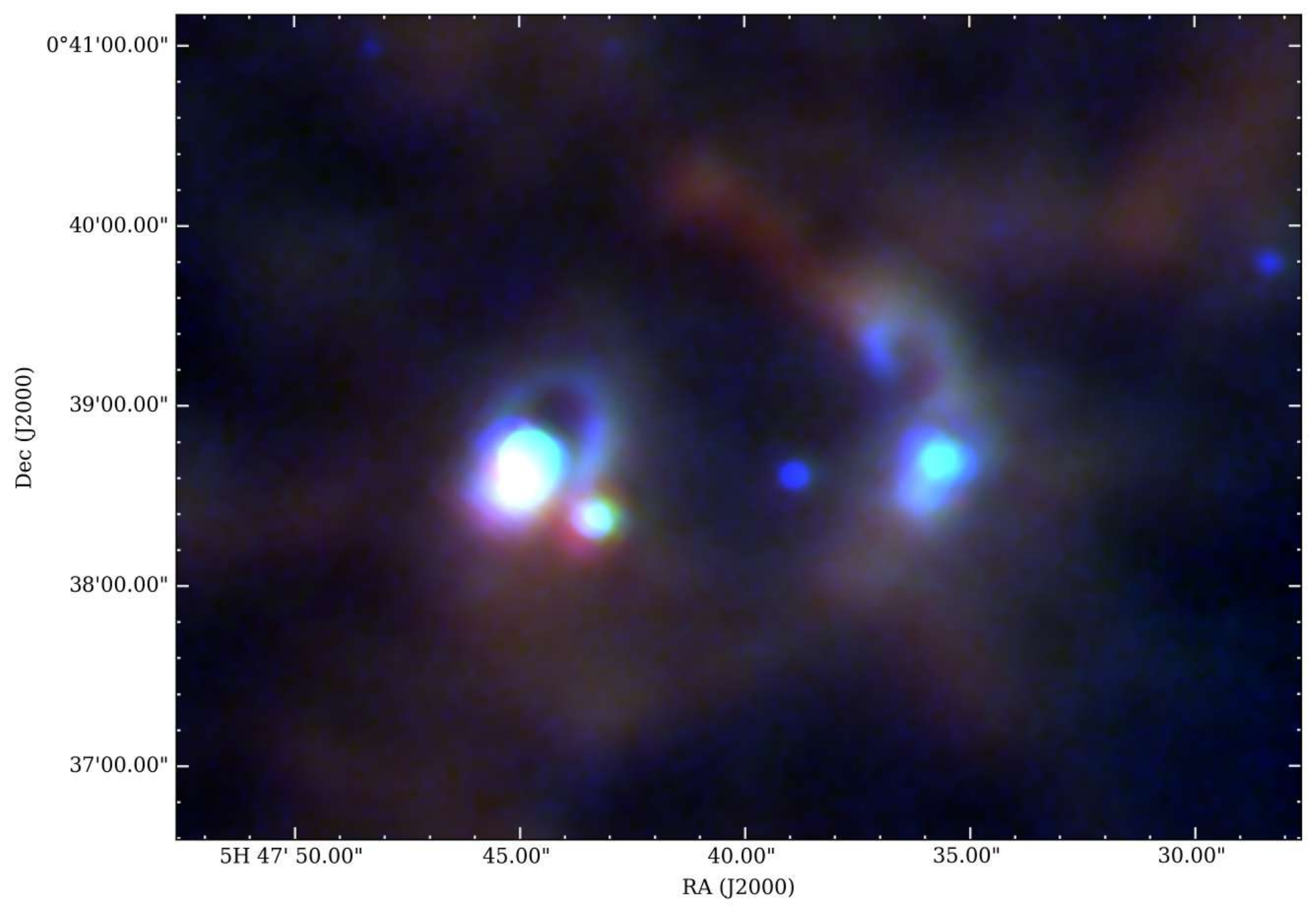}}   
  \end{minipage}
 \end{center}
  \caption{Zoomed RGB images 
          on the double centre of the NGC2071-N filament hub. The left-hand side source, midplane, is 
          IRAS 05451+0037 and the lower right-hand side one is LkH$\alpha$ 316.
          {\bf Top Left:} SDSS i, r, g images (in the RGB order). {\bf Top Right:} Pan-STARRS y, z, i filters. 
          {\bf Bottom Left:} 2MASS K$_{\rm s}$, H, J filters. {\bf Bottom Right:} Herschel 250, 160, and 100\,$\mu$m images.
          All listed in the RGB order, while Table~\ref{tab_wave} lists the nominal wavelengths of all these filters 
          and images from the shortest to the longest wavelengths.}
  \label{fig_rgb}
\end{figure*}

We have looked for these sources in archival data sets, and visualized the hub centre 
with a series of RGB images
\footnote{Images were generated with the Python package MULTICOLORFITS: https://github.com/pjcigan/multicolorfits.}
In Fig.~\ref{fig_rgb} the left-hand side source is IRAS 05451+0037 and the lower 
right-hand side one is LkH$\alpha$ 316. The used surveys and their wavelengths are 
listed in Table~\ref{tab_wave}.

\begin{table*}\small\setlength{\tabcolsep}{4.5pt}
 \caption{Summary of surveys and wavelengths providing images for Fig.~\ref{fig_rgb}. 
         SDSS: \citet{Eisenstein+2011}; Pan-STARRS: \citet{Chambers+2016}; 2MASS: \citet{Skrutskie+2006}; {\it Herschel}: \citet{Pilbratt2010}.}
 \label{tab_wave}      
 {\renewcommand{\arraystretch}{1.2}        
  \begin{tabular}{l|c|c|c|c|c|c|c|c|c|c|c|c}  
  \hline
  \hline
  Figure panel       	&\multicolumn{3}{c|}{Fig.~\ref{fig_rgb} top left}	&\multicolumn{3}{c|}{Fig.~\ref{fig_rgb} top right}	&\multicolumn{3}{c|}{Fig.~\ref{fig_rgb} bottom left}	&\multicolumn{3}{c|}{Fig.~\ref{fig_rgb} bottom right}\\
  \hline
  Survey		&SDSS g	&SDSS r	&SDSS i	&Pan-STR i	&Pan-STR z	&Pan-STR y	&2MASS J	&2MASS H	&2MASS K$_{\rm s}$	&H 100	&H 160	&H 250	\\
  \hline
  Wavelength [$\mu$m]	&0.4686	&0.6165	&0.7481	&0.7520	&0.8660	&0.9620	&1.24	&1.66	&2.16	&100	&160 	&250	\\
  \hline
  \hline  
  \end{tabular}
 }    
\end{table*}

We note that throughout the paper we are using the position of IRAS 05451+0037 given by SIMBAD, 
which is referring to \citet{Gaia2018}. This is almost identical to the one used by 
\citet{Hillenbrand+2012} based on the associated SDSS source, but different from the 
position of this same IRAS source given by \citet{Wouterloot+1988, Claussen+1996, AspinReipurth2000}. 
The location of the latter authors is $\sim$40\arcsec~west of ours, and the disagreement is
most likely due to the large beam sizes of {\it IRAS}.

\citet{AspinReipurth2000} have identified new HH objects near compact reflection nebulae --
that is our double centre in Fig.~\ref{fig_rgb}. Their HH 471 A-E jets and flows are associated with 
LkH$\alpha$ 316, and the earlier known HH 71 may as well, while HH 473-474 may be associated with 
the IRAS source as most likely exciting source. The new HH 472 seems to be associated with a 
faint and unknown source which positions are identified by SIMBAD as 2MASS J05473897+0038362. 
\citet{AspinReipurth2000} suggested that this unknown exciting source of HH 472 may 
be responsible for driving the outflow found by \citet{Fukui+1986, Iwata+1988} 
(see Sect.~\ref{sec:outflow}). 

The right-hand side group of reflection nebulosities also appear in the $I$-band 
images of \citet{AspinReipurth2000}. LkH$\alpha$ 316 is the lowest compact source, LkH$\alpha$ 316-neb 
lies $\sim$15\arcsec~north-west of it, and the 316/c component is the one $\sim$37\arcsec~north-east of them.
LkH$\alpha$ 316-neb is only a nebulosity without a star. Most probably LkH$\alpha$ 316 and LkH$\alpha$ 316-neb 
are physically associated \citep{AspinReipurth2000}, but we suggest that even 316/c may be
part of this close system, visually based on the morphology of the diffuse gas and the curved
material bridge that is best seen in the 2MASS and {\it Herschel} images (lower panels of 
Fig.~\ref{fig_rgb}). 

Furthermore, another argument strengthening the physical association between all of the three
LkH$\alpha$ 316 components is their locations in the column density map (see the contours in
Fig.~\ref{fig_100mu}). The northern nebulosity is within the ammonia core C, at the 
junction of two filaments that can also be seen in the lower right panel of Fig.~\ref{fig_rgb}. 
Summarized in Sect.~\ref{sec:props}, at such locations matterial is deposited from which 
YSOs and infrared clusters may form more efficiently, which can be further channelled 
in between 316/c and the two other nebulosities. 
The integrated intensity map of HCO$^+$(1--0) in Fig.~\ref{fig_all_mom0}, 
also suggests a connection between LkH$\alpha$ 316 and LkH$\alpha$ 316-neb,
both being on the island of strongest emission.
IRAS 05451+0037 is also found at the junction of filaments within ammonia core D, which was
originally found to drive a bipolar CO outflow \citep{Fukui+1986, Iwata+1988}. 
\citet{Hillenbrand+2012} derived a bolometric luminosity of $\sim 90\, L_\odot$ for 
IRAS 05451+0037, and presented for the first time its optical SDSS spectrum as well.
From the analysis from SDSS up to JCMT/SCUBA wavelengths and models, they concluded that 
IRAS 05451+0037 is optically faint and far-infrared bright, and its SED is consistent 
with those of Class I and flat-spectrum YSOs. They suggest that there is still likely 
a significant circumstellar envelope that is feeding the extended massive disk.

They found that our IRAS source is the brightest in the region at mid-infrared
wavelengths based on WISE images, although the nearby LkH$\alpha$ 316 and LkH$\alpha$ 316-neb 
become also quite strong by 25\,$\mu$m.

\citet{Hillenbrand+2012} saw nebulosity closely associated with the IRAS source as well
at the 2MASS K and H bands that they did not find at shorter wavelengths. 
We suggest that this gas feature can be seen at shorter than 2MASS wavelengths too
(top right panel of Fig.~\ref{fig_rgb}). At the wavelengths it is visible, it has 
a neat loop structure. 
The extent of this loop we measured from Pan-STARRS to {\it Herschel} images, 
and found to be 0.06--0.07\,pc at the distance of 400\,pc. This loop is rather thin 
and sharp, and the extent of it, from the compact source to the north-west corner, 
does not seem to vary with wavelength. 
These reflection nebulae around both hub centres indicate that there is interaction
between the point sources and the cloud material.

\subsection{CO outflow revisited}\label{sec:outflow}

The large amorphous bipolar $^{12}$CO outflow in NGC2071-N was discovered by \citet{Fukui+1986}, 
and followed up by \citet{Iwata+1988}. They defined the outflow relatively old
($\tau \sim 1.7 \times 10^5$\,yr) and suggested that its driving source is IRAS 05451+0037,
although it is off-axis to the east from the outflow lobes.
In this same study they also presented a higher resolution, though not fully calibrated CO map,
where there is indication that the lobes are extended towards the IRAS source.
They then suggested that the outflow has a U-shape with the IRAS source at the bottom of it.
The red-shifted CO lobe peaks nearly in between IRAS 05451+0037 and LkH$\alpha$ 316,
which also complicates the interpretation with the IRAS source as the origin.
\citet{Goldsmith+1992} still found the origin of this outflow less clearly identifiable.

Then \citet{AspinReipurth2000} suggested their HH 472 jet and its unknown source to be
the driving source of the outflow, as they are located between the blue and red lobes.
This molecular outflow and the HH objects are also misaligned, perhaps indicating a 
stronger or differently directed flow in the past \citep{Hillenbrand+2012}.

Interestingly, the complete loop-shaped nebulosity around IRAS 05451+0037 (Fig.~\ref{fig_rgb})
extends from the point source towards north-west, in which direction there is the apparent
centre of the outflow, however from our fine-detailed $^{13}$CO(1--0) molecular line 
data, plotting channel maps, we cannot confirm this outflow, and it is more probable that its 
SE-NW axis, interpreted at much lower resolution, corresponds to the dense features we can 
also see in column density along the SE-NW axis (e.g., Fig.~\ref{fig_fil_obj} left).

\section{Conclusions}\label{sec:concl}

We have presented {\it Herschel}, molecular line, and archival data sets, including 
100\,$\mu$m {\it Herschel} images, IRAM 30\,m, and NRO 45\,m observatons over
a newly resolved hub-filament structure in NGC2071-North featuring a double centre.   
Our main results and conclusions are summarized as follows:
\begin{enumerate}
  \item We confirm that NGC2071-North is part of the L1630 North complex at $d \sim 400$\,pc, and is 
        not affected by the Barnard's Loop (nearby in projection) which may be at a closer distance 
        ($\sim$180\,pc).
  \item The {\it Herschel} H$_2$ column density image reveals a $\sim 1.5 \times 1.5$\,pc-size 
        filamentary hub structure with curved arms, containing $\sim$500\,\msun~above 
        $A_V \sim 5$\,mag.
        The 100\,$\mu$m emission concentrates in the central part of the filament hub, at 
        IRAS 05451+0037 and the emission star LkH$\alpha$ 316, and features diffuse lobes and loops 
        around them. 
  \item We have estimated the energy balance by calculating gravitational and turbulent kinetic 
        energy densities, as well as gravitational and internal pressures within 42 spots, 
        with 25\arcsec~radius, along the filaments. From these comparisons, together with finding 
        several infall candidates in HCO$^+$ and C$^{18}$O, we conclude that gravity 
        dominates on small scales, and NGC2071-N is currently forming stars.
  \item {\it Planck} POS magnetic field lines reveal a loop structure east of NGC2071, extending 
        up to NGC2071-N. This results in east-west B-field lines in the east of the hub, and
        north-south running field lines in the west of the filament hub. This closely perpendicular
        B-field structure may be an efficient configuration for gathering material at this
        relatively isolated location.
  \item We suggest that a double centre could be formed in this hub, because the converging 
        locations of two filament pairs are offset.
        This offset is 2.3\arcmin ($0.27$\,pc) that also matches the diameter of a 
        central hub-ring that is seen in column density, traced by filaments, and in 
        HCO$^+$ {\it and} HCN, which may indicate a transition and connection between the hub 
        and the filaments.
  \item We argue that not only two, but all of the three components of the LkH$\alpha$ 316 young star
        (along with 316-neb and 316/c) are in physical association due to the location of the
        latter at the junction of two filaments.
        We find that the clear gas loop feature around IRAS 05451+0037 can already be seen at 
        shorter than 2MASS wavelengths. The extent of this loop we measured to be 0.06--0.07\,pc
        which does not seem to vary with wavelength.
  \item We have revisited the CO outflow, discovered by \citet{Fukui+1986}, and we do 
        not seem to find its lobes in our high-resolution $^{13}$CO data. 
\end{enumerate}

\section*{Acknowledgements}

We thank the anonymous reviewer for their helpful and useful suggestions and V.K.~thank K.~Pattle 
for useful discussions. 
V.K.~and D.W.-T.~acknowledge Science and Technology Facilities Council (STFC) support under grant 
number ST/R000786/1.

The present study has made use of data from the Herschel Gould Belt survey (HGBS) project 
(http://gouldbelt\-herschel.cea.fr). The HGBS is a Herschel Key Programme jointly carried out 
by SPIRE Specialist Astronomy Group 3 (SAG 3), scientists of several institutes in the PACS 
Consortium (CEA Saclay, INAF-IFSI Rome and INAF-Arcetri, KU Leuven, MPIA Heidelberg), and 
scientists of the Herschel Science Center (HSC).

Partly based on observations carried out with the IRAM 30\,m Telescope under project number 030--13. 
IRAM is supported by INSU/CNRS (France), MPG (Germany) and IGN (Spain).

The 45\,m radio telescope is operated by Nobeyama Radio Observatory, a branch of National 
Astronomical Observatory of Japan. 

Funding for SDSS-III has been provided by the Alfred P. Sloan Foundation, 
the Participating Institutions, the National Science Foundation, and the U.S. 
Department of Energy Office of Science. The SDSS-III web site is http://www.sdss3.org/.

SDSS-III is managed by the Astrophysical Research Consortium for the Participating 
Institutions of the SDSS-III Collaboration including the University of Arizona, 
the Brazilian Participation Group, Brookhaven National Laboratory, Carnegie Mellon 
University, University of Florida, the French Participation Group, the German 
Participation Group, Harvard University, the Instituto de Astrofisica de Canarias, 
the Michigan State/Notre Dame/JINA Participation Group, Johns Hopkins University, 
Lawrence Berkeley National Laboratory, Max Planck Institute for Astrophysics, 
Max Planck Institute for Extraterrestrial Physics, New Mexico State University, 
New York University, Ohio State University, Pennsylvania State University, 
University of Portsmouth, Princeton University, the Spanish Participation Group, 
University of Tokyo, University of Utah, Vanderbilt University, University of 
Virginia, University of Washington, and Yale University. 

The Pan-STARRS1 Surveys (PS1) and the PS1 public science archive have been made 
possible through contributions by the Institute for Astronomy, the University of 
Hawaii, the Pan-STARRS Project Office, the Max-Planck Society and its participating 
institutes, the Max Planck Institute for Astronomy, Heidelberg and the Max Planck 
Institute for Extraterrestrial Physics, Garching, The Johns Hopkins University, 
Durham University, the University of Edinburgh, the Queen's University Belfast, 
the Harvard-Smithsonian Center for Astrophysics, the Las Cumbres Observatory Global 
Telescope Network Incorporated, the National Central University of Taiwan, the Space 
Telescope Science Institute, the National Aeronautics and Space Administration under 
Grant No. NNX08AR22G issued through the Planetary Science Division of the NASA Science 
Mission Directorate, the National Science Foundation Grant No. AST-1238877, the 
University of Maryland, Eotvos Lorand University (ELTE), the Los Alamos National 
Laboratory, and the Gordon and Betty Moore Foundation.

\section*{Data Availability}
 
The PACS and SPIRE maps, the $Herschel$ column density and temperature maps, 
as well as the dense core lists extracted from them in the whole Orion~B region 
are publicly available at the $Herschel$ Gould Belt Survey Archive: 
http://gouldbelt-herschel.cea.fr/archives.   
Additional molecular line maps presented in this paper are available 
at http://dx.doi.org/xxxx.yyyyy. 



\bibliographystyle{mnras}
\bibliography{konyves_NGC2071N} 

\begin{thebibliography}{}
\makeatletter
\relax
\def\mn@urlcharsother{\let\do\@makeother \do\$\do\&\do\#\do\^\do\_\do\%\do\~}
\def\mn@doi{\begingroup\mn@urlcharsother \@ifnextchar [ {\mn@doi@}
  {\mn@doi@[]}}
\def\mn@doi@[#1]#2{\def\@tempa{#1}\ifx\@tempa\@empty \href
  {http://dx.doi.org/#2} {doi:#2}\else \href {http://dx.doi.org/#2} {#1}\fi
  \endgroup}
\def\mn@eprint#1#2{\mn@eprint@#1:#2::\@nil}
\def\mn@eprint@arXiv#1{\href {http://arxiv.org/abs/#1} {{\tt arXiv:#1}}}
\def\mn@eprint@dblp#1{\href {http://dblp.uni-trier.de/rec/bibtex/#1.xml}
  {dblp:#1}}
\def\mn@eprint@#1:#2:#3:#4\@nil{\def\@tempa {#1}\def\@tempb {#2}\def\@tempc
  {#3}\ifx \@tempc \@empty \let \@tempc \@tempb \let \@tempb \@tempa \fi \ifx
  \@tempb \@empty \def\@tempb {arXiv}\fi \@ifundefined
  {mn@eprint@\@tempb}{\@tempb:\@tempc}{\expandafter \expandafter \csname
  mn@eprint@\@tempb\endcsname \expandafter{\@tempc}}}

\bibitem[\protect\citeauthoryear{{Alina}, {Ristorcelli}, {Montier},
  {Abdikamalov}, {Juvela}, {Ferri{\`e}re}, {Bernard}  \& {Micelotta}}{{Alina}
  et~al.}{2019}]{Alina+2019}
{Alina} D.,  {Ristorcelli} I.,  {Montier} L.,  {Abdikamalov} E.,  {Juvela} M.,
  {Ferri{\`e}re} K.,  {Bernard} J.~P.,   {Micelotta} E.~R.,  2019, \mn@doi
  [\mnras] {10.1093/mnras/stz508}, \href
  {https://ui.adsabs.harvard.edu/abs/2019MNRAS.485.2825A} {485, 2825}

\bibitem[\protect\citeauthoryear{{Andr{\'e}} et~al.,}{{Andr{\'e}}
  et~al.}{2010}]{Andre+2010}
{Andr{\'e}} P.,  et~al., 2010, \mn@doi [A{\&}A] {10.1051/0004-6361/201014666},
  \href {http://adsabs.harvard.edu/abs/2010A%26A...518L.102A} {518, L102+}

\bibitem[\protect\citeauthoryear{{Andr{\'e}}, {Di Francesco}, {Ward-Thompson},
  {Inutsuka}, {Pudritz}  \& {Pineda}}{{Andr{\'e}} et~al.}{2014}]{Andre+2014}
{Andr{\'e}} P.,  {Di Francesco} J.,  {Ward-Thompson} D.,  {Inutsuka} S.-I.,
  {Pudritz} R.~E.,   {Pineda} J.~E.,  2014, in Protostars and Planets VI, ed.
  H. Beuther et al.. p.~27 (\mn@eprint {arXiv} {1312.6232}),
  \mn@doi{10.2458/azu_uapress_9780816531240-ch002}

\bibitem[\protect\citeauthoryear{{Andr{\'e}}, {Palmeirim}  \&
  {Arzoumanian}}{{Andr{\'e}} et~al.}{2022}]{Andre+2022}
{Andr{\'e}} P.,  {Palmeirim} P.,   {Arzoumanian} D.,  2022, \aap, \href
  {https://ui.adsabs.harvard.edu/abs/2022arXiv221004736A} {667, L1}

\bibitem[\protect\citeauthoryear{{Anthony-Twarog}}{{Anthony-Twarog}}{1982}]{Anthony-Twarog1982}
{Anthony-Twarog} B.~J.,  1982, \mn@doi [\aj] {10.1086/113204}, \href
  {https://ui.adsabs.harvard.edu/abs/1982AJ.....87.1213A} {87, 1213}

\bibitem[\protect\citeauthoryear{{Arzoumanian} et~al.,}{{Arzoumanian}
  et~al.}{2011}]{Arzoumanian+2011}
{Arzoumanian} D.,  et~al., 2011, \mn@doi [A{\&}A]
  {10.1051/0004-6361/201116596}, \href
  {http://adsabs.harvard.edu/abs/2011A%26A...529L...6A} {529, L6}

\bibitem[\protect\citeauthoryear{{Arzoumanian} et~al.,}{{Arzoumanian}
  et~al.}{2019}]{Arzoumanian+2019}
{Arzoumanian} D.,  et~al., 2019, \mn@doi [\aap] {10.1051/0004-6361/201832725},
  \href {https://ui.adsabs.harvard.edu/abs/2019A&A...621A..42A} {621, A42}

\bibitem[\protect\citeauthoryear{{Aspin} \& {Reipurth}}{{Aspin} \&
  {Reipurth}}{2000}]{AspinReipurth2000}
{Aspin} C.,  {Reipurth} B.,  2000, \mn@doi [\mnras]
  {10.1046/j.1365-8711.2000.03004.x}, \href
  {https://ui.adsabs.harvard.edu/abs/2000MNRAS.311..522A} {311, 522}

\bibitem[\protect\citeauthoryear{{Bally}}{{Bally}}{2008}]{Bally2008}
{Bally} J.,  2008, {Overview of the Orion Complex}.
p.~459

\bibitem[\protect\citeauthoryear{{Benedettini} et~al.,}{{Benedettini}
  et~al.}{2018}]{Benedettini+2018}
{Benedettini} M.,  et~al., 2018, \mn@doi [\aap] {10.1051/0004-6361/201833364},
  \href {http://adsabs.harvard.edu/abs/2018A%26A...619A..52B} {619, A52}

\bibitem[\protect\citeauthoryear{{Bertoldi} \& {McKee}}{{Bertoldi} \&
  {McKee}}{1992}]{BertoldiMcKee1992}
{Bertoldi} F.,  {McKee} C.~F.,  1992, \mn@doi [\apj] {10.1086/171638}, \href
  {http://cdsads.u-strasbg.fr/abs/1992ApJ...395..140B} {395, 140}

\bibitem[\protect\citeauthoryear{{Bohlin}, {Savage}  \& {Drake}}{{Bohlin}
  et~al.}{1978}]{Bohlin+1978}
{Bohlin} R.~C.,  {Savage} B.~D.,   {Drake} J.~F.,  1978, \mn@doi [\apj]
  {10.1086/156357}, \href {http://adsabs.harvard.edu/abs/1978ApJ...224..132B}
  {224, 132}

\bibitem[\protect\citeauthoryear{{Bonnor}}{{Bonnor}}{1956}]{Bonnor1956}
{Bonnor} W.~B.,  1956, \mnras, \href
  {http://adsabs.harvard.edu/abs/1956MNRAS.116..351B} {116, 351}

\bibitem[\protect\citeauthoryear{{Carter} et~al.,}{{Carter}
  et~al.}{2012}]{Carter+2012}
{Carter} M.,  et~al., 2012, \mn@doi [\aap] {10.1051/0004-6361/201118452}, \href
  {https://ui.adsabs.harvard.edu/abs/2012A&A...538A..89C} {538, A89}

\bibitem[\protect\citeauthoryear{{Chambers} et~al.,}{{Chambers}
  et~al.}{2016}]{Chambers+2016}
{Chambers} K.~C.,  et~al., 2016, arXiv e-prints, \href
  {https://ui.adsabs.harvard.edu/abs/2016arXiv161205560C} {p. arXiv:1612.05560}

\bibitem[\protect\citeauthoryear{{Chen}, {King}  \& {Li}}{{Chen}
  et~al.}{2016}]{Chen+2016}
{Chen} C.-Y.,  {King} P.~K.,   {Li} Z.-Y.,  2016, \mn@doi [\apj]
  {10.3847/0004-637X/829/2/84}, \href
  {https://ui.adsabs.harvard.edu/abs/2016ApJ...829...84C} {829, 84}

\bibitem[\protect\citeauthoryear{{Claussen}, {Wilking}, {Benson}, {Wootten},
  {Myers}  \& {Terebey}}{{Claussen} et~al.}{1996}]{Claussen+1996}
{Claussen} M.~J.,  {Wilking} B.~A.,  {Benson} P.~J.,  {Wootten} A.,  {Myers}
  P.~C.,   {Terebey} S.,  1996, \mn@doi [\apjs] {10.1086/192330}, \href
  {https://ui.adsabs.harvard.edu/abs/1996ApJS..106..111C} {106, 111}

\bibitem[\protect\citeauthoryear{{Cutri} et~al.,}{{Cutri}
  et~al.}{2003}]{Cutri+2003}
{Cutri} R.~M.,  et~al., 2003, VizieR Online Data Catalog, \href
  {https://ui.adsabs.harvard.edu/abs/2003yCat.2246....0C} {p. II/246}

\bibitem[\protect\citeauthoryear{{Dale} \& {Bonnell}}{{Dale} \&
  {Bonnell}}{2011}]{DaleBonnell2011}
{Dale} J.~E.,  {Bonnell} I.,  2011, \mn@doi [\mnras]
  {10.1111/j.1365-2966.2011.18392.x}, \href
  {http://adsabs.harvard.edu/abs/2011MNRAS.414..321D} {414, 321}

\bibitem[\protect\citeauthoryear{{Dewangan}, {Luna}, {Ojha}, {Anandarao},
  {Mallick}  \& {Mayya}}{{Dewangan} et~al.}{2015}]{Dewangan+2015}
{Dewangan} L.~K.,  {Luna} A.,  {Ojha} D.~K.,  {Anandarao} B.~G.,  {Mallick}
  K.~K.,   {Mayya} Y.~D.,  2015, \mn@doi [\apj] {10.1088/0004-637X/811/2/79},
  \href {https://ui.adsabs.harvard.edu/abs/2015ApJ...811...79D} {811, 79}

\bibitem[\protect\citeauthoryear{{Dhabal}, {Mundy}, {Rizzo}, {Storm}  \&
  {Teuben}}{{Dhabal} et~al.}{2018}]{Dhabal+2018}
{Dhabal} A.,  {Mundy} L.~G.,  {Rizzo} M.~J.,  {Storm} S.,   {Teuben} P.,  2018,
  \mn@doi [\apj] {10.3847/1538-4357/aaa76b}, \href
  {https://ui.adsabs.harvard.edu/abs/2018ApJ...853..169D} {853, 169}

\bibitem[\protect\citeauthoryear{{Dunham}, {Rosolowsky}, {Evans}, {Cyganowski}
  \& {Urquhart}}{{Dunham} et~al.}{2011}]{Dunham+2011}
{Dunham} M.~K.,  {Rosolowsky} E.,  {Evans} Neal~J. I.,  {Cyganowski} C.,
  {Urquhart} J.~S.,  2011, \mn@doi [\apj] {10.1088/0004-637X/741/2/110}, \href
  {https://ui.adsabs.harvard.edu/abs/2011ApJ...741..110D} {741, 110}

\bibitem[\protect\citeauthoryear{{Eisenstein} et~al.,}{{Eisenstein}
  et~al.}{2011}]{Eisenstein+2011}
{Eisenstein} D.~J.,  et~al., 2011, \mn@doi [\aj] {10.1088/0004-6256/142/3/72},
  \href {https://ui.adsabs.harvard.edu/abs/2011AJ....142...72E} {142, 72}

\bibitem[\protect\citeauthoryear{{Evans}}{{Evans}}{1999}]{Evans1999}
{Evans} Neal~J. I.,  1999, \mn@doi [\araa] {10.1146/annurev.astro.37.1.311},
  \href {https://ui.adsabs.harvard.edu/abs/1999ARA&A..37..311E} {37, 311}

\bibitem[\protect\citeauthoryear{{Fiorellino} et~al.,}{{Fiorellino}
  et~al.}{2021}]{Fiorellino+2021}
{Fiorellino} E.,  et~al., 2021, \mn@doi [\mnras] {10.1093/mnras/staa3420},
  \href {https://ui.adsabs.harvard.edu/abs/2021MNRAS.500.4257F} {500, 4257}

\bibitem[\protect\citeauthoryear{{Fukui}, {Sugitani}, {Takaba}, {Iwata},
  {Mizuno}, {Ogawa}  \& {Kawabata}}{{Fukui} et~al.}{1986}]{Fukui+1986}
{Fukui} Y.,  {Sugitani} K.,  {Takaba} H.,  {Iwata} T.,  {Mizuno} A.,  {Ogawa}
  H.,   {Kawabata} K.,  1986, \mn@doi [\apjl] {10.1086/184803}, \href
  {https://ui.adsabs.harvard.edu/abs/1986ApJ...311L..85F} {311, L85}

\bibitem[\protect\citeauthoryear{{Gaia Collaboration}}{{Gaia
  Collaboration}}{2018}]{Gaia2018}
{Gaia Collaboration} 2018, VizieR Online Data Catalog, \href
  {https://ui.adsabs.harvard.edu/abs/2018yCat.1345....0G} {p. I/345}

\bibitem[\protect\citeauthoryear{{Gibb}}{{Gibb}}{2008}]{Gibb2008}
{Gibb} A.~G.,  2008, {Star Formation in NGC 2068, NGC 2071, and Northern
  L1630}.
p.~693

\bibitem[\protect\citeauthoryear{{Goldsmith}, {Margulis}, {Snell}  \&
  {Fukui}}{{Goldsmith} et~al.}{1992}]{Goldsmith+1992}
{Goldsmith} P.~F.,  {Margulis} M.,  {Snell} R.~L.,   {Fukui} Y.,  1992, \mn@doi
  [\apj] {10.1086/170960}, \href
  {https://ui.adsabs.harvard.edu/abs/1992ApJ...385..522G} {385, 522}

\bibitem[\protect\citeauthoryear{{Greenfield} et~al.,}{{Greenfield}
  et~al.}{2013}]{Greenfield+2013}
{Greenfield} P.,  et~al., 2013, {Astropy: Community Python library for
  astronomy} (\mn@eprint {ascl} {1304.002})

\bibitem[\protect\citeauthoryear{{Hacar}, {Clark}, {Heitsch}, {Kainulainen},
  {Panopoulou}, {Seifried}  \& {Smith}}{{Hacar} et~al.}{2022}]{Hacar+2022}
{Hacar} A.,  {Clark} S.,  {Heitsch} F.,  {Kainulainen} J.,  {Panopoulou} G.,
  {Seifried} D.,   {Smith} R.,  2022, arXiv e-prints, \href
  {https://ui.adsabs.harvard.edu/abs/2022arXiv220309562H} {p. arXiv:2203.09562}

\bibitem[\protect\citeauthoryear{{Hatchell}, {Richer}, {Fuller}, {Qualtrough},
  {Ladd}  \& {Chandler}}{{Hatchell} et~al.}{2005}]{Hatchell+2005}
{Hatchell} J.,  {Richer} J.~S.,  {Fuller} G.~A.,  {Qualtrough} C.~J.,  {Ladd}
  E.~F.,   {Chandler} C.~J.,  2005, \mn@doi [\aap]
  {10.1051/0004-6361:20041836}, \href
  {http://adsabs.harvard.edu/abs/2005A%26A...440..151H} {440, 151}

\bibitem[\protect\citeauthoryear{{He} et~al.,}{{He} et~al.}{2015}]{He+2015}
{He} Y.-X.,  et~al., 2015, \mn@doi [\mnras] {10.1093/mnras/stv732}, \href
  {https://ui.adsabs.harvard.edu/abs/2015MNRAS.450.1926H} {450, 1926}

\bibitem[\protect\citeauthoryear{{Hillenbrand}, {Knapp}, {Padgett}, {Rebull}
  \& {McGehee}}{{Hillenbrand} et~al.}{2012}]{Hillenbrand+2012}
{Hillenbrand} L.~A.,  {Knapp} G.~R.,  {Padgett} D.~L.,  {Rebull} L.~M.,
  {McGehee} P.~M.,  2012, \mn@doi [\aj] {10.1088/0004-6256/143/2/37}, \href
  {https://ui.adsabs.harvard.edu/abs/2012AJ....143...37H} {143, 37}

\bibitem[\protect\citeauthoryear{{Iwata}, {Fukui}  \& {Ogawa}}{{Iwata}
  et~al.}{1988}]{Iwata+1988}
{Iwata} T.,  {Fukui} Y.,   {Ogawa} H.,  1988, \mn@doi [\apj] {10.1086/166009},
  \href {https://ui.adsabs.harvard.edu/abs/1988ApJ...325..372I} {325, 372}

\bibitem[\protect\citeauthoryear{{Johnstone}, {Di Francesco}  \&
  {Kirk}}{{Johnstone} et~al.}{2004}]{Johnstone+2004}
{Johnstone} D.,  {Di Francesco} J.,   {Kirk} H.,  2004, \mn@doi [ApJ]
  {10.1086/423737}, \href {http://adsabs.harvard.edu/abs/2004ApJ...611L..45J}
  {611, L45}

\bibitem[\protect\citeauthoryear{{Kirk, H.}, {Johnstone}  \& {Di
  Francesco}}{{Kirk, H.} et~al.}{2006}]{KirkH+2006}
{Kirk, H.} {Johnstone} D.,   {Di Francesco} J.,  2006, \mn@doi [ApJ]
  {10.1086/503193}, \href {http://adsabs.harvard.edu/abs/2006ApJ...646.1009K}
  {646, 1009}

\bibitem[\protect\citeauthoryear{{Kirk} et~al.,}{{Kirk}
  et~al.}{2016a}]{HKirk+2016a}
{Kirk} H.,  et~al., 2016a, \mn@doi [\apj] {10.3847/0004-637X/817/2/167}, \href
  {http://adsabs.harvard.edu/abs/2016ApJ...817..167K} {817, 167}

\bibitem[\protect\citeauthoryear{{Kirk} et~al.,}{{Kirk}
  et~al.}{2016b}]{KirkH+2016a}
{Kirk} H.,  et~al., 2016b, \mn@doi [\apj] {10.3847/0004-637X/817/2/167}, \href
  {http://adsabs.harvard.edu/abs/2016ApJ...817..167K} {817, 167}

\bibitem[\protect\citeauthoryear{{Koch} \& {Rosolowsky}}{{Koch} \&
  {Rosolowsky}}{2015}]{KochRosolowsky2015}
{Koch} E.~W.,  {Rosolowsky} E.~W.,  2015, \mn@doi [\mnras]
  {10.1093/mnras/stv1521}, \href
  {http://adsabs.harvard.edu/abs/2015MNRAS.452.3435K} {452, 3435}

\bibitem[\protect\citeauthoryear{{K{\"o}nyves} et~al.,}{{K{\"o}nyves}
  et~al.}{2015}]{Konyves+2015}
{K{\"o}nyves} V.,  et~al., 2015, \mn@doi [\aap] {10.1051/0004-6361/201525861},
  \href {http://adsabs.harvard.edu/abs/2015A%26A...584A..91K} {584, A91}

\bibitem[\protect\citeauthoryear{{K{\"o}nyves} et~al.,}{{K{\"o}nyves}
  et~al.}{2020}]{Konyves+2020}
{K{\"o}nyves} V.,  et~al., 2020, \mn@doi [\aap] {10.1051/0004-6361/201834753},
  \href {https://ui.adsabs.harvard.edu/abs/2020A&A...635A..34K} {635, A34}

\bibitem[\protect\citeauthoryear{{Ladjelate} et~al.,}{{Ladjelate}
  et~al.}{2020}]{Ladjelate+2020}
{Ladjelate} B.,  et~al., 2020, \mn@doi [\aap] {10.1051/0004-6361/201936442},
  \href {https://ui.adsabs.harvard.edu/abs/2020A&A...638A..74L} {638, A74}

\bibitem[\protect\citeauthoryear{{Lallement}, {Vergely}, {Valette},
  {Puspitarini}, {Eyer}  \& {Casagrande}}{{Lallement}
  et~al.}{2014}]{Lallement+2014}
{Lallement} R.,  {Vergely} J.-L.,  {Valette} B.,  {Puspitarini} L.,  {Eyer} L.,
    {Casagrande} L.,  2014, \mn@doi [\aap] {10.1051/0004-6361/201322032}, \href
  {http://adsabs.harvard.edu/abs/2014A%26A...561A..91L} {561, A91}

\bibitem[\protect\citeauthoryear{{Li}, {Fang}, {Henning}  \&
  {Kainulainen}}{{Li} et~al.}{2013}]{Li+2013}
{Li} H.-b.,  {Fang} M.,  {Henning} T.,   {Kainulainen} J.,  2013, \mn@doi
  [\mnras] {10.1093/mnras/stt1849}, \href
  {http://adsabs.harvard.edu/abs/2013MNRAS.436.3707L} {436, 3707}

\bibitem[\protect\citeauthoryear{{Lyo} et~al.,}{{Lyo} et~al.}{2021}]{Lyo+2021}
{Lyo} A.~R.,  et~al., 2021, \mn@doi [\apj] {10.3847/1538-4357/ac0ce9}, \href
  {https://ui.adsabs.harvard.edu/abs/2021ApJ...918...85L} {918, 85}

\bibitem[\protect\citeauthoryear{{Marsh} et~al.,}{{Marsh}
  et~al.}{2016}]{Marsh+2016}
{Marsh} K.~A.,  et~al., 2016, \mn@doi [\mnras] {10.1093/mnras/stw301}, \href
  {http://adsabs.harvard.edu/abs/2016MNRAS.459..342M} {459, 342}

\bibitem[\protect\citeauthoryear{{Megeath} et~al.,}{{Megeath}
  et~al.}{2012}]{Megeath+2012}
{Megeath} S.~T.,  et~al., 2012, \mn@doi [\aj] {10.1088/0004-6256/144/6/192},
  \href {http://adsabs.harvard.edu/abs/2012AJ....144..192M} {144, 192}

\bibitem[\protect\citeauthoryear{{Menten}, {Reid}, {Forbrich}  \&
  {Brunthaler}}{{Menten} et~al.}{2007}]{Menten+2007}
{Menten} K.~M.,  {Reid} M.~J.,  {Forbrich} J.,   {Brunthaler} A.,  2007,
  \mn@doi [\aap] {10.1051/0004-6361:20078247}, \href
  {https://ui.adsabs.harvard.edu/abs/2007A&A...474..515M} {474, 515}

\bibitem[\protect\citeauthoryear{{Miville-Desch{\^e}nes}, {Murray}  \&
  {Lee}}{{Miville-Desch{\^e}nes} et~al.}{2017}]{MAMD+2017}
{Miville-Desch{\^e}nes} M.-A.,  {Murray} N.,   {Lee} E.~J.,  2017, \mn@doi
  [\apj] {10.3847/1538-4357/834/1/57}, \href
  {https://ui.adsabs.harvard.edu/abs/2017ApJ...834...57M} {834, 57}

\bibitem[\protect\citeauthoryear{{Myers}}{{Myers}}{2009}]{Myers2009}
{Myers} P.~C.,  2009, \mn@doi [\apj] {10.1088/0004-637X/700/2/1609}, \href
  {http://adsabs.harvard.edu/abs/2009ApJ...700.1609M} {700, 1609}

\bibitem[\protect\citeauthoryear{{Myers}, {Mardones}, {Tafalla}, {Williams}  \&
  {Wilner}}{{Myers} et~al.}{1996}]{Myers+1996}
{Myers} P.~C.,  {Mardones} D.,  {Tafalla} M.,  {Williams} J.~P.,   {Wilner}
  D.~J.,  1996, \mn@doi [\apjl] {10.1086/310146}, \href
  {https://ui.adsabs.harvard.edu/abs/1996ApJ...465L.133M} {465, L133}

\bibitem[\protect\citeauthoryear{{Nakamura} \& {Li}}{{Nakamura} \&
  {Li}}{2008}]{NakamuraLi2008}
{Nakamura} F.,  {Li} Z.-Y.,  2008, \mn@doi [\apj] {10.1086/591641}, \href
  {https://ui.adsabs.harvard.edu/abs/2008ApJ...687..354N} {687, 354}

\bibitem[\protect\citeauthoryear{{Ochsendorf}, {Brown}, {Bally}  \&
  {Tielens}}{{Ochsendorf} et~al.}{2015}]{Ochsendorf+2015}
{Ochsendorf} B.~B.,  {Brown} A.~G.~A.,  {Bally} J.,   {Tielens} A.~G.~G.~M.,
  2015, \mn@doi [\apj] {10.1088/0004-637X/808/2/111}, \href
  {http://adsabs.harvard.edu/abs/2015ApJ...808..111O} {808, 111}

\bibitem[\protect\citeauthoryear{{Onishi}, {Mizuno}, {Kawamura}, {Ogawa}  \&
  {Fukui}}{{Onishi} et~al.}{1998}]{Onishi+1998}
{Onishi} T.,  {Mizuno} A.,  {Kawamura} A.,  {Ogawa} H.,   {Fukui} Y.,  1998,
  \mn@doi [ApJ] {10.1086/305867}, \href
  {http://adsabs.harvard.edu/abs/1998ApJ...502..296O} {502, 296}

\bibitem[\protect\citeauthoryear{{Palmeirim} et~al.,}{{Palmeirim}
  et~al.}{2013}]{Palmeirim+2013}
{Palmeirim} P.,  et~al., 2013, \mn@doi [\aap] {10.1051/0004-6361/201220500},
  \href {http://adsabs.harvard.edu/abs/2013A%26A...550A..38P} {550, A38}

\bibitem[\protect\citeauthoryear{{Panopoulou}, {Clark}, {Hacar}, {Heitsch},
  {Kainulainen}, {Ntormousi}, {Seifried}  \& {Smith}}{{Panopoulou}
  et~al.}{2022}]{Panopoulou+2022}
{Panopoulou} G.~V.,  {Clark} S.~E.,  {Hacar} A.,  {Heitsch} F.,  {Kainulainen}
  J.,  {Ntormousi} E.,  {Seifried} D.,   {Smith} R.~J.,  2022, \mn@doi [\aap]
  {10.1051/0004-6361/202142281}, \href
  {https://ui.adsabs.harvard.edu/abs/2022A&A...657L..13P} {657, L13}

\bibitem[\protect\citeauthoryear{{Pattle} et~al.,}{{Pattle}
  et~al.}{2015}]{Pattle+2015}
{Pattle} K.,  et~al., 2015, \mn@doi [\mnras] {10.1093/mnras/stv376}, \href
  {https://ui.adsabs.harvard.edu/abs/2015MNRAS.450.1094P} {450, 1094}

\bibitem[\protect\citeauthoryear{{Peretto}, {Andr{\'e}}  \&
  {Belloche}}{{Peretto} et~al.}{2006}]{Peretto+2006}
{Peretto} N.,  {Andr{\'e}} P.,   {Belloche} A.,  2006, \mn@doi [A{\&}A]
  {10.1051/0004-6361:20053324}, \href
  {http://adsabs.harvard.edu/abs/2006A%26A...445..979P} {445, 979}

\bibitem[\protect\citeauthoryear{{Peretto} et~al.,}{{Peretto}
  et~al.}{2012}]{Peretto+2012}
{Peretto} N.,  et~al., 2012, \mn@doi [\aap] {10.1051/0004-6361/201118663},
  \href {http://adsabs.harvard.edu/abs/2012A%26A...541A..63P} {541, A63}

\bibitem[\protect\citeauthoryear{{Peretto} et~al.,}{{Peretto}
  et~al.}{2013}]{Peretto+2013}
{Peretto} N.,  et~al., 2013, \mn@doi [\aap] {10.1051/0004-6361/201321318},
  \href {http://adsabs.harvard.edu/abs/2013A%26A...555A.112P} {555, A112}

\bibitem[\protect\citeauthoryear{{Pilbratt} et~al.,}{{Pilbratt}
  et~al.}{2010}]{Pilbratt2010}
{Pilbratt} G.~L.,  et~al., 2010, \mn@doi [A{\&}A]
  {10.1051/0004-6361/201014759}, \href
  {http://cdsads.u-strasbg.fr/abs/2010A%26A...518L...1P} {518, L1+}

\bibitem[\protect\citeauthoryear{{Pineda} et~al.,}{{Pineda}
  et~al.}{2022}]{Pineda+2022}
{Pineda} J.~E.,  et~al., 2022, arXiv e-prints, \href
  {https://ui.adsabs.harvard.edu/abs/2022arXiv220503935P} {p. arXiv:2205.03935}

\bibitem[\protect\citeauthoryear{{Planck Collaboration} et~al.,}{{Planck
  Collaboration} et~al.}{2016}]{PlanckXXXV2016}
{Planck Collaboration} et~al., 2016, \mn@doi [\aap]
  {10.1051/0004-6361/201525896}, \href
  {https://ui.adsabs.harvard.edu/abs/2016A&A...586A.138P} {586, A138}

\bibitem[\protect\citeauthoryear{{Rayner} et~al.,}{{Rayner}
  et~al.}{2017}]{Rayner+2017}
{Rayner} T.~S.~M.,  et~al., 2017, \mn@doi [\aap] {10.1051/0004-6361/201630039},
  \href {http://adsabs.harvard.edu/abs/2017A%26A...607A..22R} {607, A22}

\bibitem[\protect\citeauthoryear{{Rygl}, {Wyrowski}, {Schuller}  \&
  {Menten}}{{Rygl} et~al.}{2013}]{Rygl+2013}
{Rygl} K.~L.~J.,  {Wyrowski} F.,  {Schuller} F.,   {Menten} K.~M.,  2013,
  \mn@doi [\aap] {10.1051/0004-6361/201219574}, \href
  {https://ui.adsabs.harvard.edu/abs/2013A&A...549A...5R} {549, A5}

\bibitem[\protect\citeauthoryear{{Sadavoy}, {Shirley}, {Di Francesco},
  {Henning}, {Currie}, {Andr{\'e}}  \& {Pezzuto}}{{Sadavoy}
  et~al.}{2015}]{Sadavoy+2015}
{Sadavoy} S.~I.,  {Shirley} Y.,  {Di Francesco} J.,  {Henning} T.,  {Currie}
  M.~J.,  {Andr{\'e}} P.,   {Pezzuto} S.,  2015, \mn@doi [\apj]
  {10.1088/0004-637X/806/1/38}, \href
  {https://ui.adsabs.harvard.edu/abs/2015ApJ...806...38S} {806, 38}

\bibitem[\protect\citeauthoryear{{Sault}, {Teuben}  \& {Wright}}{{Sault}
  et~al.}{1995}]{Sault+1995}
{Sault} R.~J.,  {Teuben} P.~J.,   {Wright} M.~C.~H.,  1995, in {Shaw} R.~A.,
  {Payne} H.~E.,   {Hayes} J.~J.~E.,  eds,  Astronomical Society of the Pacific
  Conference Series Vol. 77, Astronomical Data Analysis Software and Systems
  IV. p.~433 (\mn@eprint {arXiv} {astro-ph/0612759})

\bibitem[\protect\citeauthoryear{{Schlafly} et~al.,}{{Schlafly}
  et~al.}{2014}]{Schlafly+2014}
{Schlafly} E.~F.,  et~al., 2014, \mn@doi [\apj] {10.1088/0004-637X/786/1/29},
  \href {https://ui.adsabs.harvard.edu/abs/2014ApJ...786...29S} {786, 29}

\bibitem[\protect\citeauthoryear{{Schneider} et~al.,}{{Schneider}
  et~al.}{2010}]{Schneider+2010}
{Schneider} N.,  et~al., 2010, \mn@doi [\aap] {10.1051/0004-6361/201014627},
  \href {http://adsabs.harvard.edu/abs/2010A%26A...518L..83S} {518, L83}

\bibitem[\protect\citeauthoryear{{Schneider} et~al.,}{{Schneider}
  et~al.}{2012}]{Schneider+2012}
{Schneider} N.,  et~al., 2012, \mn@doi [\aap] {10.1051/0004-6361/201118566},
  \href {http://adsabs.harvard.edu/abs/2012A%26A...540L..11S} {540, L11}

\bibitem[\protect\citeauthoryear{{Shimajiri} et~al.,}{{Shimajiri}
  et~al.}{2017}]{Shimajiri+2017}
{Shimajiri} Y.,  et~al., 2017, \mn@doi [\aap] {10.1051/0004-6361/201730633},
  \href {https://ui.adsabs.harvard.edu/abs/2017A&A...604A..74S} {604, A74}

\bibitem[\protect\citeauthoryear{{Skrutskie} et~al.,}{{Skrutskie}
  et~al.}{2006}]{Skrutskie+2006}
{Skrutskie} M.~F.,  et~al., 2006, \mn@doi [\aj] {10.1086/498708}, \href
  {https://ui.adsabs.harvard.edu/abs/2006AJ....131.1163S} {131, 1163}

\bibitem[\protect\citeauthoryear{{Snell}}{{Snell}}{1989}]{Snell1989}
{Snell} R.~L.,  1989, {Molecular Outflows}.
p.~231, \mn@doi{10.1007/BFb0114872}

\bibitem[\protect\citeauthoryear{{Soler}}{{Soler}}{2019}]{Soler2019}
{Soler} J.~D.,  2019, \mn@doi [\aap] {10.1051/0004-6361/201935779}, \href
  {https://ui.adsabs.harvard.edu/abs/2019A&A...629A..96S} {629, A96}

\bibitem[\protect\citeauthoryear{{Soler} \& {Hennebelle}}{{Soler} \&
  {Hennebelle}}{2017}]{SolerHennebelle2017}
{Soler} J.~D.,  {Hennebelle} P.,  2017, \mn@doi [\aap]
  {10.1051/0004-6361/201731049}, \href
  {https://ui.adsabs.harvard.edu/abs/2017A&A...607A...2S} {607, A2}

\bibitem[\protect\citeauthoryear{{Soler}, {Hennebelle}, {Martin},
  {Miville-Desch{\^e}nes}, {Netterfield}  \& {Fissel}}{{Soler}
  et~al.}{2013}]{Soler+2013}
{Soler} J.~D.,  {Hennebelle} P.,  {Martin} P.~G.,  {Miville-Desch{\^e}nes}
  M.~A.,  {Netterfield} C.~B.,   {Fissel} L.~M.,  2013, \mn@doi [\apj]
  {10.1088/0004-637X/774/2/128}, \href
  {https://ui.adsabs.harvard.edu/abs/2013ApJ...774..128S} {774, 128}

\bibitem[\protect\citeauthoryear{{Sousbie}}{{Sousbie}}{2011}]{Sousbie2011}
{Sousbie} T.,  2011, \mn@doi [\mnras] {10.1111/j.1365-2966.2011.18394.x}, \href
  {http://adsabs.harvard.edu/abs/2011MNRAS.414..350S} {414, 350}

\bibitem[\protect\citeauthoryear{{Tahani}, {Plume}, {Brown}  \&
  {Kainulainen}}{{Tahani} et~al.}{2018}]{Tahani+2018}
{Tahani} M.,  {Plume} R.,  {Brown} J.~C.,   {Kainulainen} J.,  2018, \mn@doi
  [\aap] {10.1051/0004-6361/201732219}, \href
  {https://ui.adsabs.harvard.edu/abs/2018A&A...614A.100T} {614, A100}

\bibitem[\protect\citeauthoryear{{Tenorio-Tagle}}{{Tenorio-Tagle}}{1982}]{Tenorio-Tagle1982}
{Tenorio-Tagle} G.,  1982, in {Roger} R.~S.,  {Dewdney} P.~E.,  eds,
  Astrophysics and Space Science Library Vol. 93, Regions of Recent Star
  Formation. pp 1--13, \mn@doi{10.1007/978-94-009-7778-5\_1}

\bibitem[\protect\citeauthoryear{{Traficante}, {Fuller}, {Billot},
  {Duarte-Cabral}, {Merello}, {Molinari}, {Peretto}  \&
  {Schisano}}{{Traficante} et~al.}{2017}]{Traficante+2017}
{Traficante} A.,  {Fuller} G.~A.,  {Billot} N.,  {Duarte-Cabral} A.,  {Merello}
  M.,  {Molinari} S.,  {Peretto} N.,   {Schisano} E.,  2017, \mn@doi [\mnras]
  {10.1093/mnras/stx1375}, \href
  {https://ui.adsabs.harvard.edu/abs/2017MNRAS.470.3882T} {470, 3882}

\bibitem[\protect\citeauthoryear{{Trevi{\~n}o-Morales}
  et~al.,}{{Trevi{\~n}o-Morales} et~al.}{2019}]{Trevino-Morales+2019}
{Trevi{\~n}o-Morales} S.~P.,  et~al., 2019, \mn@doi [\aap]
  {10.1051/0004-6361/201935260}, \href
  {https://ui.adsabs.harvard.edu/abs/2019A&A...629A..81T} {629, A81}

\bibitem[\protect\citeauthoryear{{Wouterloot}, {Walmsley}  \&
  {Henkel}}{{Wouterloot} et~al.}{1988}]{Wouterloot+1988}
{Wouterloot} J.~G.~A.,  {Walmsley} C.~M.,   {Henkel} C.,  1988, \aap, \href
  {https://ui.adsabs.harvard.edu/abs/1988A&A...203..367W} {203, 367}

\bibitem[\protect\citeauthoryear{{Zucker}, {Speagle}, {Schlafly}, {Green},
  {Finkbeiner}, {Goodman}  \& {Alves}}{{Zucker} et~al.}{2019}]{Zucker+2019}
{Zucker} C.,  {Speagle} J.~S.,  {Schlafly} E.~F.,  {Green} G.~M.,  {Finkbeiner}
  D.~P.,  {Goodman} A.~A.,   {Alves} J.,  2019, \mn@doi [\apj]
  {10.3847/1538-4357/ab2388}, \href
  {https://ui.adsabs.harvard.edu/abs/2019ApJ...879..125Z} {879, 125}

\makeatother
\end{thebibliography}



\appendix

\section{Averaged Spectra}\label{app:spectra}
\begin{figure*}
\subfloat{\includegraphics[width=0.165\textwidth]{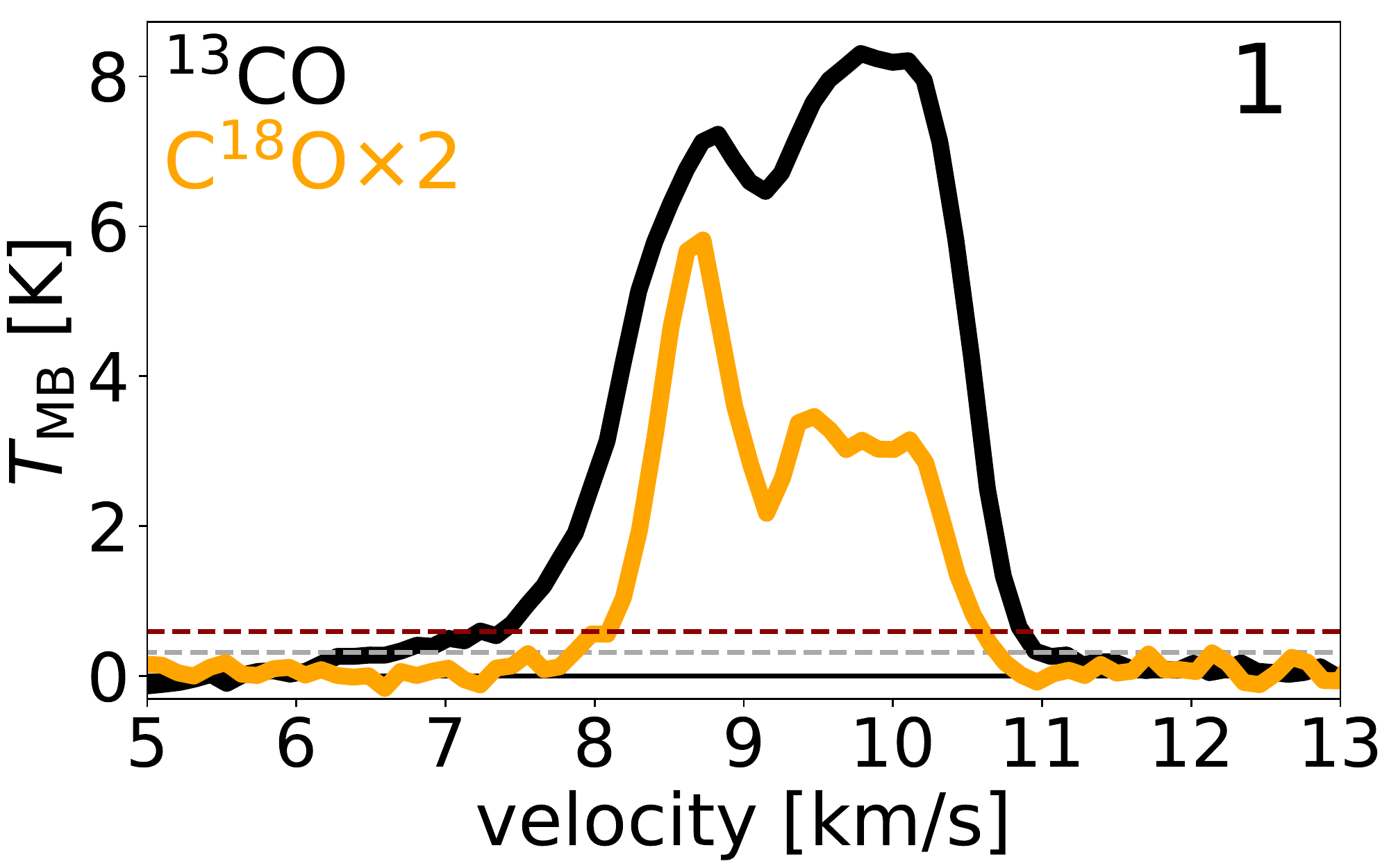}} 
\subfloat{\includegraphics[width=0.165\textwidth]{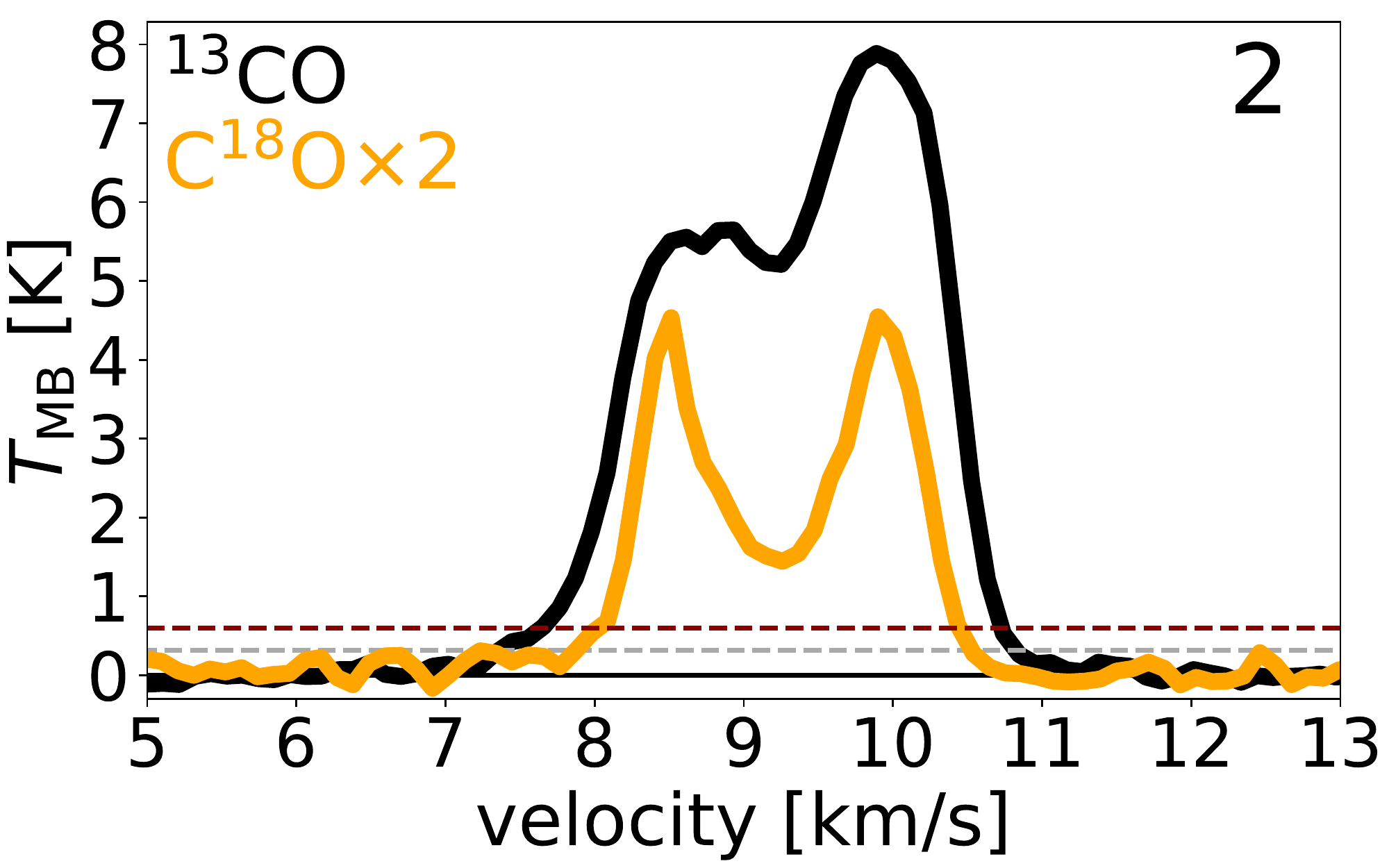}} 
\subfloat{\includegraphics[width=0.165\textwidth]{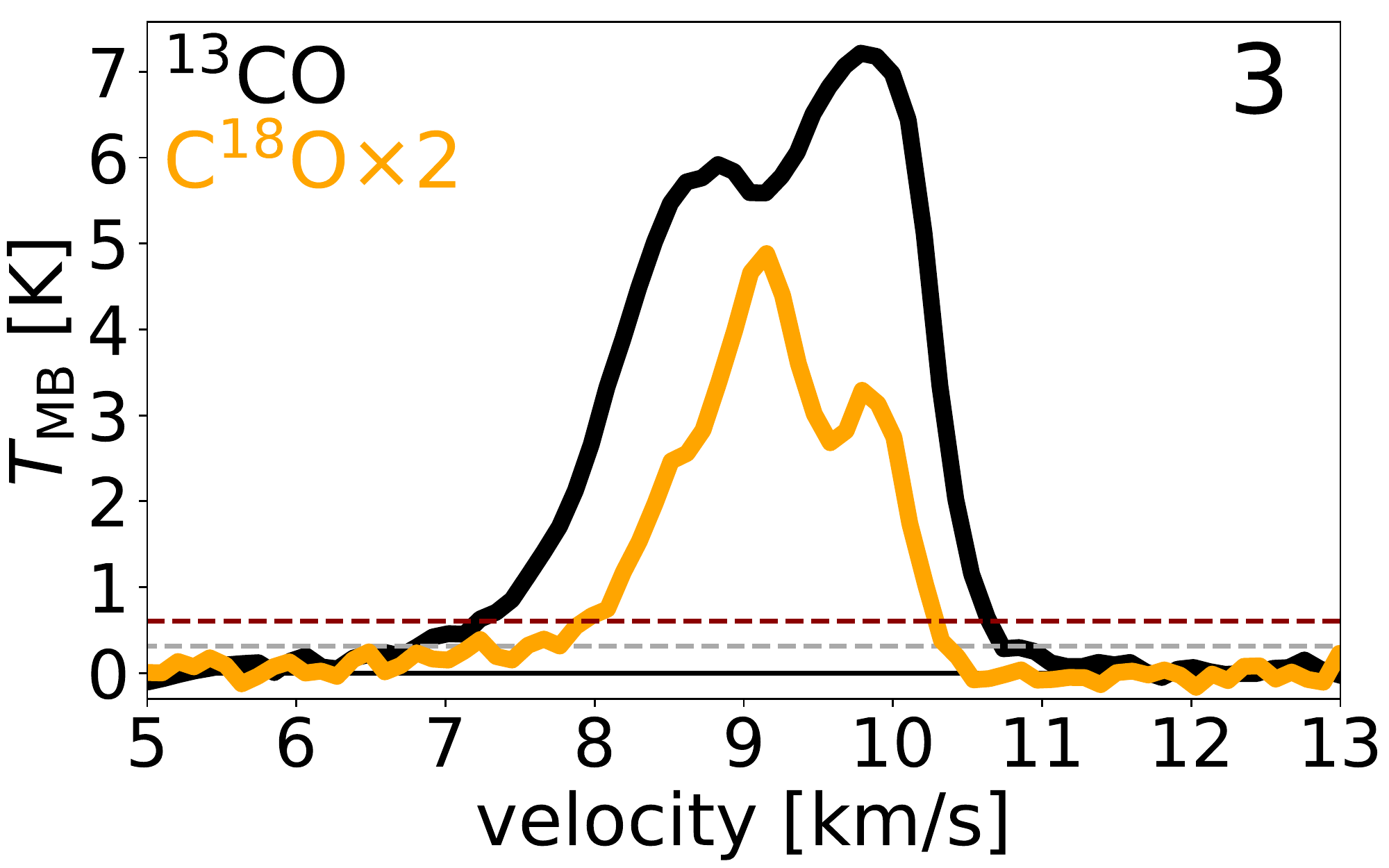}} 
\subfloat{\includegraphics[width=0.165\textwidth]{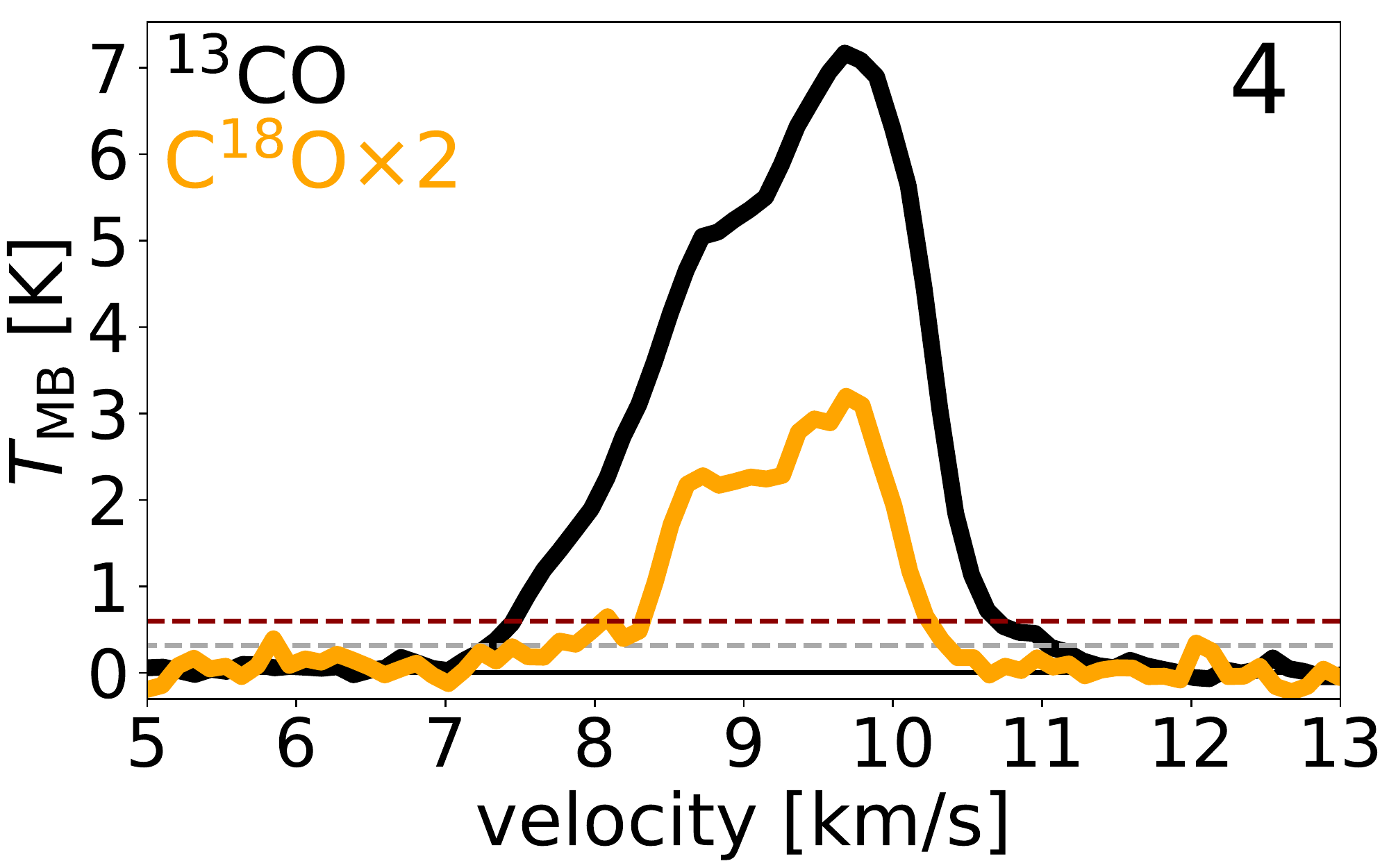}}
\subfloat{\includegraphics[width=0.165\textwidth]{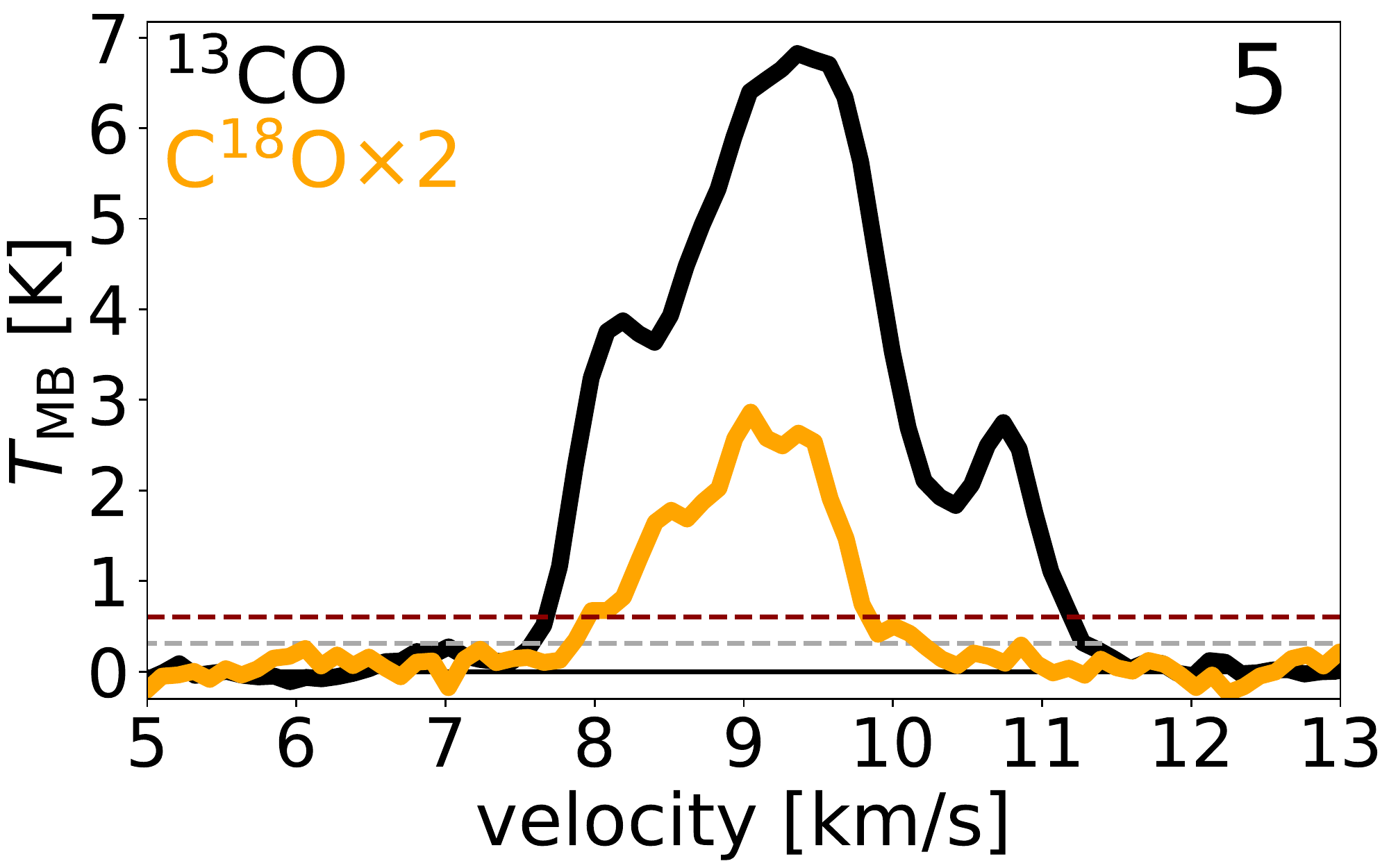}} 
\subfloat{\includegraphics[width=0.165\textwidth]{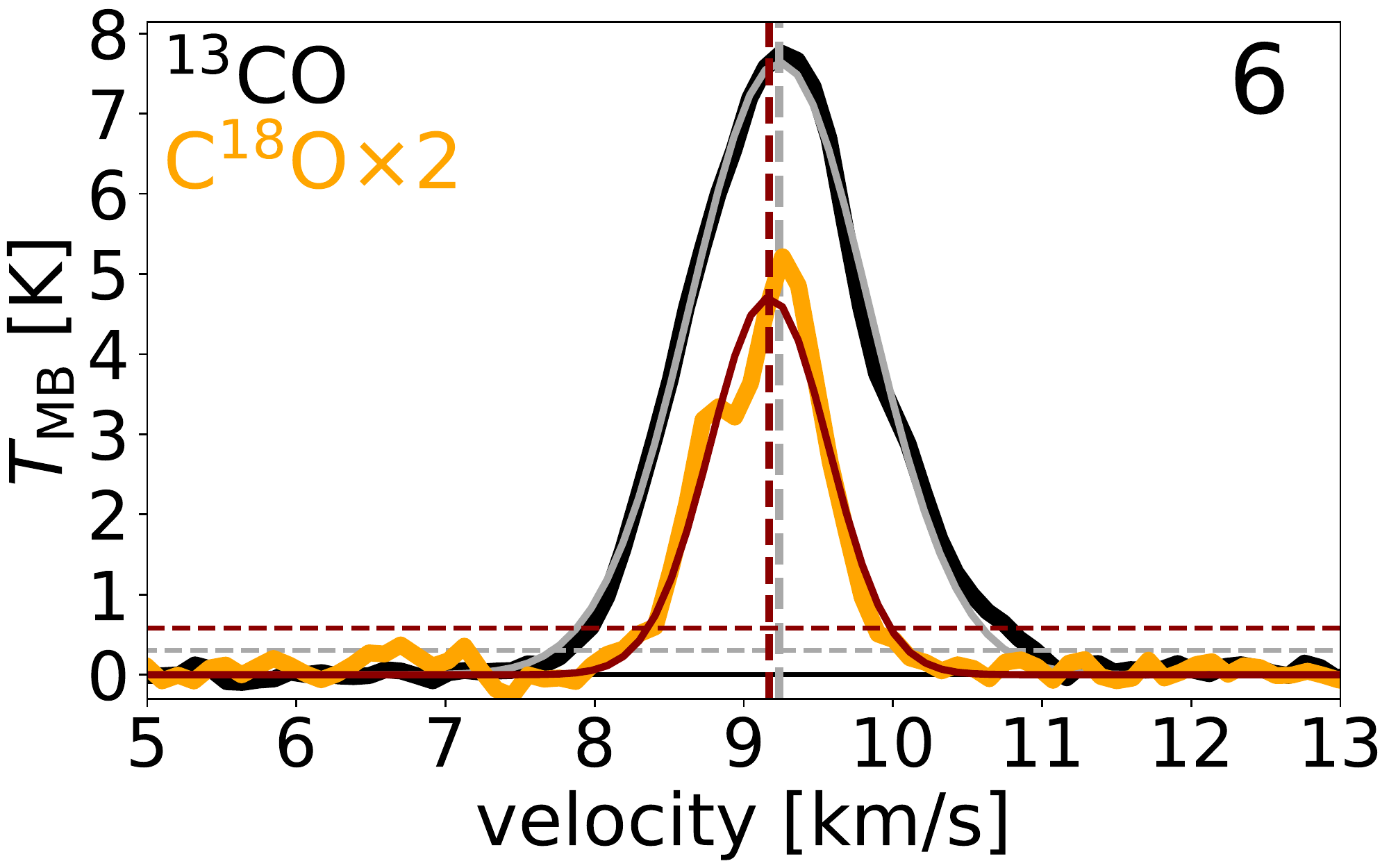}}\\ 
\subfloat{\includegraphics[width=0.165\textwidth]{./figs/13CO_C18O_spectra_cNo7.pdf}} 
\subfloat{\includegraphics[width=0.165\textwidth]{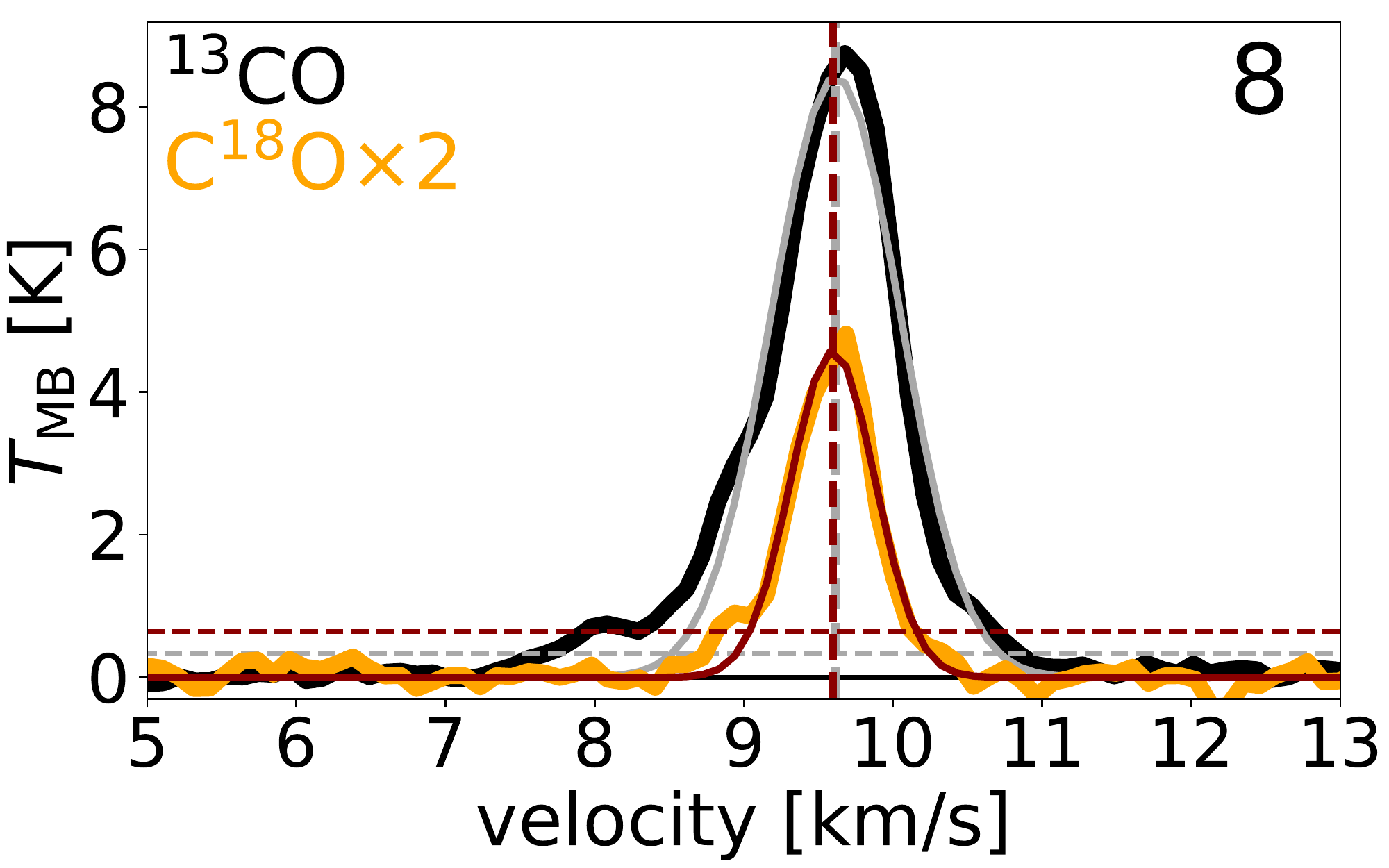}} 
\subfloat{\includegraphics[width=0.165\textwidth]{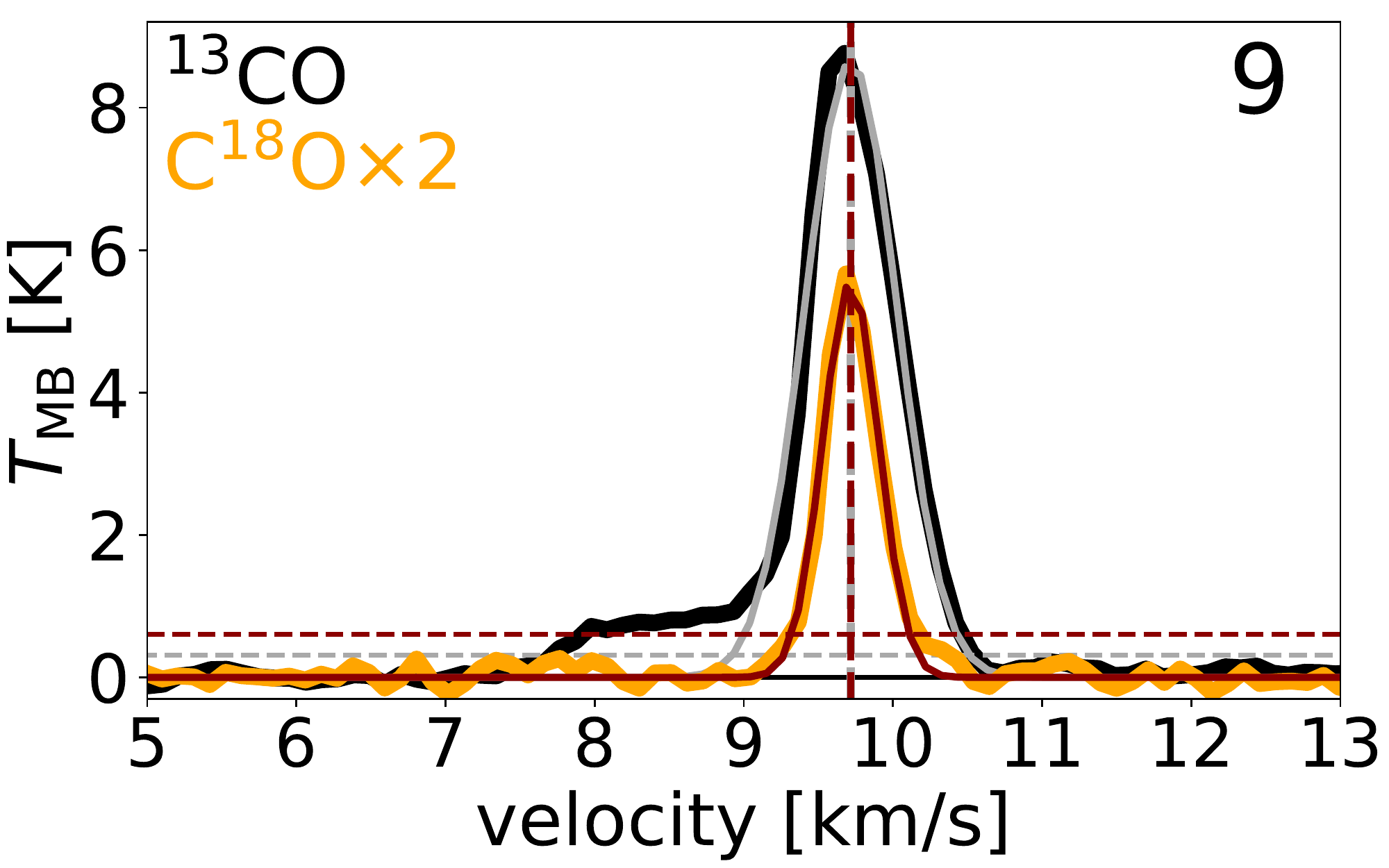}} 
\subfloat{\includegraphics[width=0.165\textwidth]{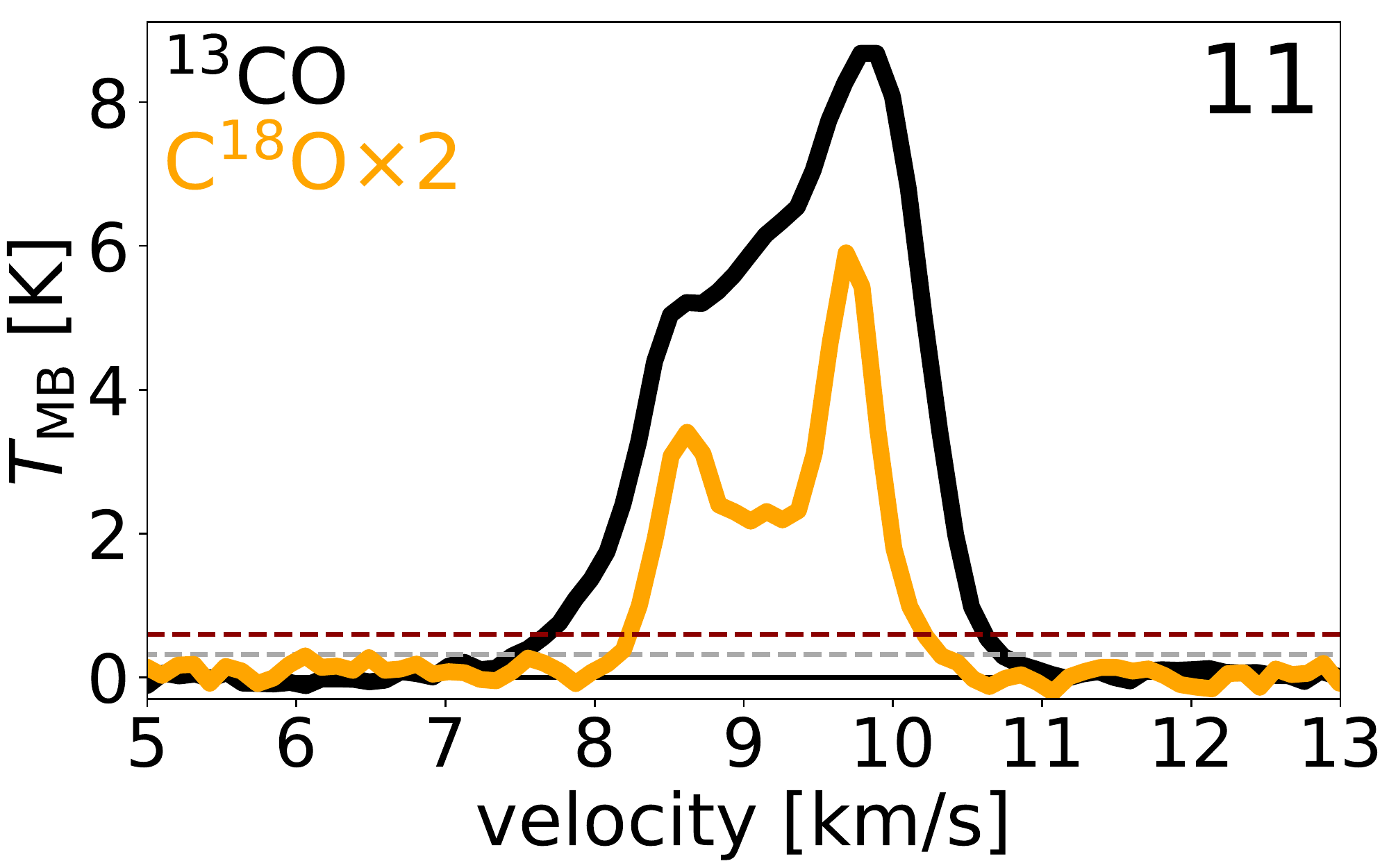}}
\subfloat{\includegraphics[width=0.165\textwidth]{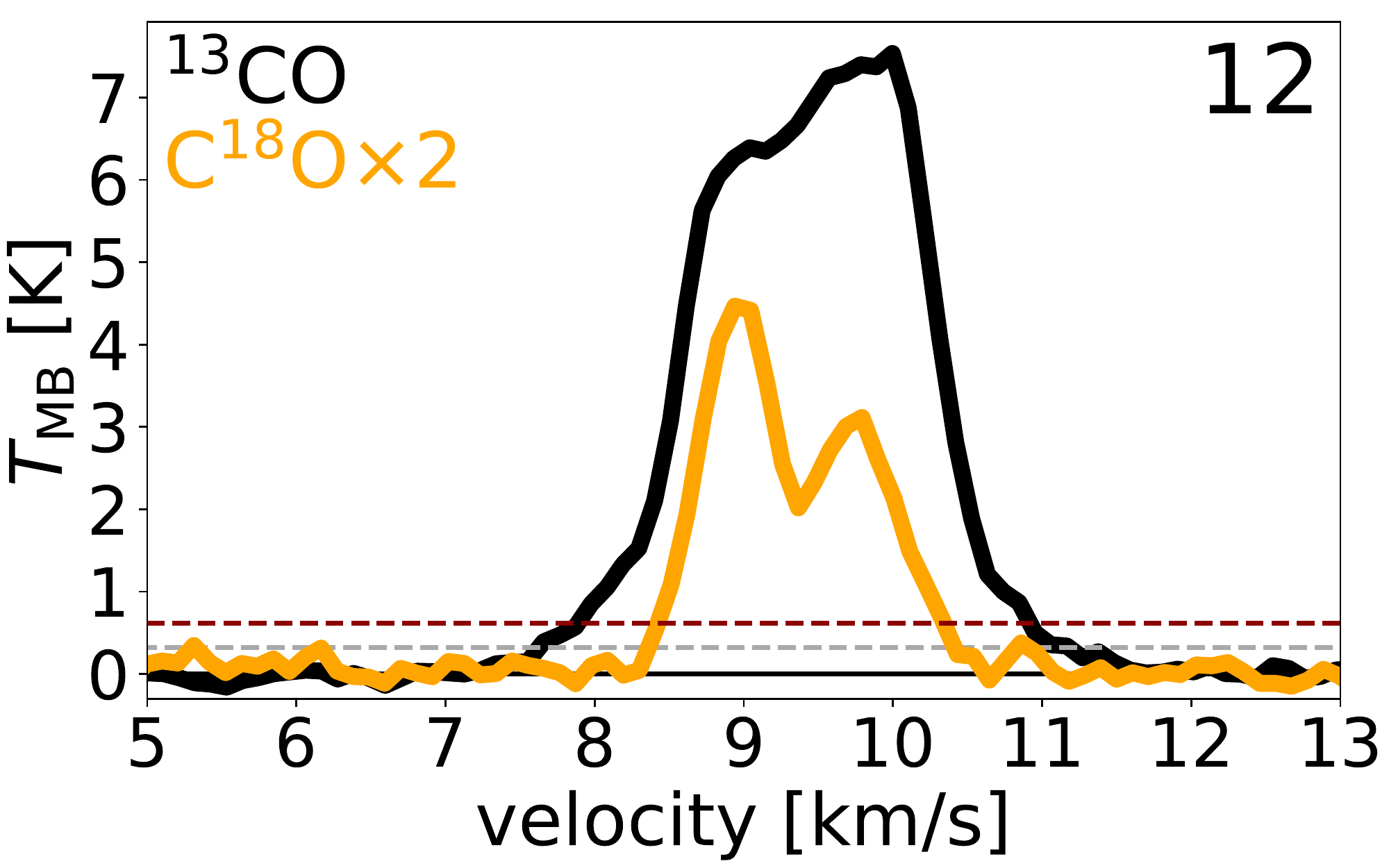}} 
\subfloat{\includegraphics[width=0.165\textwidth]{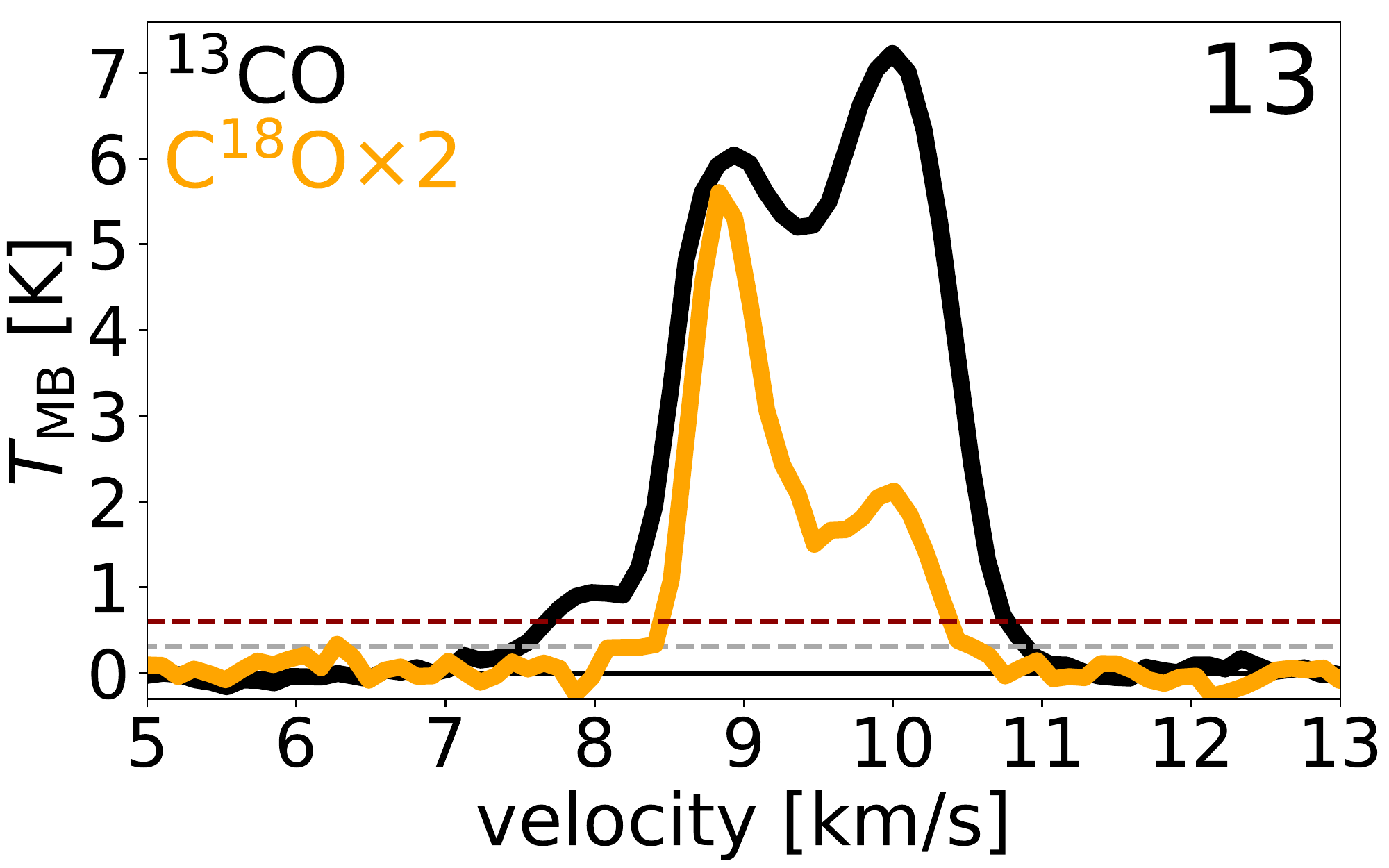}}\\
 
\subfloat{\includegraphics[width=0.165\textwidth]{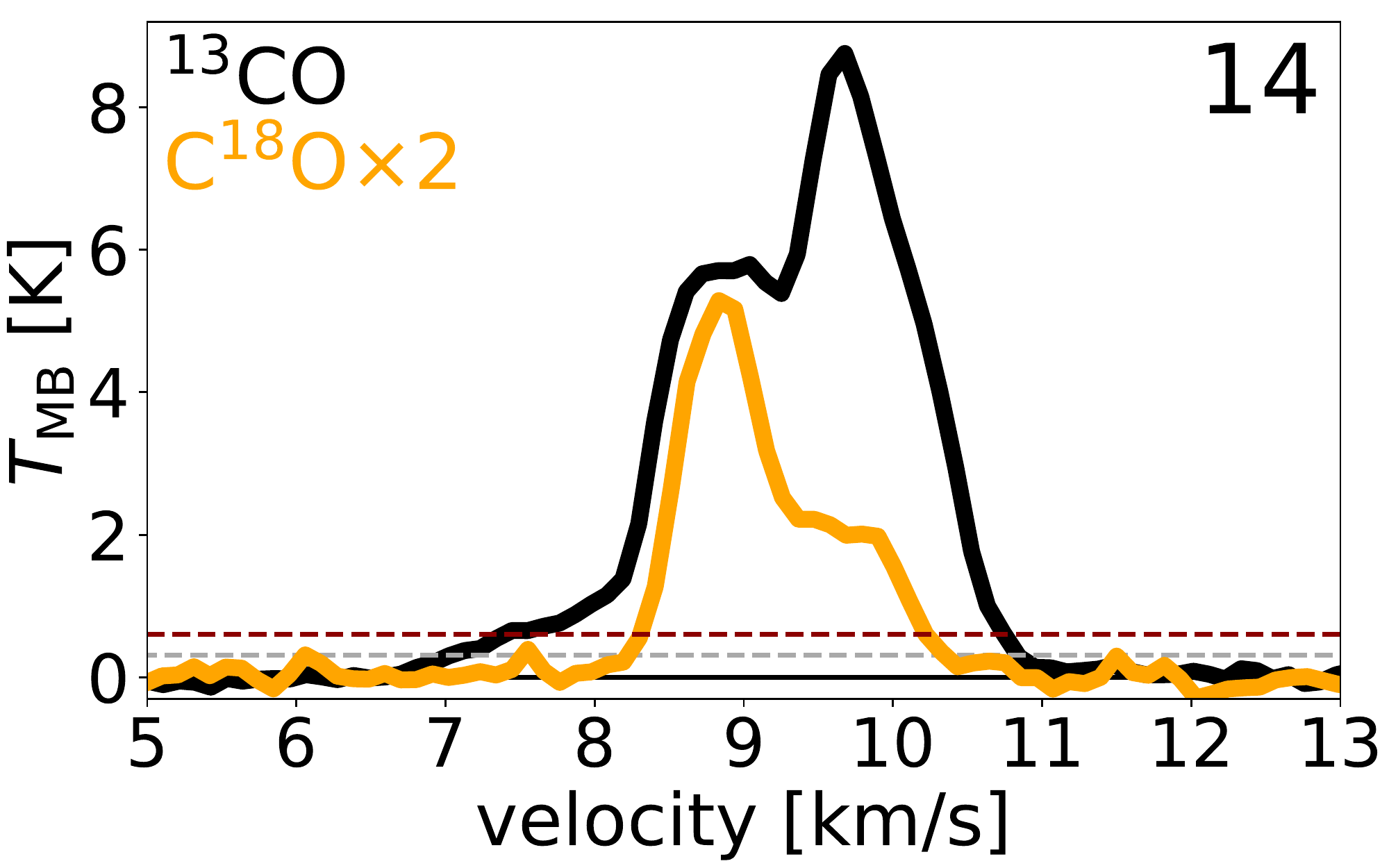}} 
\subfloat{\includegraphics[width=0.165\textwidth]{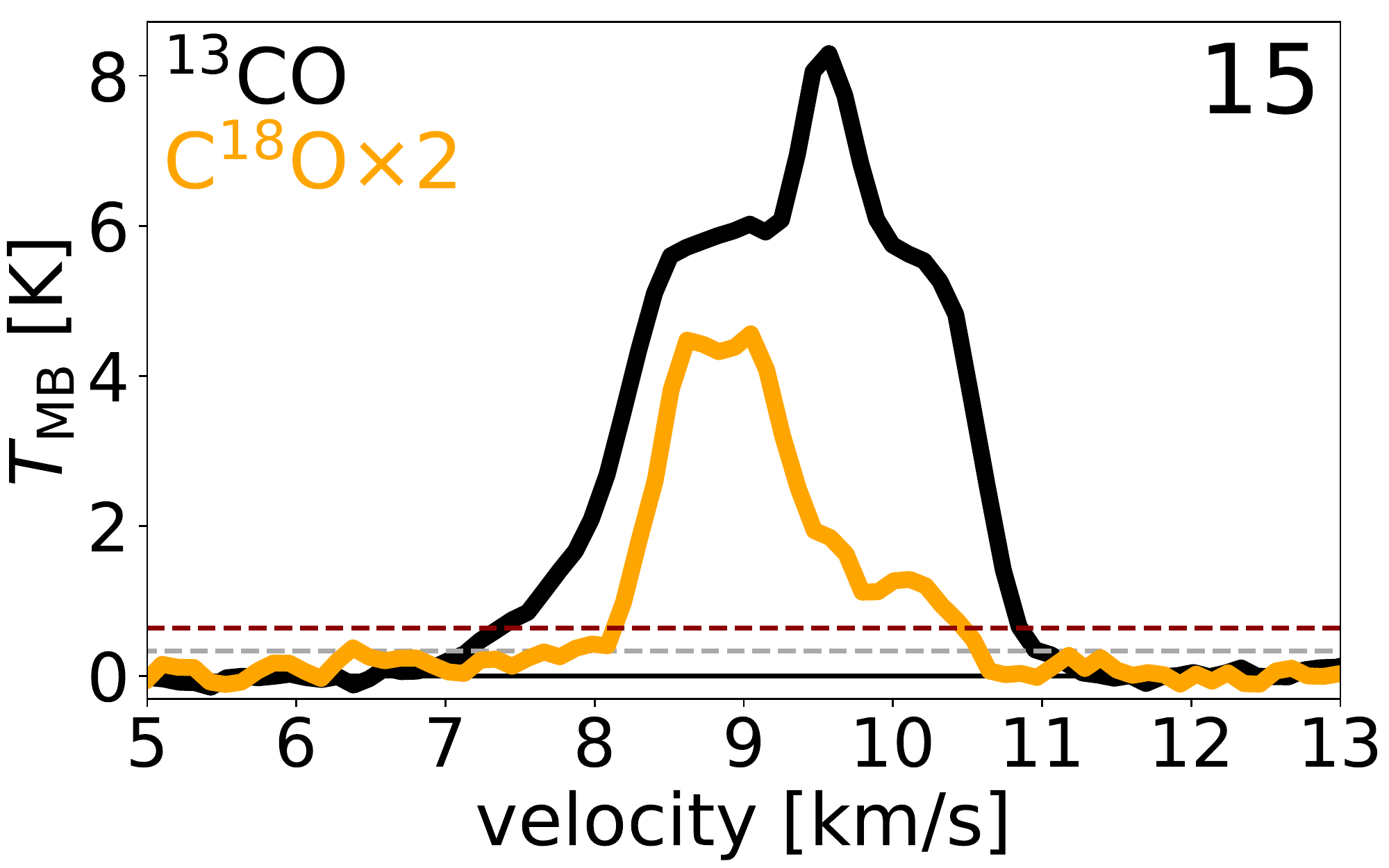}} 
\subfloat{\includegraphics[width=0.165\textwidth]{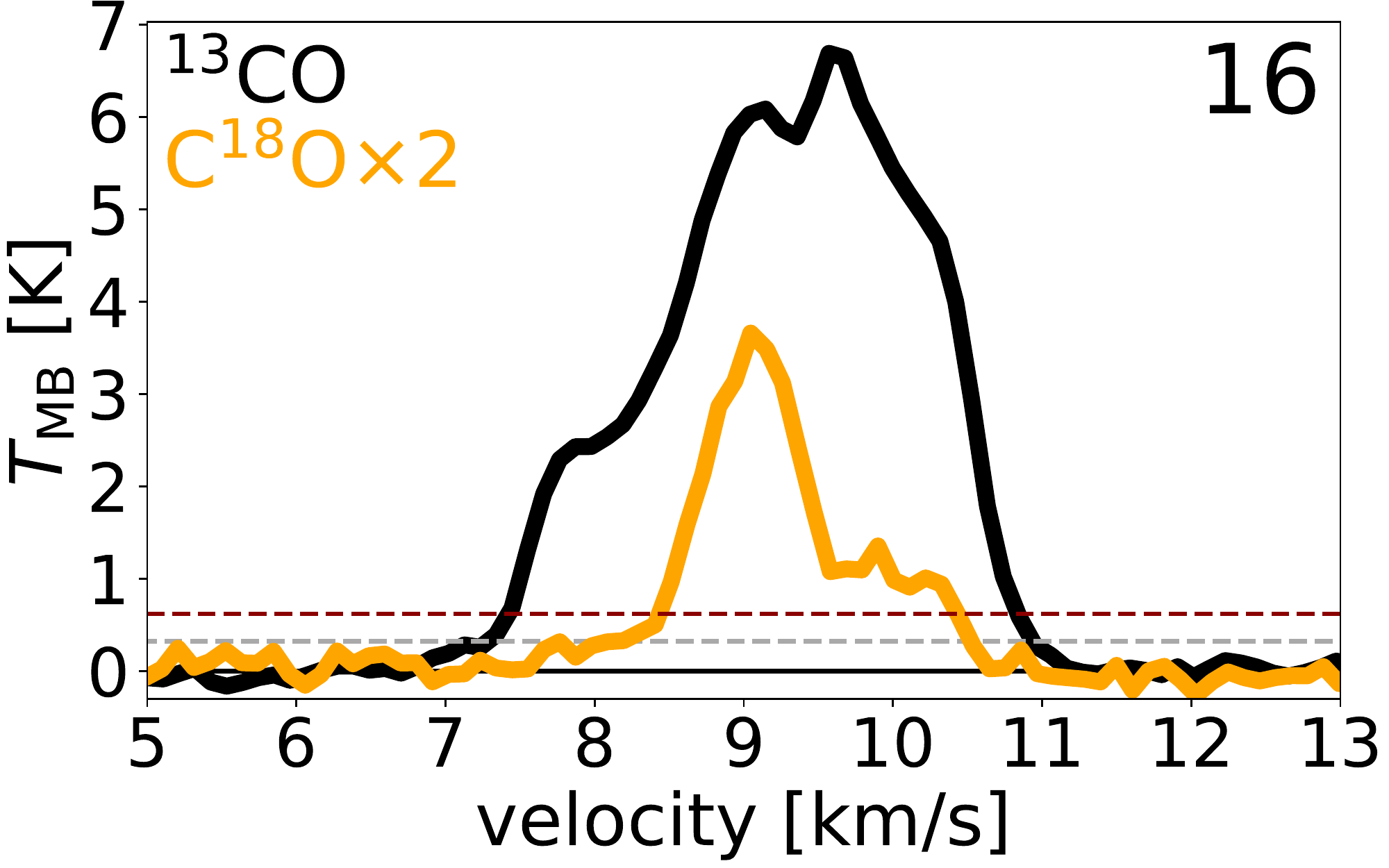}} 
\subfloat{\includegraphics[width=0.165\textwidth]{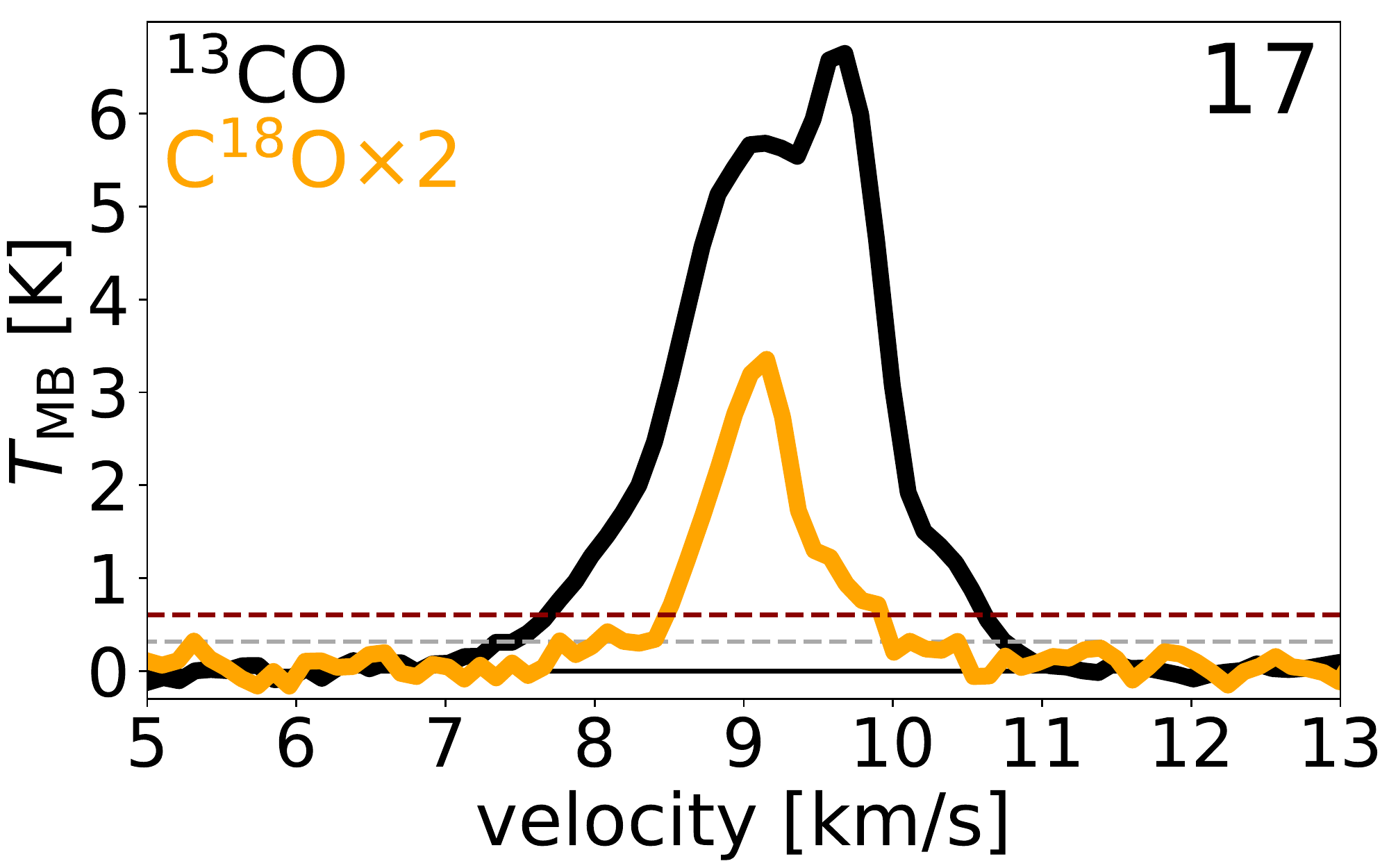}} 
\subfloat{\includegraphics[width=0.165\textwidth]{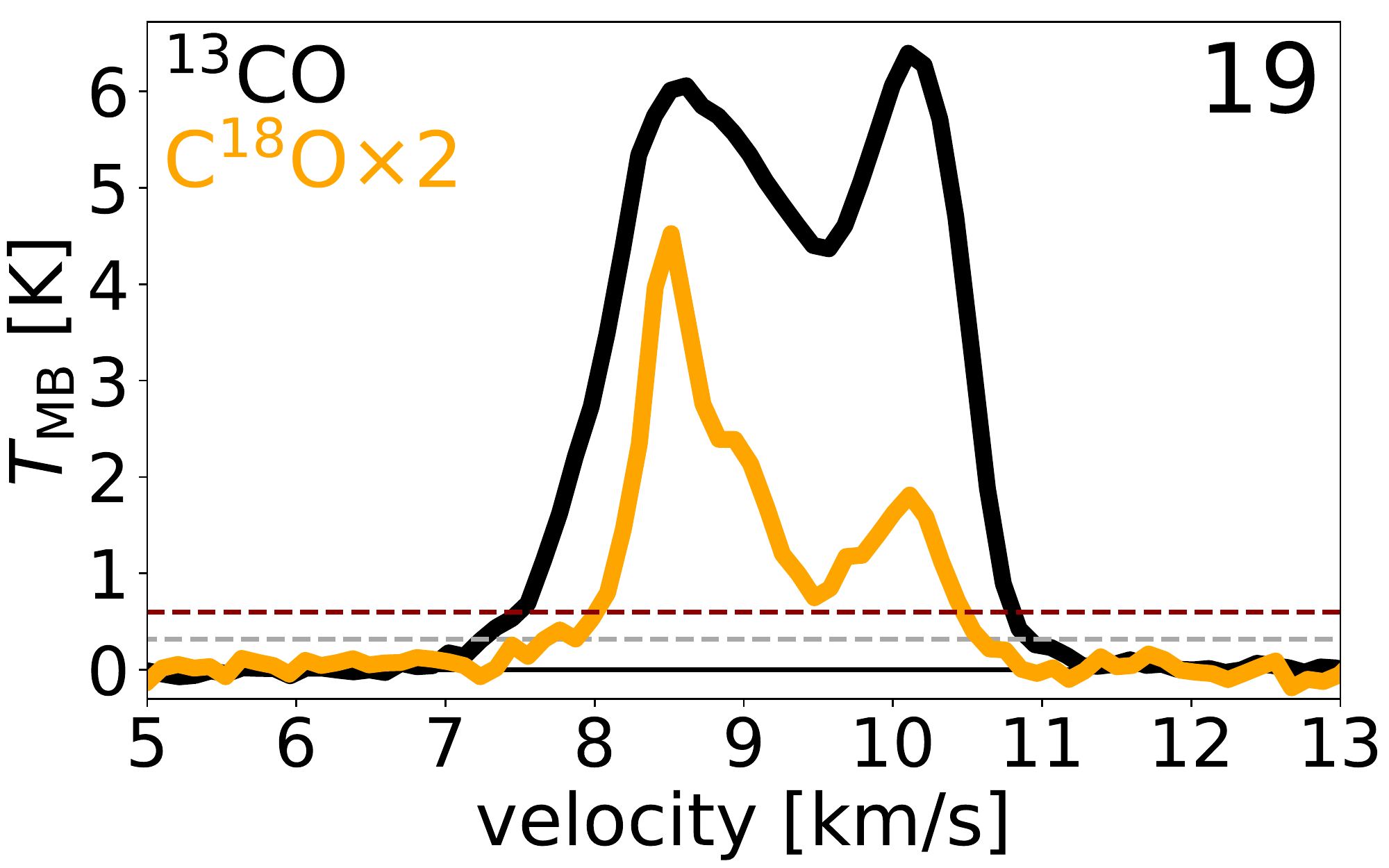}} 
\subfloat{\includegraphics[width=0.165\textwidth]{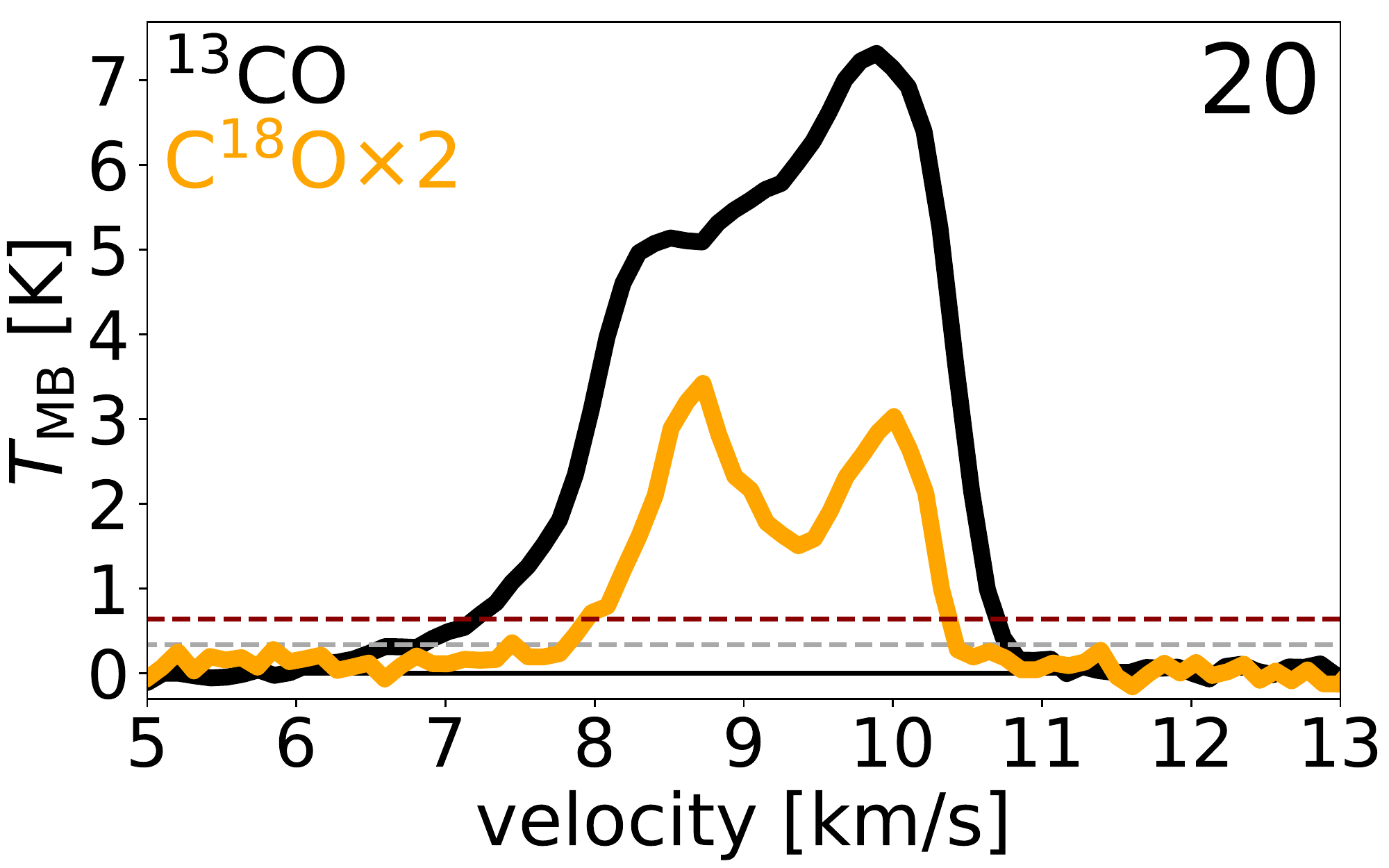}}\\ 
\subfloat{\includegraphics[width=0.165\textwidth]{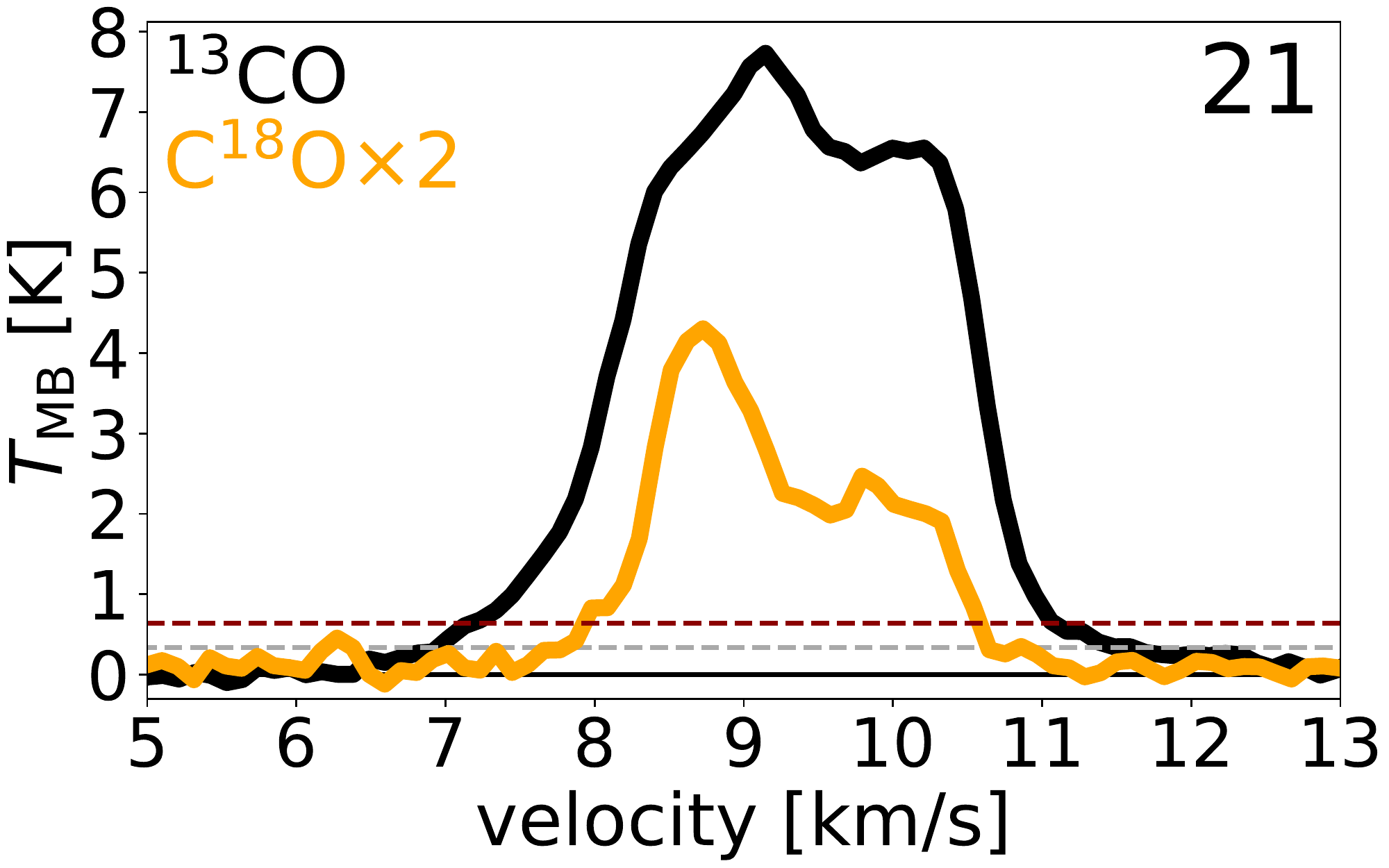}} 
\subfloat{\includegraphics[width=0.165\textwidth]{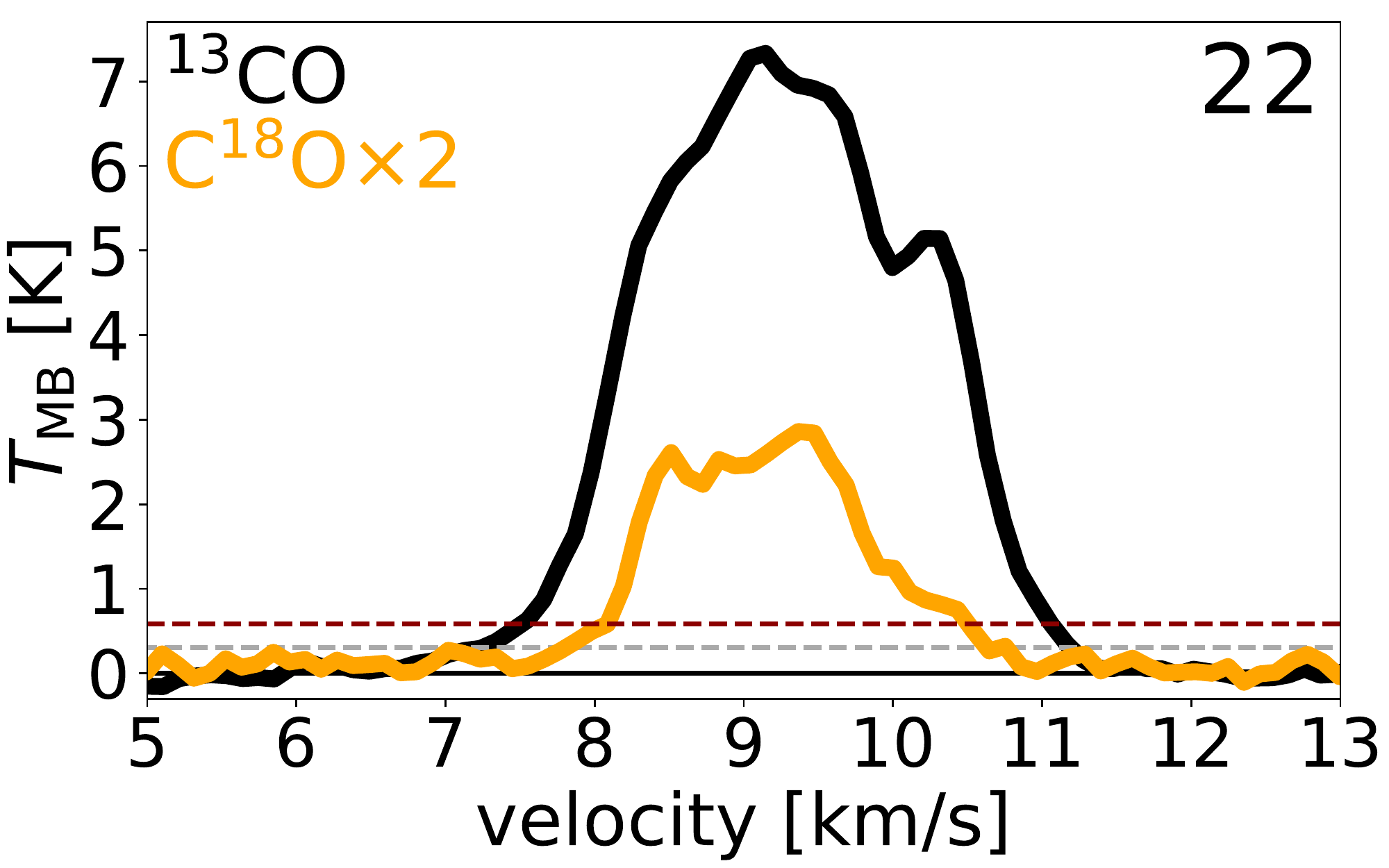}} 
\subfloat{\includegraphics[width=0.165\textwidth]{./figs/13CO_C18O_spectra_noFit_cNo23.pdf}} 
\subfloat{\includegraphics[width=0.165\textwidth]{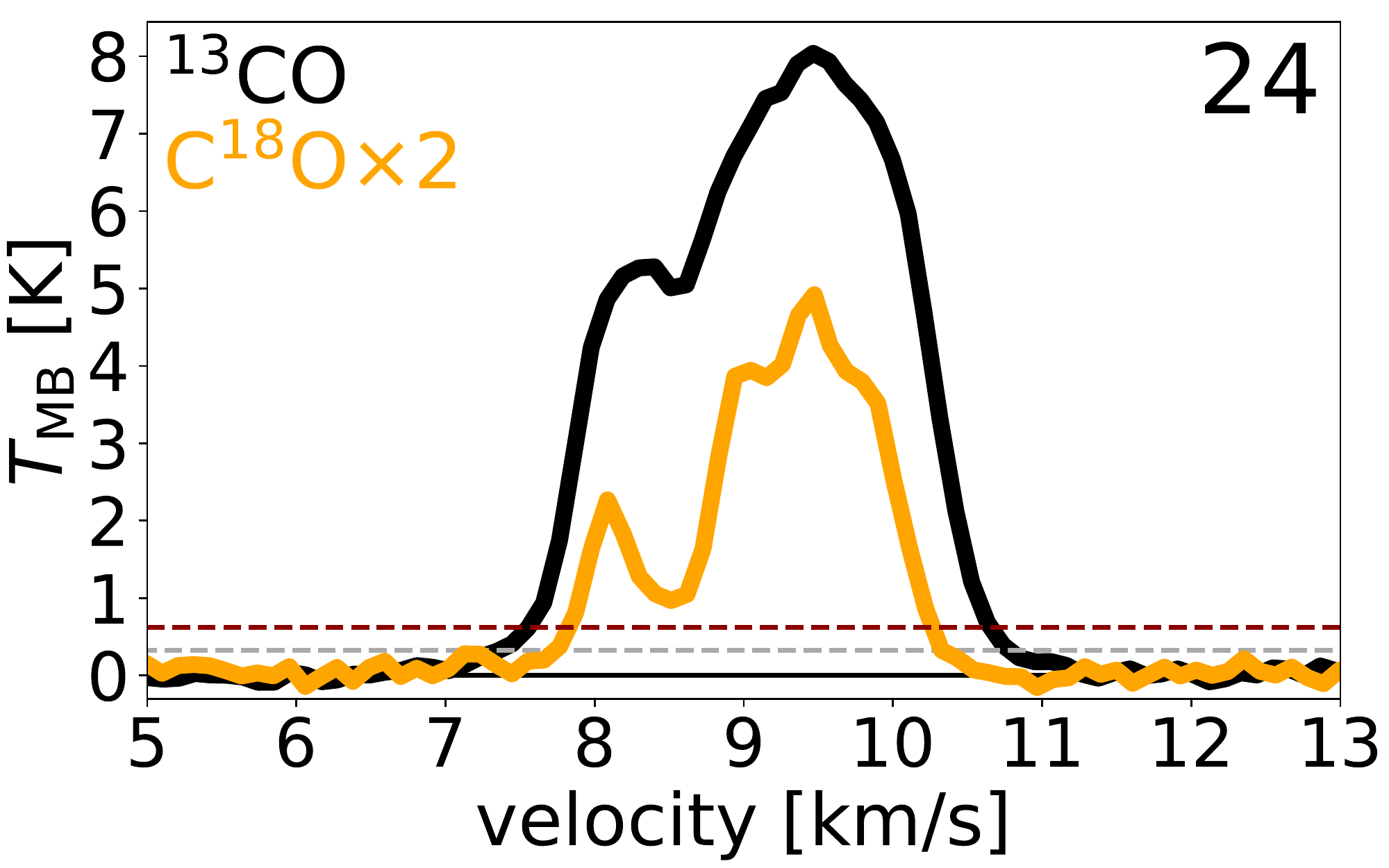}} 
\subfloat{\includegraphics[width=0.165\textwidth]{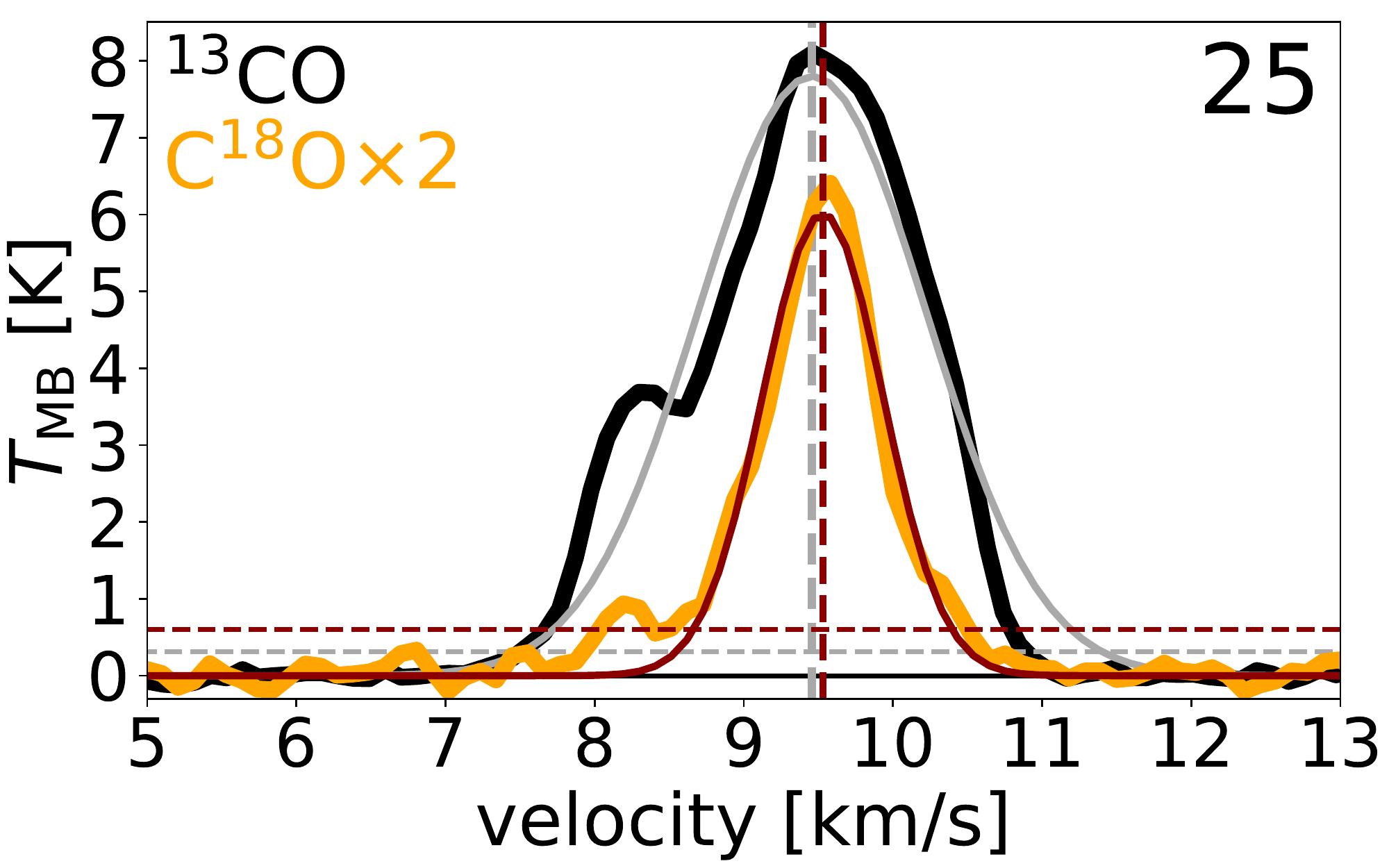}} 
\subfloat{\includegraphics[width=0.165\textwidth]{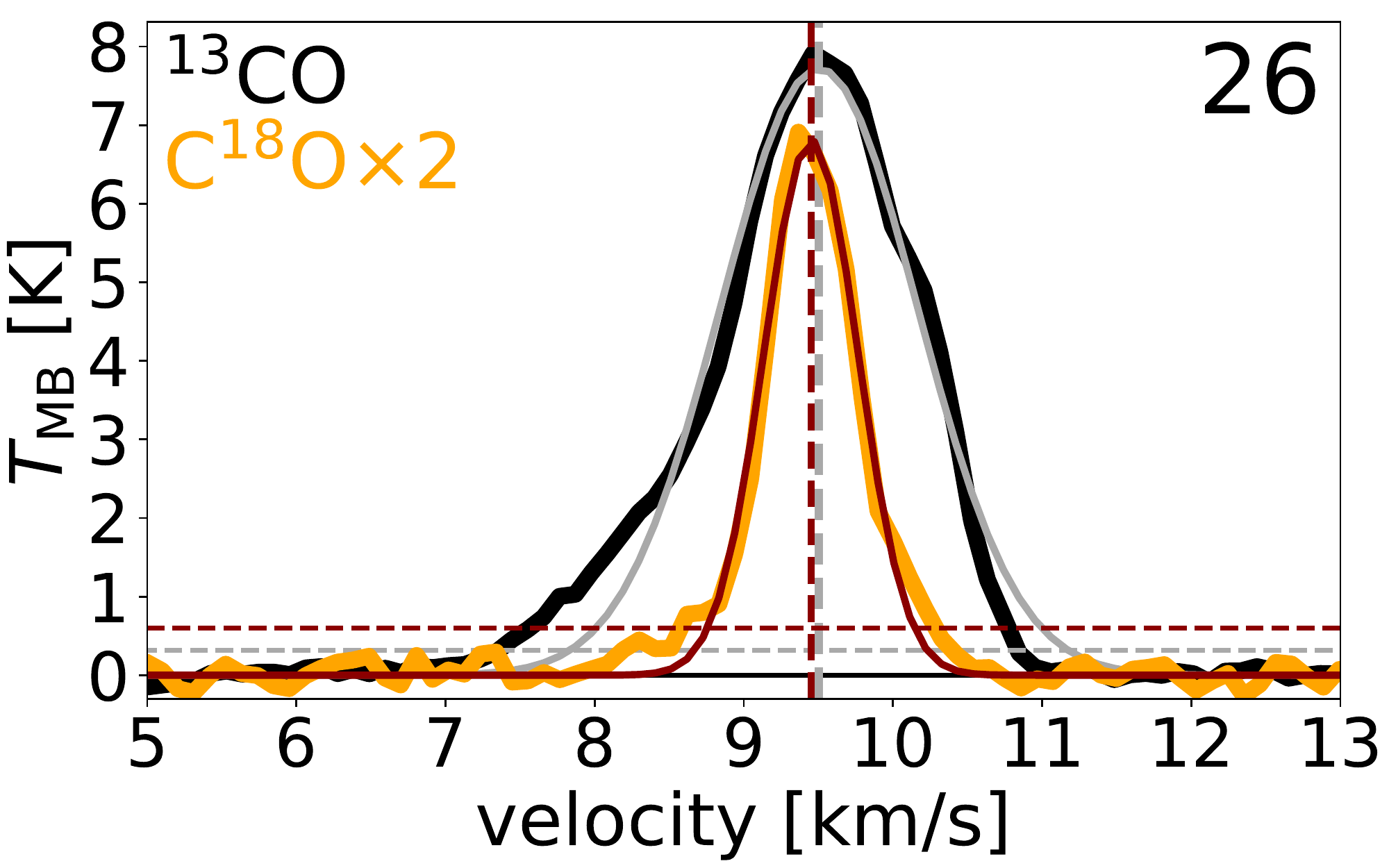}}\\
 
\subfloat{\includegraphics[width=0.165\textwidth]{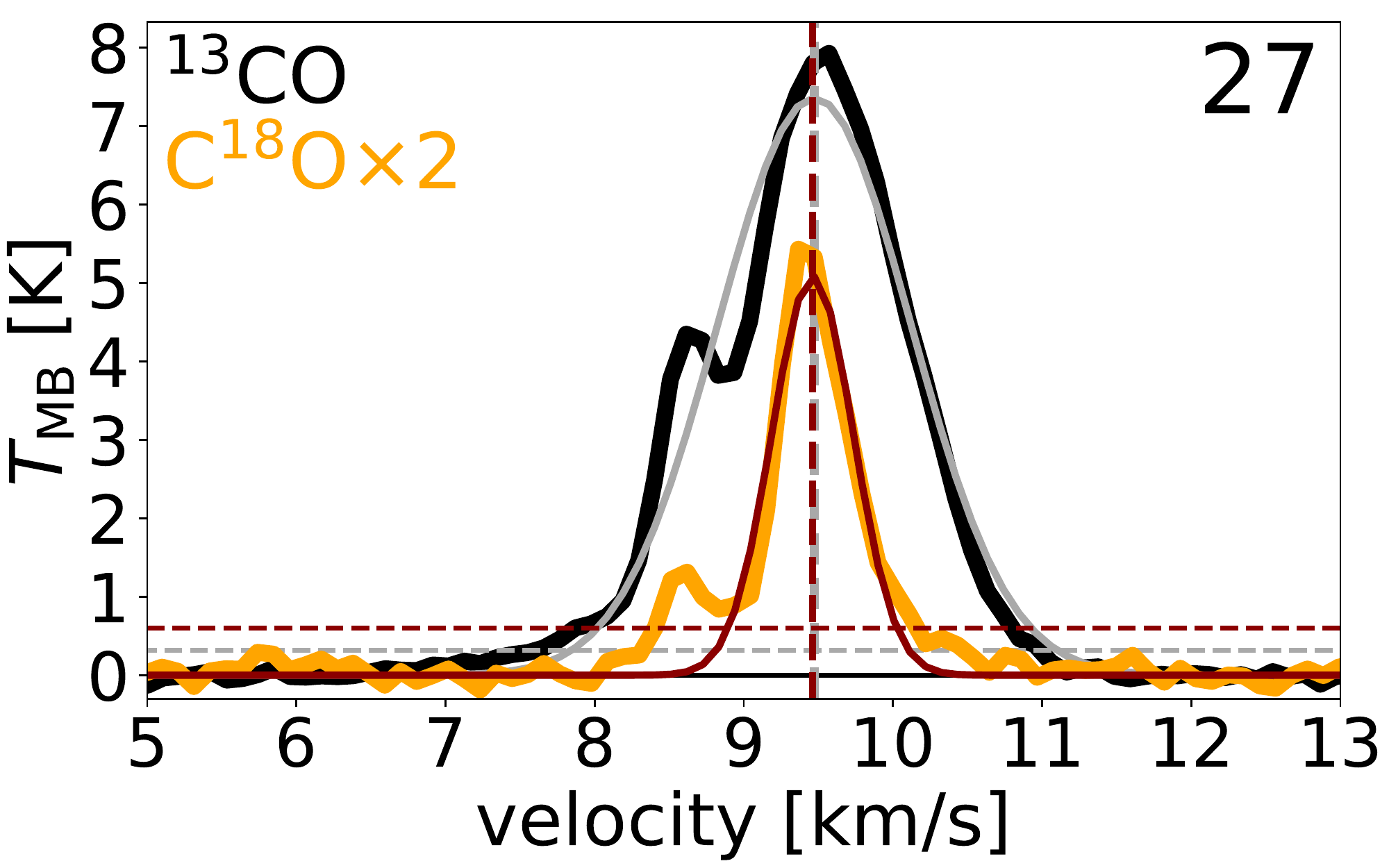}} 
\subfloat{\includegraphics[width=0.165\textwidth]{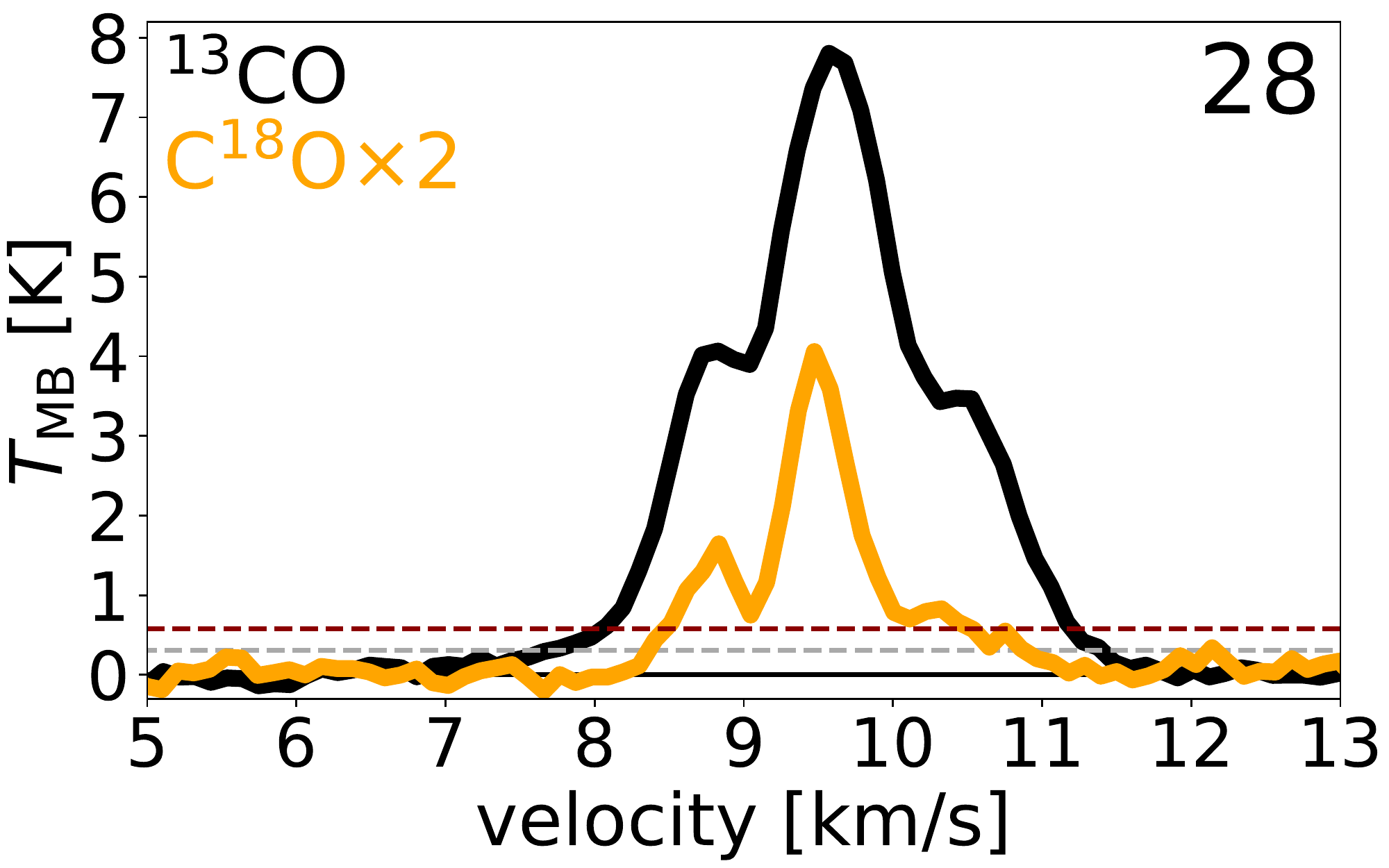}} 
\subfloat{\includegraphics[width=0.165\textwidth]{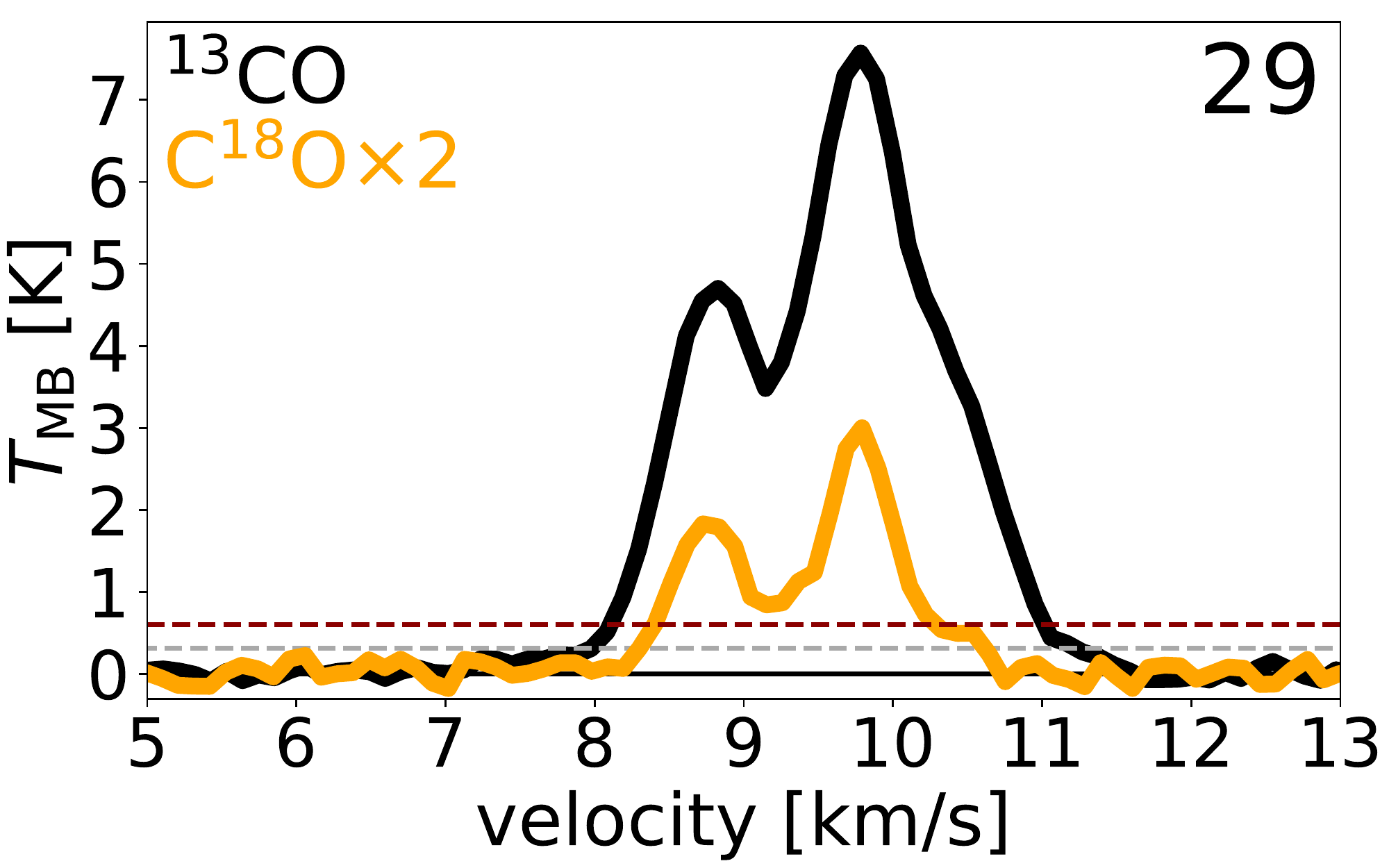}} 
\subfloat{\includegraphics[width=0.165\textwidth]{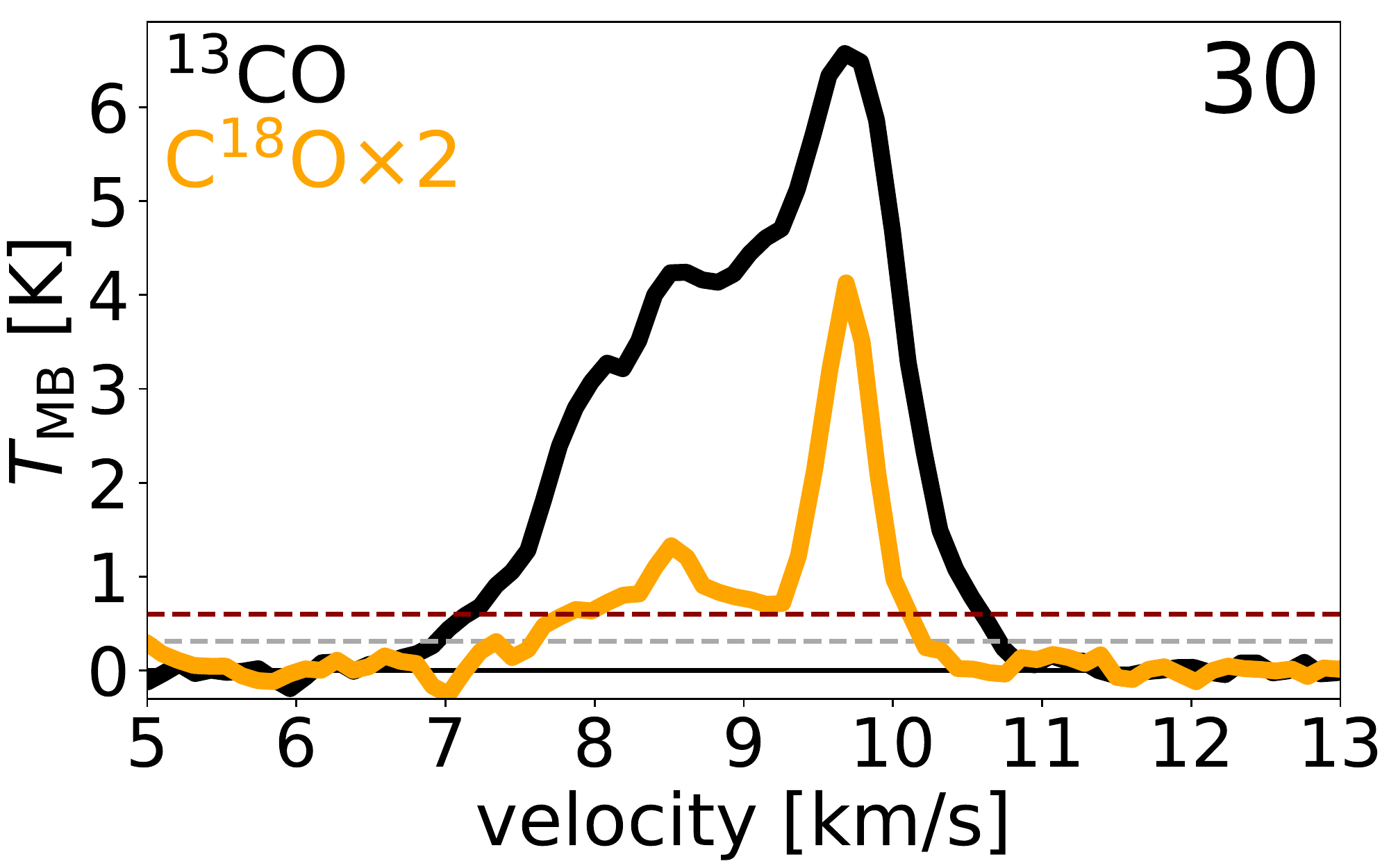}}
\subfloat{\includegraphics[width=0.165\textwidth]{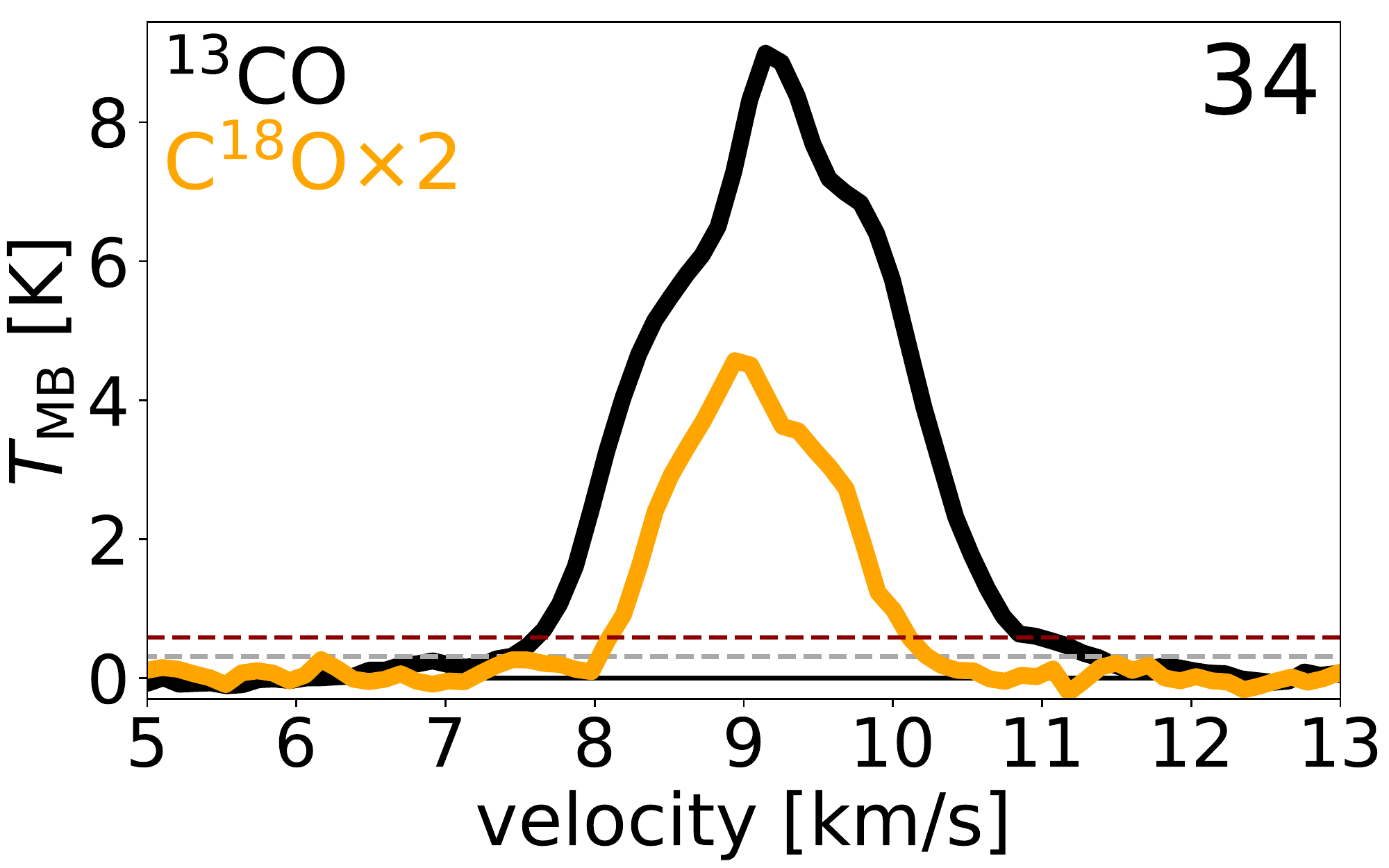}} 
\subfloat{\includegraphics[width=0.165\textwidth]{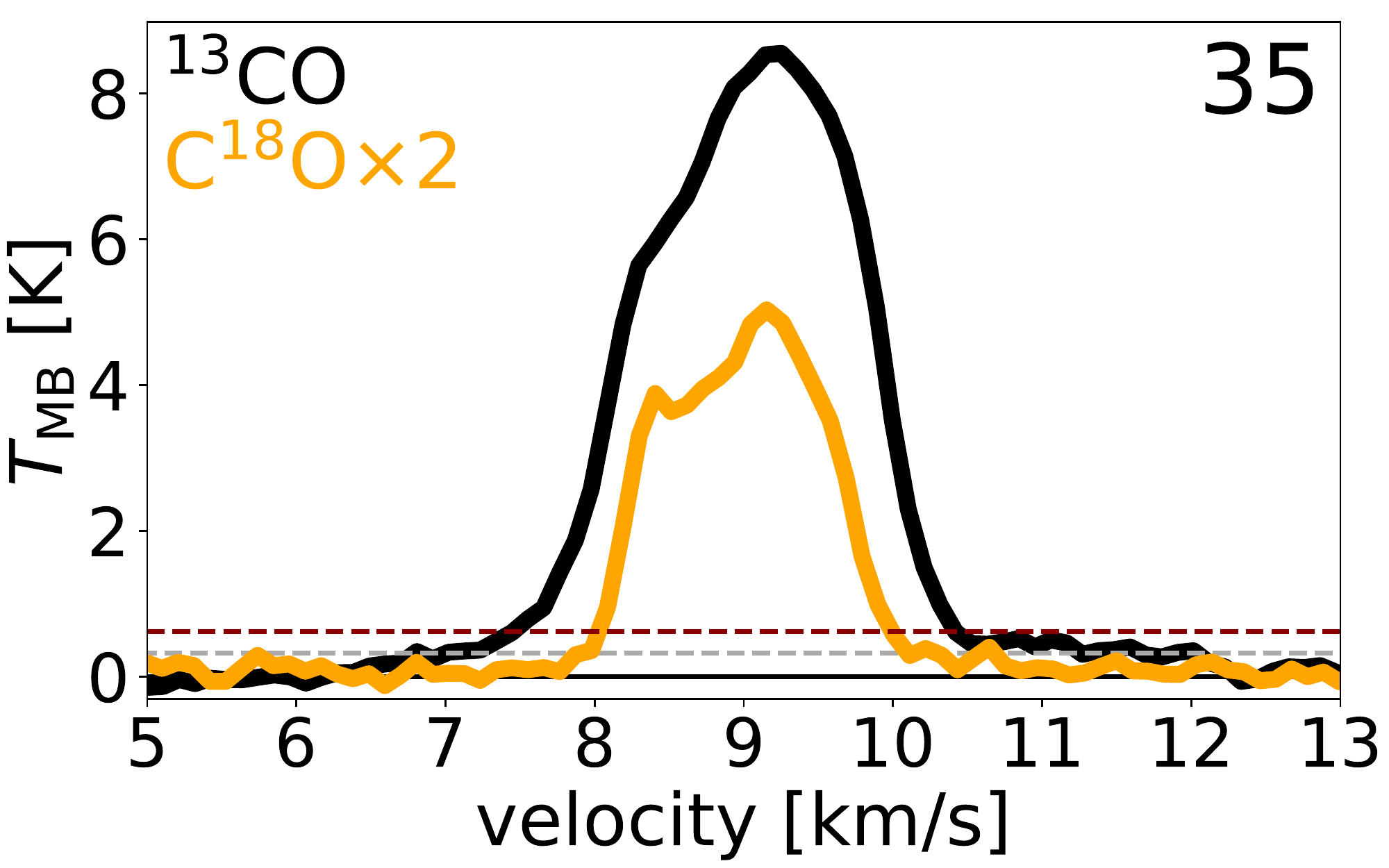}}\\ 
\subfloat{\includegraphics[width=0.165\textwidth]{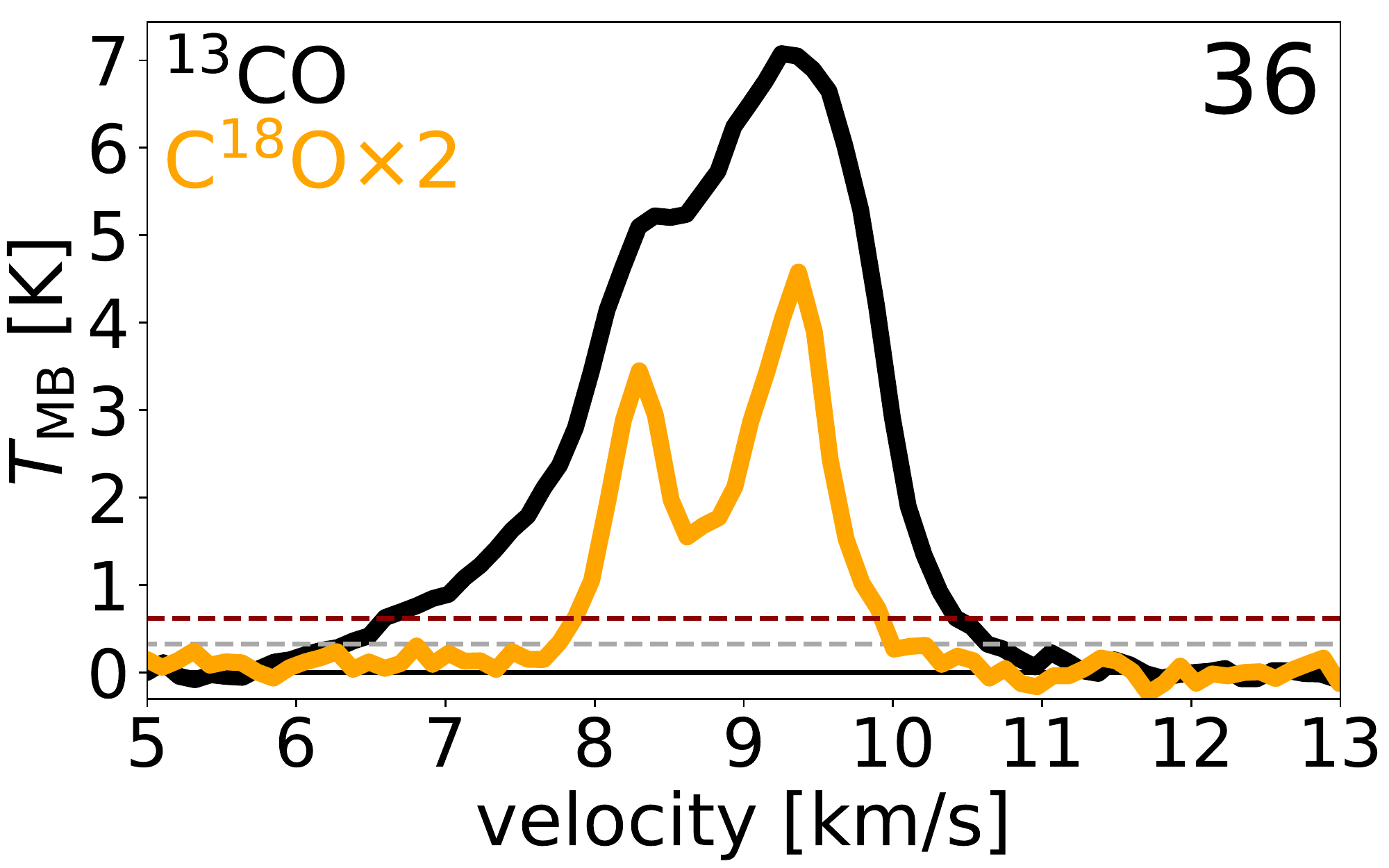}} 
\subfloat{\includegraphics[width=0.165\textwidth]{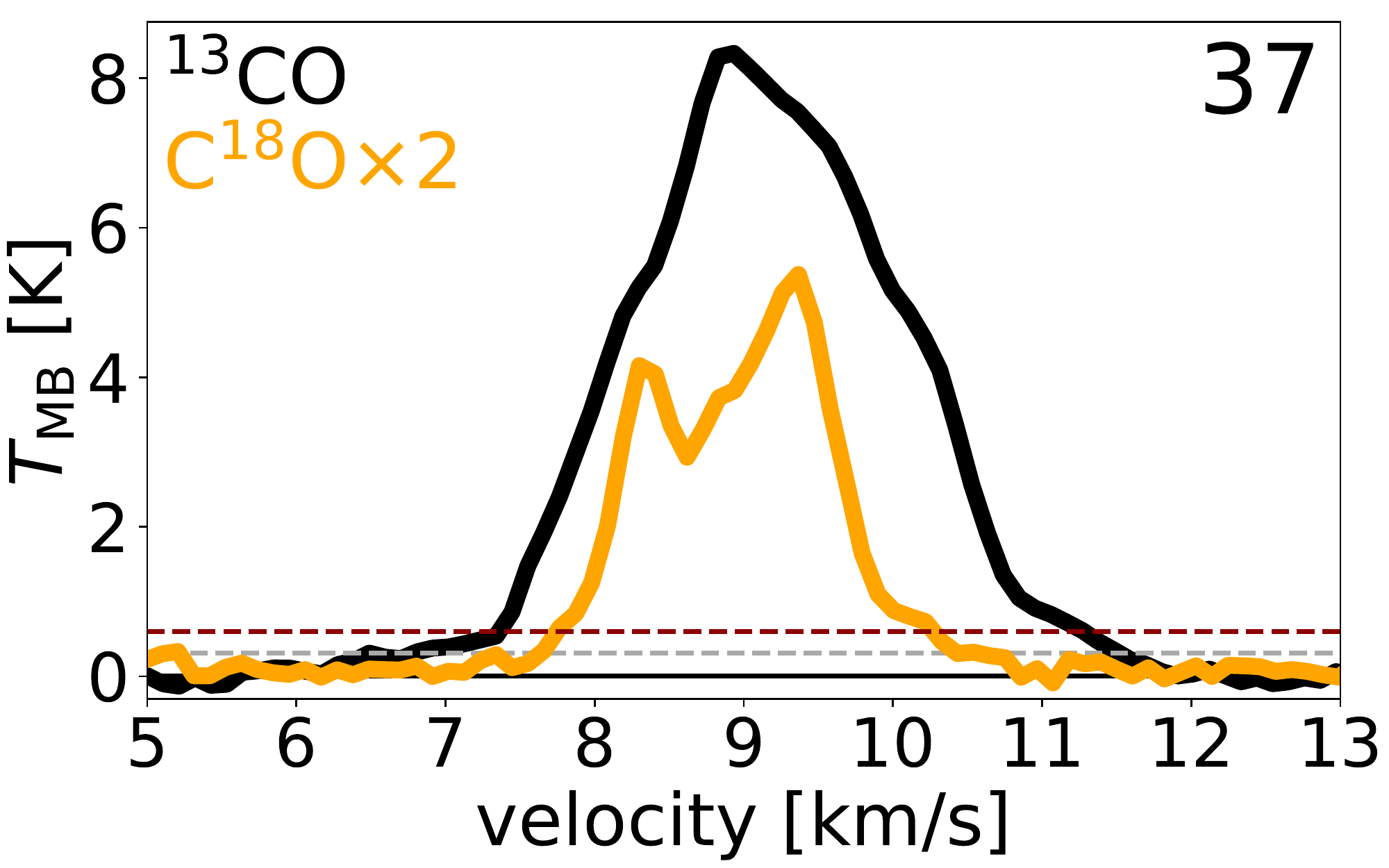}} 
\subfloat{\includegraphics[width=0.165\textwidth]{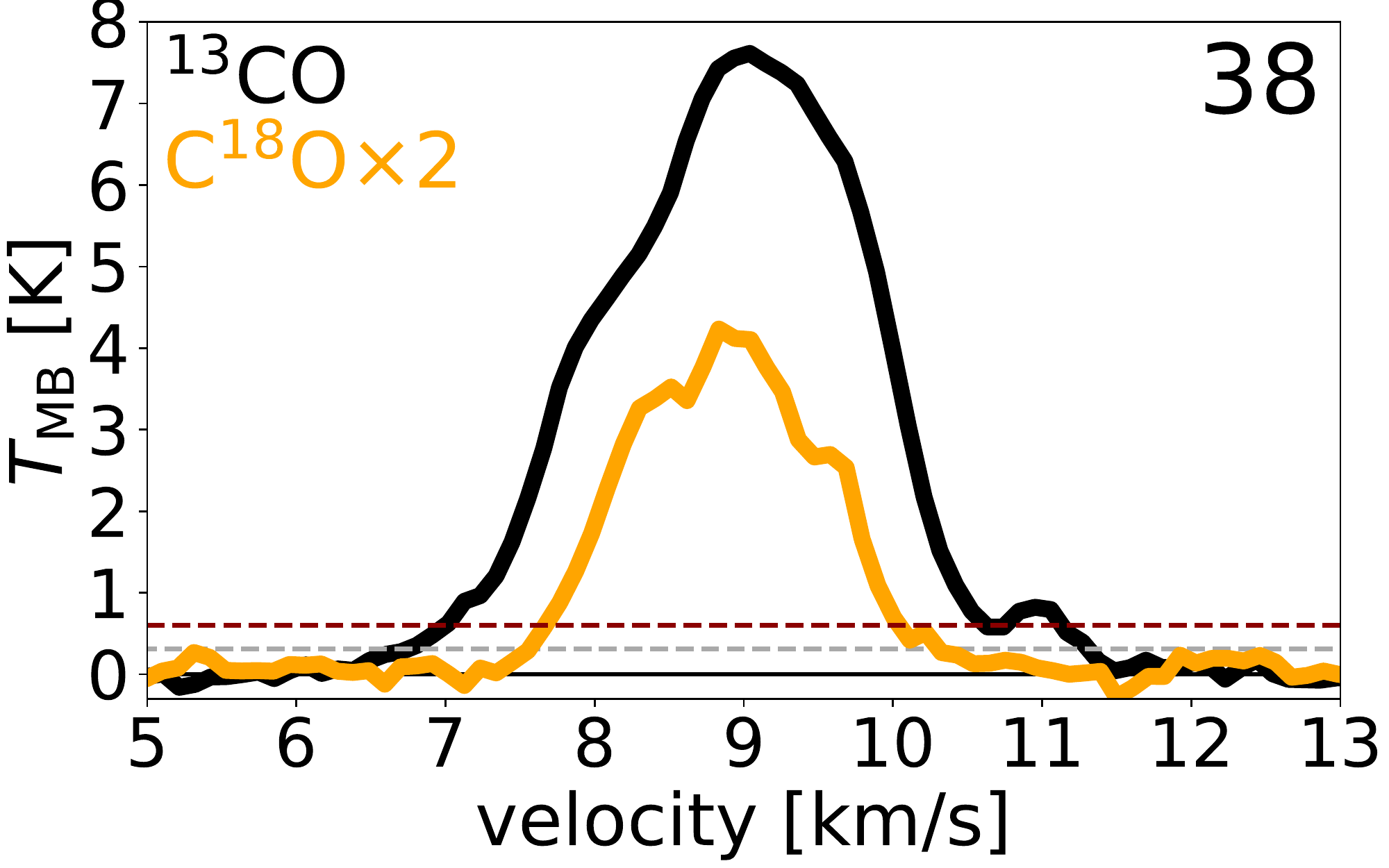}} 
\subfloat{\includegraphics[width=0.165\textwidth]{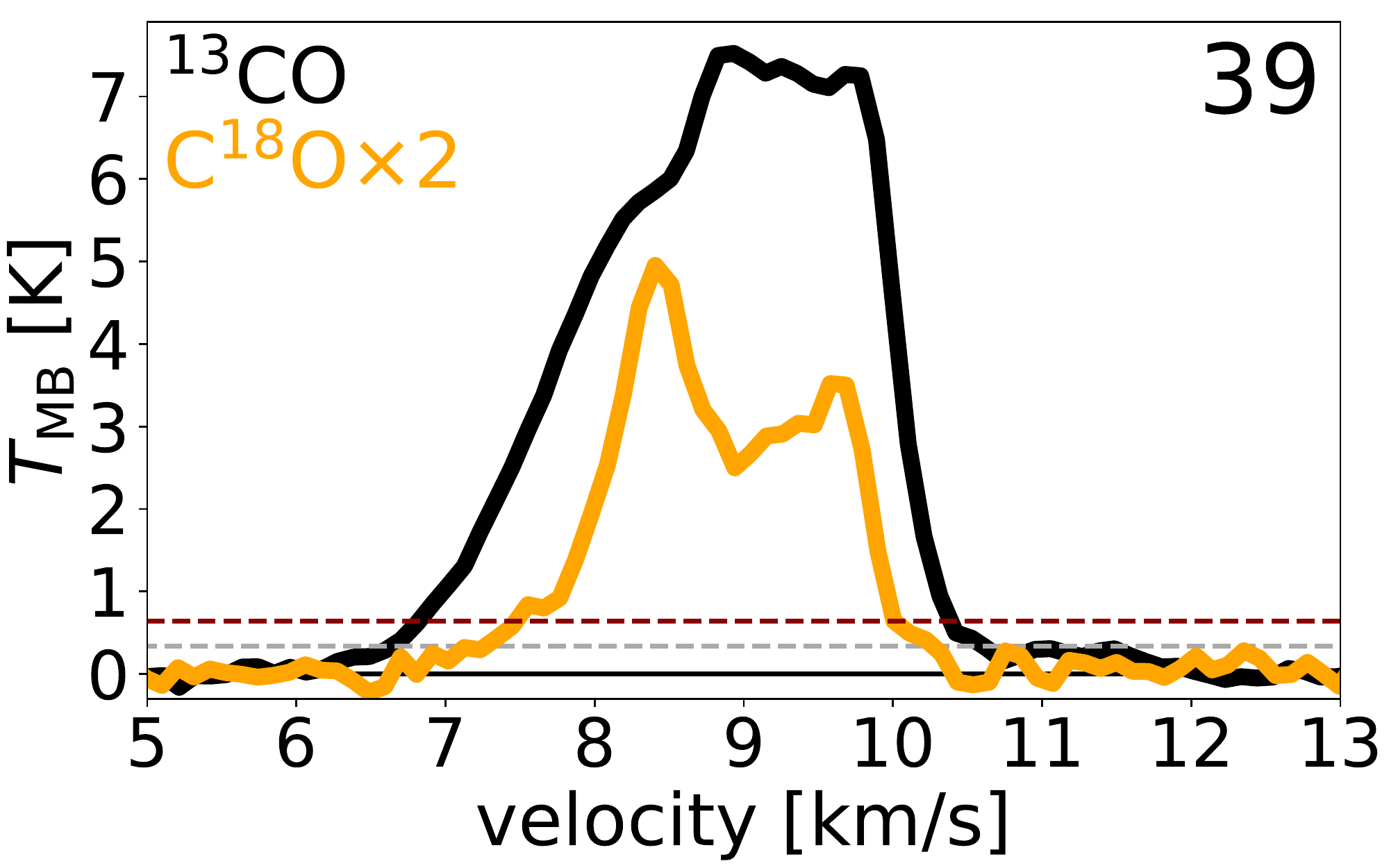}} 

\caption{$^{13}$CO(1--0) (black) and C$^{18}$O(1--0) (orange) spectra averaged over the analysis spots where observed (see Fig.~\ref{fig_circ}). 
         When it is reasonable, a Gaussian fit and its peak position are also included. 
         Dashed horizontal lines mark 6$\sigma$ rms thresholds; grey for the black spectra, red for the overplotted profiles.   
         For details, see Sects.~\ref{sec:molec} and \ref{sec:energy}. Derived properties from these profiles are listed in Table~\ref{tab_uCalcT}. 
         } 
\label{fig_13CO_C18O_spectra}
\end{figure*}

\begin{figure*}
\subfloat{\includegraphics[width=0.165\textwidth]{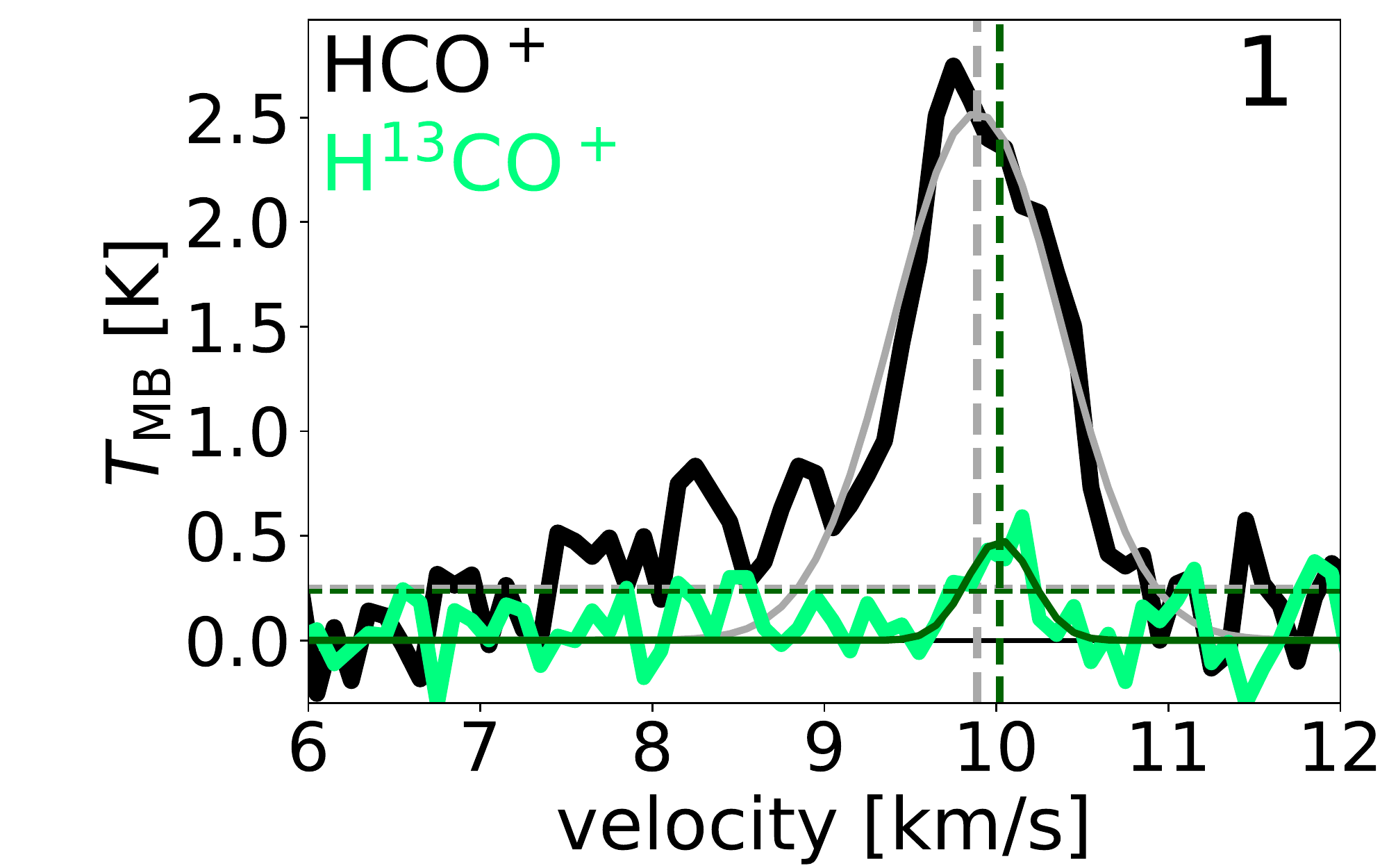}} 
\subfloat{\includegraphics[width=0.165\textwidth]{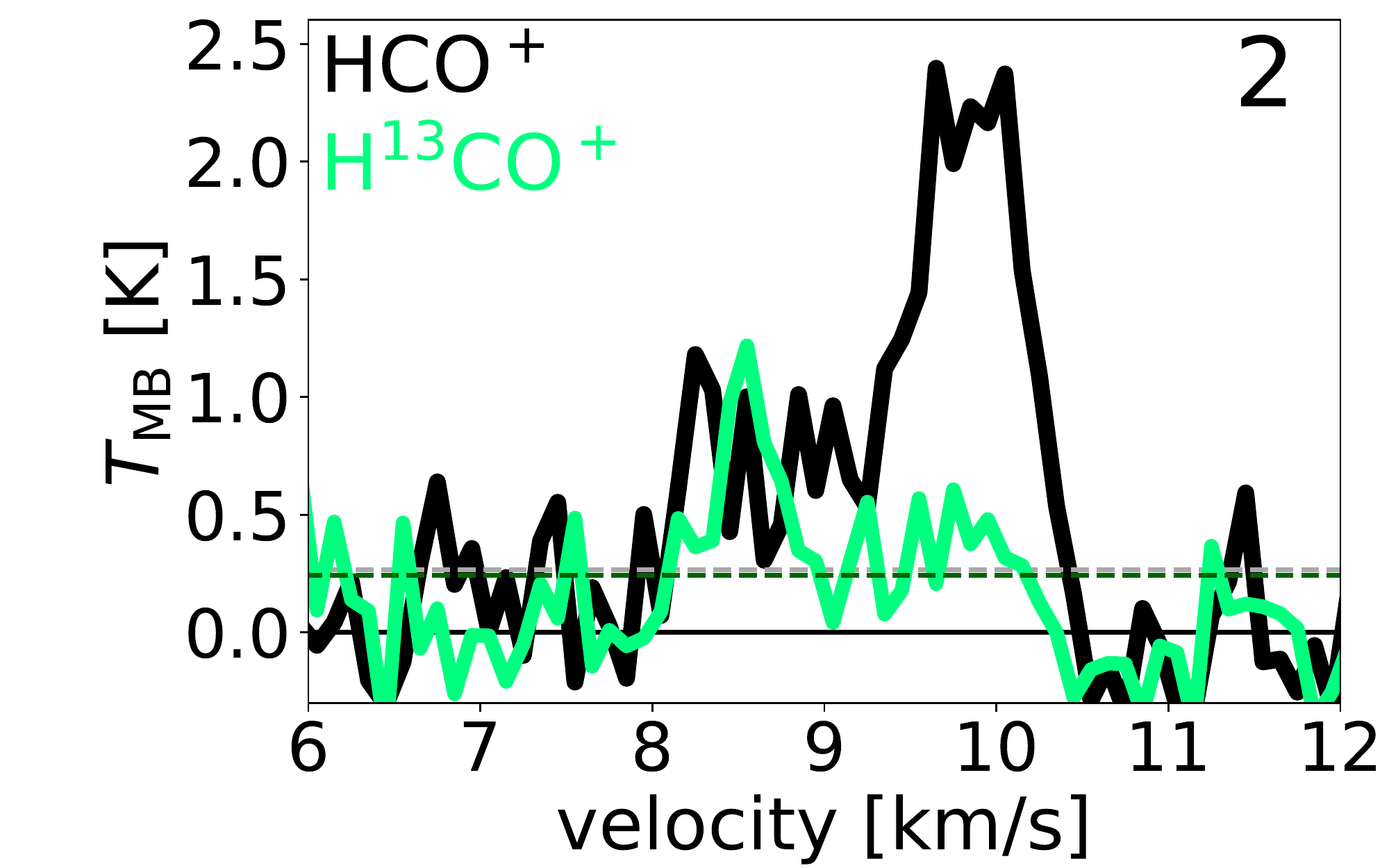}} 
\subfloat{\includegraphics[width=0.165\textwidth]{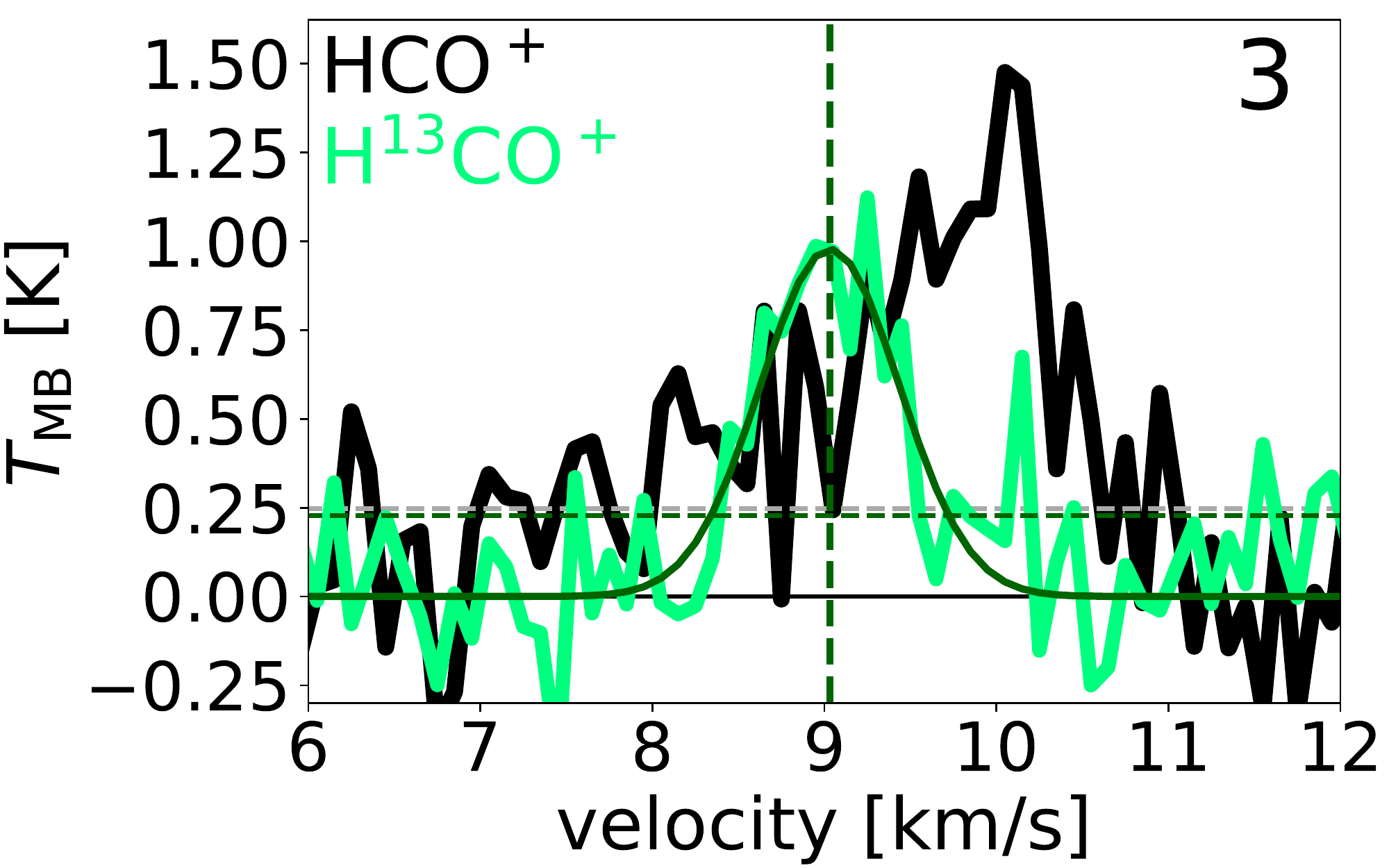}} 
\subfloat{\includegraphics[width=0.165\textwidth]{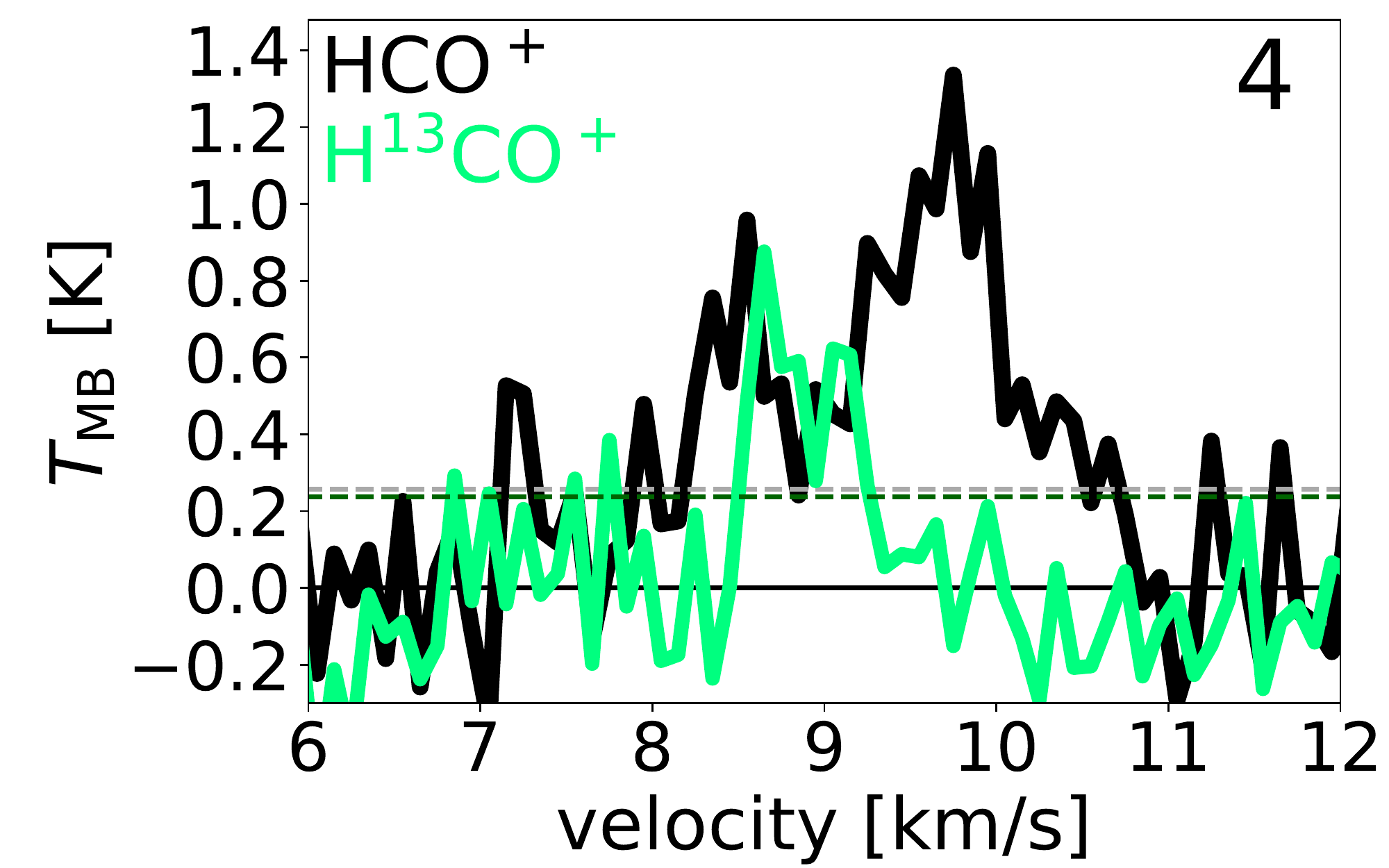}}
\subfloat{\includegraphics[width=0.165\textwidth]{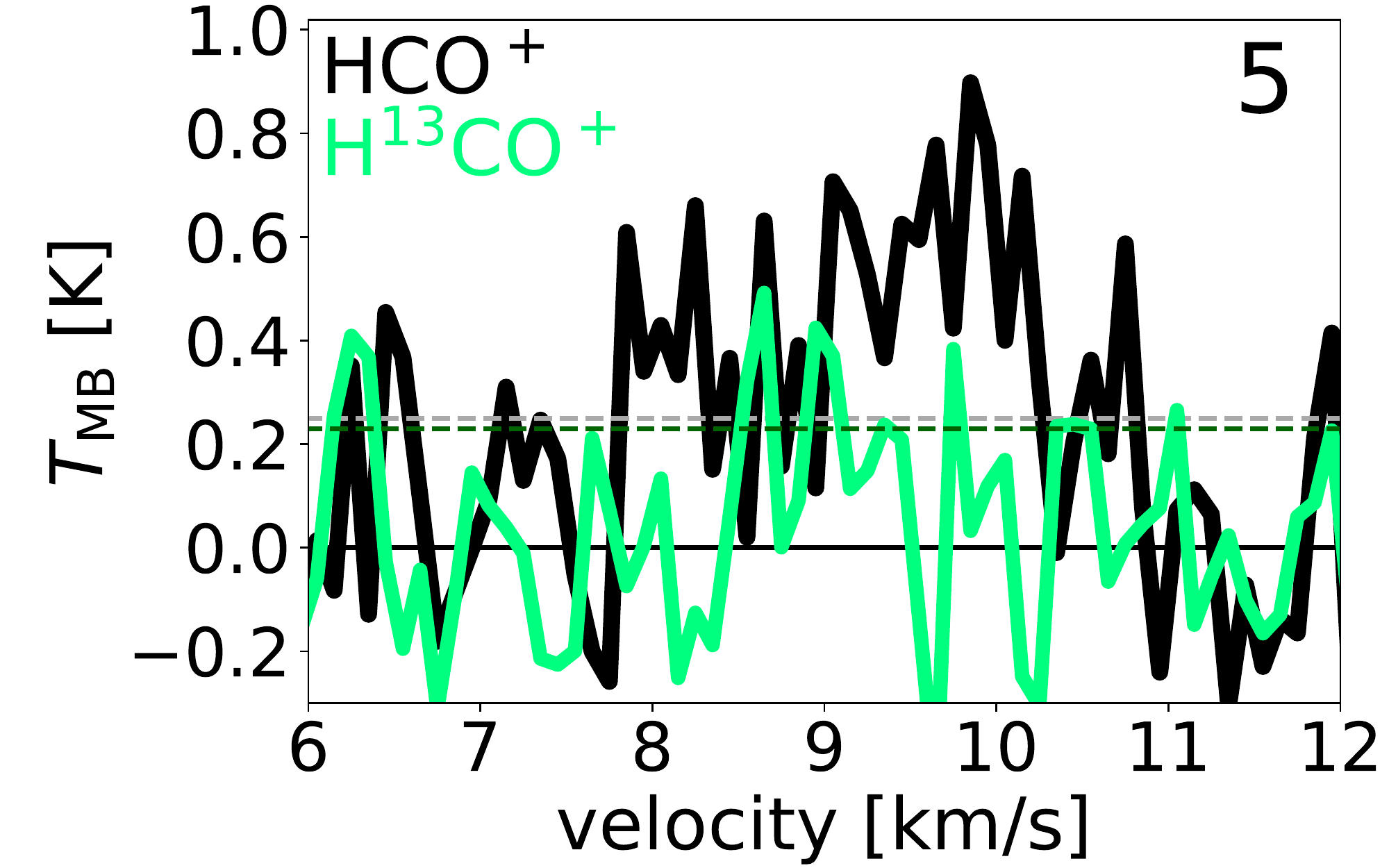}} 
\subfloat{\includegraphics[width=0.165\textwidth]{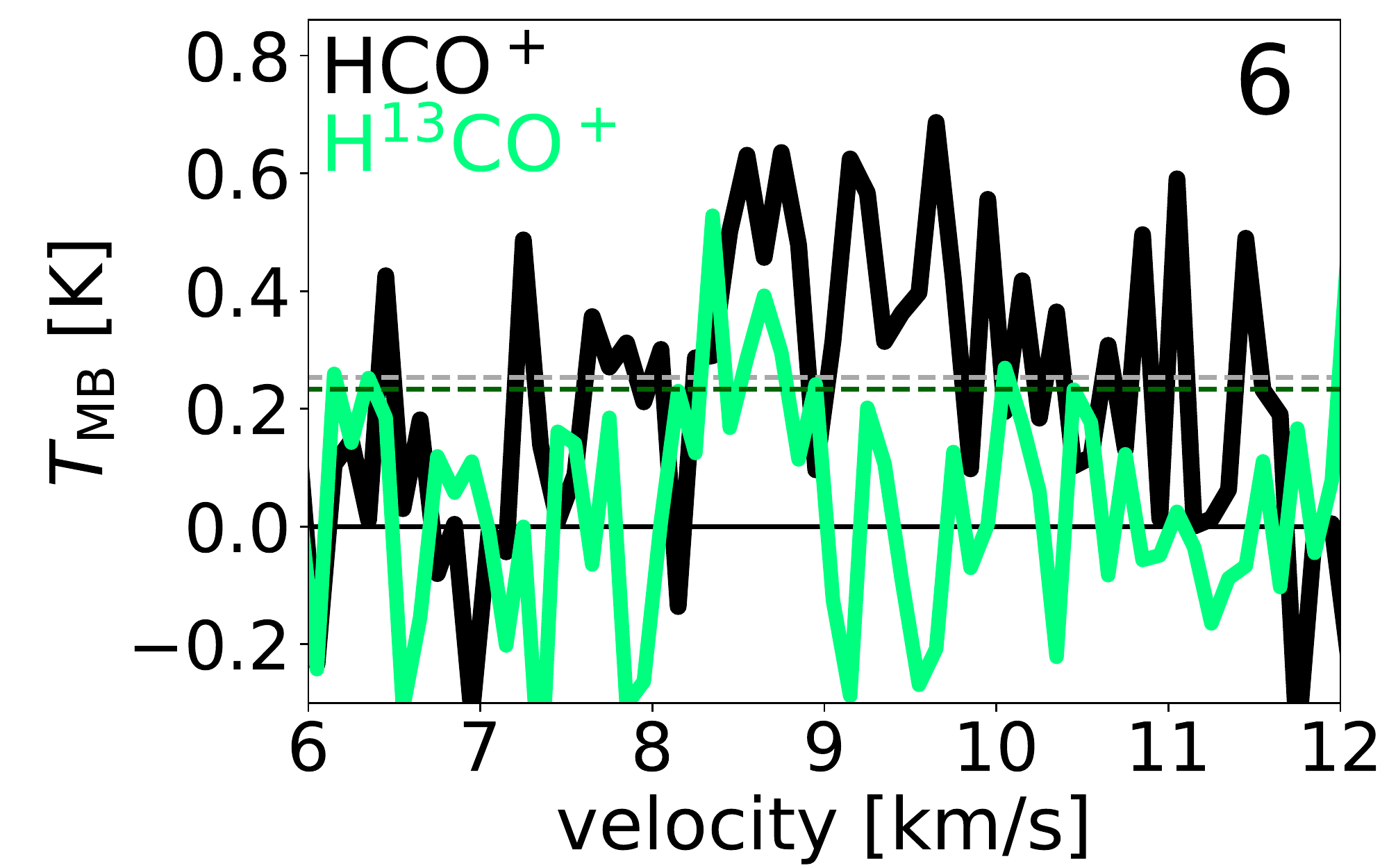}}\\ 
\subfloat{\includegraphics[width=0.165\textwidth]{./figs/HCO+_H13CO+_spectra_noFit_cNo7.pdf}} 
\subfloat{\includegraphics[width=0.165\textwidth]{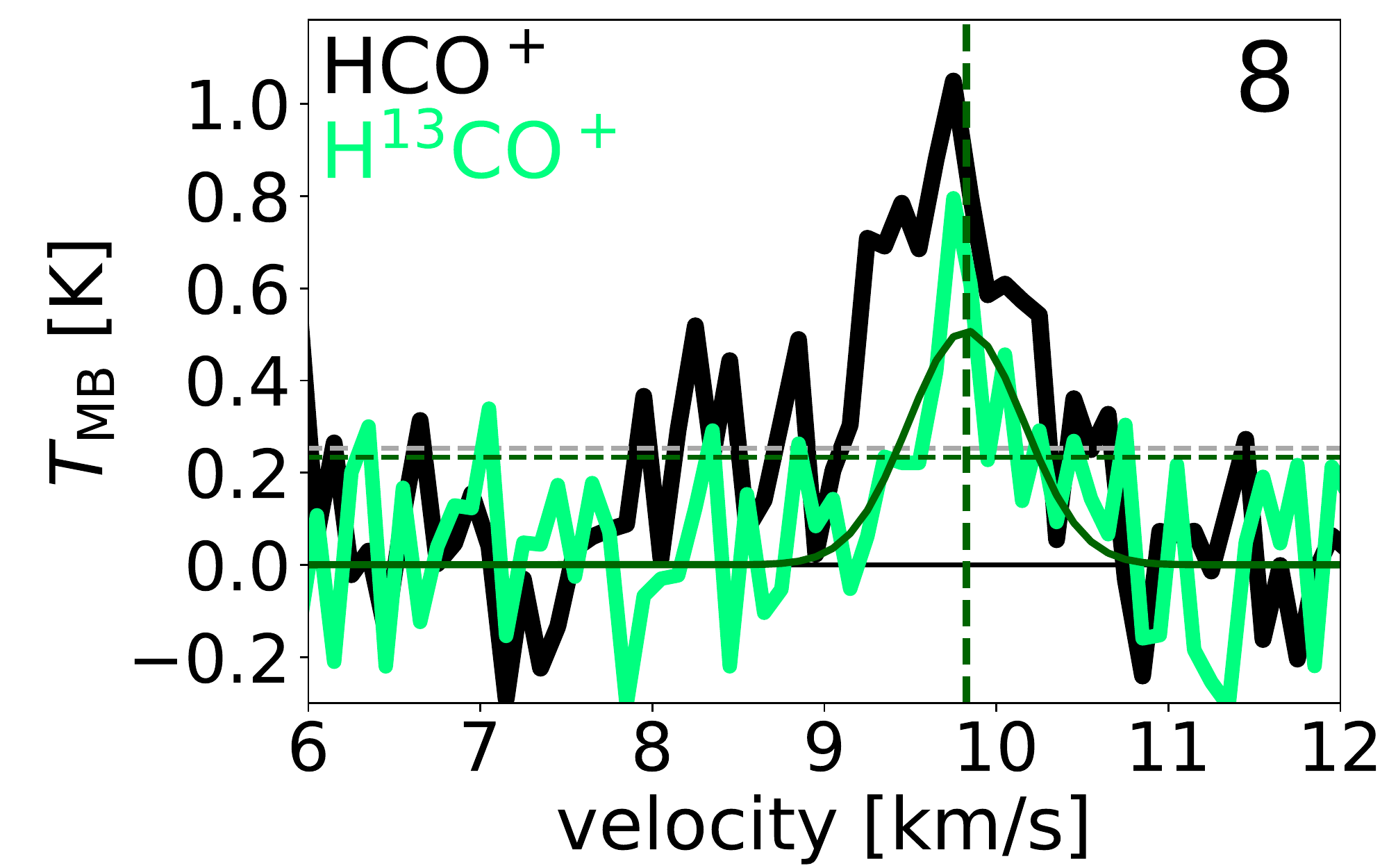}} 
\subfloat{\includegraphics[width=0.165\textwidth]{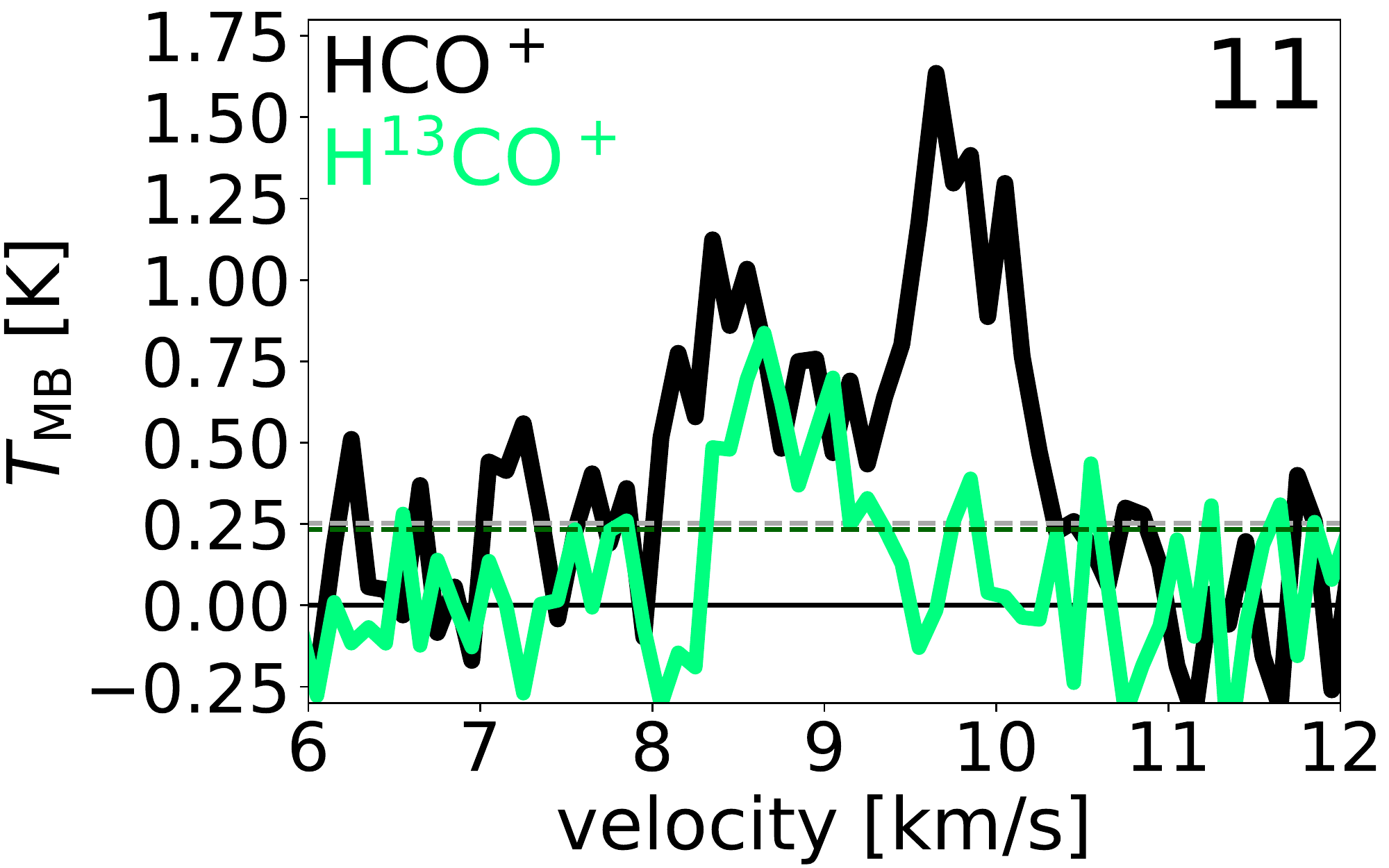}}
\subfloat{\includegraphics[width=0.165\textwidth]{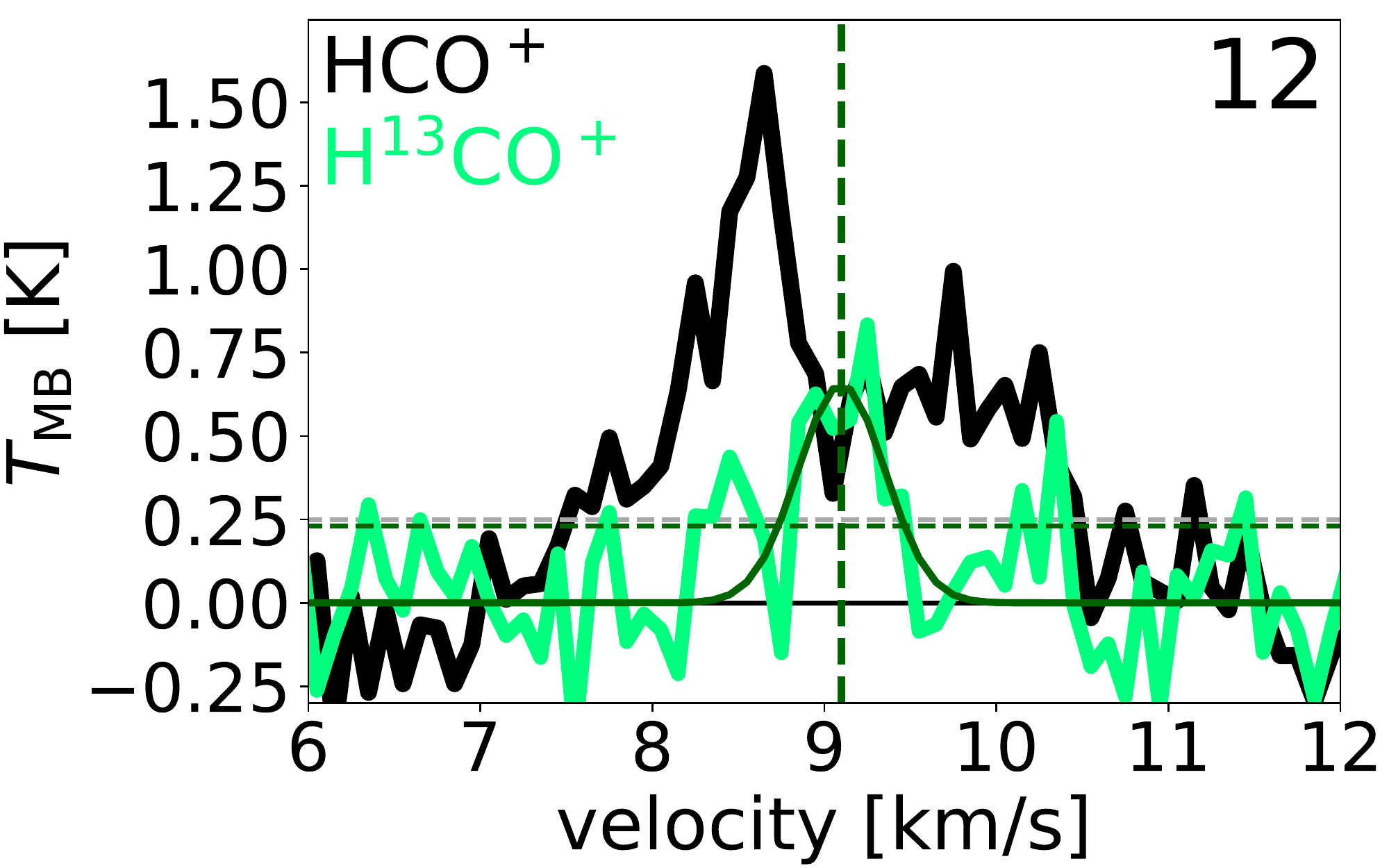}}
 \subfloat{\includegraphics[width=0.165\textwidth]{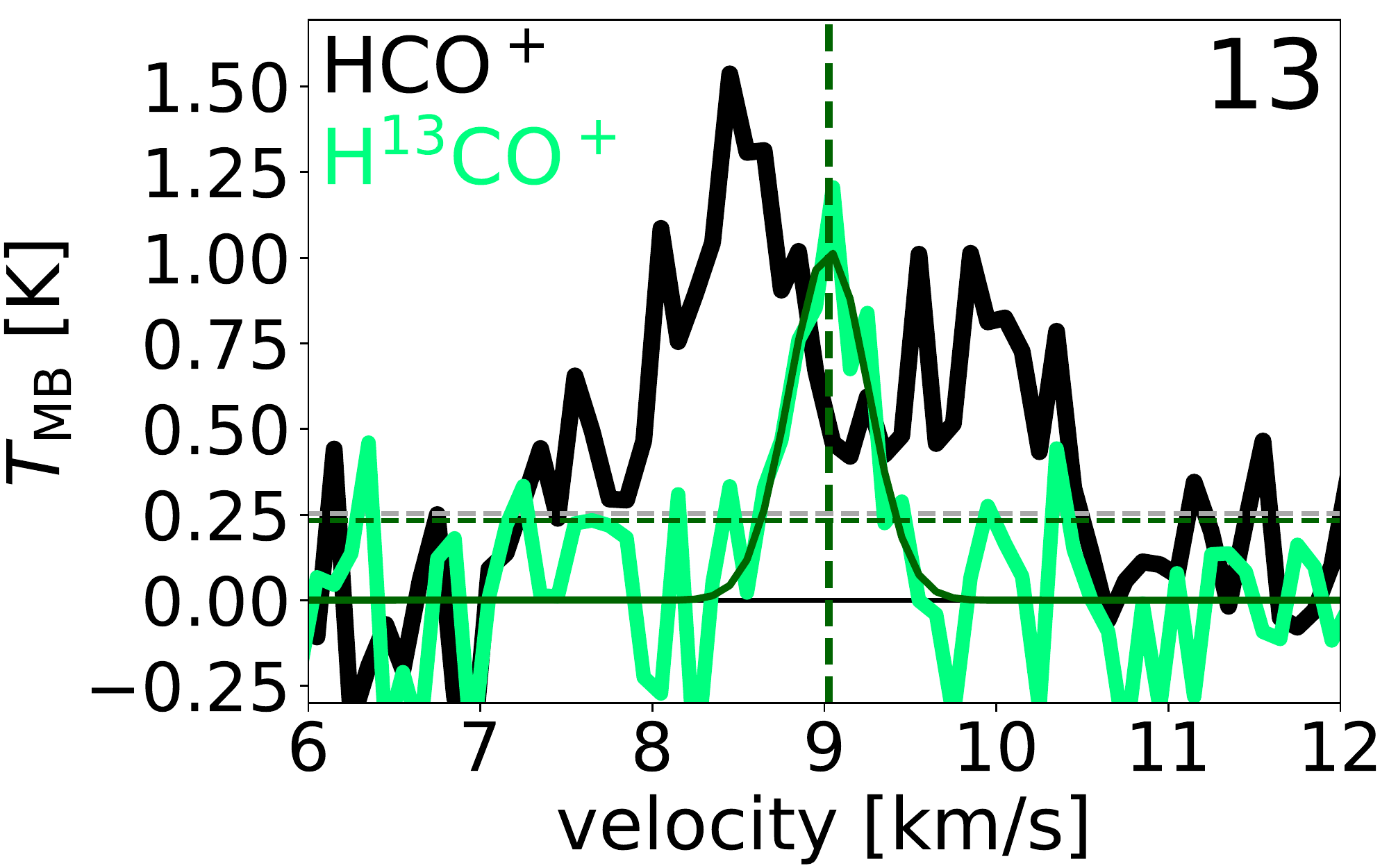}} 
\subfloat{\includegraphics[width=0.165\textwidth]{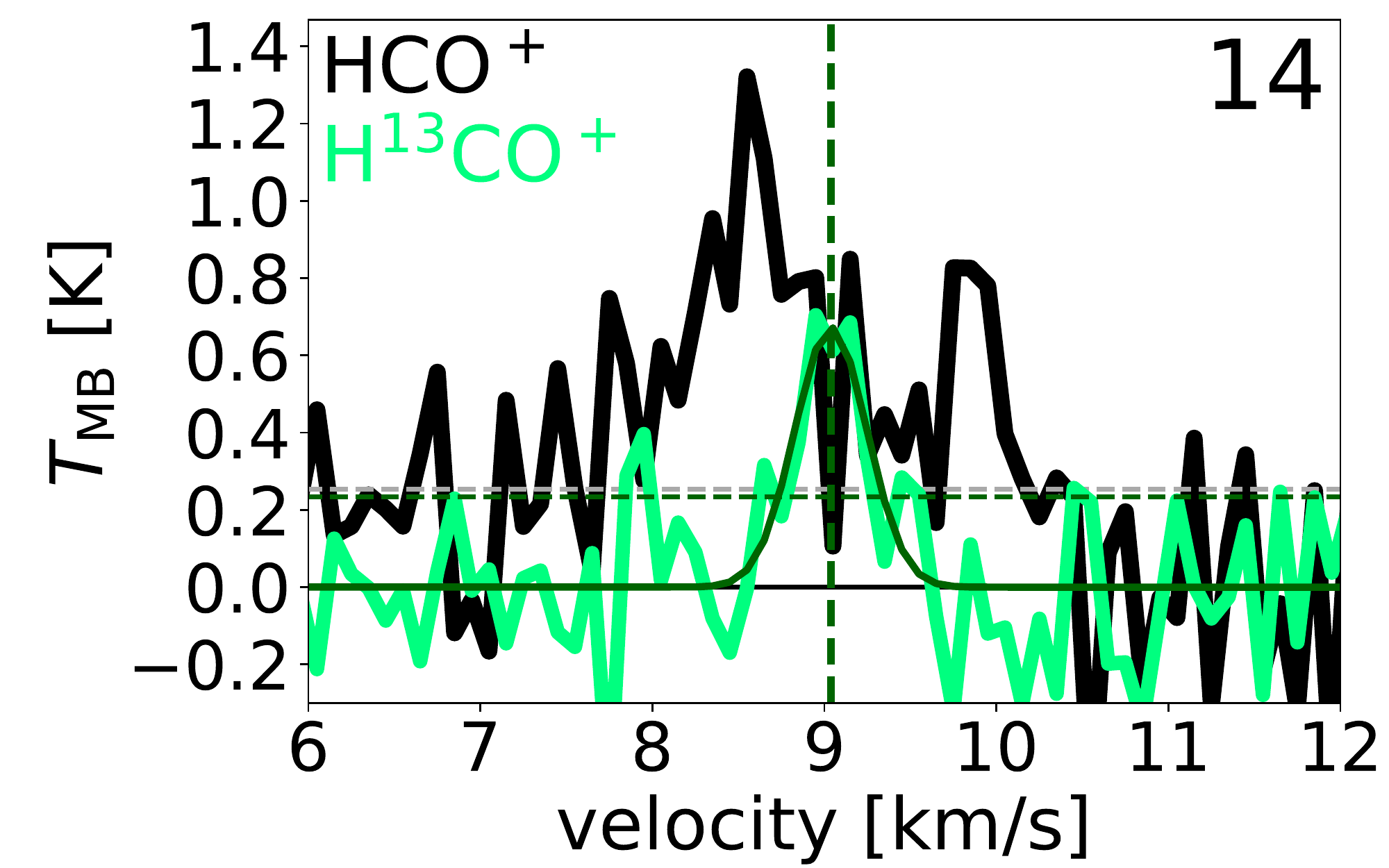}}\\
 
\subfloat{\includegraphics[width=0.165\textwidth]{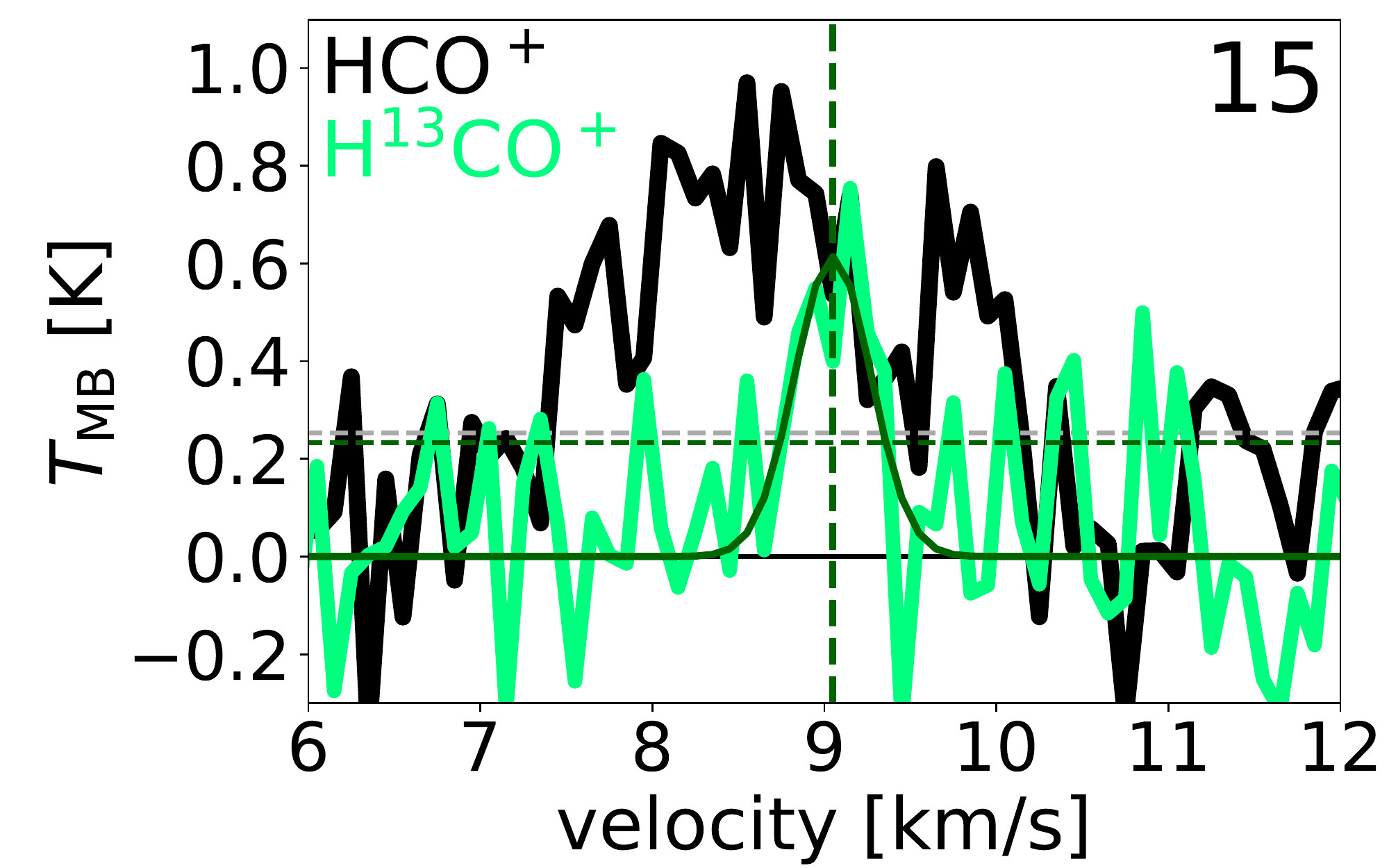}} 
\subfloat{\includegraphics[width=0.165\textwidth]{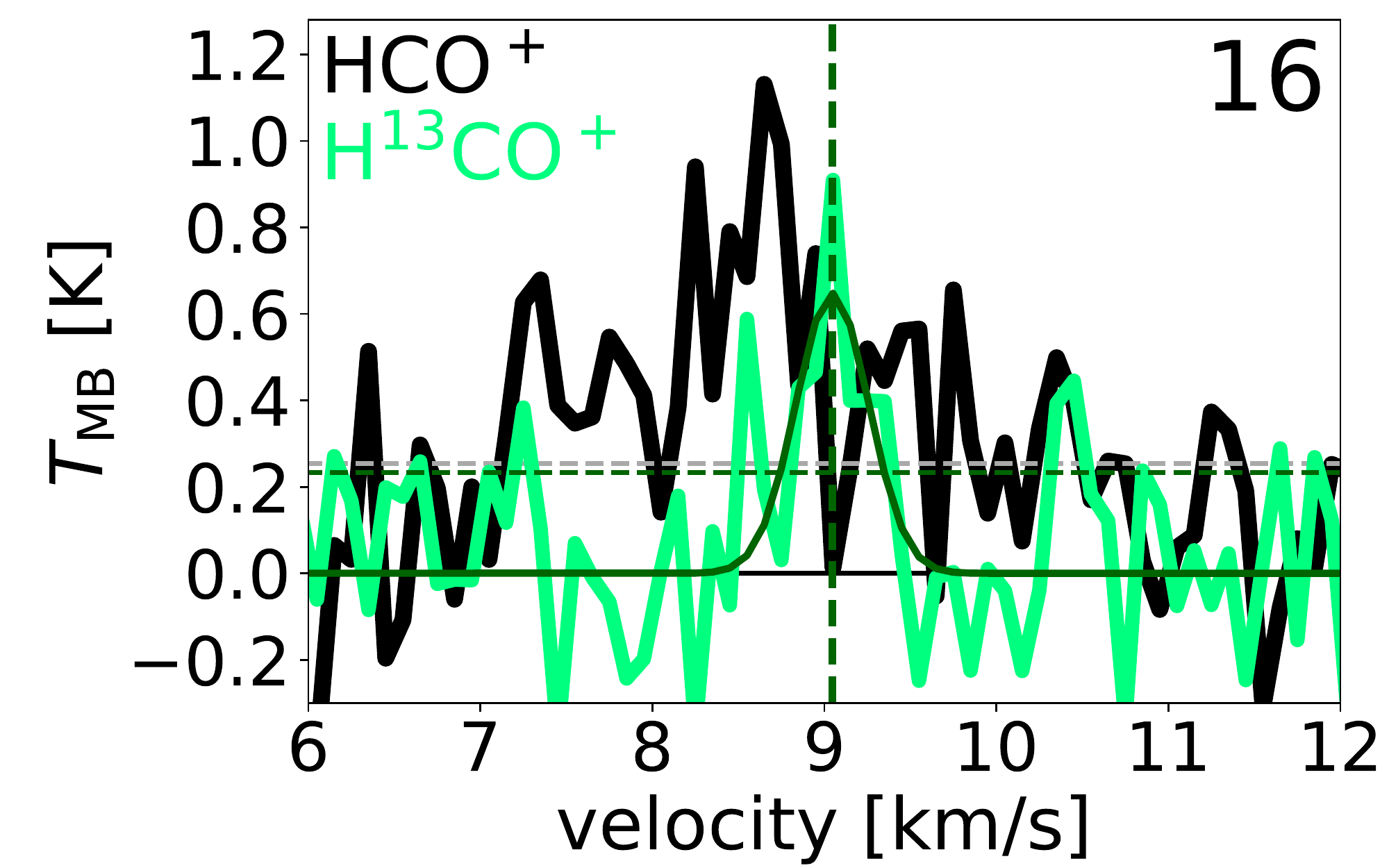}} 
\subfloat{\includegraphics[width=0.165\textwidth]{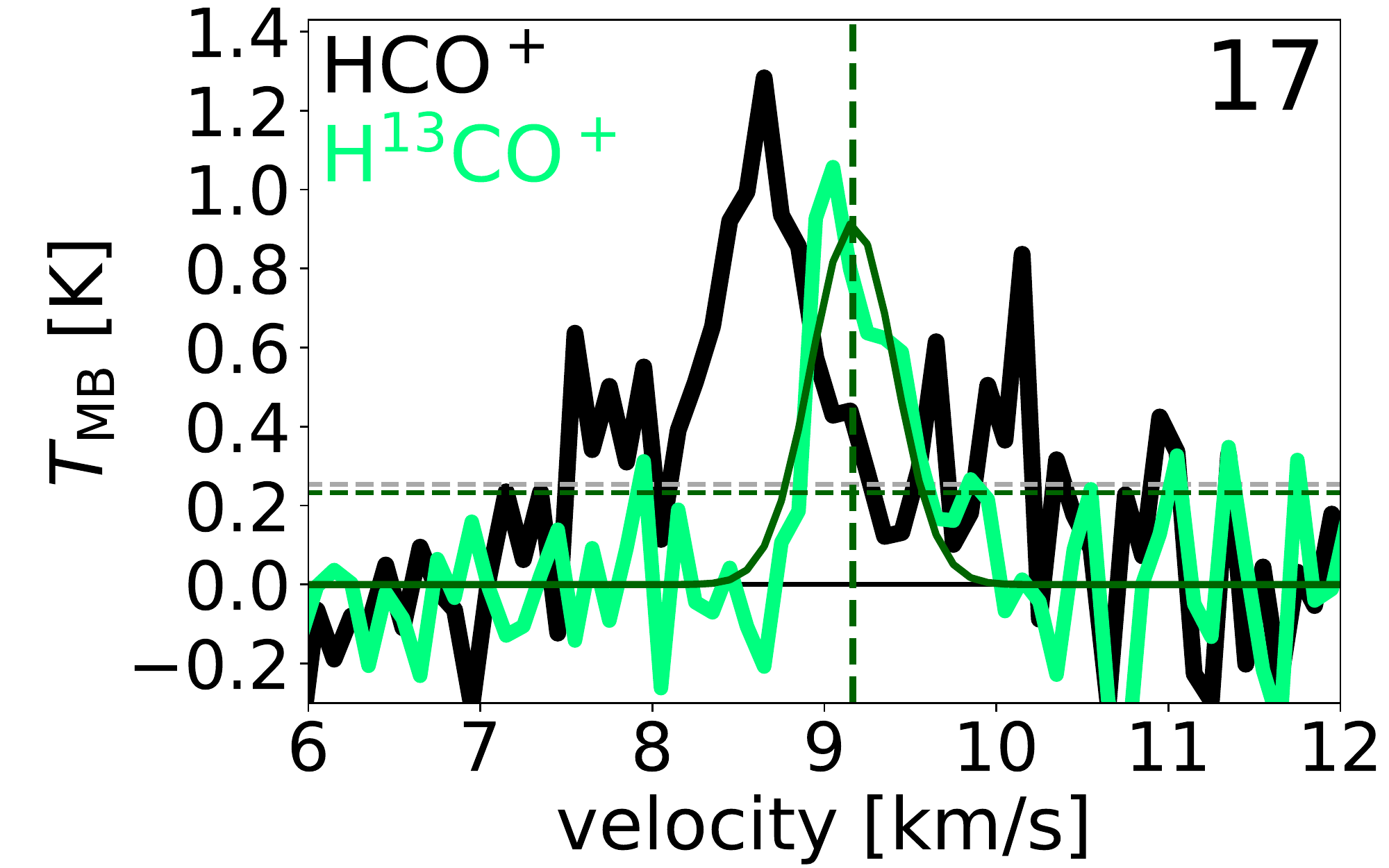}} 
\subfloat{\includegraphics[width=0.165\textwidth]{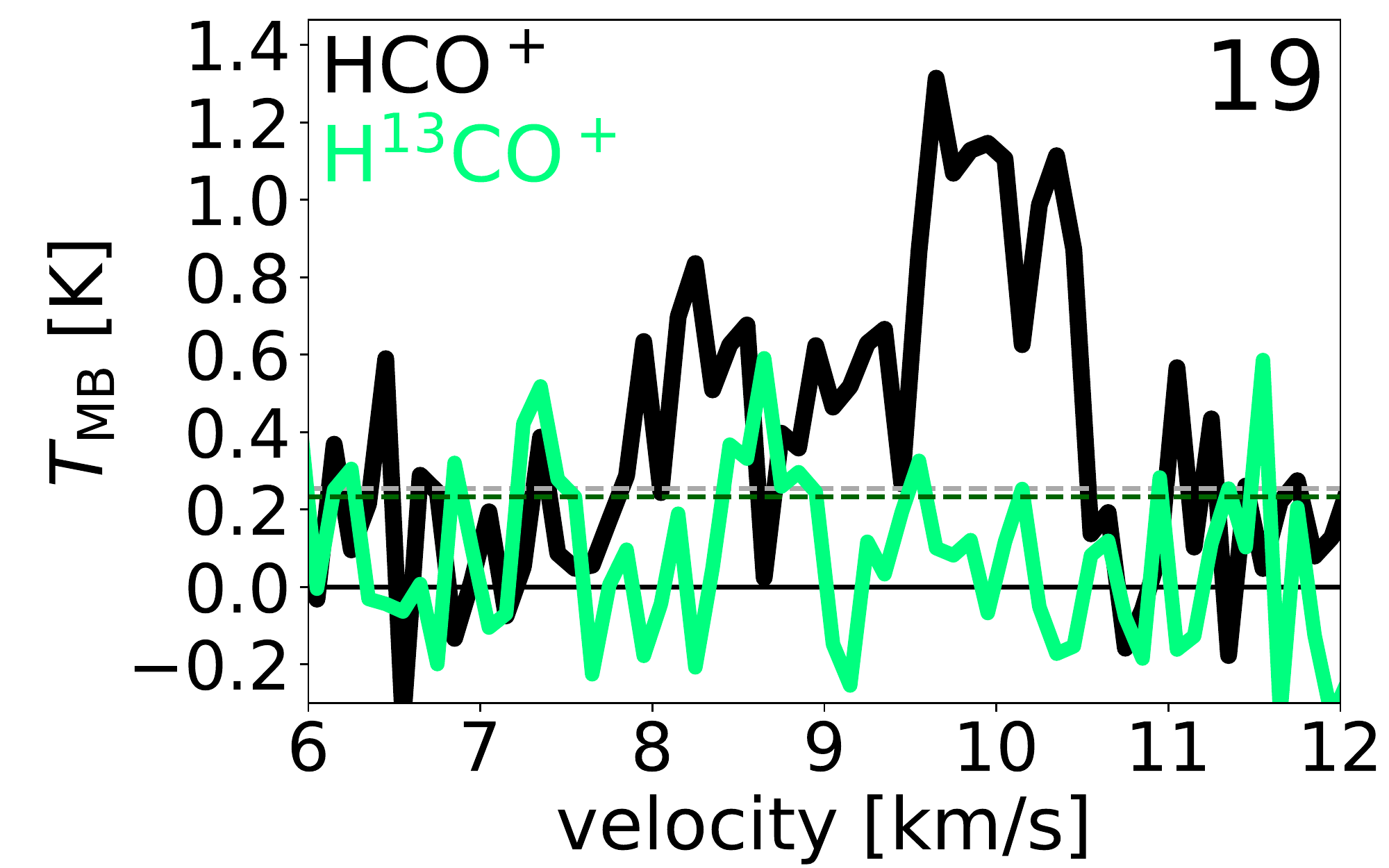}} 
\subfloat{\includegraphics[width=0.165\textwidth]{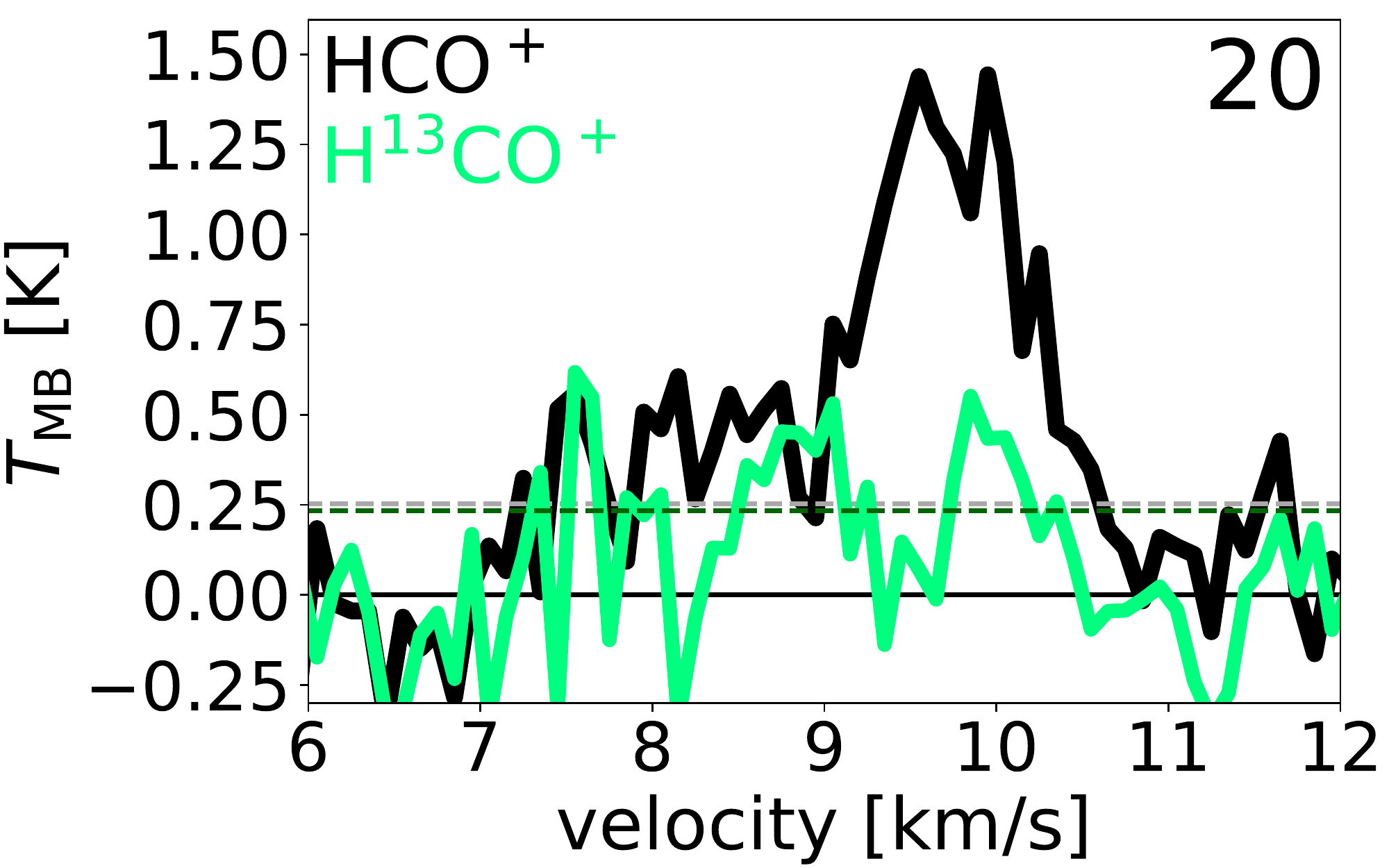}} 
\subfloat{\includegraphics[width=0.165\textwidth]{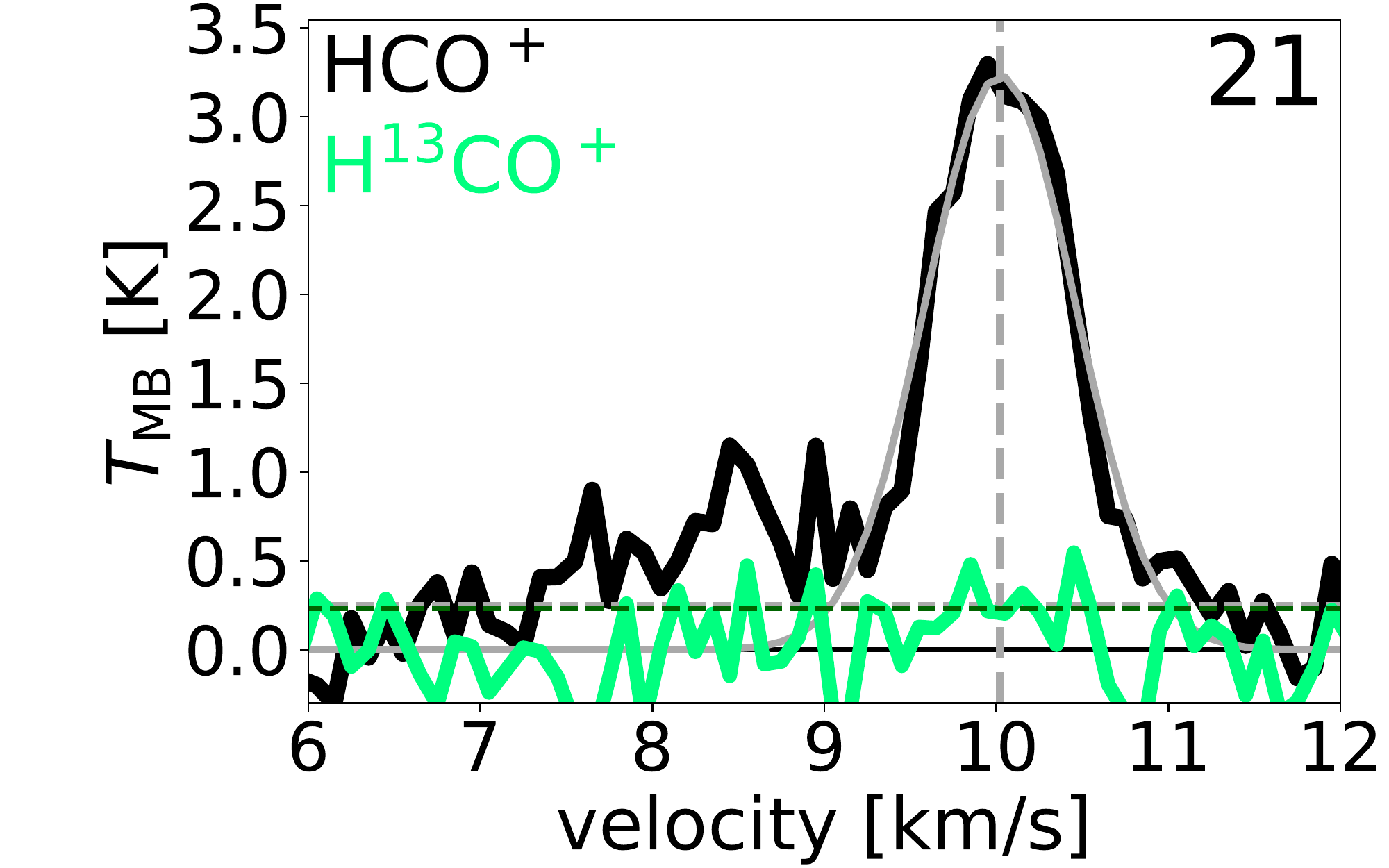}}\\ 
\subfloat{\includegraphics[width=0.165\textwidth]{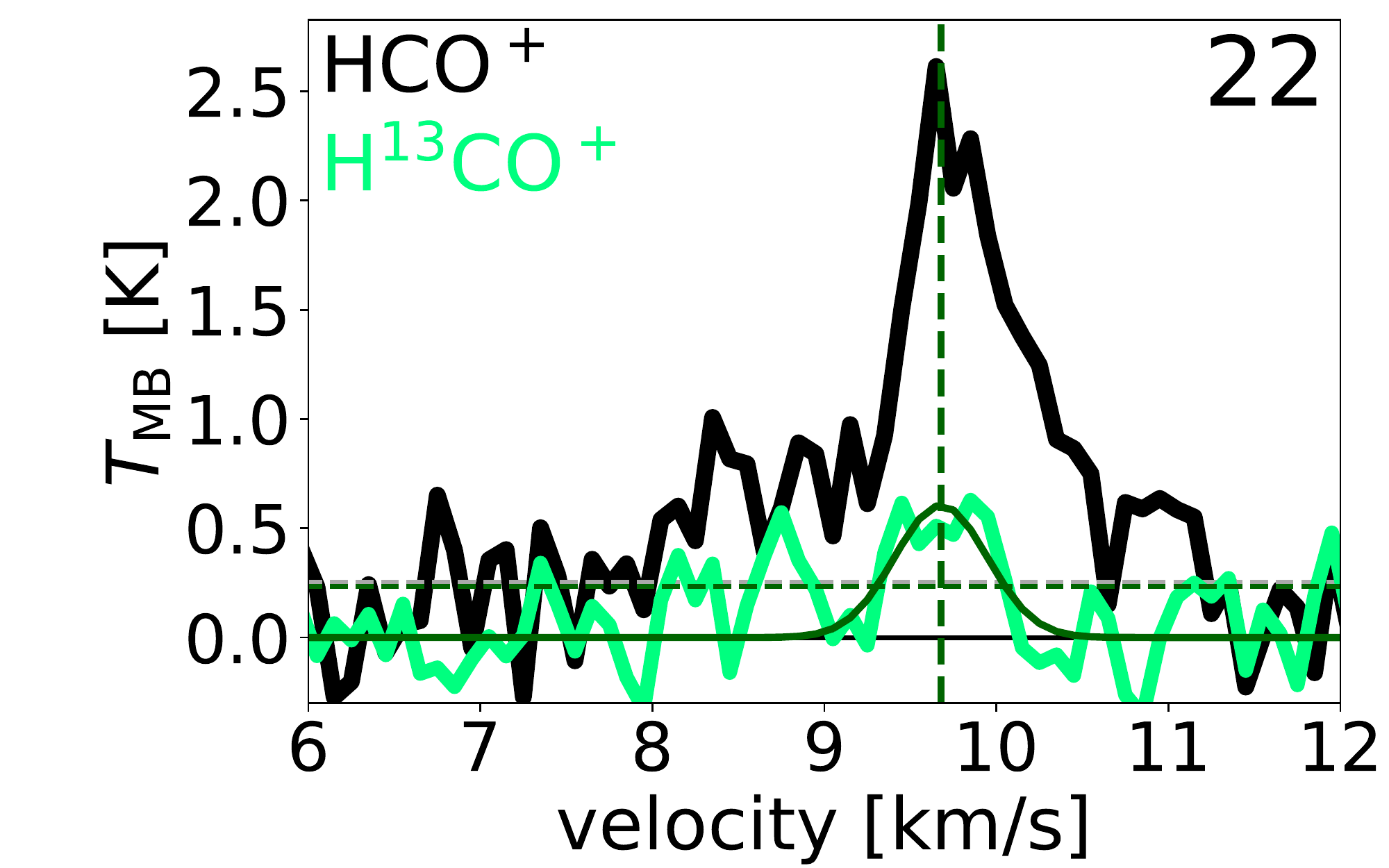}} 
\subfloat{\includegraphics[width=0.165\textwidth]{./figs/HCO+_H13CO+_spectra_onlyH13COFit_cNo23.pdf}} 
\subfloat{\includegraphics[width=0.165\textwidth]{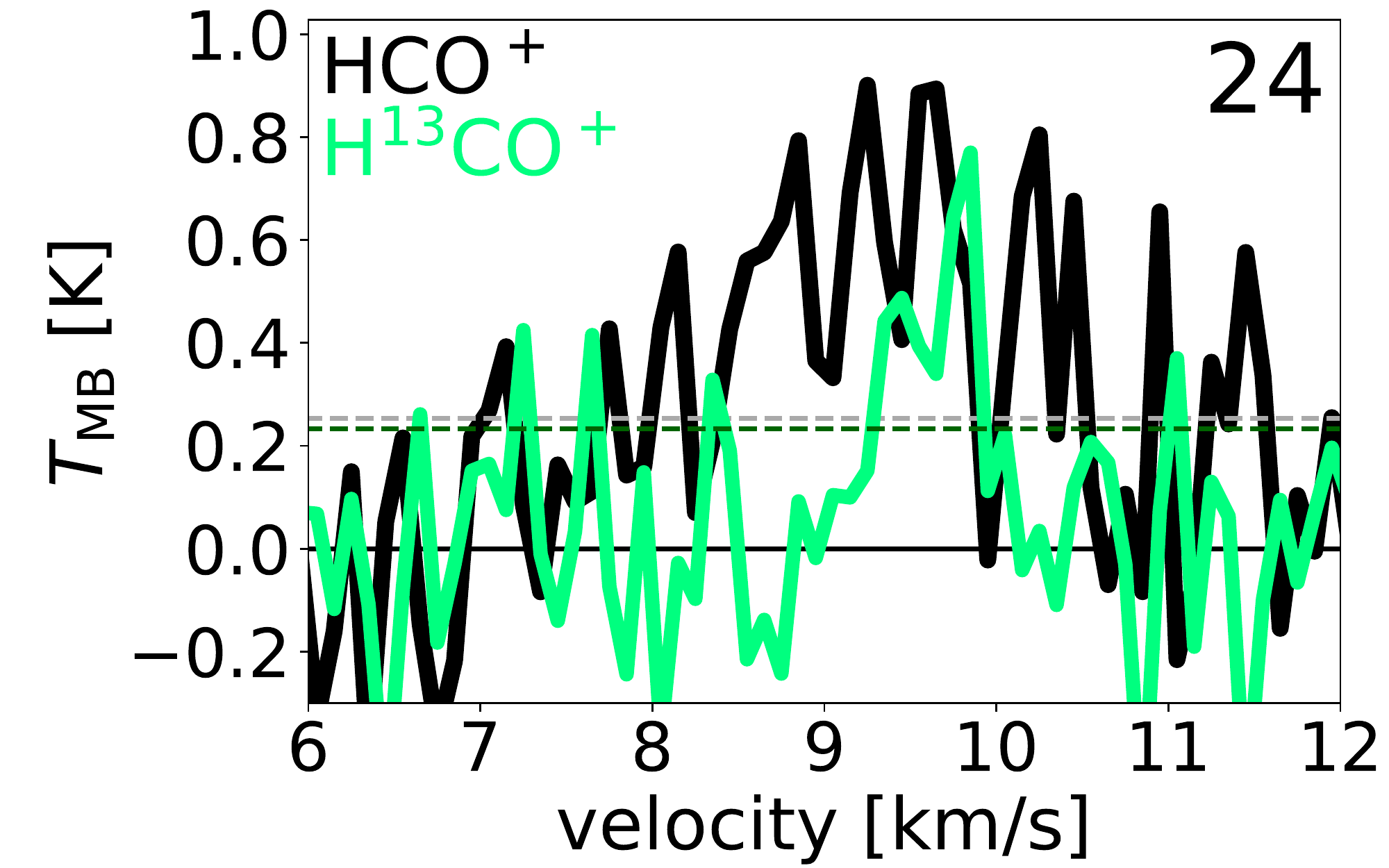}} 
\subfloat{\includegraphics[width=0.165\textwidth]{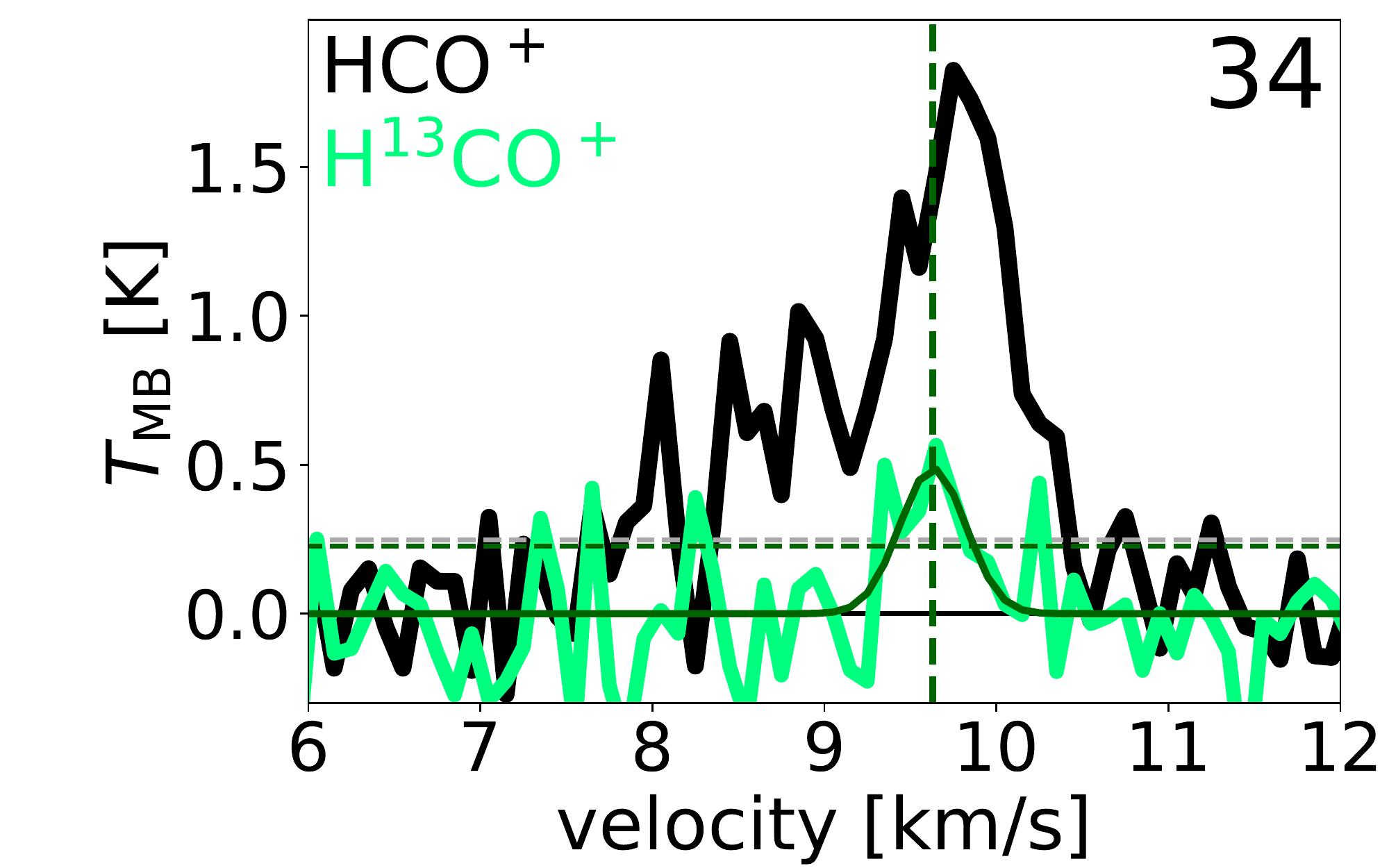}} 
\subfloat{\includegraphics[width=0.165\textwidth]{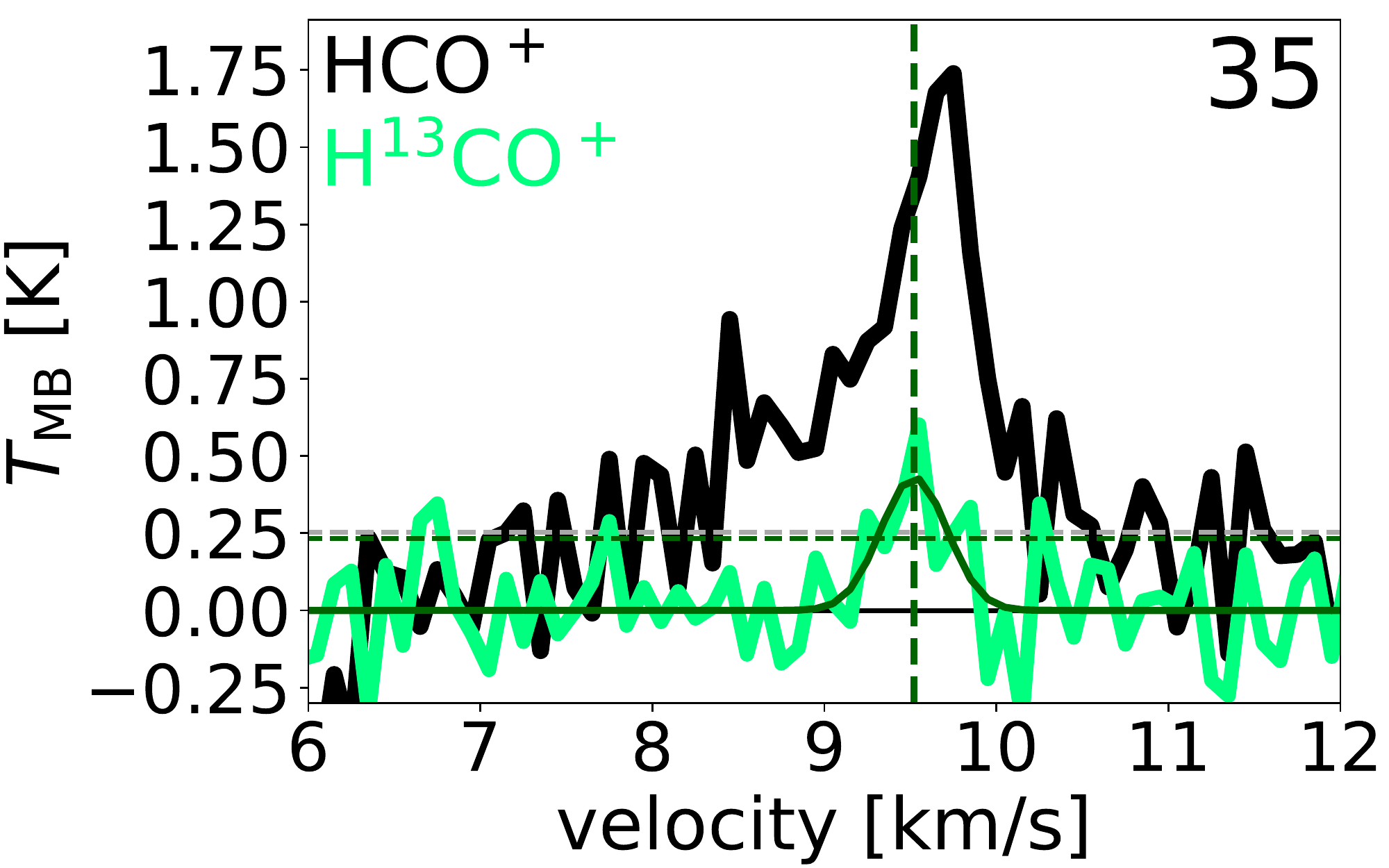}} 
\subfloat{\includegraphics[width=0.165\textwidth]{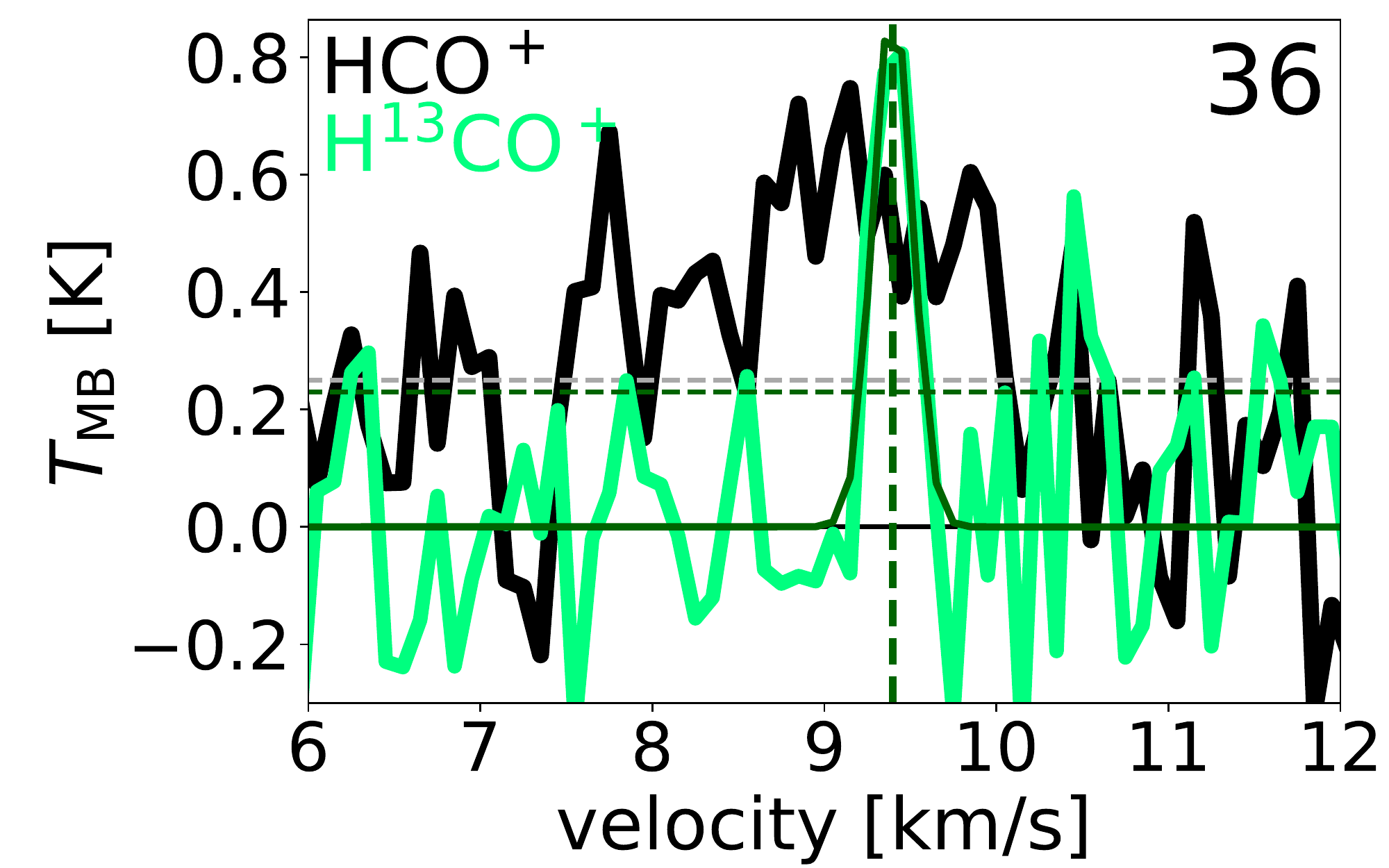}}\\
 
\subfloat{\includegraphics[width=0.165\textwidth]{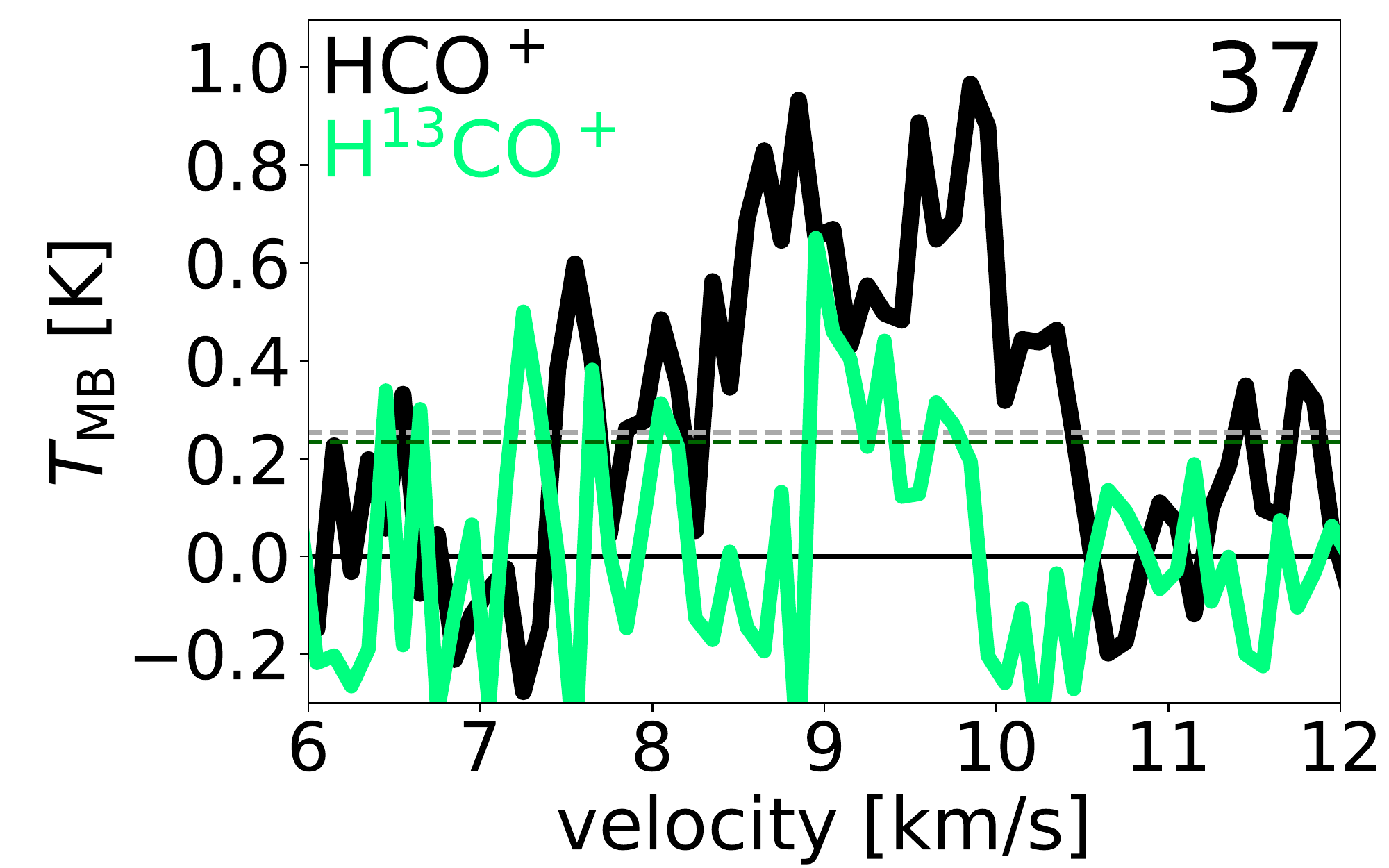}} 
\subfloat{\includegraphics[width=0.165\textwidth]{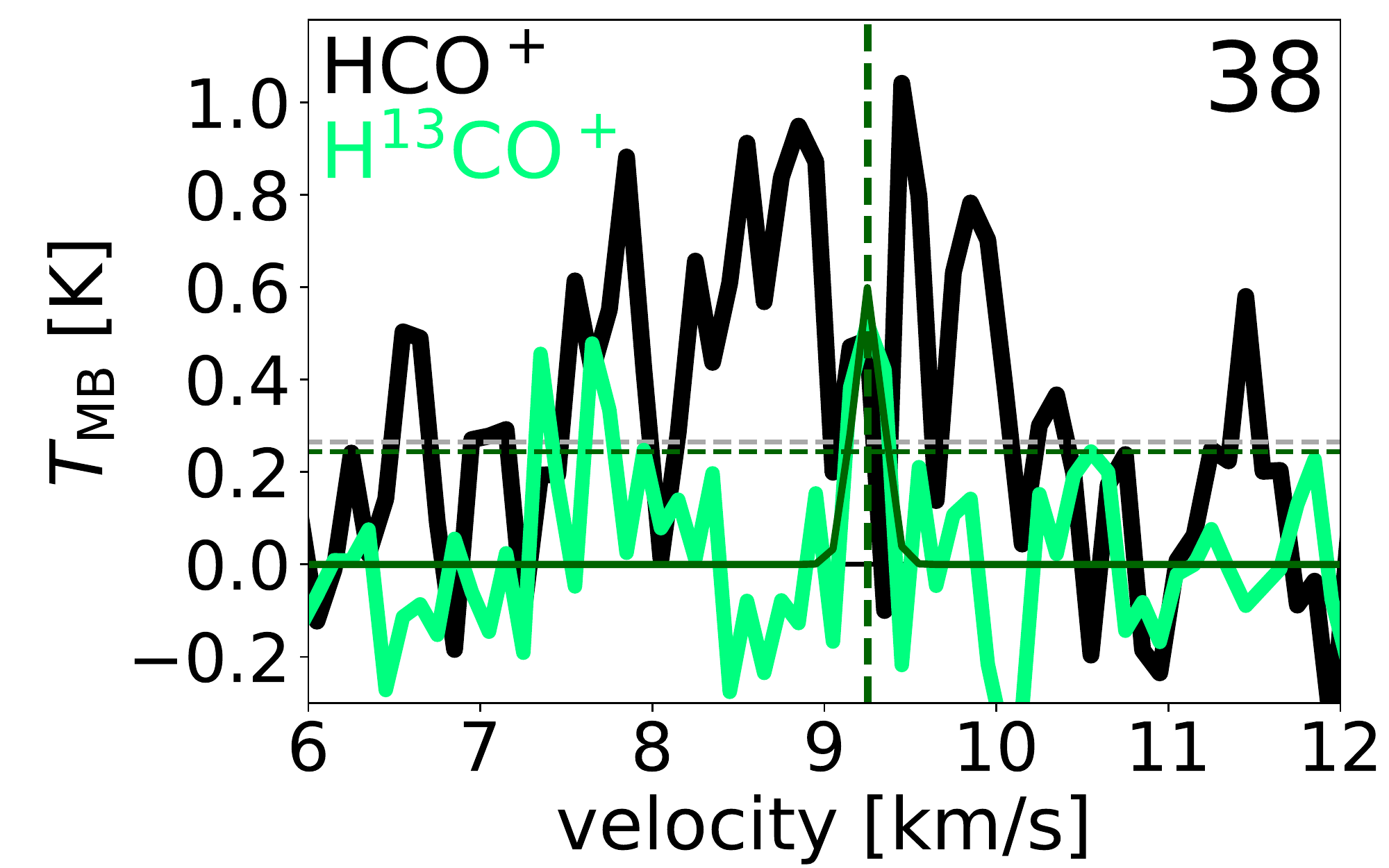}} 
\subfloat{\includegraphics[width=0.165\textwidth]{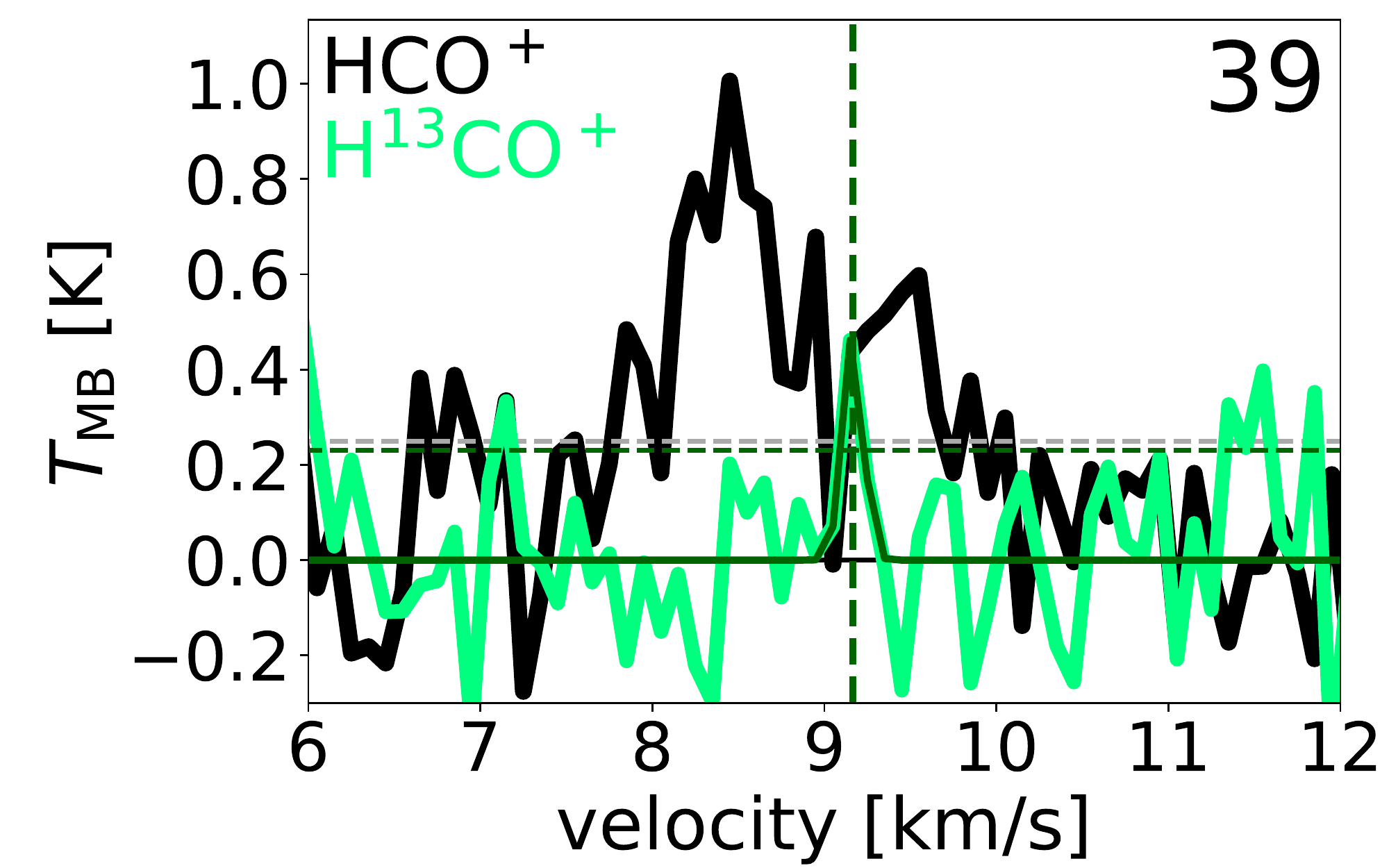}} 
\subfloat{\includegraphics[width=0.165\textwidth]{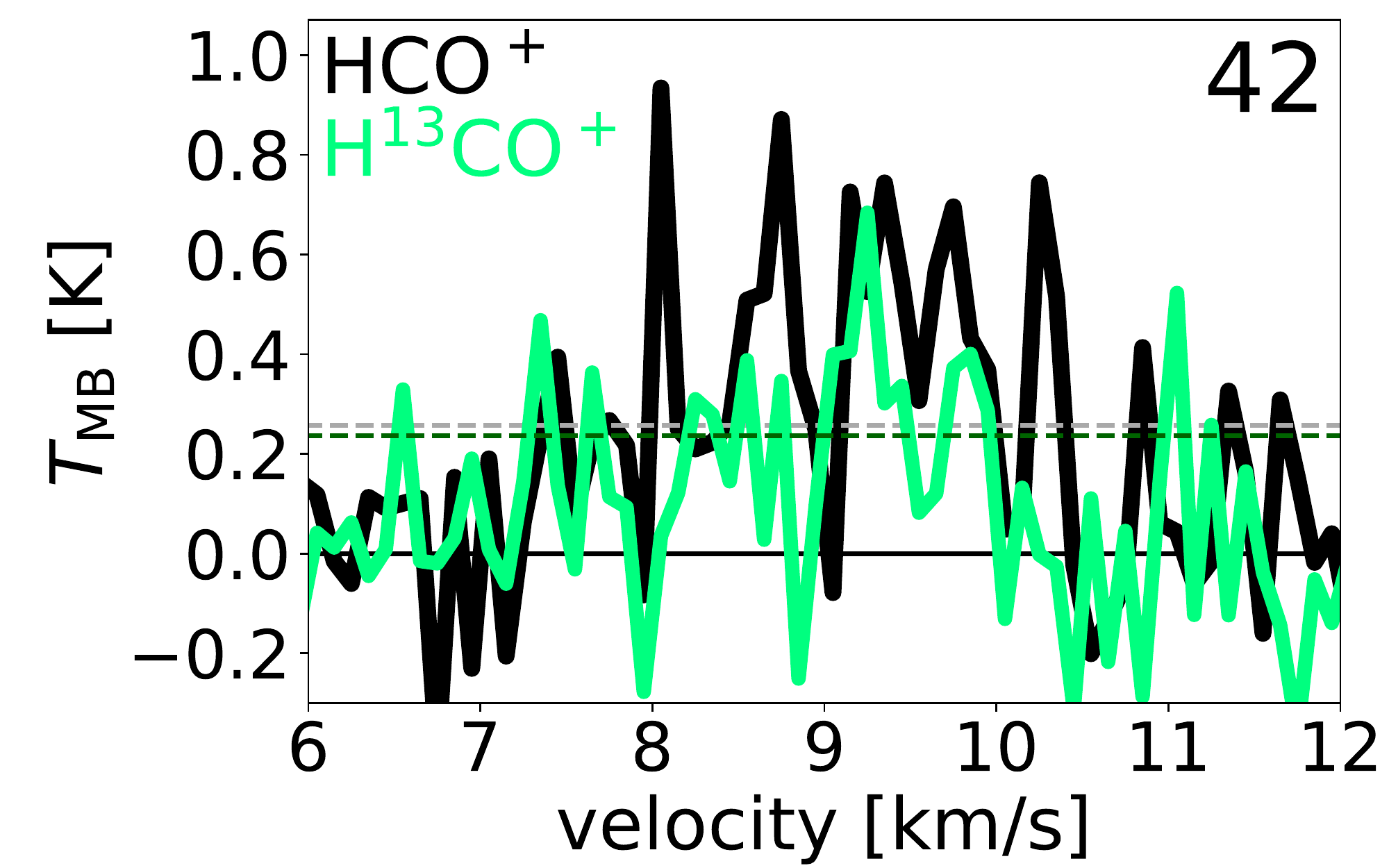}} 

\caption{HCO$^+$(1--0) (black) and H$^{13}$CO$^+$(1--0) (light green) spectra averaged over the analysis spots where observed (see Fig.~\ref{fig_circ}). 
         When it is reasonable, a Gaussian fit and its peak position are also included. 
         Dashed horizontal lines mark 3$\sigma$ rms thresholds; grey for the black spectra, green for the overplotted profiles.   
         For details, see Sects.~\ref{sec:molec} and \ref{sec:energy}. Derived properties from these profiles are listed in Table~\ref{tab_uCalcT}. 
         } 
\label{fig_HCO+_H13CO+_spectra}
\end{figure*}

\begin{figure*}
\subfloat{\includegraphics[width=0.249\textwidth]{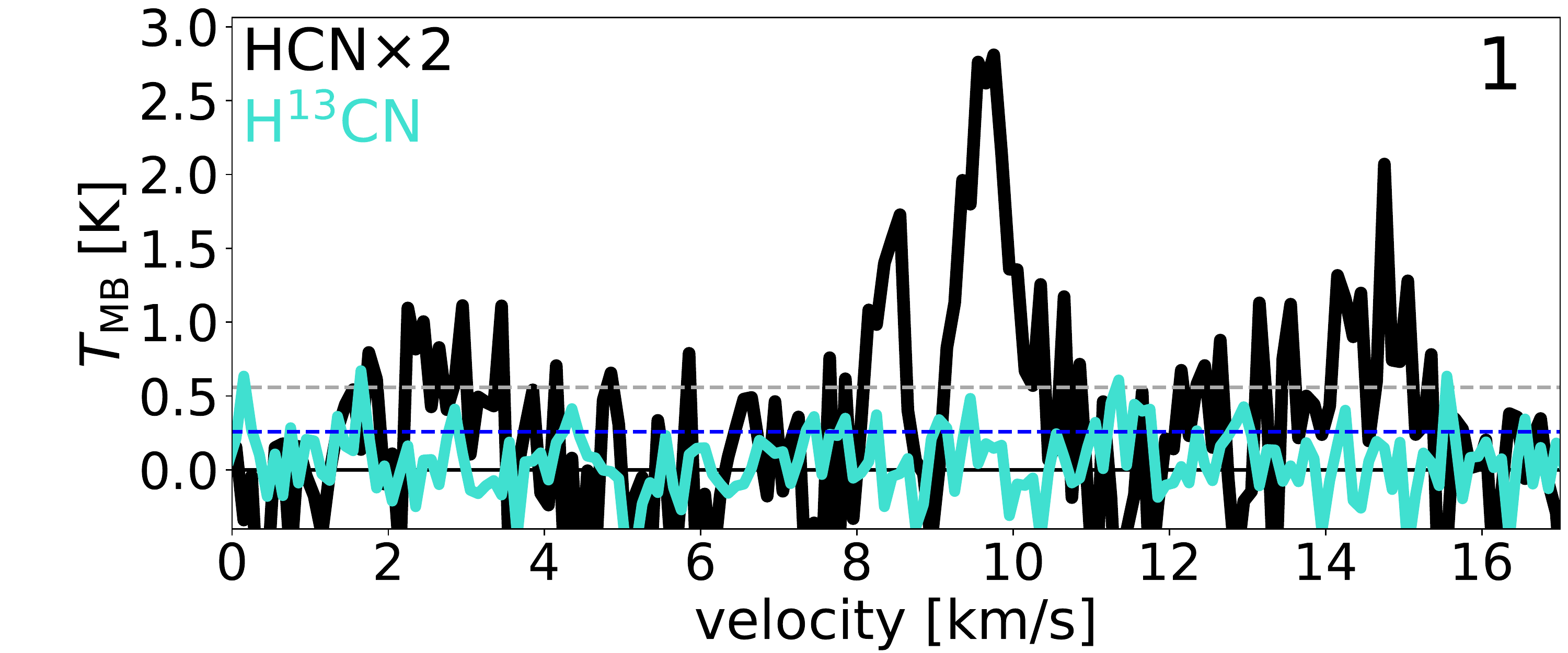}} 
\subfloat{\includegraphics[width=0.249\textwidth]{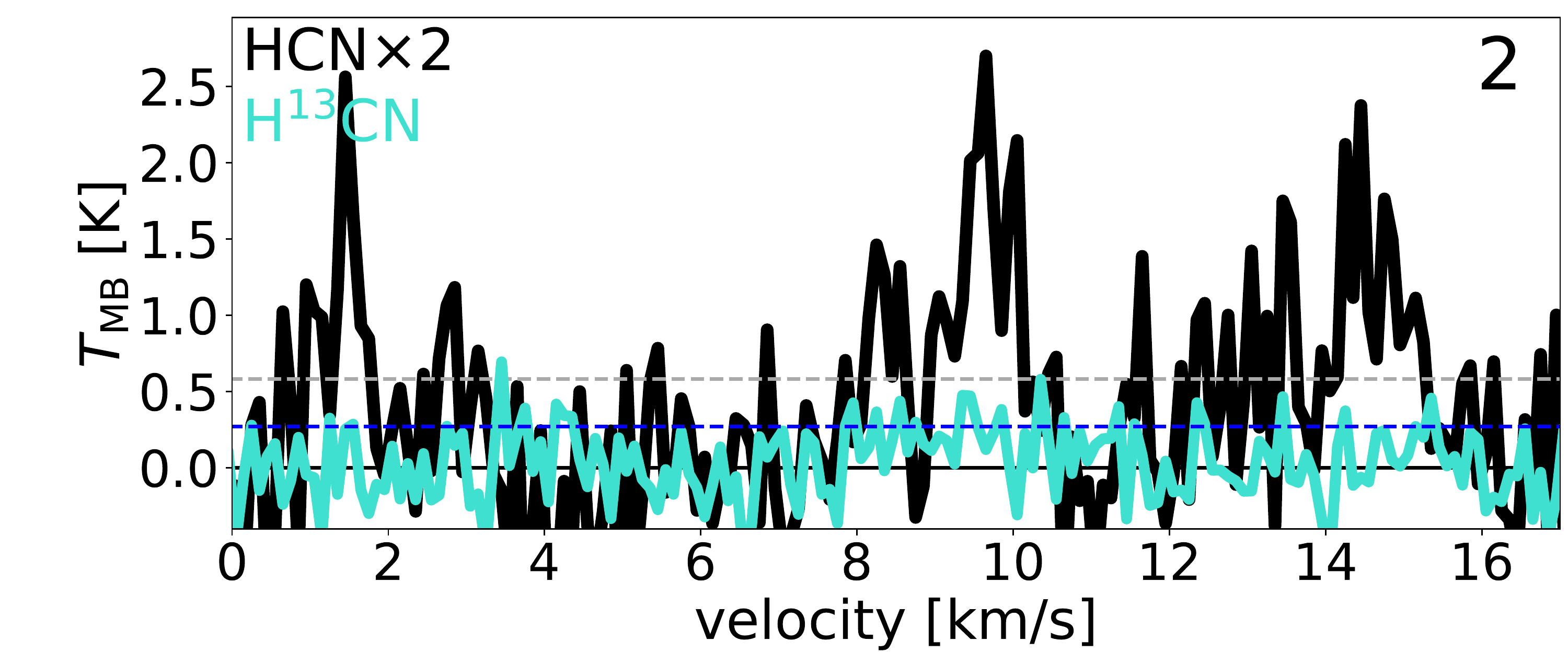}} 
\subfloat{\includegraphics[width=0.249\textwidth]{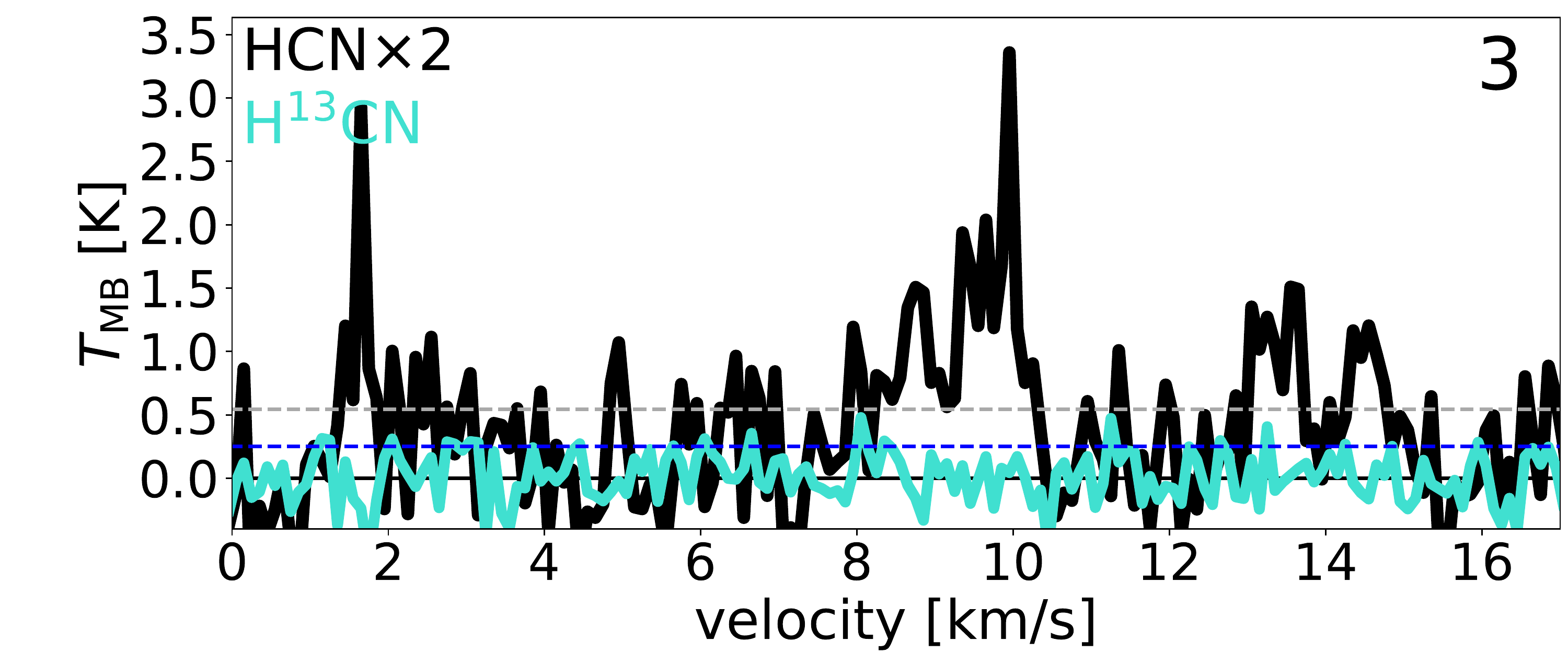}}
\subfloat{\includegraphics[width=0.249\textwidth]{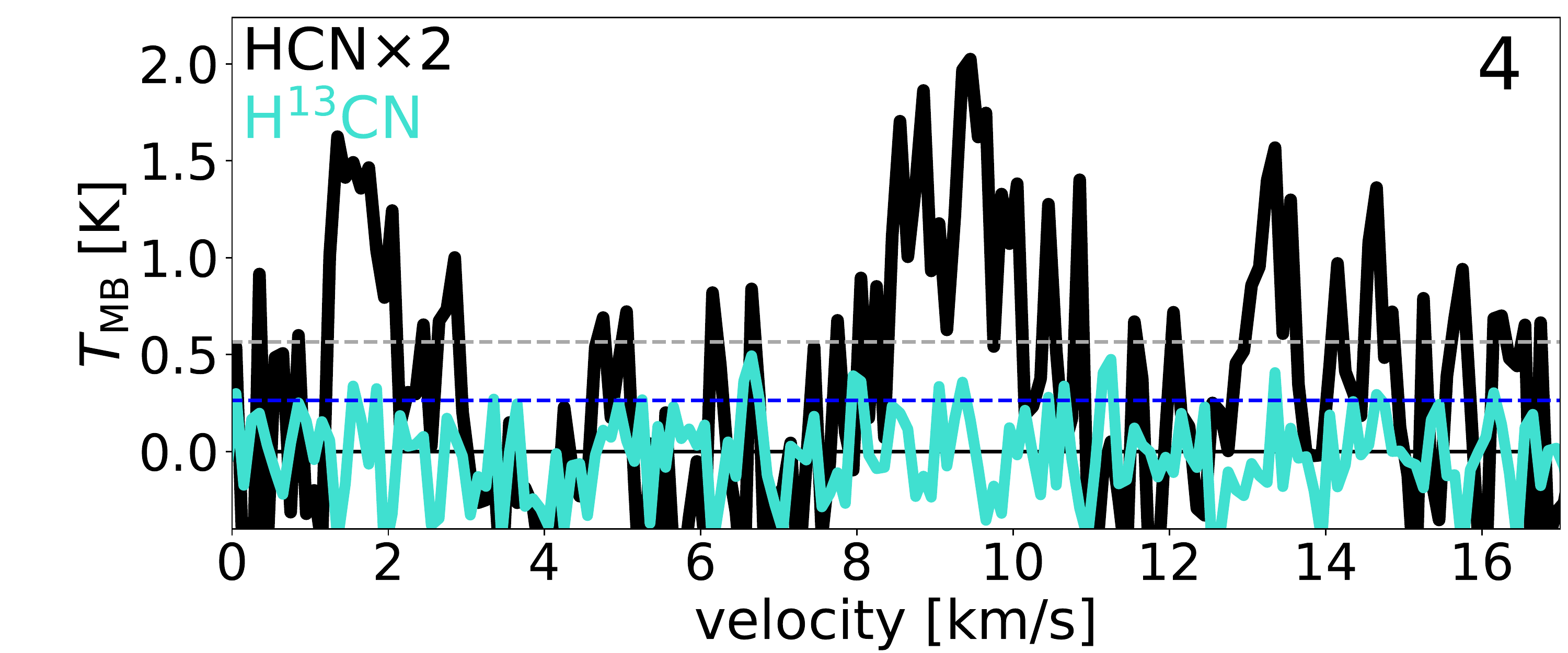}}\\
\subfloat{\includegraphics[width=0.249\textwidth]{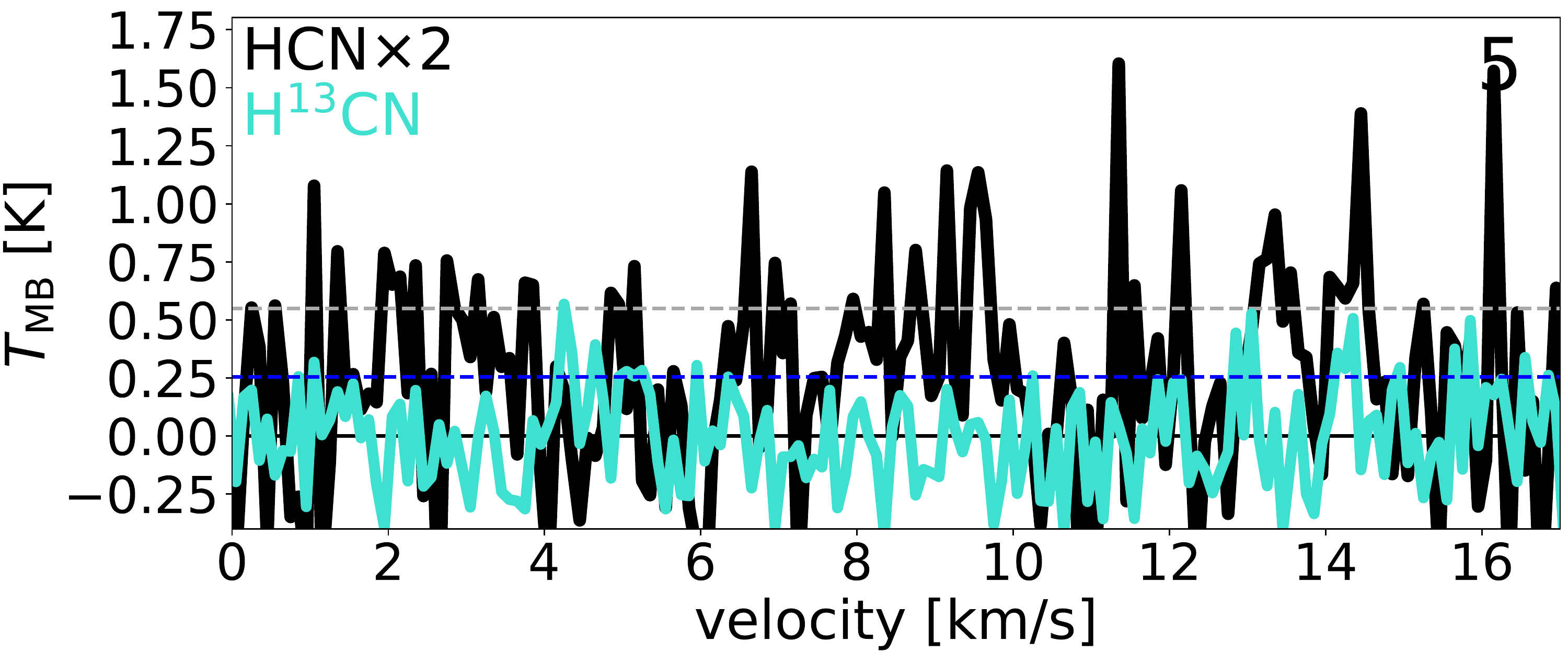}} 
\subfloat{\includegraphics[width=0.249\textwidth]{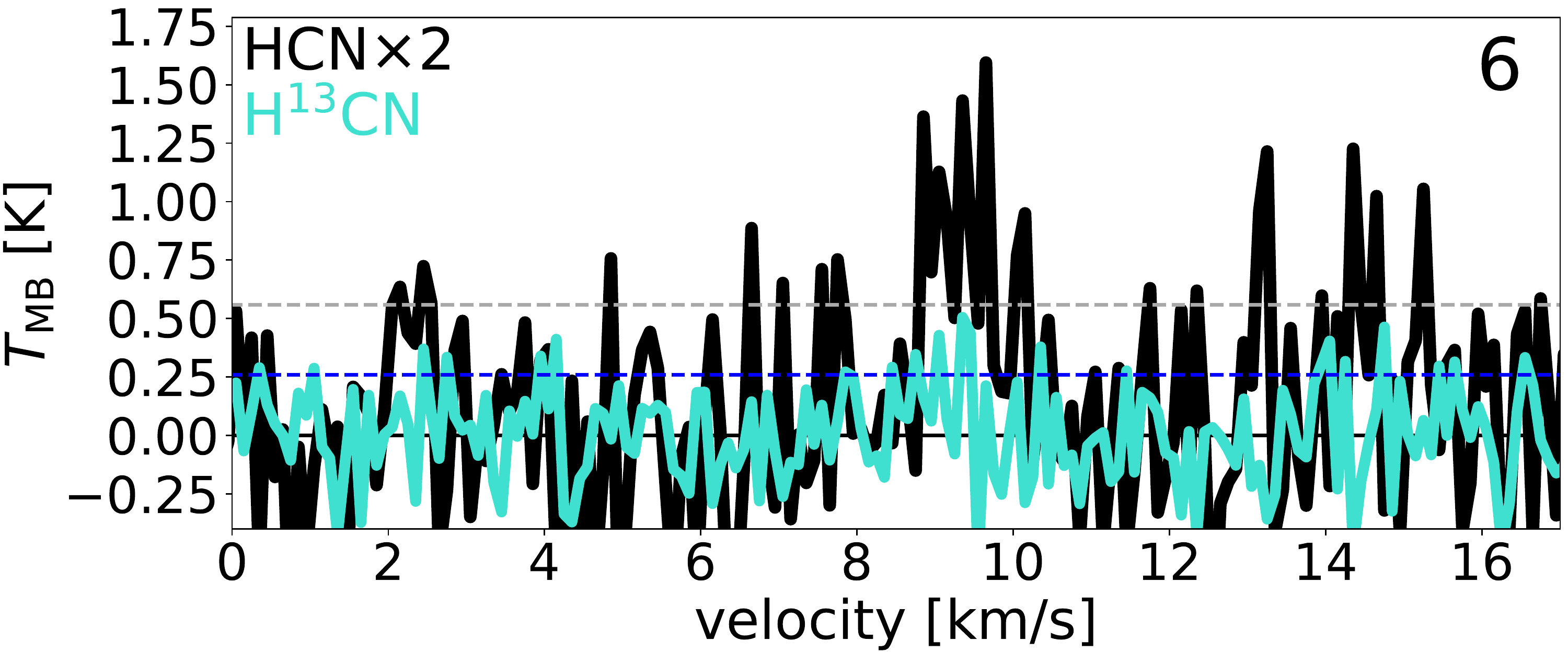}} 
\subfloat{\includegraphics[width=0.249\textwidth]{./figs/HCN_H13CN_spectra_noFit_cNo7.pdf}} 
\subfloat{\includegraphics[width=0.249\textwidth]{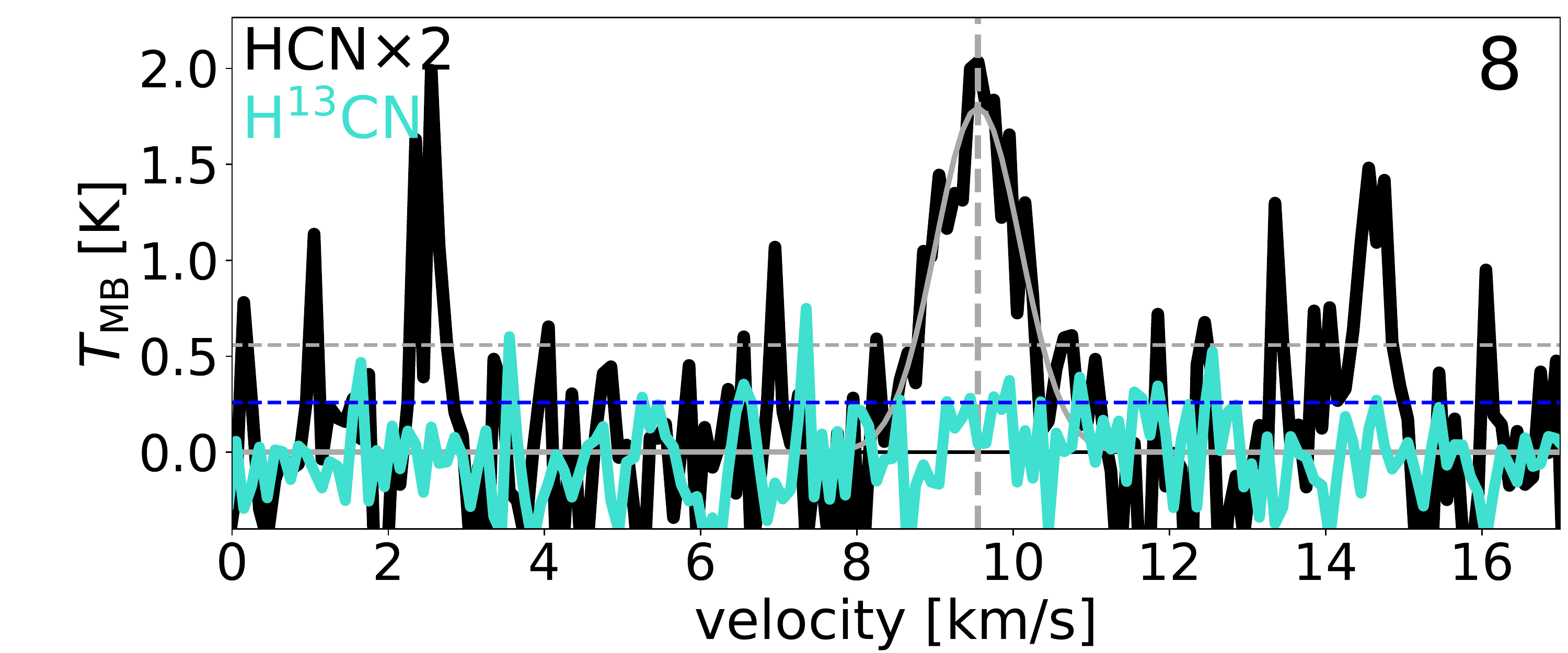}}\\ 

\subfloat{\includegraphics[width=0.249\textwidth]{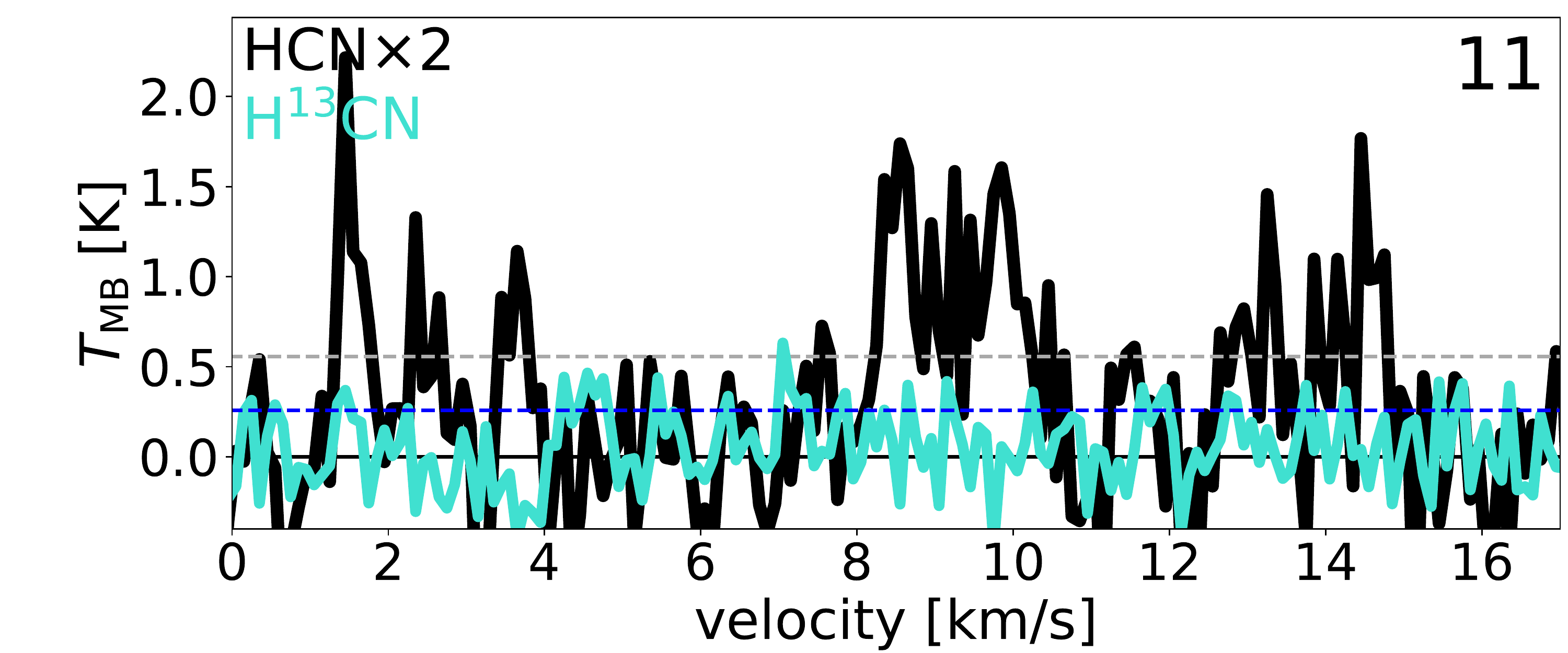}}
\subfloat{\includegraphics[width=0.249\textwidth]{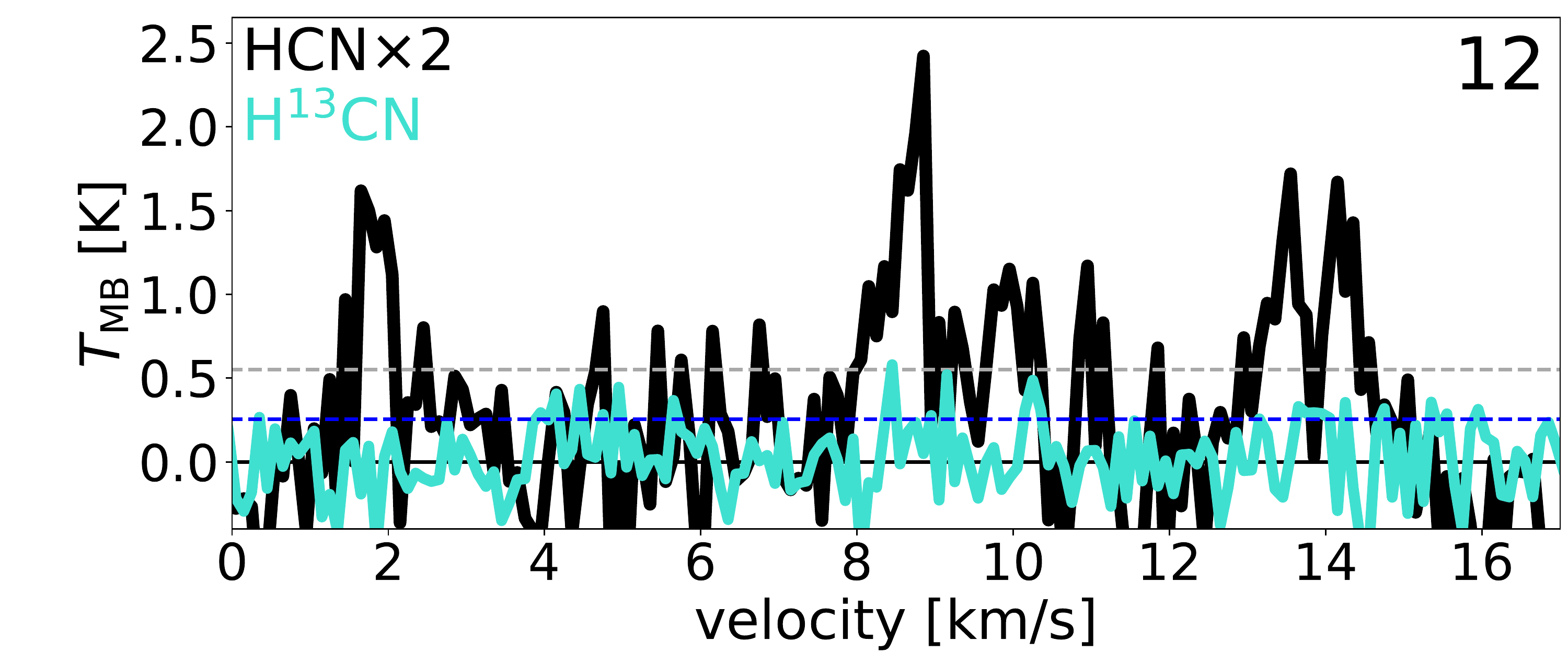}}
\subfloat{\includegraphics[width=0.249\textwidth]{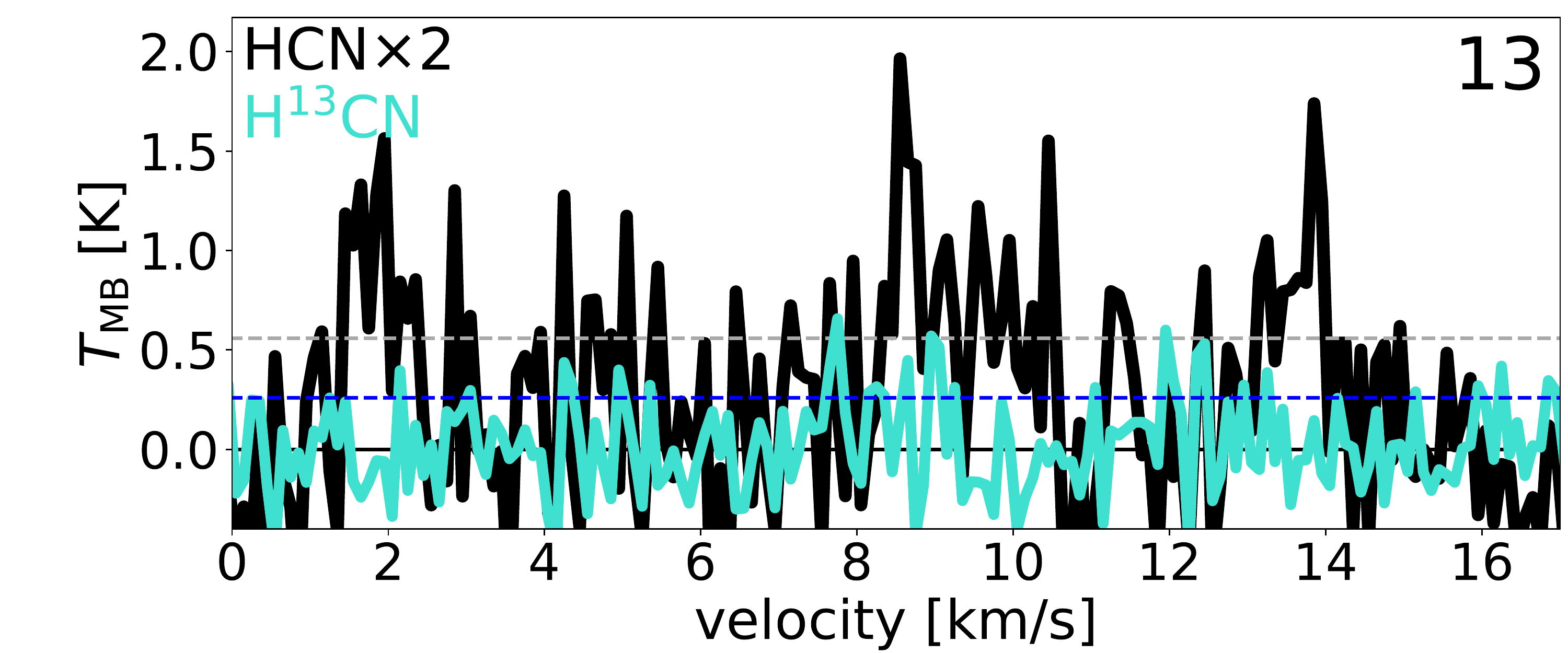}} 
\subfloat{\includegraphics[width=0.249\textwidth]{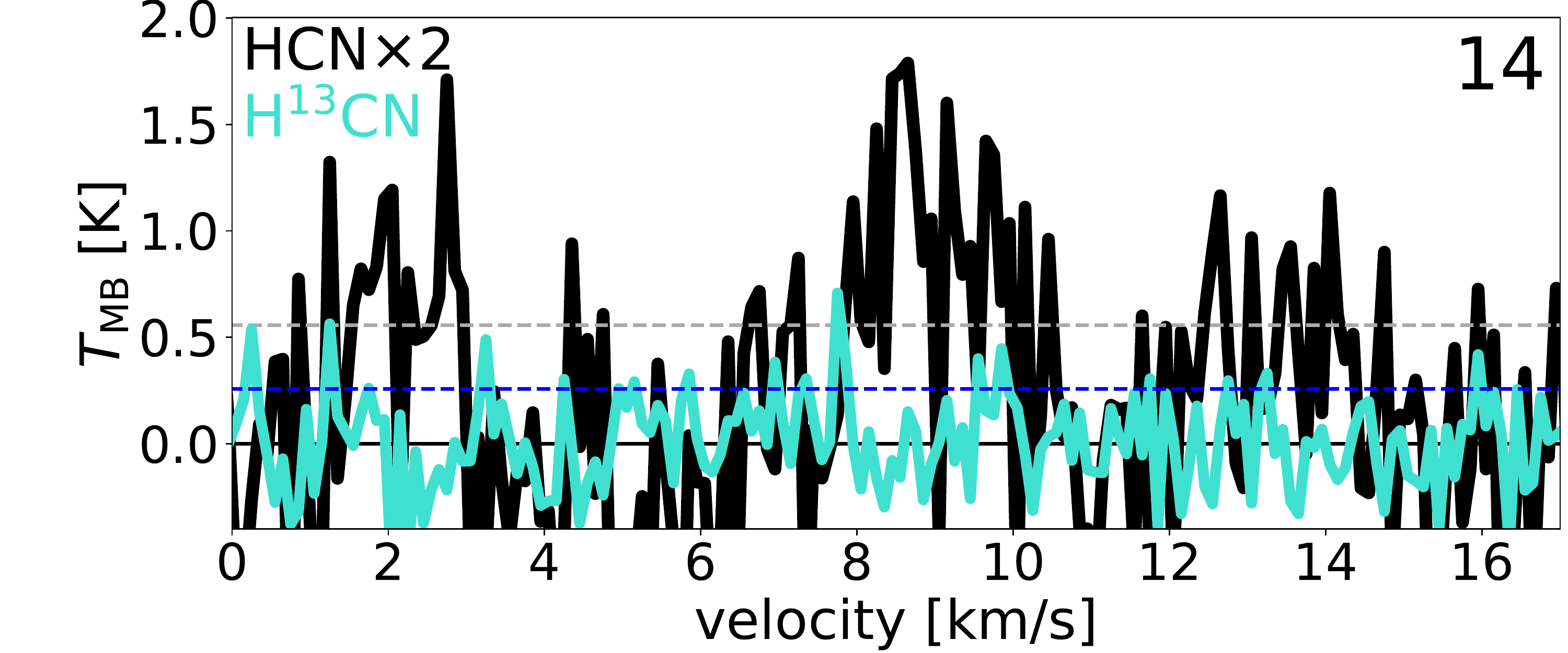}}\\ 
\subfloat{\includegraphics[width=0.249\textwidth]{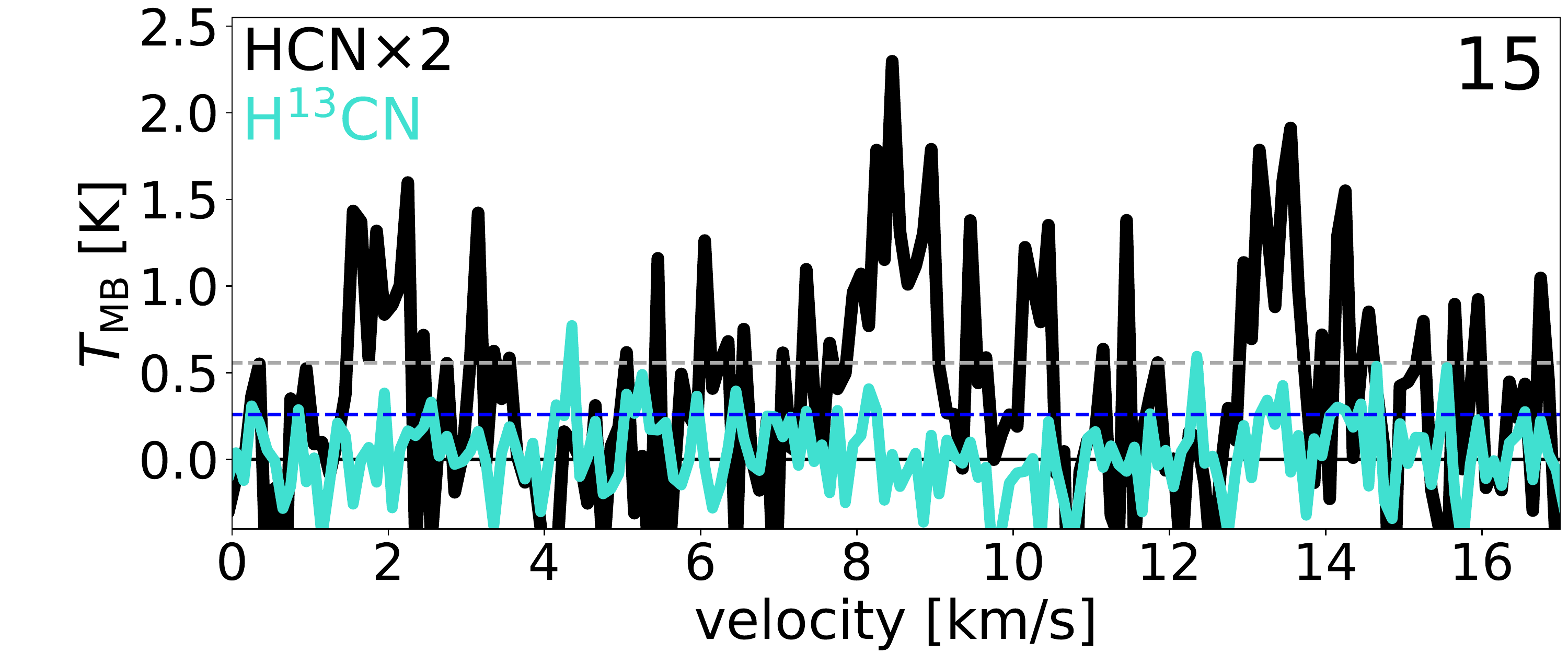}} 
\subfloat{\includegraphics[width=0.249\textwidth]{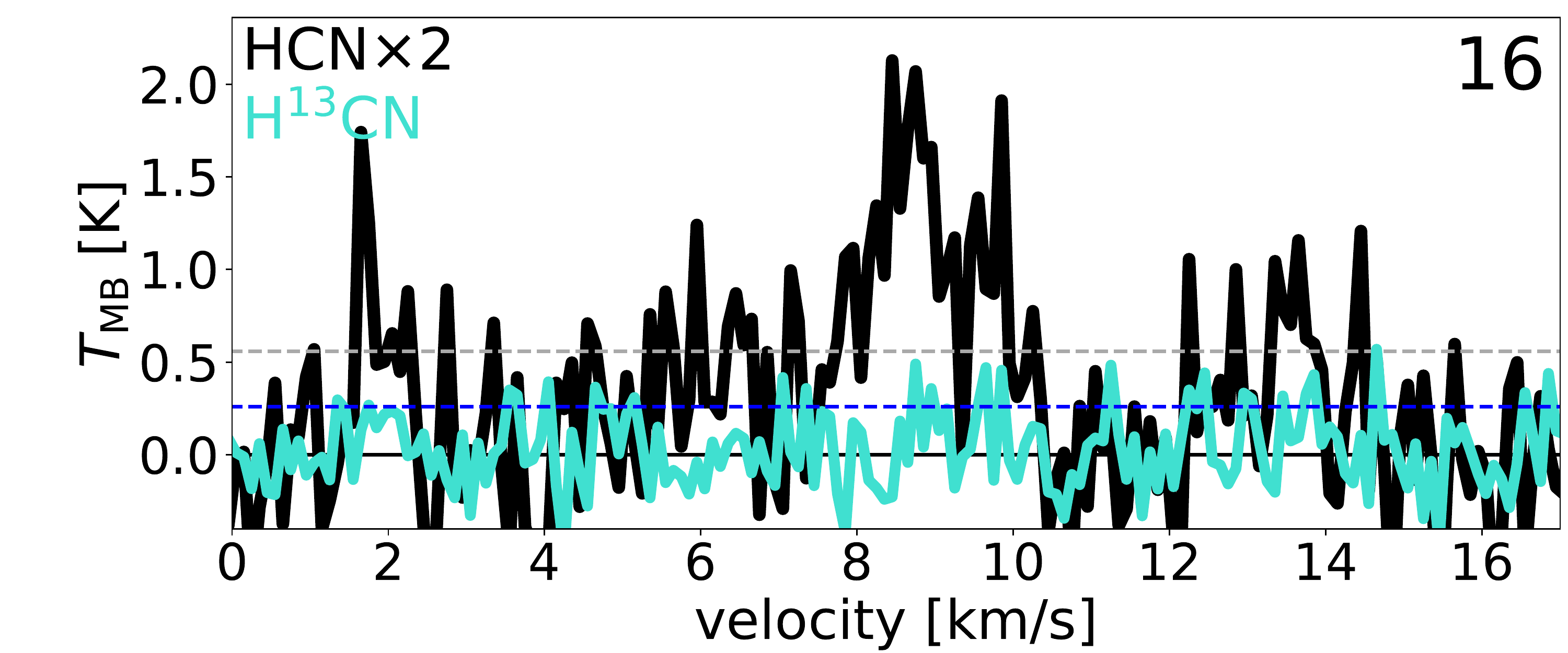}}
\subfloat{\includegraphics[width=0.249\textwidth]{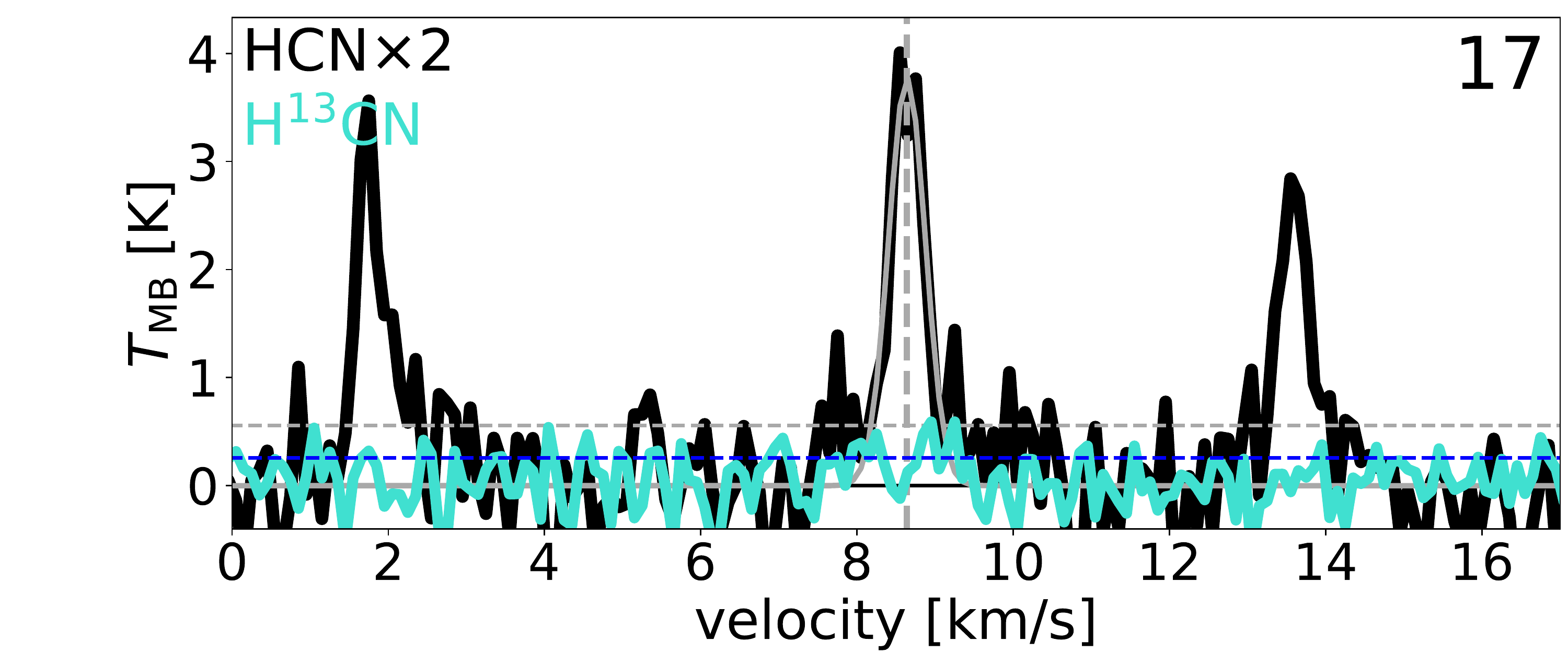}} 
\subfloat{\includegraphics[width=0.249\textwidth]{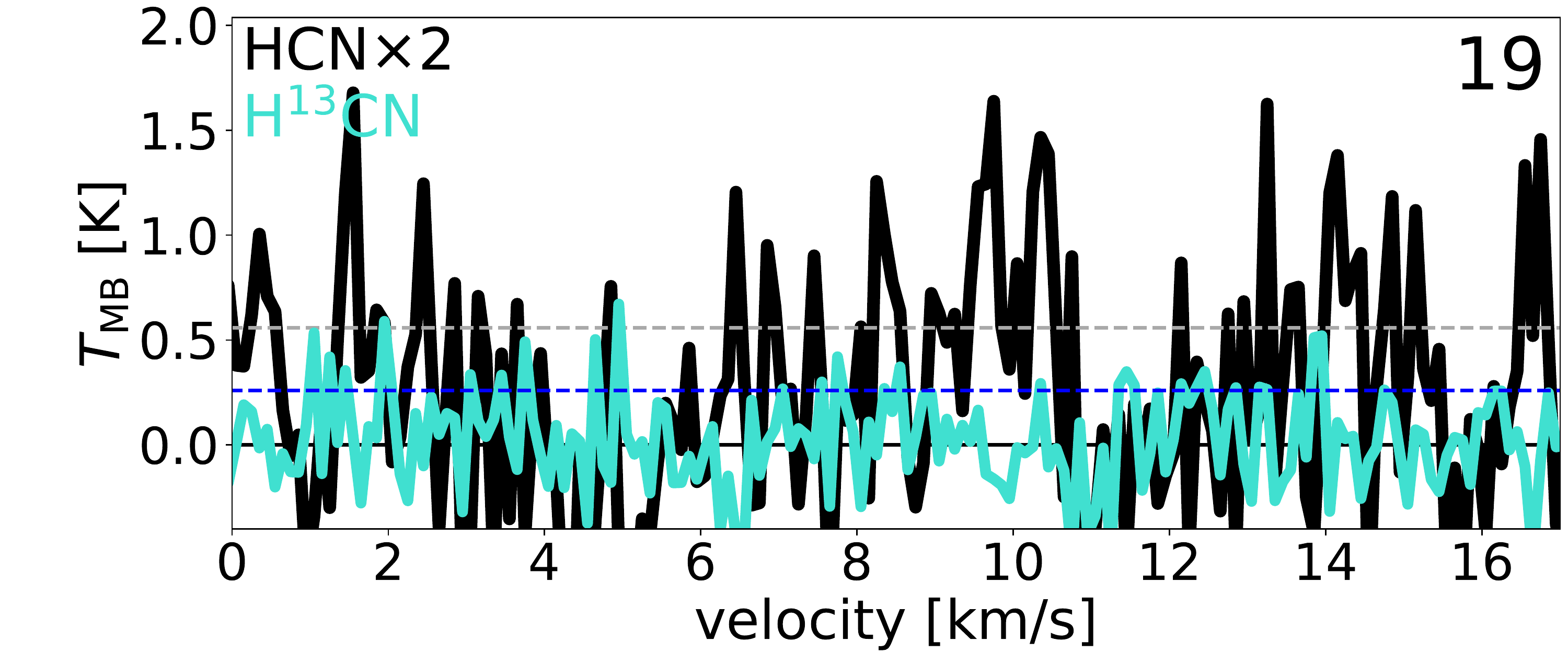}}\\
 
\subfloat{\includegraphics[width=0.249\textwidth]{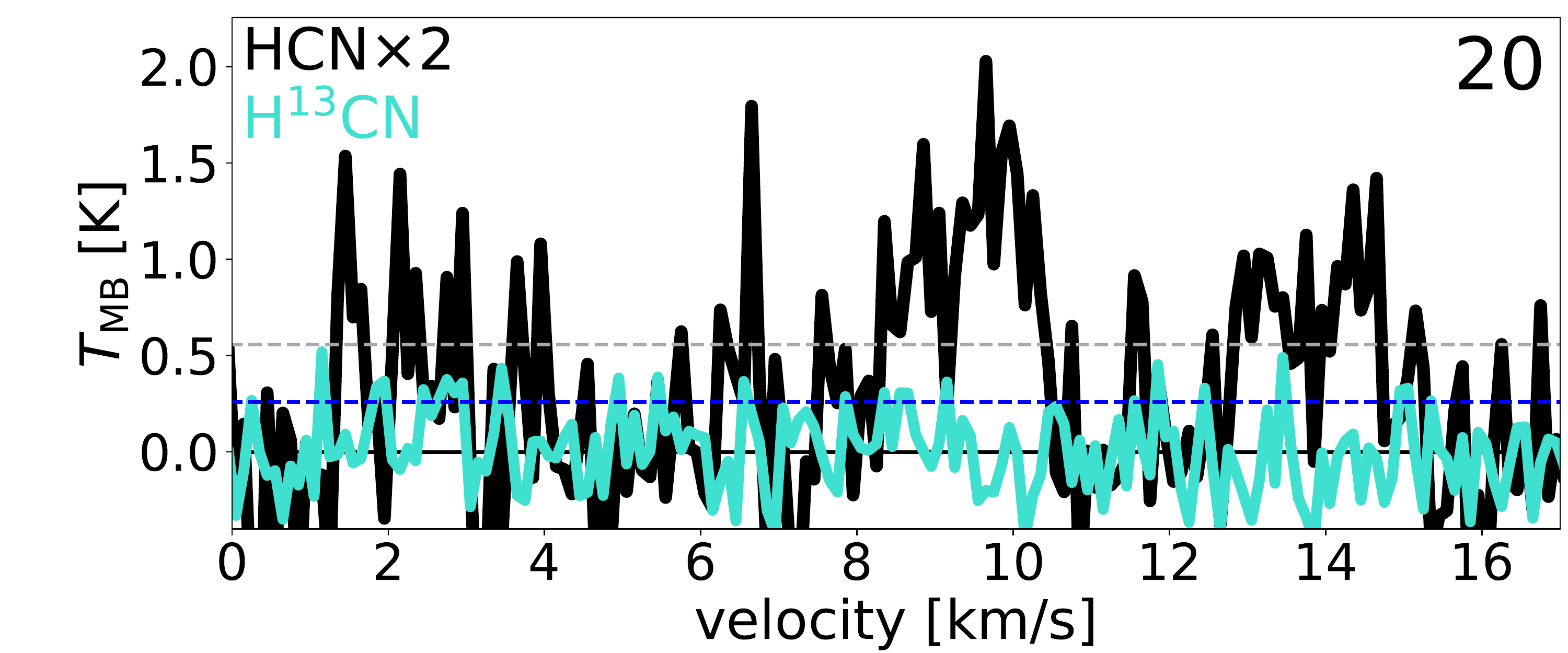}} 
\subfloat{\includegraphics[width=0.249\textwidth]{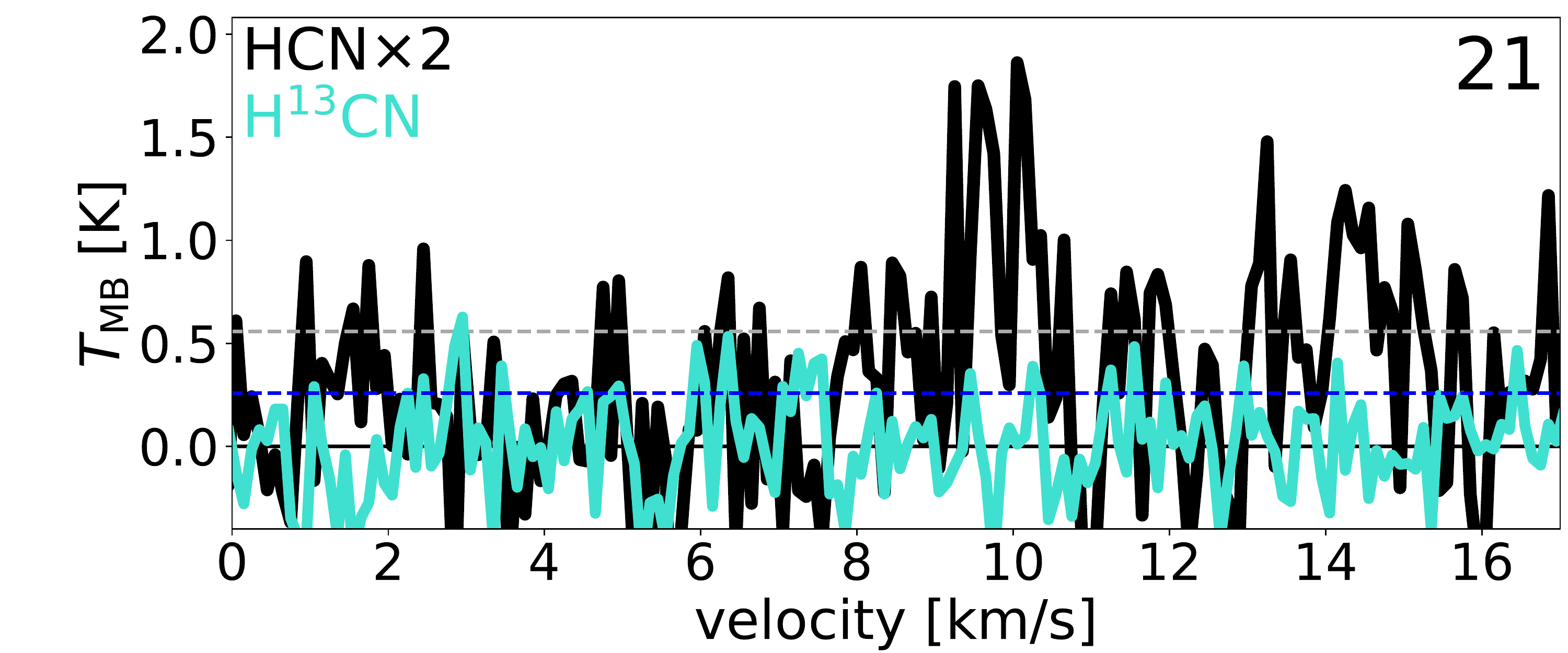}} 
\subfloat{\includegraphics[width=0.249\textwidth]{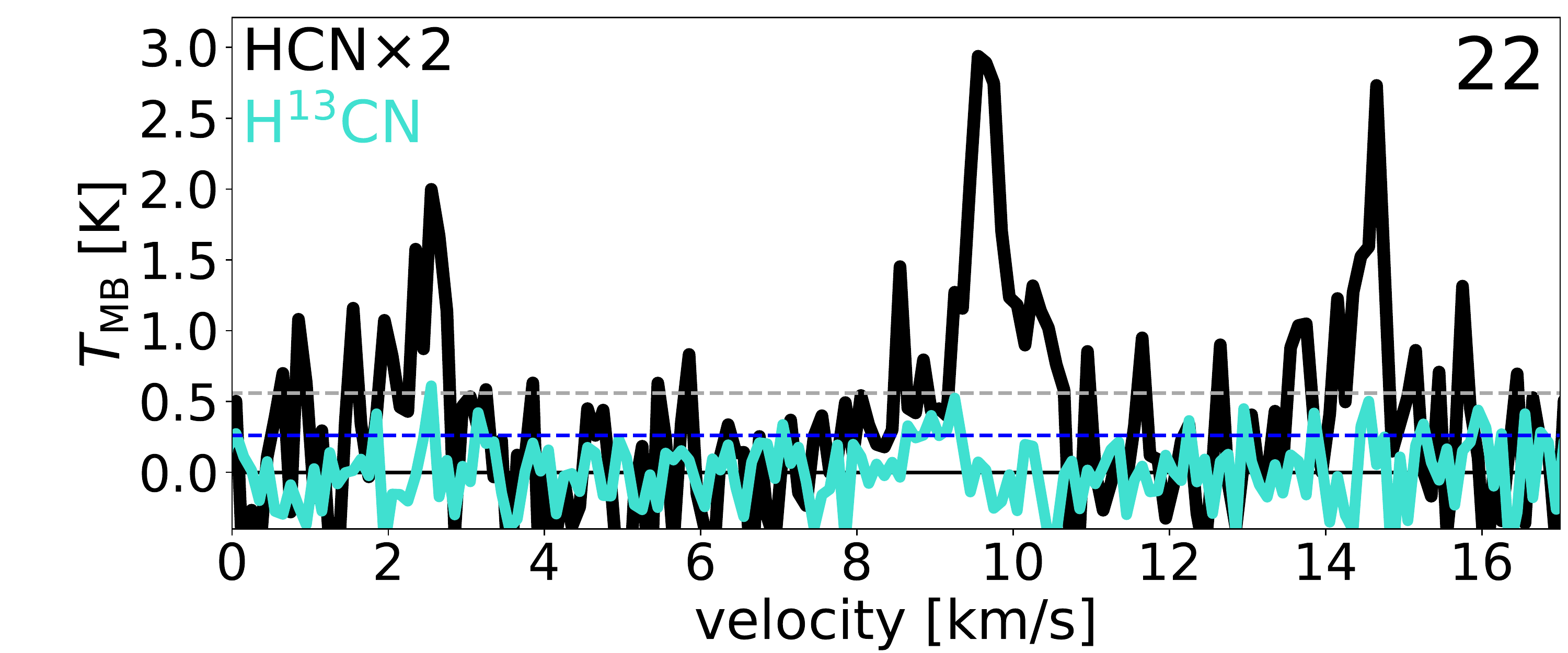}} 
\subfloat{\includegraphics[width=0.249\textwidth]{./figs/HCN_H13CN_spectra_cNo23.pdf}}\\ 
\subfloat{\includegraphics[width=0.249\textwidth]{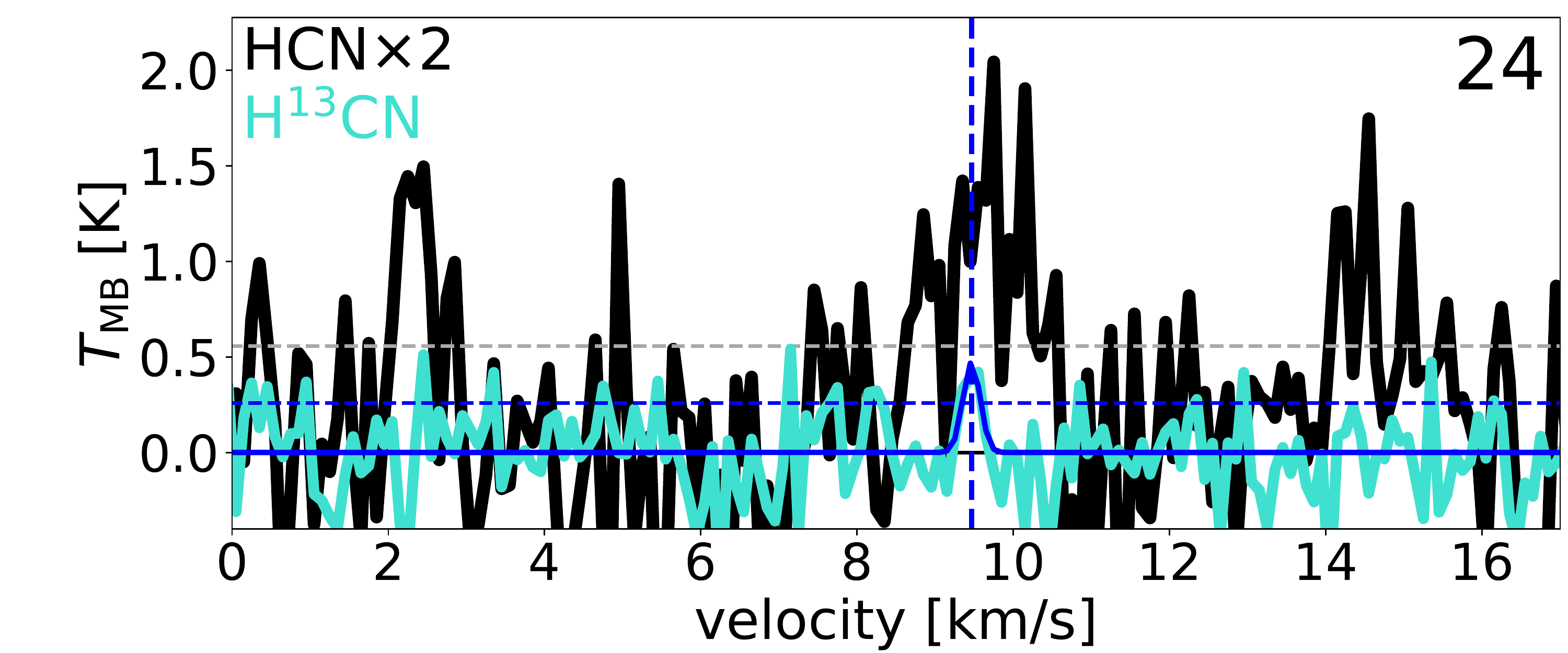}} 
\subfloat{\includegraphics[width=0.249\textwidth]{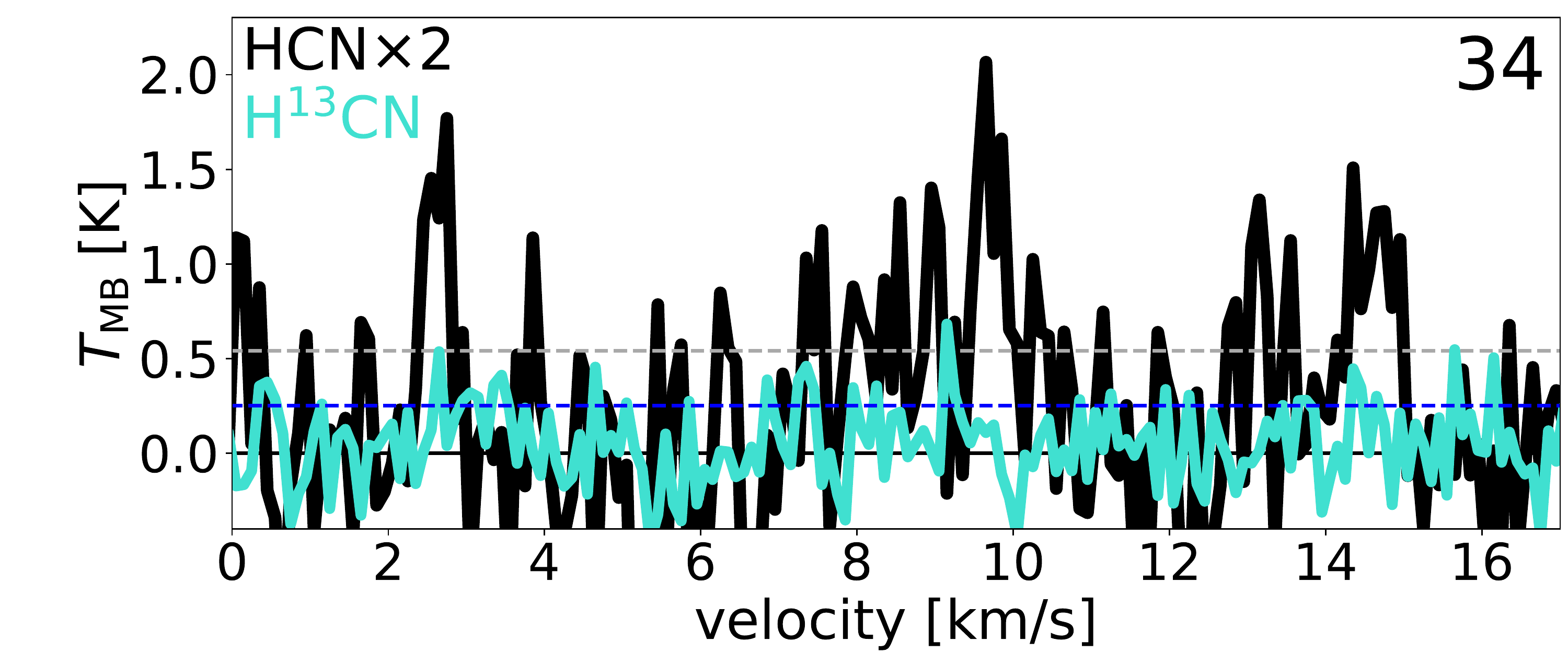}} 
\subfloat{\includegraphics[width=0.249\textwidth]{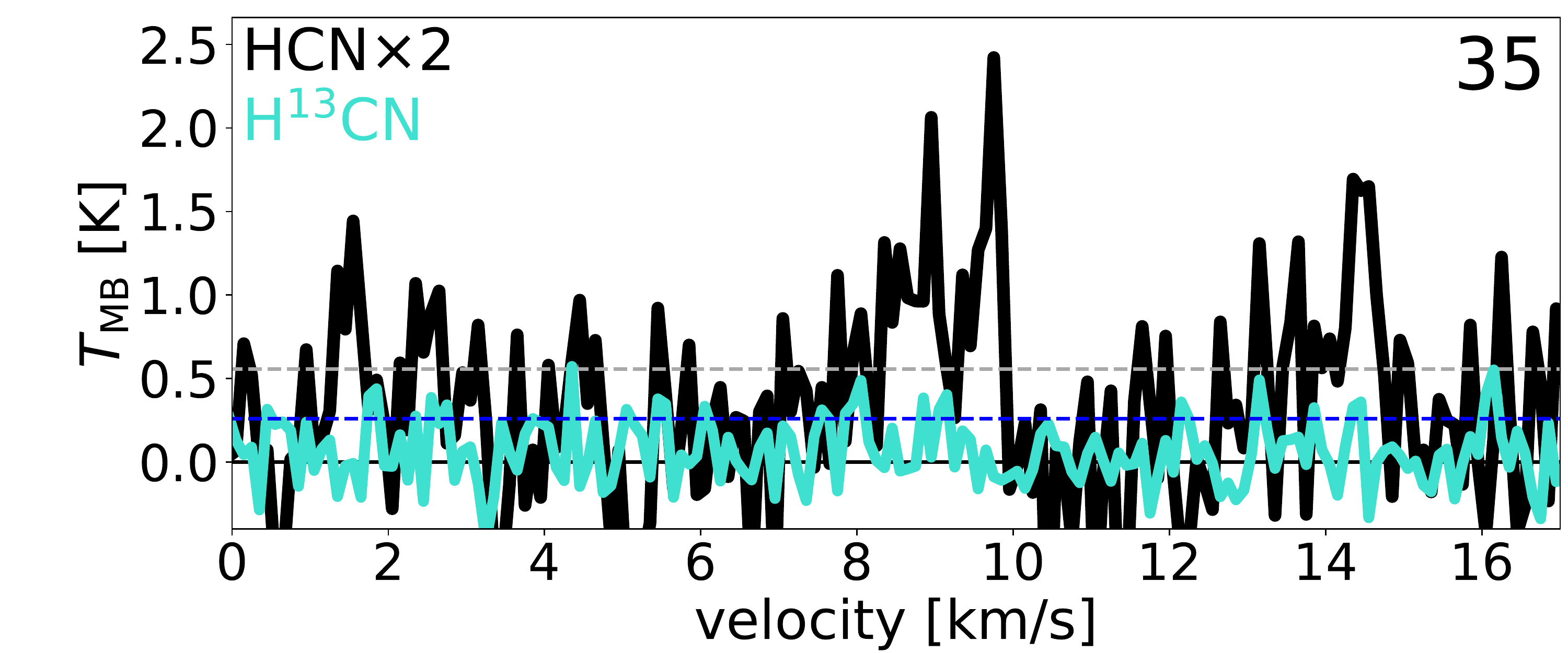}} 
\subfloat{\includegraphics[width=0.249\textwidth]{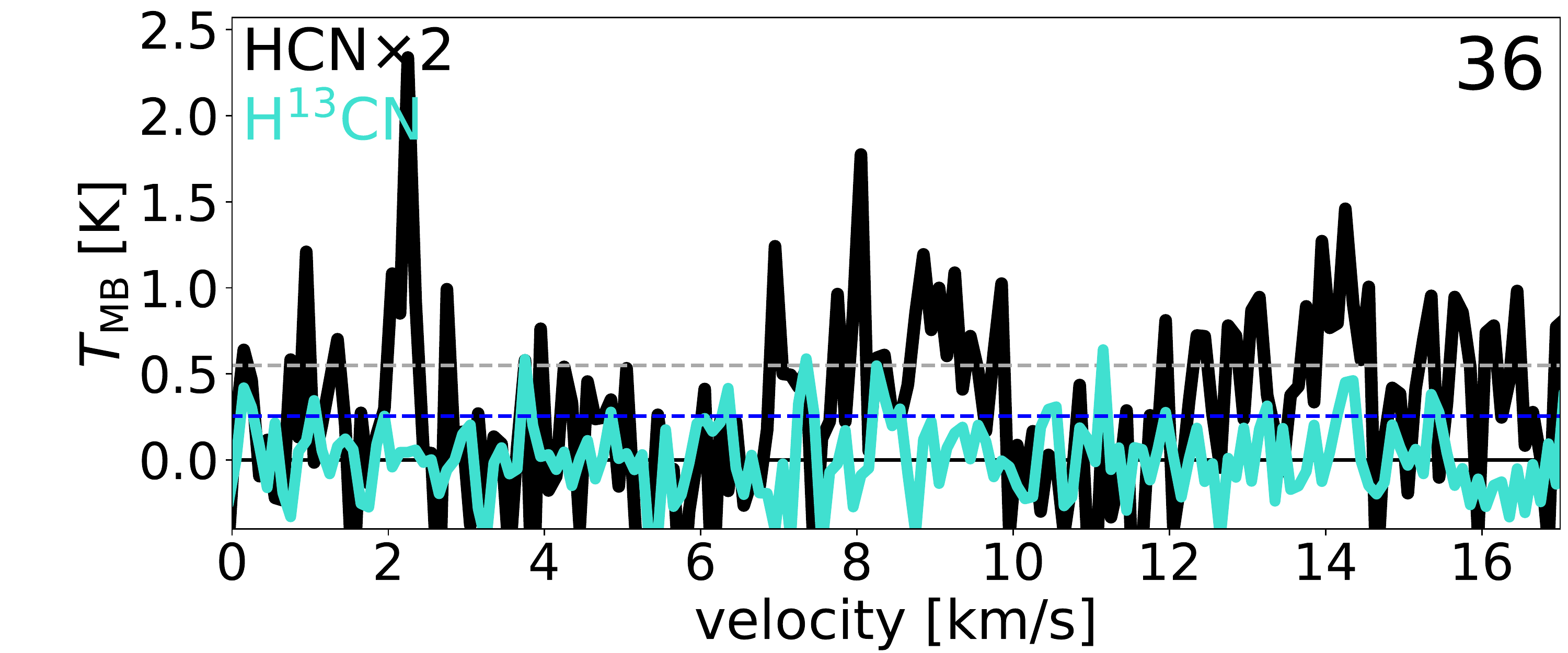}}\\
 
\subfloat{\includegraphics[width=0.249\textwidth]{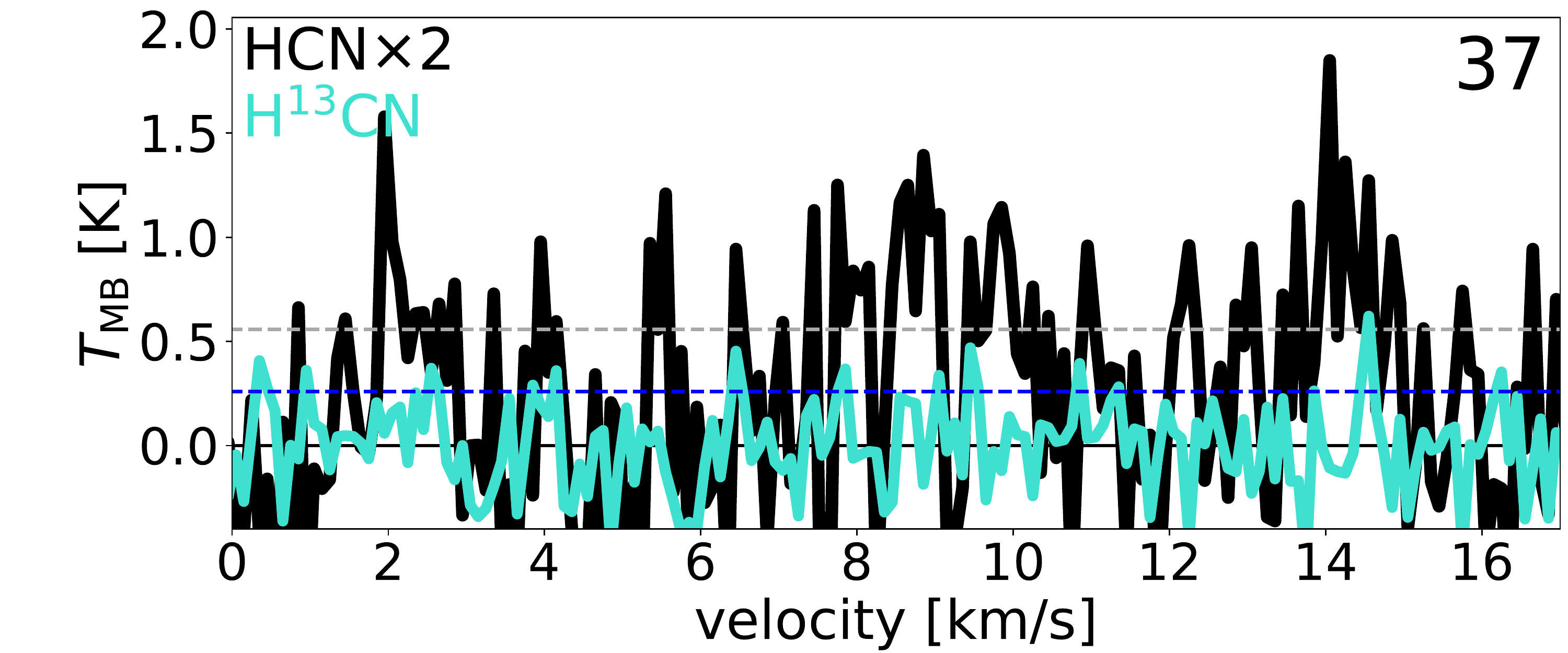}} 
\subfloat{\includegraphics[width=0.249\textwidth]{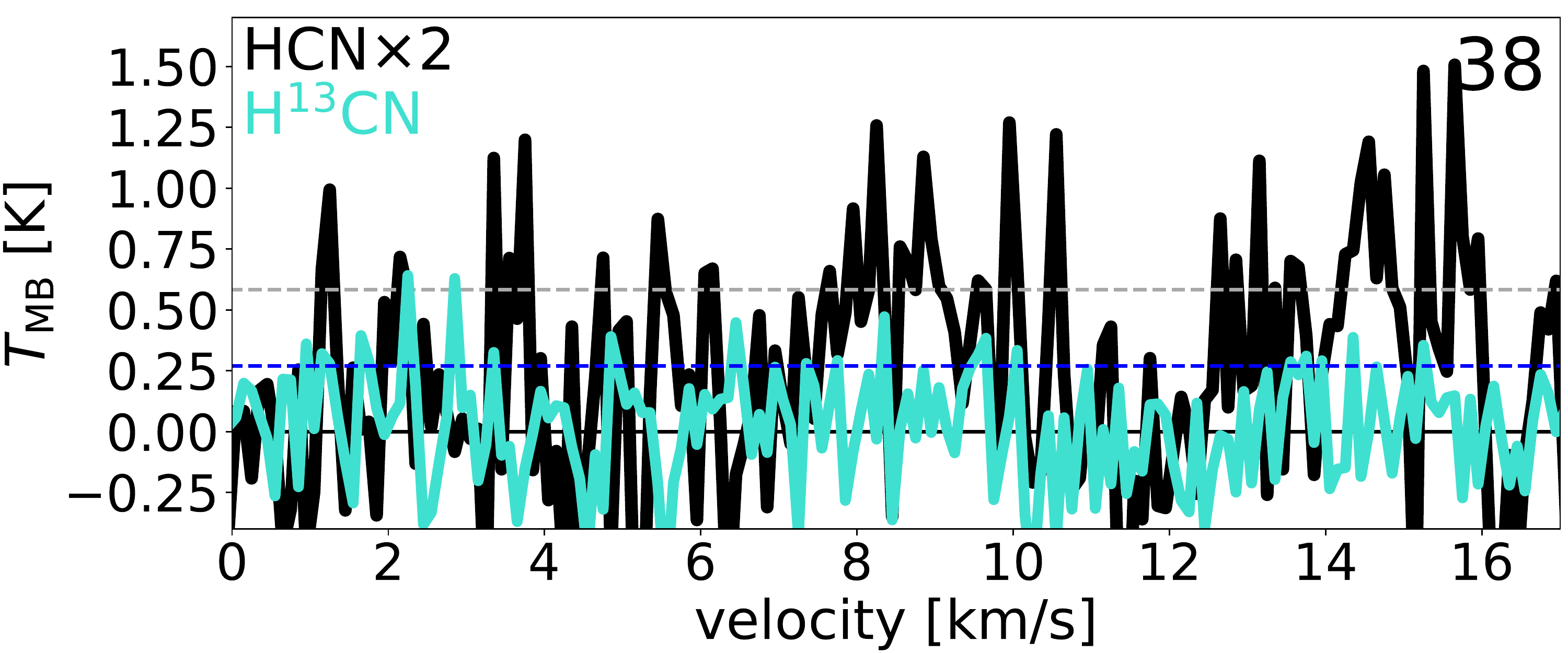}}
\subfloat{\includegraphics[width=0.249\textwidth]{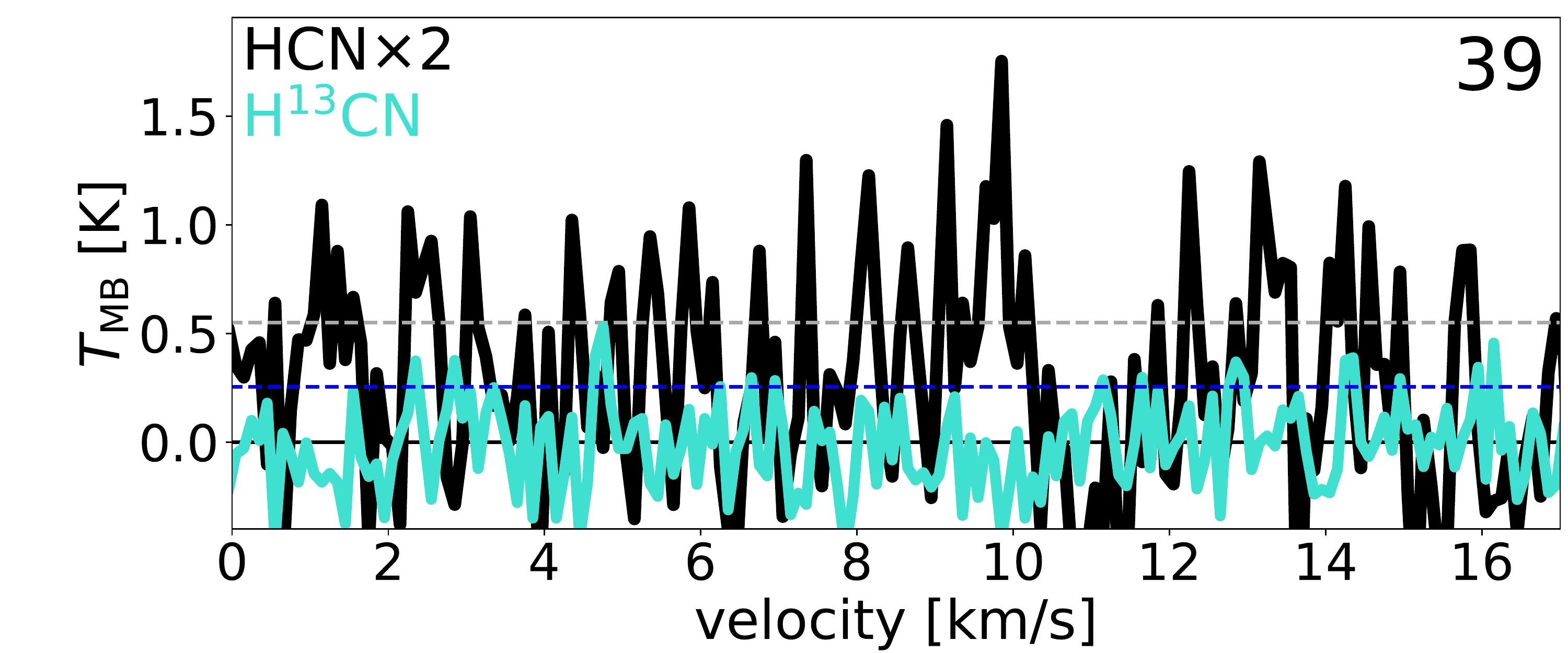}} 
\subfloat{\includegraphics[width=0.249\textwidth]{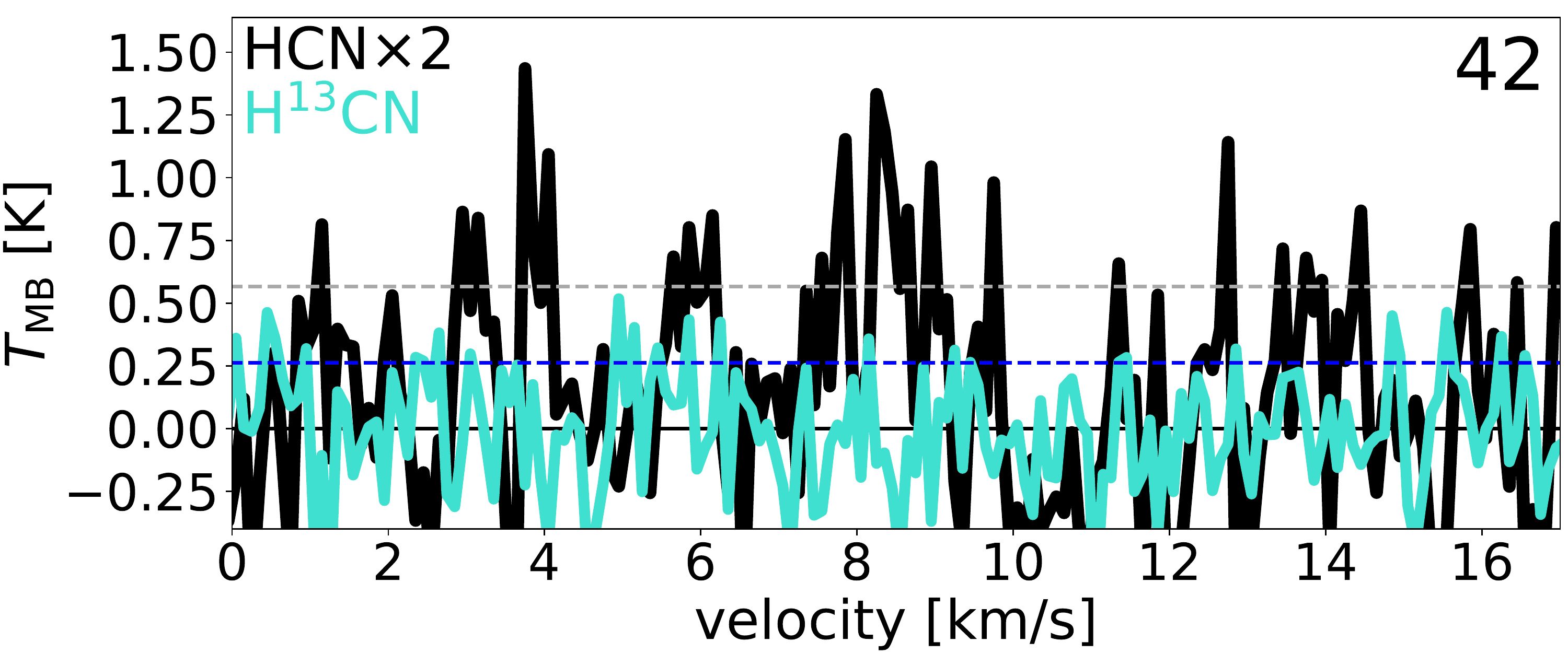}} 

\caption{HCN(1--0) (black) and H$^{13}$CN(1--0) (cyan) spectra averaged over the analysis spots where observed (see Fig.~\ref{fig_circ}). 
         When it is reasonable, a Gaussian fit and its peak position are also included. 
         Dashed horizontal lines mark 3$\sigma$ rms thresholds; grey for the black spectra, blue for the overplotted profiles.   
         For details, see Sects.~\ref{sec:molec} and \ref{sec:energy}. Derived properties from these profiles are listed in Table~\ref{tab_uCalcT}. 
         } 
\label{fig_HCN_H13CN_spectra}
\end{figure*}

\section{Integrated Intensity Maps}\label{app:mom0}

\begin{figure*}
\subfloat{\includegraphics[width=0.48\textwidth]{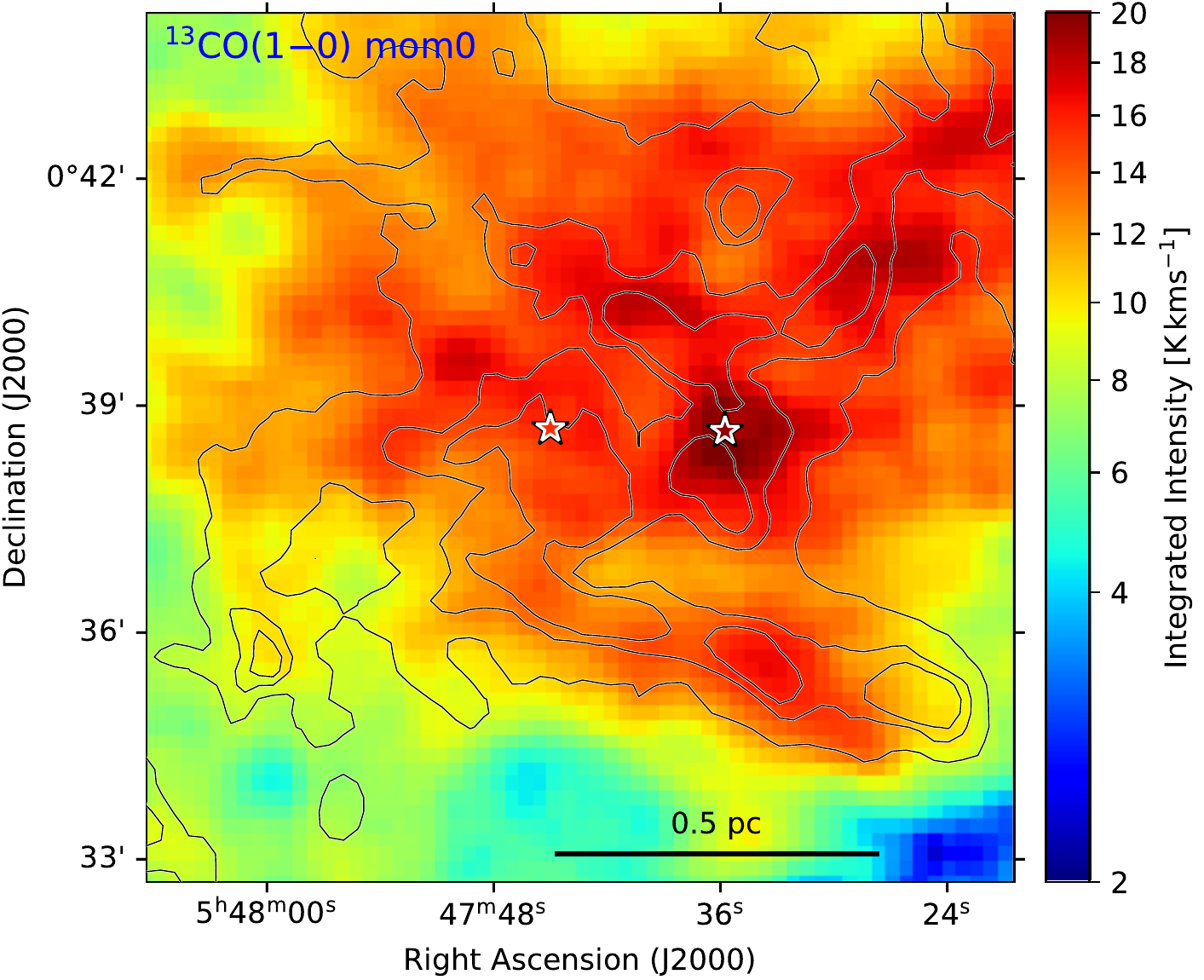}} 
\subfloat{\includegraphics[width=0.48\textwidth]{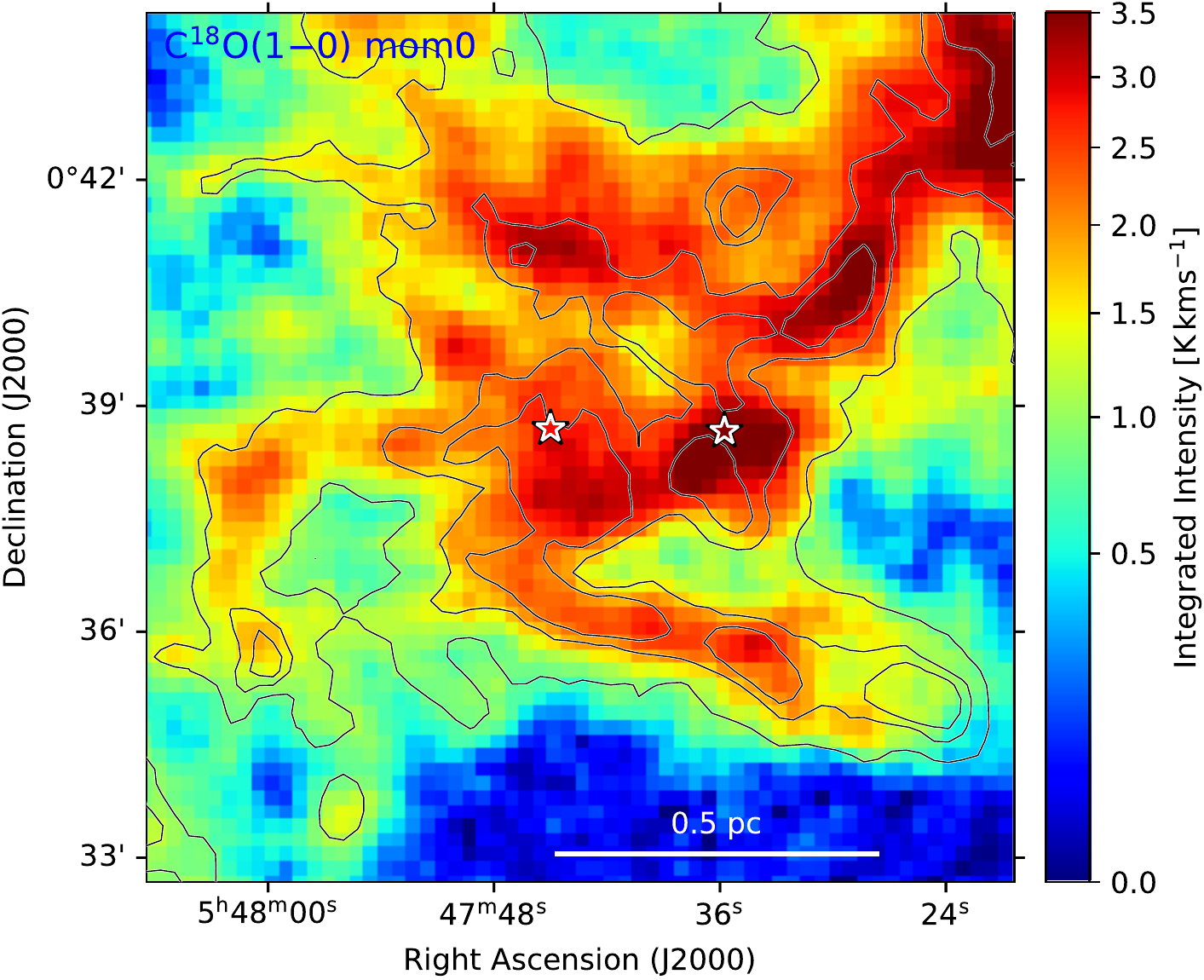}}\\ 
\subfloat{\includegraphics[width=0.48\textwidth]{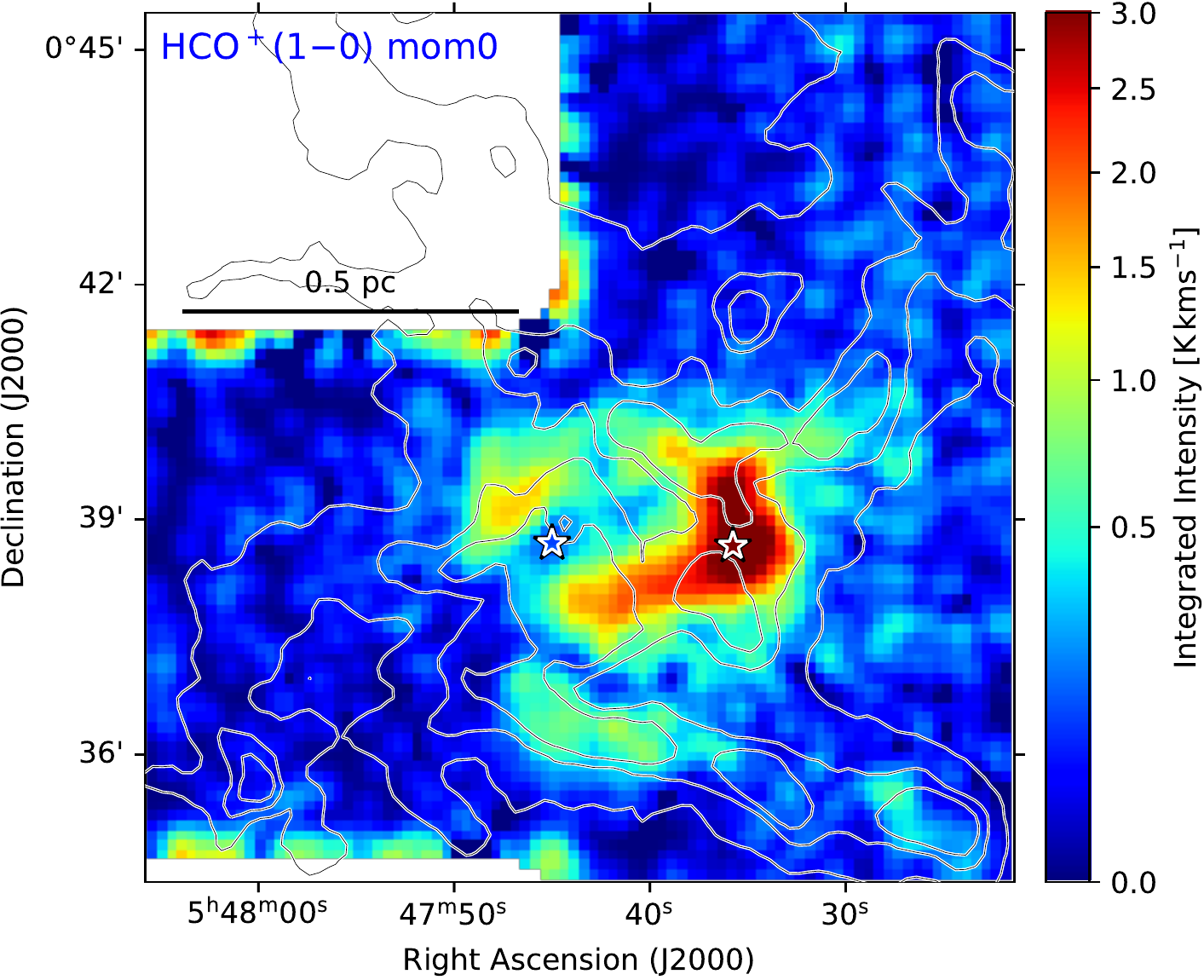}} 
\subfloat{\includegraphics[width=0.48\textwidth]{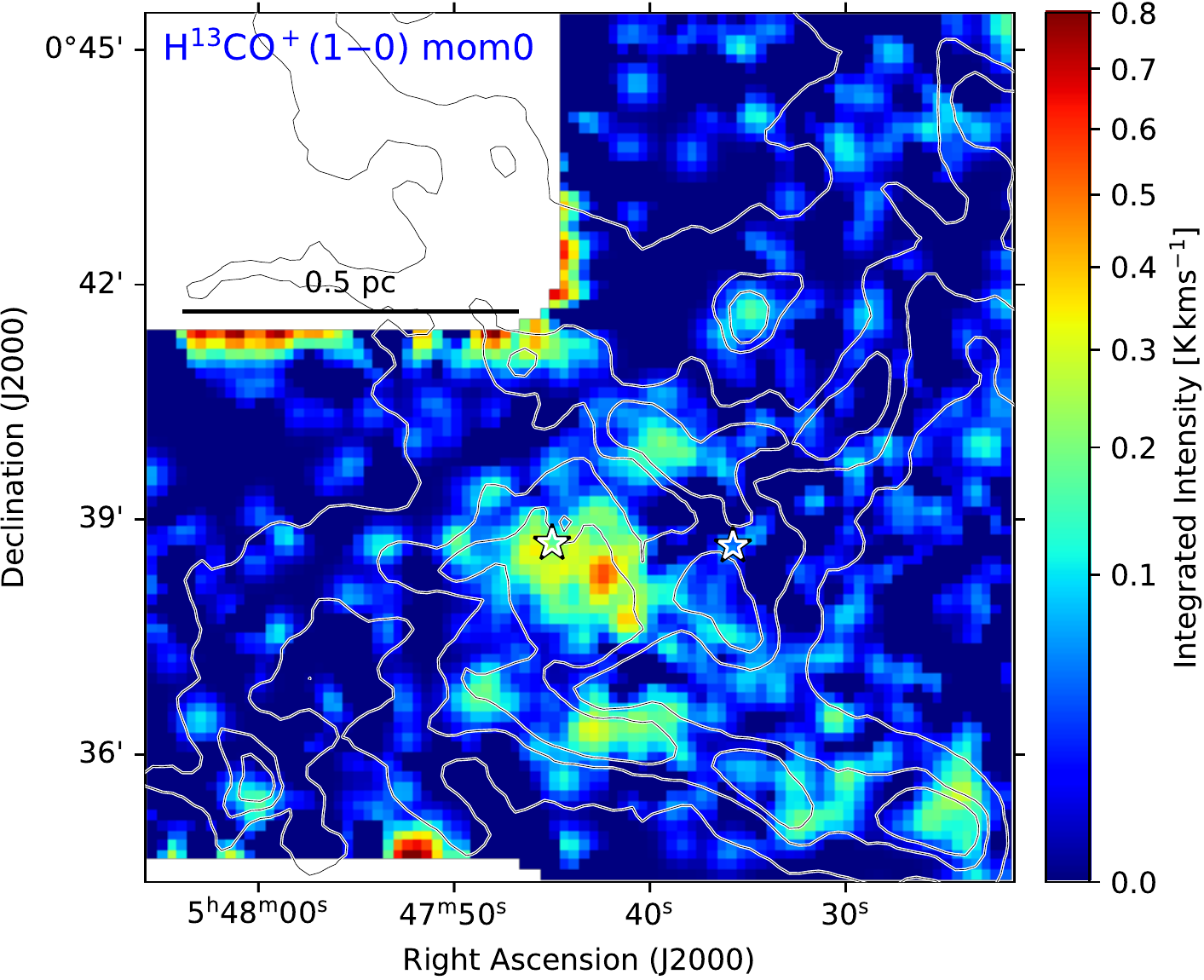}}\\
\subfloat{\includegraphics[width=0.48\textwidth]{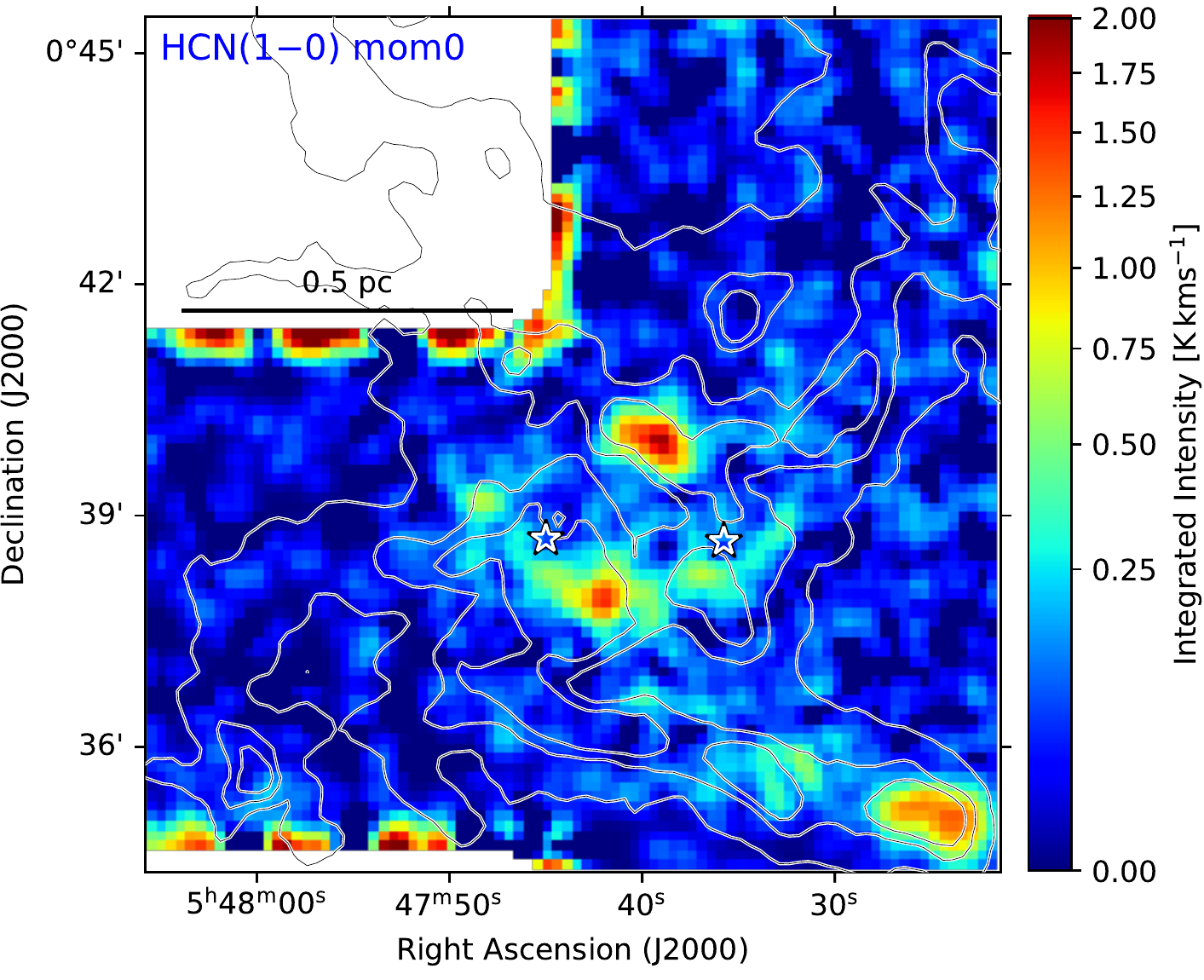}} 
\subfloat{\includegraphics[width=0.48\textwidth]{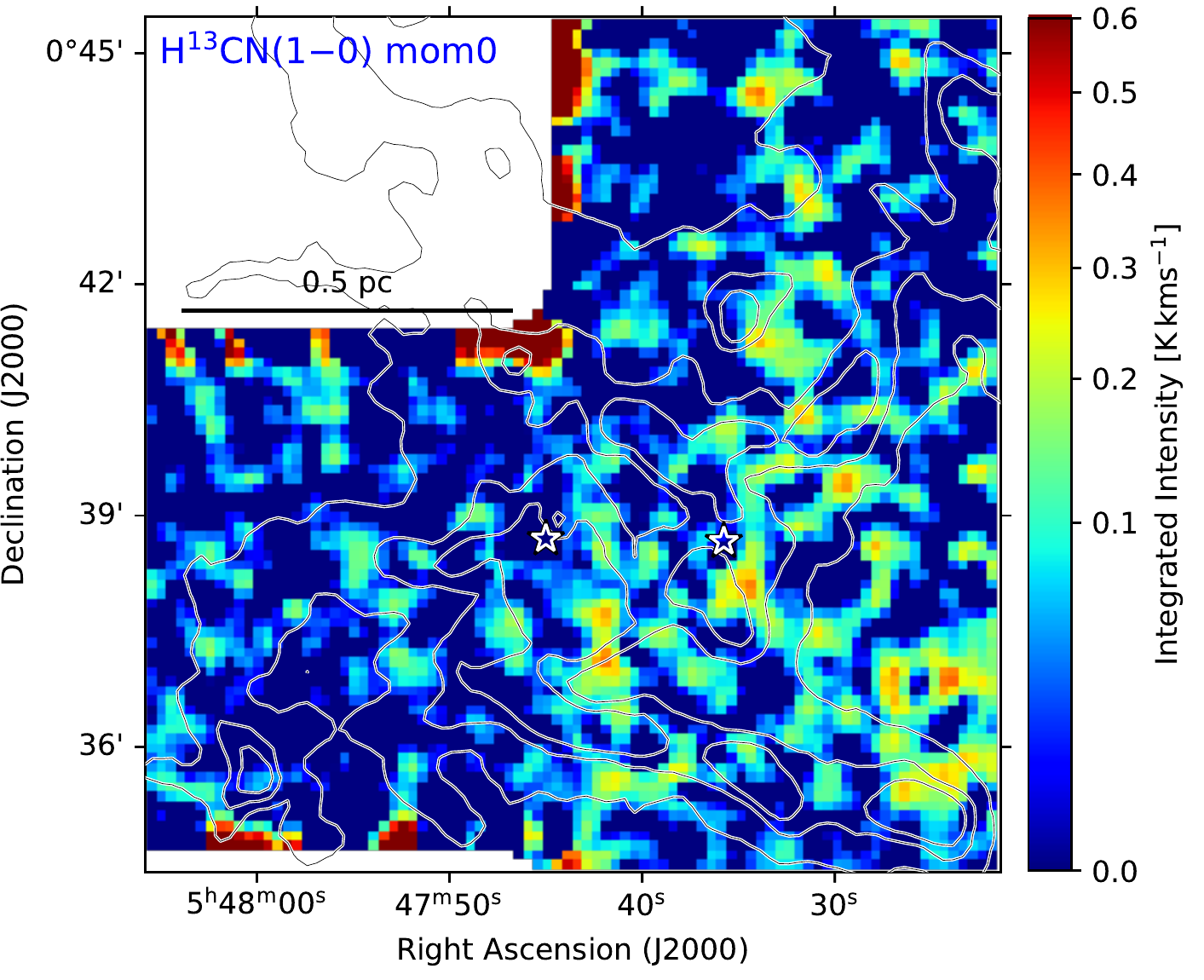}}\\ 
\caption{Integrated intensity (moment 0) maps of the hub centre (in main beam temperature) in the molecular lines indicated
         in the upper left corner.
         Higher intensity ``strings of beads'' at the map edges are artifacts.
         In all panels smoothed column density contours are overplotted, and the left and right white stars mark the 
         locations of IRAS 05451+0037 and LkH$\alpha$ 316, resp.
         } 
\label{fig_all_mom0}
\end{figure*}


\bsp	
\label{lastpage}
\end{document}